%% file: master.tex
\documentclass[12pt,a4paper,fleqn]{book}
\usepackage{epsfig,latexsym,fancyheadings,amssymb,feynmp,axodraw}
\usepackage[english]{babel}

\catcode`@=11
\def\citer{\@ifnextchar
[{\@tempswatrue\@citexr}{\@tempswafalse\@citexr[]}}

\def\@citexr[#1]#2{\if@filesw\immediate\write\@auxout{\string\citation{#2}}\fi
  \def\@citea{}\@cite{\@for\@citeb:=#2\do
    {\@citea\def\@citea{--\penalty\@m}\@ifundefined
       {b@\@citeb}{{\bf ?}\@warning
       {Citation `\@citeb' on page \thepage \space undefined}}%
\hbox{\csname b@\@citeb\endcsname}}}{#1}}
\catcode`@=12

\newcommand{\str}{\vphantom{\bigg(}}

\newcommand{\lra}{\longrightarrow}

\newcommand{\beq}{\begin{eqnarray}}
\newcommand{\eeq}{\end{eqnarray}}
\newcommand{\be}{\begin{equation}}
\newcommand{\ee}{\end{equation}}
\newcommand{\<}{\langle}
\newcommand{\non}{\nonumber}

\newcommand{\qq}{$q\bar{q}$}

\renewcommand{\slash}[1]{#1\!\!\!/}

\renewcommand{\>}{\rangle}

\newcommand{\lesim}{\makebox[0pt][l]{\raisebox{2pt}{$\scriptstyle{<}$}}
\raisebox{-2pt}{$\scriptstyle{\sim}$}}
\newcommand{\gesim}{\makebox[0pt][l]{\raisebox{2pt}{$\scriptstyle{>}$}}
\raisebox{-2pt}{$\scriptstyle{\sim}$}}
\newcommand{\grgl}{\hbox
  to-0.5pt{\lower2.5pt\hbox{$\sim$}\hss}{\raise3pt\hbox{$>$}}}
\newcommand{\klgl}{\hbox
  to-0.5pt{\lower2.5pt\hbox{$\sim$}\hss}{\raise3pt\hbox{$<$}}}
\newcommand{\s}{\newline \vspace*{-3.5mm}}

\def\fmfL(#1,#2,#3)#4{\put(#1,#2){\makebox(0,0)[#3]{#4}}}

\begin{document}

\pagenumbering{roman}
\begin{titlepage}
  \begin{center}
 \sc{
  {\huge
  Higgs Particles in the \\ Standard Model \\ and \\ 
  Supersymmetric Theories \\[3pt]
  }}
  \vspace{4.8cm}
    {\large Dissertation}\\
    \vspace{0.5cm}
    {\large zur Erlangung des Doktorgrades}\\
    \vspace{0.5cm}
    {\large des Fachbereichs Physik}\\
    \vspace{0.5cm}
    {\large der Universit\"at Hamburg}\\
  \vspace{3.8cm}
    {\large vorgelegt von}\\
    \vspace{0.5cm}
    {\large Milada Margarete M\"uhlleitner}\\
    \vspace{0.5cm}
    {\large aus Aalen}\\
    \vfill
    {\large Hamburg}\\
    \vspace{0.5cm}
    {\large 2000}   
  \end{center}
\end{titlepage}
\pagestyle{empty}

\newpage


{\phantom H}
\vfill
\begin{tabular}{ll}
Gutachter der Dissertation: & 
Prof.~Dr.~P.M.~Zerwas \\
& Prof.~Dr.~B.A.~Kniehl \\
\\
Gutachter der Disputation: & 
Prof.~Dr.~P.M.~Zerwas \\
& Prof.~Dr.~J.~Bartels \\
\\
Datum der Disputation: & 
13.~Juli 2000 \\
\\
Dekan des Fachbereichs Physik \\
und Vorsitzender des \\
Promotionsausschusses: &
Prof.~Dr.~F.\,--\,W.~B\"u{\ss}er
\end{tabular}
\pagestyle{empty}
\newpage


\pagestyle{empty}

{\phantom H}
\vspace{4cm}
\begin{center}
{\Large F\"ur meinen Vater}
\end{center}
\newpage


\pagestyle{empty}
\begin{center}
\sc{{\large Abstract}}
\end{center}
This thesis presents a theoretical analysis of the properties of the
Higgs bosons in the Stan\-dard Model (SM) and the minimal supersymmetric
extension (MSSM), which can be in\-ves\-ti\-ga\-ted at the LHC and $e^+e^-$
linear colliders. The final goal is the reconstruction of the Higgs
potential and thus the verification of the Higgs mechanism.  MSSM
Higgs boson production processes at future $\gamma\gamma$ colliders
are calculated in several decay channels.  Heavy scalar and
pseudoscalar Higgs bosons can be discovered in the $b\bar{b}$
final state in the investigated mass range 200 to 800~GeV for moderate
and large values of $\tan\beta$. The $\tau^+\tau^-$ channel provides a
heavy Higgs boson discovery potential for large values of $\tan\beta$.
%
Several mechanisms that can be exploited at $e^+e^-$ linear colliders
for the measurement of the lifetime of a SM Higgs boson in the
intermediate mass range are analysed. In the $WW$ mode, the lifetime
of Higgs scalars with masses below $\sim 160$~GeV can be determined
with an error less than 10\%. The reconstruction of the Higgs
potential requires the measurement of the Higgs self-couplings.  The
SM and MSSM trilinear Higgs self-couplings are accessible in double
and triple Higgs production. A theoretical analysis is presented in
the relevant channels at the LHC and $e^+e^-$ linear colliders. For
high luminosities, the SM trilinear Higgs self-coupling can be
measured with an accuracy of 20\% at a 500~GeV $e^+e^-$ linear
collider. The MSSM coupling among three light Higgs bosons has to be
extracted from continuum production. The other trilinear Higgs
couplings are measurable in a restricted range of the MSSM parameter
space. At the LHC, the $Hhh$ coupling can be probed in resonant
decays.
\newline
\begin{center}
\sc{{\large Zusammenfassung}}
\end{center}
Diese Doktorarbeit pr\"asentiert eine theoretische Analyse der
Eigenschaften von Higgsteilchen im Standard Modell (SM) und der
minimalen supersymmetrischen Erweiterung (MSSM), die am LHC und
$e^+e^-$ Linearcollidern untersucht werden k\"onnen. Das Ziel ist, das
Higgspotential zu rekonstruieren und dadurch den Higgsmechanismus zu
\"uberpr\"ufen.  F\"ur die MSSM Higgs-Boson Produktion an
zuk\"unftigen $\gamma\gamma$-Beschleunigern werden Prozesse in
verschiedenen Zerfallskan\"alen berechnet. Schwere skalare und
pseudoskalare Higgs-Bosonen k\"onnen im $b\bar{b}$ Endzustand im
gesamten untersuchten Massenbereich von 200 bis 800~GeV f\"ur mittlere
und gro{\ss}e Werte von $\tan\beta$ entdeckt werden. Der
$\tau^+\tau^-$ Kanal erlaubt die Entdeckung von schweren Higgs-Bosonen
f\"ur gro{\ss}e Werte von $\tan\beta$. Es werden mehrere Mechanismen
untersucht, die an $e^+e^-$ Linearbeschleunigern f\"ur die
Lebensdauerbe\-stim\-mung eines Higgs-Bosons im intermedi\"aren
Massenbereich verwendet werden k\"onnen. Im $WW$-Kanal kann die
Lebensdauer von Higgs-Skalaren, die leichter als $\sim 160$~GeV sind,
mit einem Fehler von weniger als 10\% bestimmt werden. Die
Rekonstruktion des Higgspotentials erfordert die Messung der
Higgs-Selbstkopplungen. Die trilinearen Higgs-Kopplungen des SM und
des MSSM sind in der Produktion von zwei und drei Higgsteilchen
zug\"anglich. Es wird eine theoretische Analyse in den am LHC und an
$e^+e^-$ Linearbeschleunigern relevanten Kan\"alen durchgef\"uhrt.
Bei hohen Luminosit\"aten kann die trilineare Higgs-Selbstkopplung des
SM mit einer Genauigkeit von 20\% an einem 500~GeV $e^+e^-$
Linearbeschleuniger gemessen werden. Die MSSM Kopplung zwischen drei
leichten Higgs-Bosonen mu{\ss} in der Kontinuumsproduktion ermittelt
werden. Die \"ubrigen trilinearen Kopplungen sind in einem
eingeschr\"ankten Parameterbereich des MSSM-Parameterraumes
me{\ss}bar.  Am LHC kann die $Hhh$-Kopplung in resonanten Zerf\"allen
untersucht werden.  
\pagestyle{empty}
\newpage


\pagestyle{fancy}
\setcounter{page}{1}

\tableofcontents 

\input{chapter1}

\input{chapter2}

\input{chapter3}

\input{chapter4}
\input{chapter5}
\input{chapter6}

\input{app.tex}
\input{lit}

\cleardoublepage


\pagestyle{empty}
\begin{center}
{\sc{{\large Acknowledgements}}}
\end{center}

\noindent
I would like to thank my advisor P.M.~Zerwas for the suggestion of the
project, the continuous support of this work, his enthusiasm and
encouragement. I have profited from his experience and numerous
helpful discussions. \s

\noindent
I am grateful to A.~Djouadi, W.~Kilian and M.~Spira for a pleasant
collaboration and their patience in answering my questions. I would
like to thank W.~Kilian for proof-reading parts of the draft. Special
thanks go to M.~Spira for the careful proof-reading of the draft and
for countless clarifying discussions.\s

\noindent
I would like to take the opportunity to thank R.~Lafaye, D.J.~Miller
and S.~Moretti for an enjoyable working atmosphere. Furthermore, I am
indebted to D.J.~Miller for advice in the English language.\s

\noindent
During the first stage of my work I profited a lot from T.~Plehn's
experience.\s

\noindent
I would like to thank all my friends at DESY, in particular O.~B\"ar,
Y.~Schr\"oder, M.~Spira, A.~Brandenburg, D.J.~Miller, U.~Nierste,
B.~Pl\"umper, M.~Maniatis and M.~Weber, for the pleasant working
environment.\s

\noindent
O.~B\"ar is due specific thanks for supporting me at all stages of the
thesis and fortifying me in the realization of my ideas.\s

\noindent
My special thanks go to my parents. To my mother who has
always believed in me. To my father who had his special great manner
in encouraging me to go my way.\s

\noindent
Financial support by Deutsches Elektronen-Synchrotron (DESY)
is gratefully acknowledged. 

%
\cleardoublepage

\end{document}

%% file: chapter1.tex
\chapter{Introduction}
\pagenumbering{arabic}
The Standard Model of particle physics (SM) combines the electroweak
and strong interactions based on the local $SU(3)\times SU(2)_L \times
U(1)_Y$ gauge group. The electroweak gauge fields $W^\pm$, $Z$ and the
photon field $A$ correspond to the four generators of the non-abelian
$SU(2)_L \times U(1)_Y$ gauge group whereas the eight generators of
the colour group $SU(3)$ are associated with the equivalent number of
gluons. The interaction between matter and gauge fields is
incorporated in the theory by minimal substitution which replaces the
partial derivatives in the Lagrangean with the covariant ones
including the couplings related to the various gauge groups. At this
stage matter and gauge fields are still massless in contradiction to
the experimental observation. Introducing explicit mass terms,
however, would violate the gauge invariance and lead to a
non-renormalizable theory with an infinite number of parameters to be
adjusted by the experiment. Another difficulty arises due to the
violation of the unitarity bounds at high energies by the scattering
amplitudes of the massive $W^\pm/Z$ bosons and fermions. \s

A way out is provided by introducing an additional weak isodoublet
scalar field \citer{wsalam,smpub}. One of the four degrees of freedom
corresponds to a physical particle, the so-called Higgs boson. The
self-interaction of the scalar field leads to an infinite number of
degenerate ground states with non-zero vacuum expectation value (VEV)
$v=(\sqrt{2}G_F)^{-1/2}\approx 246$~GeV. By choosing one of them as
physical ground state, the $SU(2)_L\times U(1)_Y$ symmetry is hidden
with the $U(1)_{em}$ symmetry left over. The Higgs couplings to
other particles are defined by the constraint set by unitarity.
Through the interaction with the scalar field in the ground state, the
electroweak gauge bosons and the fundamental matter particles acquire
their masses. The non-vanishing field strength, essential for the
non-zero particle masses, is induced by the typical minimax form of
the Higgs potential. After the "spontaneous symmetry breaking" the
three Goldstone bosons among the four degrees of freedom of the Higgs
doublet are absorbed to provide the longitudinal modes for the massive
$Z$ and $W^\pm$ gauge bosons. Since all the Higgs couplings are
predetermined, the parameters describing the Higgs particle are
entirely fixed by its mass. This is the only unknown parameter in the
SM Higgs sector \cite{gunion,smpub}.\s

%
The Standard Model is in very good agreement with electroweak
precision tests at LEP, SLC, Tevatron and HERA. In some cases it has
been tested to an accuracy better than 0.1\%. Up to now, the only
deviation from the Standard Model has been the experimental indication
of massive neutrinos, which can be embedded in the SM by introducing a
mass term for neutrinos in the Yukawa Lagrangean. This strongly
limits possible forms of new physics. Models that preserve the SM
structure (extended by the neutrino mass term) are to be favoured. In
this respect a promising candidate for possible extensions is
supersymmetry \cite{gunion,susy,susyprod}. \s

In view of the impressing success of the SM one may ask of course why
to consider extensions to new physics. The reason is that the SM
implies many unanswered questions. The extrapolation of the model to
high energies indicates a unification of all three gauge couplings at
mass scales of the order of $M_{GUT} \sim 10^{14}-10^{16}$~GeV
\cite{gut} (Grand Unified Theories GUT's). If one assumes new physics
to set in at the GUT scale, the SM will be valid in
a mass range between the relative small weak scale of the order of
$(\sqrt{2}G_F)^{-1/2}\approx 246$~GeV set by the Higgs mechanism and
$M_{GUT}$ given by the grand unification scale. Quantum fluctuations,
however, lead to large corrections of the Higgs mass at high scales.
In order to keep the Higgs mass stable in the presence of the large
GUT scale, the considerable corrections have to be absorbed in the
mass counterterms leading to a finetuning of the Higgs parameters. \s

A stabilization of the separation between the electroweak and the GUT
scale is provided by the introduction of supersymmetry (SUSY) which
incorporates the most general symmetry of the $S$-matrix. Within the
framework of supersymmetry the masses of the scalar particles remain
moderate even in the presence of high energy scales. This is due to
supersymmetry representing a connection between fermionic and bosonic
degrees of freedom so that quantum corrections arising from the two
particle types cancel each other. Even in softly broken supersymmetry
the leading quadratic singularities vanish and are replaced by
logarithmic divergences. Furthermore, SUSY GUT's lead to a prediction
of the electroweak mixing angle \cite{sinw} that is in very good
agreement with present high-precision measurements of
$\sin^2\theta_W$. A minimal realization of supersymmetry is given by
the Minimal Supersymmetric extension of the Standard Model, the MSSM
\cite{gunion,susy,susyprod}. \s

The Higgs mechanism described above is a basic ingredient of the
electroweak sector both of the Standard Model and the MSSM. So far,
however, it has not been verified experimentally. In order to
establish this most important aspect for the consistent formulation of
an electroweak theory experimentally, three steps have to be
performed.
\begin{enumerate}
\item First of all the Higgs particle(s) must be discovered. 
\item In a second step, the Higgs couplings to the fermions and
      gauge particles have to be determined via the partial decay widths 
      and the Higgs production cross sections. 
\item In order to finally 
      reconstruct the Higgs potential, the trilinear and quadrilinear 
      Higgs self-couplings have to be measured.  
\end{enumerate}
The discovery of a scalar particle alone will not reveal the mechanism
by which the fundamental particle masses are generated. Only the
knowledge of the Higgs self-interaction will allow for the
determination of the explicit form of the Higgs potential.\s

In this work the procedure for the experimental verification of the
Higgs sector will be analysed. For this task the picture of the Higgs
particle will be evolved in the three steps described in the previous
paragraph. For reasons resumed above not only the SM but also a
supersymmetric extension, the MSSM which involves five physical Higgs
particles \cite{gunion,susy,susyprod}, will be considered. The
theoretical foundations will be developed and the experimental
implications at the Large Hadron Collider \cite{atlascms}, which is
constructed for a c.m.\ energy of 14~TeV, and at high energy $e^+e^-$
linear colliders \cite{accomando} will be discussed. The total
integrated luminosity for the LHC is $\int {\cal L}= 100$~fb$^{-1}$
after 3 years. $e^+e^-$ linear colliders are planned to run in the
energy range between $500$~GeV and 1~TeV, possibly extending later up
to 5~TeV. For the TESLA design this corresponds to an integrated
luminosity of $\int {\cal L}= 300$~fb$^{-1}$ per year at $500$~GeV and
of $500$~fb$^{-1}$ for a c.m.\ energy of 800~GeV \cite{accomando}. \s

The structure of the thesis which reflects the experimental scheme to
be pursued is as follows. Chapter~2 will start with the setting of the
scene by describing the SM Higgs couplings and the MSSM Higgs sector
from the theoretical point of view. \s

The third Chapter will be devoted to the Higgs boson search at $pp$
colliders and in $e^+e^-$ collisions. First, the production mechanisms
will be reviewed before turning on to the per\-spec\-ti\-ves for the Higgs
discovery in the diverse decay channels at the two collider types. The
final picture will display a parameter region in which the discovery
of the heavy MSSM Higgs bosons will be difficult if not impossible.
Therefore the production and discovery potential in the $\gamma\gamma$
mode at $e^+e^-$ linear colliders will be analysed. Using
Weizs\"acker-Williams photons for the production of Higgs bosons
provides too small a number of events. The rate is sufficiently large,
however, if the photons are produced through the backscattering of
laser light from high-energy electron/positron beams. This mechanism
\cite{ggprop} allows for the production of real photons with high
energy, luminosity \cite{ggvgl} and monochromaticity and a high degree
of polarization of the scattered photons. Several decay channels will
be examined with regard to a sufficiently large signal to background
ratio. Next-to-leading order corrections will be included and where
necessary the resummation of large logarithms will thoroughly be taken
into account. \s

Once the Higgs boson will be discovered and its mass will be
determined, the next step is the measurement of the Higgs couplings to
the gauge and matter particles via the Higgs decays. Chapter~4 will
present the branching ratios for the various decay channels of the SM
Higgs scalar and the MSSM Higgs bosons. The knowledge of the branching
ratios and the decay widths allows for the extraction of the lifetime
of the Higgs bosons, which together with the mass forms the basic
characteristics of particles. Therefore, Chapter~4 will continue with
evolving the theoretical set-up for the determination of the SM and
MSSM Higgs lifetimes in a model-independent way. Some representative
examples at $e^+e^-$ colliders will be given.\s

%
The first Chapters describe the determination of the necessary
ingredients, {\it i.e.} the mass and lifetime of the Higgs states and
the Yukawa and gauge-Higgs boson couplings, from a theoretical and
experimental point of view, enabling finally the measurement of the
Higgs self-couplings. Chapter~5 deals with this task which is
essential in order to establish experimentally the Higgs sector
harboring the mechanism for the generation of particle masses. Special
emphasis will hence be on this part of the thesis. The plethora of
mechanisms which allow the access to the trilinear Higgs
self-couplings in $pp$ and $e^+e^-$ collisions will be described. The
final picture unveils a theoretically conclusive way of determining
all possible trilinear Higgs self-couplings in the SM and the MSSM.
The size of the cross sections for these processes will be discussed
including a short sideview on the ones involving self-couplings among
four Higgs particles. The phenomenological implications, in particular
the per\-spec\-ti\-ves for the measurement of the trilinear Higgs
self-couplings will be examined by reviewing existing background
studies. \s

The thesis will be concluded with a summary of the salient features of
the investigation of the Higgs sector in the SM and the MSSM.

%% file: chapter2.tex
\chapter{The SM and the MSSM Higgs sector\label{chap:higgssect}}

\section{The Higgs particle of the Standard Model}
The dynamics of the $SU(2)_L$ Higgs doublet field $\Phi$, introduced
in order to ensure unitarity and to provide a mechanism for the
generation of masses without violating gauge principles, is described
by the Lagrangean
\beq
{\cal L}_\Phi = (D_\mu \Phi)^\dagger (D^\mu \Phi) - V(\Phi)
\label{philagr}
\eeq
where $V$ denotes the Higgs self-interaction potential
\beq
V = \lambda\left[ \Phi^\dagger \Phi - \frac{1}{2}v^2 \right]^2
\label{selfpot}
\eeq
with a minimum at $\langle \Phi \rangle_0 = (0,v/\sqrt{2})$. By
introducing the Higgs field in the unitary gauge
\beq
\Phi = \frac{1}{\sqrt{2}}\left( \begin{array}{c} 0 \\ v+H 
\end{array} \right)
\label{unitary}
\eeq
the potential Eq.~(\ref{selfpot}) can be cast into the form
\beq
V_H = \frac{1}{2} (2\lambda v^2) H^2 + \lambda v H^3 + \frac{\lambda}{4}
H^4
\eeq
where the Higgs mass $M_H$ and the Higgs self-interactions can be read
off directly. Apparently, the Higgs mass
\beq
M_H = \sqrt{2\lambda} v
\eeq
is related to the quadrilinear coupling $\lambda$. The trilinear Higgs 
self-coupling can be expressed as 
\beq
\lambda_{HHH} = 3 M_H^2/M_Z^2
\eeq
in units of $\lambda_0 = M_Z^2/v$ and the self-coupling among four Higgs
bosons in units of $\lambda_0^2$ is given by
\beq
\lambda_{HHHH} = 3 M_H^2/M_Z^4
\eeq
where $\lambda_0 = 33.8$~GeV, numerically.  For a typical energy scale
$M_Z$ and a Higgs mass $M_H=110$~GeV, the trilinear Higgs
self-coupling equals to $\lambda_{HHH} \lambda_0/ M_Z = 1.6$. In
contrast, the quadrilinear coupling $\lambda_{HHHH} \lambda_0^2 = 0.6$
is suppressed compared to the trilinear coupling by a factor of about
the weak gauge coupling. Evidently, in the SM the Higgs self-couplings
are uniquely defined by the mass of the Higgs boson.\s

The covariant derivative in (\ref{philagr}) is given by 
\beq
D_\mu = i\partial_\mu + gT_a W_\mu^a - g'\frac{Y}{2} B_\mu
\label{covder}
\eeq
and $(D_\mu \Phi)^\dagger (D^\mu \Phi)$ hence describes the kinetic
Higgs term and the interaction between Higgs and gauge bosons. $T_a$
($a=1,2,3$) denote the isospin-generators of the $SU(2)_L$ gauge group
and $Y$ corresponds to the $U(1)_Y$ hypercharge-generator. $g$ and
$g'$ are the electroweak couplings and $W_\mu^a$ and $B_\mu$ are the
gauge fields associated with the two symmetry groups, respectively.
After introducing the physical Higgs field (\ref{unitary})
and transforming the electroweak eigenstates $W^a_\mu$, $B_\mu$ to the
mass eigenstates, the kinetic term in Eq.~(\ref{philagr}) yields the
mass terms for the electroweak gauge bosons $W^\pm$, $Z$ and the
photon field $A$
\beq
M_W = \frac{1}{2}gv, \quad
M_Z = \frac{1}{2}\sqrt{g^2+g'^2} v, \quad
M_A = 0
\eeq
as well as the Higgs-gauge boson interaction strengths 
\beq
\lambda_{HVV} = 2(\sqrt{2}G_F)^{1/2} M_V^2\;, \qquad 
\lambda_{HHVV} = 2(\sqrt{2}G_F) M_V^2\;, \qquad [V = W,Z]
\label{higgsgauge}
\eeq
\indent The interaction between the Higgs boson and fermions must
respect the $SU(2)_L \times U(1)_Y$ gauge symmetry and maintain the
renormalizability of the theory. The operators which fulfill these
conditions are combined in the Yukawa Lagrangean
\beq
{\cal L}_{yuk} = - \bar{E}_R C_E \Phi^\dagger L_L + 
\bar{U}_R C_U \Phi^T_i \epsilon_{ij} Q'_{Lj} - 
\bar{D}'_R C_D \Phi^\dagger Q'_L
\label{yuklagr}
\eeq
with 
\beq
E = (e,\mu,\tau)^T, \qquad U = (u,c,t)^T, \qquad 
D' = (d',s',b')^T
\eeq
and
\beq
Q'_L = 
\left( \begin{array}{c} U \\ D' \end{array} \right)_L, \qquad
L_L = 
\left( \begin{array}{c} N \\ E \end{array} \right)_L, \quad
\mathrm{where} \quad 
N_L = (\nu_{eL},\nu_{\mu L},\nu_{\tau L})^T
\eeq
The right-handed fermions denoted by the index $R$ behave as singlets
under $SU(2)_L$-trans\-for\-ma\-tions whereas the left-handed fermions
denoted by the index $L$ are combined in isospin-doublets.  The prime
indicates that the corresponding quarks are given in the electroweak
basis which is connected to the basis of the mass eigenstates via the
unitary Cabibbo-Kobayashi-Maskawa matrix $V$ \cite{kobay}, $D' = V D$.
The $3\times 3$ matrices $C_E,C_U,C_D$ are unitary. By appropriate
change of the basis $(E,U,D')$ with a constant $U(3)$-matrix and
making use of the unitarity of the matrices $C_i$, Eq.~(\ref{yuklagr})
provides the fermion mass terms $m_f$ after expanding $\Phi$ around
the VEV, cf.~Eq.~(\ref{unitary}). They are related to the Higgs
fermion interaction coefficient via
\beq
\lambda_{Hff} = (\sqrt{2} G_F)^{1/2} m_f
\label{higgsf}
\eeq
\indent Taking into account the recent indication of massive neutrinos
one possibility to include them in the theory is the addition of an
operator containing right-handed Dirac neutrinos to the Yukawa
Lagrangean:
\beq
\bar{N}_R C_N \Phi_i^T \epsilon_{ij} L_{Lj}'
\eeq
and changing the mass eigenstates $E$ to electroweak eigenstates $E'$,
\beq
E \to E'
\eeq
This Yukawa Lagrangean leads then to a Dirac mass term for neutrinos
and furthermore encounters a "Cabibbo-Kobayashi-Maskawa" matrix $V_E$
for charged leptons transforming the electroweak eigenstates $E'$ to mass
eigenstates $E$:
\beq
E' = V_E E
\eeq
There are also other possibilities of implementing massive neutrinos in
the theory.

\section{Supersymmetry \label{susysec}}

\subsection{The Minimal Supersymmetric Standard Model}
In the minimal realization of a supersymmetric theory, the MSSM, a
minimal number of supersymmetric particles is introduced as partners
to the SM particles. The scalar partners of quarks and leptons are
called squarks and sleptons with the degrees of freedom corresponding
to the degrees of freedom of the SM states. The fermions and their
supersymmetric partners are combined in chiral superfields. For
reasons described below, the MSSM which includes the same $SU(3)\times
SU(2)_L \times U(1)_Y$ gauge symmetry as the SM contains two $SU(2)_L$
Higgs doublets. They are associated with $SU(2)_L$ doublets of
Majorana fermion fields, the higgsinos. The electroweak gauge bosons
and the gluons also acquire SUSY partners with spin 1/2, which are
called gauginos and gluinos, respectively. Charginos and neutralinos
are the physical mass eigenstates of the higgsinos and gauginos and
are given by linear combinations of these fields.  Gauge bosons and
their supersymmetric partners are described by vector superfields. The
particles in a superfield differ only by spin 1/2.  Since
supersymmetry relates fermions and bosons, the particles in a
supermultiplet have equal masses and residual quantum numbers in an
unbroken supersymmetric theory.\s
%

The superfields are introduced in order to build up the supersymmetric
Lagrangean, and the space-time integral over the Lagrange density is
extended to a superspace integral which involves two more dimensions.
They correspond to two-component Grassmann variables, denoted by
$\theta$ and $\bar{\theta}$. The superfields can be expanded in terms
of these Grassmann variables developing the component fields of the
supermultiplets as coefficients in these finite series. The so-called
$F$ term is the one proportional to $\theta\theta$ in the expansion of
the chiral superfield. The $D$ term represents the contribution
proportional $\theta^2\bar{\theta}^2$ in the vector superfield
expansion \cite{susy}. Starting with a theory without gauge fields, the
supermultiplets can be described using chiral superfields $\Phi_i$
only. Since a renormalizable theory must not contain higher orders
than three in $\Phi_i$, the Lagrangean for chiral superfields is given
by
\beq
{\cal L} = \sum_i (\bar{\Phi}_i \Phi_i)_D + ({\cal W} + \bar{{\cal W}})_F
\eeq
with the superpotential 
\beq
{\cal W} = \sum_{ijk} \left( m_{ij}\Phi_i\Phi_j + 
\lambda_{ijk} \Phi_i\Phi_j\Phi_k\right)
\label{superpot}
\eeq
The products $\Phi^2$ and $\Phi^3$ are chiral again and the real
product $\bar{\Phi} \Phi$ forms a vector field. The subscripts $D$ and
$F$ denote the prescriptions for taking the $D$ and $F$ terms in the
expansion of the superfield products, respectively. The analytical
superpotential ${\cal W}$ must not contain complex conjugate
superfields. Neither will the scalar potential contain complex
conjugate component Higgs fields. As anticipated, two Higgs doublets
are required in order to give masses to both the up- and down-type
quarks. Integrating out the Grassmann variables and using the
Euler-Lagrange equations
\beq
F_i = -\frac{\partial \bar{{\cal W}} (\bar{A})}{\partial \bar{A}_i}
\eeq
the scalar potential is given by 
\beq
{\cal V} = \sum_i |F_i|^2
\eeq
where the Yukawa terms included in ${\cal W}$ have been dropped. $A$
is a complex scalar component field in the expansion of the scalar
superfield. \s

Including a non-abelian gauge sector the Lagrangean for a minimal
supersymmetric gauge theory in the Wess-Zumino gauge reads
\cite{zumino}
\beq
{\cal L} &=&
\sum_{ij} \left[ \bar{\Phi}_i \left( e^{2gV^a T^a} \right)_{ij} \Phi_j 
\right]_D \non\\ 
&+& \left[ {\scriptstyle \frac{1}{4 C(R)}} Tr(W^\alpha W_\alpha) 
+ {\cal W}(\Phi_i) \right]_F + h.c.
\label{gaugel}
\eeq
The non-abelian field strength superfield $W_\alpha$ is given by
\beq
W_\alpha = -\frac{1}{8g}\bar{D}^2 e^{-2gV} \left( D_\alpha e^{2gV}\right)
\eeq
$D_\alpha$ is the covariant derivative acting on superspace. $g$
denotes the non-abelian gauge coupling and $V$ reads
\beq
V = V^a T^a
\eeq
where $V^a$ denotes a vector supermultiplet and $T^a$ are the
generators of the Lie-Algebra. They are normalized to
\beq
Tr[T^a T^b] = C(R) \delta^{ab}
\eeq
The factor $C$ depends on the representation $R$. Integrating out
the Grassmann fields the scalar potential for a non-abelian gauge
theory can be read off from (\ref{gaugel}):
\beq
{\cal V} = \sum_i |F_i|^2 + \frac{1}{2} \sum_a (D^a)^2
\label{gaugepot}
\eeq
The auxiliary fields $F_i$ and $D^a$ are given by the Euler-Lagrange
equations,
\beq
F_i &=& -\frac{\partial \bar{{\cal W}} (\bar{A})}{\partial \bar{A}_i} \non
\\
D^a &=& - \sum_{ij} g\bar{A}_i (T^a)_{ij} A_j
\eeq
$A_i$ are scalar fields that belong to the the fundamental
representation of a general non-abelian $SU(N)$ gauge group. A $U(1)$
gauge multiplet can be included in the scalar potential by adding
\beq
\frac{1}{2} (D')^2
\eeq
to Eq.~(\ref{gaugepot}). $D'$ is defined by
\beq
D' = \sum_i g' \frac{Y}{2} \bar{A}_i A_i + \xi
\label{defdp}
\eeq
$g'$ denotes the $U(1)$ gauge coupling and $Y$ the generator of the
abelian group. The constant Fayet-Iliopoulos term $\xi$
\cite{fayetterm} can in many models argued to be small
\cite{smallterm} and will be set equal to zero in the following.  \s


The superpotential as given in (\ref{superpot}) implies lepton and
baryon number violating interactions which may lead to proton decay at
tree level by exchanging a s-down-quark. In the SM, no such problem
occurs since the corresponding interactions do not arise due to gauge
invariance unless the dimension of the operators is larger than six.
In order to avoid the problem in the MSSM, the corresponding couplings
must either be suppressed or the interactions must be forbidden by a
symmetry. The $\Bbb{Z}_2$ symmetry under which the Grassmann variables
in the Lagrangean change sign fulfills this condition. The symmetry is
called $R$ parity and characterized by a multiplicative quantum number
which can be defined as
\beq
R = (-1)^{(3B+L+2S)}
\eeq
where $B$ is the baryon number, $L$ the lepton number and $S$ the spin
of the particle. The SM particles have $R$ parity $+1$ whereas their
SUSY partners have $R$ parity $-1$. From the phenomenological point of
view, the assumption of $R$ parity conservation implies that SUSY
partners can only be pair produced in collisions of SM particles.
Furthermore, a theory with $R$ parity conservation contains a
lightest SUSY particle (LSP) which is stable.\s

So far no SUSY particle with the same mass as its corresponding SM
partner has been observed. In order to explain the experiments
the SUSY particles must be assumed to be heavier than their SM
counterparts.  Hence supersymmetry has to be broken. For the time
being, supersymmetry is usually broken by introducing explicit soft
mass terms \cite{girardello} and regarding the MSSM as an effective
low energy theory \cite{hall}. The mass terms are called soft because
they are chosen such that they do not develop quadratic divergences
again.  The form of the soft-breaking terms is restricted by the
requirement of gauge invariance as well as weak-scale $R$ parity
invariance and stable scalar masses. A further constraint is set by
the experimental bounds.  In generic notation they are given by
\begin{itemize}
\item $-m_{ij}^2 S_i^*S_j$ \\[0.1cm] 
  where $S_i$ denote scalar component fields and $m_{ij}^2$,
  $i,j=1,...,n$, scalar mass matrices for squarks and sleptons with
  $n$ generations. There are two masses for the Higgs scalars and a
  complex mass term mixing the scalar components of the two Higgs doublets:
  \\[0.1cm]
  $\phantom{-}\tilde{m}_1^2|H_1|^2+\tilde{m}_2^2|H_2|^2-[B\mu H_1H_2+h.c.]$ 
  \\[0.1cm]
  The Higgs fields $H_1$ and $H_2$ are contracted using
  $\epsilon_{ij}$ ($\epsilon_{12}=1$) and $\mu$ is called mixing
  parameter.
\item $-\frac{1}{2} [m'_i\lambda_i\lambda_i + h.c.]$\\
  for Majorana fermions $\lambda_i$ with three real gaugino masses 
  $m'_i$, $i=1,2,3$.
\item $-[A_{ijk} C_i C_j C_k + h.c.] $\\[0.1cm]
  which involves scalar fields $C_i$. $A_{ijk}$
  [$i,j,k=1,2,3$] are complex trilinear couplings respecting $R$
  parity. Non-zero trilinear couplings $A_{ijk}$ lead to a mixing of
  the left- and right-handed sfermions.
\end{itemize}

\subsection{The MSSM Higgs sector}
The MSSM Higgs potential can be derived using
Eqs.~(\ref{gaugepot})--(\ref{defdp}) \cite{gunion,susy,susyprod}. The
contribution $V_D$ due to the $D$ term
\beq
V_D = \frac{1}{2} \sum_a (D^a)^2 + \frac{1}{2} (D')^2
\eeq
for two Higgs doublets $H_1$, $H_2$ with hypercharges $Y=-1$ and $Y+1$,
respectively, is given by
\beq
D_{U(1)_Y} &=& -\frac{g'}{2} \left( |H_2|^2 - |H_1|^2 \right) \\[0.2cm]
D_{SU(2)_L}^a &=& -\frac{g}{2} \left( H_1^{i*} \sigma_{ij}^a H_1^j + 
                  H_2^{i*} \sigma_{ij}^a H_2^j \right) 
\eeq 
where $T^a=\sigma^a/2$ and $\sigma^a$ ($a=1,2,3$) are the Pauli matrices.
The second contribution to the potential arises from the derivative of
the superpotential with respect to the chiral superfields. The
superpotential ${\cal W}$ conserving baryon and lepton number and
respecting the $SU(3)\times SU(2)_L \times U(1)_Y$ symmetry is given by
\beq {\cal W} = -\epsilon_{ij} \mu \hat{H}_1^i \hat{H}_2^j +
\epsilon_{ij} [ \lambda_L \hat{H}^i_1 \hat{L}^{j}_L \hat{\bar{E}}_R +
\lambda_D \hat{H}^i_1 \hat{Q}_L^j \hat{\bar{D}}_R + \lambda_U \hat{H}_2^j
\hat{Q}_L^i \hat{\bar{U}}_R ] 
\eeq 
$\hat{H}_k$ ($k=1,2$) are the superfields containing the Higgs fields
$H_k$ and their SUSY partners. The $SU(2)_L$ doublet superfield
$\hat{L}_L$ includes the left-handed electron and neutrino and the
corresponding scalar particles. The superfield $\hat{\bar{E}}_R$
contains the right-handed anti-electron and its scalar partner.
$\hat{Q}_L$ includes an $SU(2)_L$ doublet of quarks and the associated
SUSY partners.  $\hat{\bar{U}}_R$ ($\hat{\bar{D}}_R$) consists of the
right-handed up- (down-) antiquark and its supersymmetric counterpart.
For the sake of simplicity the superpotential is given only for the
first generation. The supersymmetric potential can then be cast into
the form
\beq 
V = |\mu|^2 \left(
  |H_1|^2 + |H_2|^2 \right) + \frac{g^2+g'^2}{8} \left( |H_1|^2 -
  |H_2|^2 \right)^2 + \frac{g^2}{2} |H_1^*H_2|^2 
\eeq
The minimum of the potential is given by $\< H_1^0 \> = \< H_2^0 \> =
0$ with $V=0$. So far no violation of supersymmetry has been applied.
Including the possible soft SUSY breaking terms, the potential is
given by
\beq 
V_H &=& m_{11}^2 |H_1|^2 + m_{22}^2 |H_2|^2 - m_{12}^2
\epsilon_{ij} \left( H_1^i H_2^j + h.c. \right) + \frac{g^2+g'^2}{8}
\left( |H_1|^2 - |H_2|^2 \right)^2 
\non \\
&& + \frac{g^2}{2} |H_1^* H_2|^2 
\label{mssmpot}
\eeq 
where the squared masses are 
\beq
m_{11}^2 &=& |\mu|^2 + \tilde{m}_1^2 \non\\
m_{22}^2 &=& |\mu|^2 + \tilde{m}_2^2 
\label{masspar} \\
m_{12}^2 &=& \mu B \non
\eeq 
%
%
with $\tilde{m}_1^2$, $\tilde{m}_2^2$ and $\mu B$ being the soft SUSY
breaking mass parameters. $V_H$ manifestly conserves CP invariance as
any complex phase in $\mu B$ can be absorbed by redefining the Higgs
fields. The electroweak symmetry is broken for non-vanishing vacuum
expectation values of the neutral components of the Higgs doublets,
\beq
\<H_1\> = \left( 
\begin{array}{c}
\frac{v_1}{\sqrt{2}} \\
0
\end{array}
\right) \quad , \qquad 
\<H_2\> = \left( 
\begin{array}{c}
0 \\
\frac{v_2}{\sqrt{2}}
\end{array}
\right)
\eeq
By appropriate choice of the phases of the Higgs fields, $v_1$ and
$v_2$ can always be adjusted real and positive. The relation of the
VEVs $v_1$ and $v_2$ defines the angle $\beta$,
\beq
\tan\beta = \frac{v_2}{v_1}
\eeq
For positive $v_1$, $v_2$, $0\le\beta \le \pi/2$.  The potential acquires a
stable minimum if the following conditions are fulfilled:
\beq
m_{11}^2 + m_{22}^2 &>& 2\,|m_{12}^2| \non\\
m_{11}^2 m_{22}^2 &<& |m_{12}^2|^2
\eeq
\indent The MSSM Higgs potential may be compared to a general self-interaction
potential of two Higgs doublets $\varphi_i$ in a CP-conserving theory
\beq
V^{2HDM}[\varphi_1,\varphi_2] &=& 
m_{11}^2 \varphi_1^\dagger \varphi_1 + m_{22}^2 \varphi_2^\dagger 
\varphi_2 - [m_{12}^2 \varphi_1^\dagger \varphi_2 + {\rm h.c.}]
\non \\ 
& + & 
\textstyle{\frac{1}{2}} \lambda_1 (\varphi_1^\dagger \varphi_1)^2 + 
\textstyle{\frac{1}{2}} \lambda_2 (\varphi_2^\dagger \varphi_2)^2 +
\lambda_3 (\varphi_1^\dagger \varphi_1)(\varphi_2^\dagger \varphi_2) +
\lambda_4 (\varphi_1^\dagger \varphi_2)(\varphi_2^\dagger \varphi_1) 
\non\\
&+& \left\{ 
\textstyle{\frac{1}{2}} \lambda_5 (\varphi_1^\dagger \varphi_2)^2
+ [\lambda_6 (\varphi_1^\dagger \varphi_1) + \lambda_7
(\varphi_2^\dagger \varphi_2)]\varphi_1^\dagger \varphi_2 + 
{\rm h.c.} \right\}
\eeq
The seven couplings $\lambda_i$ and the three mass parameters
$m_{11}^2$, $m_{22}^2$, $m_{12}^2$ are real. By applying the relations
between the fields $\varphi_i$ and the Higgs doublets $H_1$ and $H_2$
with opposite hypercharge \cite{susyprod}
\beq
( \varphi_1 )^j = \epsilon_{ij} H_1^{i*}\;, 
\qquad 
( \varphi_2 )^j = H_2^j
\eeq
one will find that in the MSSM the $\lambda$ parameters at tree level
are uniquely defined in terms of the electroweak gauge couplings
\cite{gunion,susy,susyprod,haber}
\beq
\lambda_1 \!&=&\! \lambda_2 \;=\; \textstyle{\frac{1}{4}} (g^2 + g'^2) 
\non\\
\lambda_3 \!&=&\! \textstyle{\frac{1}{4}} (g^2 - g'^2) \non\\
\lambda_4 \!&=&\! -\textstyle{\frac{1}{2}} g^2 \non\\
\lambda_5 \!&=&\! \lambda_6 \;=\; \lambda_7 \;=\; 0
\eeq
The two complex $SU(2)_L$ Higgs doublets have eight degrees of
freedom. After the electroweak symmetry breaking three of them will be
absorbed to provide longitudinal degrees of freedom to the electroweak
gauge bosons. The remaining five degrees of freedom correspond to the
equivalent number of physical Higgs states, two charged Higgs bosons
$H^\pm$, a CP-odd neutral Higgs boson $A$ and two CP-even neutral
Higgs bosons $h$ and $H$ \cite{gunion,susy,susyprod}. Expanding the
Higgs doublets $H_i$ around their vacuum expectation values leads to
the mass matrix $M_{ij}^2$ for the component fields $\eta_i$ of the
Higgs doublets
\beq
M_{ij}^2 = \left.
\frac{\partial^2 V_H}{\partial\eta_i \partial\eta_j}
\right|_{min}
\label{matrix}
\eeq
with the minimum given by $\< 0|\eta_i|0 \>=0$. Diagonalizing the
mass matrix (\ref{matrix}) the masses for the physical Higgs states
are obtained in terms of the parameters in the Higgs potential $V_H$.
The parameters change when top and stop-loop radiative
corrections are included. In the one-loop leading $m_t^4$
approximation, where $m_t$ denotes the top mass, they can be
approximated by
\beq
\epsilon \approx \frac{3G_F m_t^4}{\sqrt{2} \pi^2 \sin^2\beta} \ln
\frac{\tilde{M}^2}{m_t^2}
\eeq
with the common squark-mass value $\tilde{M}$ setting the scale of
supersymmetry breaking.  Provided that stop mixing effects are
moderate at the SUSY scale, they can be implemented by shifting
$\tilde{M}^2$ in $\epsilon$
\beq
\begin{array}{clrcl}
\tilde{M}^2 \to \tilde{M}^2 + \Delta \tilde{M}^2 &:& \Delta \tilde{M}^2 
&=& \hat{A}^2 [1-\hat{A}^2/(12\tilde{M}^2)] \\[0.1cm]
&&\hat{A} &=& A - \mu \cot\beta
\end{array}
\eeq
where $A$ and $\mu$ correspond to the trilinear coupling in the top
sector and the higgsino mass parameter in the superpotential,
respectively. Using the mass $M_A$ and tan $\beta$ as input parameters
the charged Higgs mass and the masses for the neutral CP-even Higgs
bosons as well as the mixing angle $\alpha$ in the neutral
sector are given by

\vspace{-0.5cm}
{\footnotesize \beq
M_{H^\pm}^2 \!\!&=&\!\!  M_A^2 + M_W^2 \non\\
M_{h,H}^2 \!\!&=&\!\! \textstyle{\frac{1}{2}}
\left[ M_A^2+M_Z^2+\epsilon \mp
\sqrt{(M_A^2+M_Z^2+\epsilon)^2- 4M_A^2 M_Z^2 \cos^2 2\beta
   - 4\epsilon( M_A^2 \sin^2 \beta + M_Z^2 \cos^2 \beta)} \right]
\non \\
\tan 2\alpha \!\!&=&\!\! \tan 2\beta
 \frac{M_A^2 + M_Z^2}{M_A^2 - M_Z^2 +\epsilon/\cos 2\beta} \qquad
\mbox{with} \qquad  - \frac{\pi}{2} \leq \alpha \leq 0
\label{mass}
\eeq}
\hspace{-0.3cm} in the $\epsilon$ approximation. Evidently, at tree
level, where $\epsilon = 0$, the parameters of the Higgs potential can
be expressed by the electroweak and two additional parameters which
are usually chosen as $M_A$ and $\tan\beta$. [The mass $M_A$ is
related to $m_{12}$ via $M_A^2=m_{12}^2/\sin\beta\cos\beta$. The
masses $m_{11},m_{22}$ can be eliminated by minimizing the potential.]
The Born masses fulfill the following conditions
\beq
\begin{array}{rllll}
M_H &>& M_A, M_Z &>& M_h \\
M_{H^\pm} &>& M_A, M_W & & \\
M_h &<& M_Z |\cos2\beta| & & 
\end{array}
\label{massconstr}
\eeq
When radiative corrections are taken into account the upper bound of
the light neutral Higgs mass $M_h$ will extend to about 130~GeV
\cite{okada,carena}. In contrast, the masses of the heavy CP-even and
CP-odd neutral Higgs bosons $H$, $A$, and the charged Higgs particles
$H^\pm$ may vary in the mass range from the order of the electroweak
symmetry scale $v$ up to about 1~TeV.\s

After introducing the physical Higgs states the self-couplings among
the Higgs bosons can be derived from (\ref{mssmpot}).
Respecting CP-invariance there are six trilinear couplings
among the neutral Higgs bosons $h$ and $H$. In units of $\lambda_0=
M_Z^2/v$, $v=\sqrt{v_1^2+v_2^2}$, they may be written as
\cite{okada,djouadi1,djouadi2}
\beq
\lambda_{hhh} &=& 3 \cos2\alpha \sin (\beta+\alpha) 
+ 3 \frac{\epsilon}{M_Z^2} \frac{\cos \alpha}{\sin\beta} \cos^2\alpha  
\non \\
\lambda_{Hhh} &=& 2\sin2 \alpha \sin (\beta+\alpha) -\cos 2\alpha \cos(\beta
+ \alpha) + 3 \frac{\epsilon}{M_Z^2} \frac{\sin \alpha}{\sin\beta}
\cos^2\alpha \non \\
\lambda_{HHh} &=& -2 \sin 2\alpha \cos (\beta+\alpha) - \cos 2\alpha \sin(\beta
+ \alpha) + 3 \frac{\epsilon}{M_Z^2} \frac{\cos \alpha}{\sin\beta}
\sin^2\alpha \non \\
\lambda_{HHH} &=& 3 \cos 2\alpha \cos (\beta+\alpha) 
+ 3 \frac{\epsilon}{M_Z^2} \frac{\sin \alpha}{\sin\beta} \sin^2 \alpha
\non \\
\lambda_{hAA} &=& \cos 2\beta \sin(\beta+ \alpha)+ 
\frac{\epsilon}{M_Z^2} 
\frac{\cos \alpha}{\sin\beta} \cos^2\beta \non \\
\lambda_{HAA} &=& - \cos 2\beta \cos(\beta+ \alpha) + 
\frac{\epsilon}{M_Z^2} 
\frac{\sin \alpha}{\sin\beta} \cos^2\beta
\label{coup}
\eeq
In the decoupling limit $M_A^2 \sim M_H^2 \sim M^2_{H^\pm} \gg v^2/2$,
\beq
\sin(\beta-\alpha) \to 1 \\
\cos(\beta-\alpha) \to 0
\eeq
and the trilinear Higgs couplings approach the values  
{\small \beq
\lambda_{hhh} &\lra& 3 M_h^2/M_Z^2 \non\\
\lambda_{Hhh} &\lra& -3 \sqrt{ 
\left( \frac{M_h^2}{M_Z^2}-\frac{\epsilon}{M_Z^2}\sin^2\beta \right)
\left( 1 - \frac{M_h^2}{M_Z^2} + \frac{\epsilon}{M_Z^2}\sin^2\beta 
\right) } - \frac{3\epsilon}{M_Z^2}\sin\beta\cos\beta \non\\
\lambda_{HHh} &\lra& 2 - \frac{3 M_h^2}{M_Z^2} + 
\frac{3\epsilon}{M_Z^2} \non\\
\lambda_{HHH} &\lra& 3 \sqrt{ 
\left( \frac{M_h^2}{M_Z^2}-\frac{\epsilon}{M_Z^2}\sin^2\beta \right)
\left( 1 - \frac{M_h^2}{M_Z^2} + \frac{\epsilon}{M_Z^2}\sin^2\beta 
\right) } -
\frac{3\epsilon}{M_Z^2} \frac{\cos^3\beta}{\sin\beta} \non\\
\lambda_{hAA} &\lra& - \frac{M_h^2}{M_Z^2} + \frac{\epsilon}{M_Z^2} 
\non\\
\lambda_{HAA} &\lra& \sqrt{ 
\left( \frac{M_h^2}{M_Z^2}-\frac{\epsilon}{M_Z^2}\sin^2\beta \right)
\left( 1 - \frac{M_h^2}{M_Z^2} + \frac{\epsilon}{M_Z^2}\sin^2\beta 
\right) } -
\frac{\epsilon}{M_Z^2}\frac{\cos^3\beta}{\sin\beta}
\eeq} 
\noindent \hspace{-0.3cm} with $M_h^2 = M_Z^2 \cos^2 2\beta + 
\epsilon\sin^2\beta$.  Evidently, the self-coupling of the light 
CP-even neutral Higgs boson $h$ reaches the SM value in the 
decoupling limit.
\begin{table}[t]
\begin{center}$
\begin{array}{lllll}\hline
\Phi & & g_{\Phi\bar{u}u} & g_{\Phi\bar{d}d} & g_{\Phi VV} \\[0.1cm] \hline
SM & H & 1 & 1 & 1 \\[0.1cm] \hline  
MSSM &h& \cos\alpha/\sin\beta & -\sin\alpha/\cos\beta & \sin(\beta-\alpha) \\
&H& \sin\alpha/\sin\beta & \cos\alpha/\cos\beta & \cos(\beta-\alpha) \\ 
&A& 1/\tan\beta & \tan\beta & 0 \\[0.1cm] \hline
\end{array}
$\end{center}
\caption{Yukawa and gauge boson [$V=W,Z$] Higgs couplings in the MSSM 
normalized to the SM couplings. }
\label{hcoup}
\end{table}

\subsection{The Higgs-Yukawa and the Higgs-gauge couplings}

\begin{figure}[ht]
\unitlength 1cm
\begin{center}
\begin{picture}(10,11)
\put(-2.8,4){\epsfig{figure=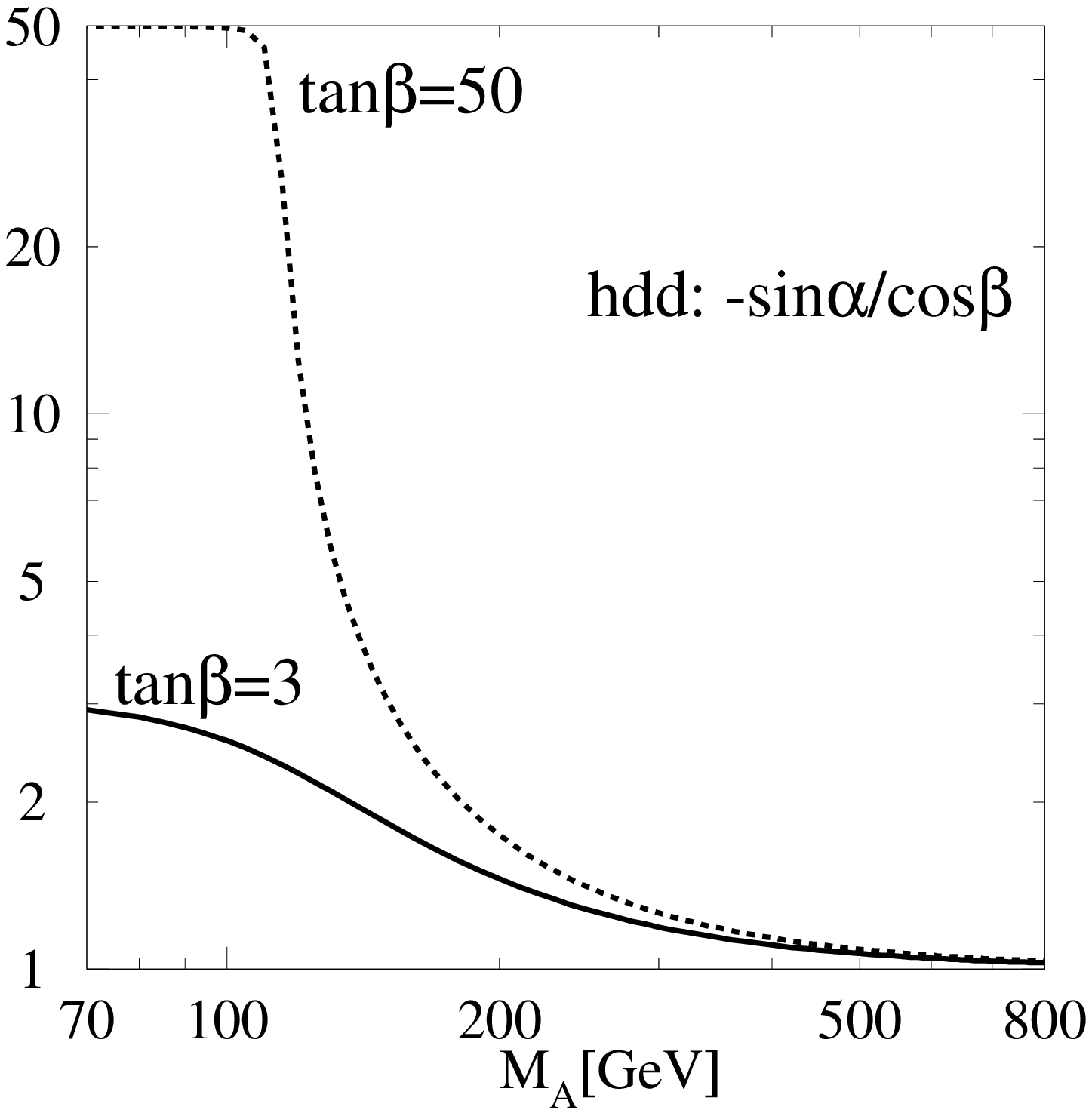,width=7.5cm}}
\put(5.3,4){\epsfig{figure=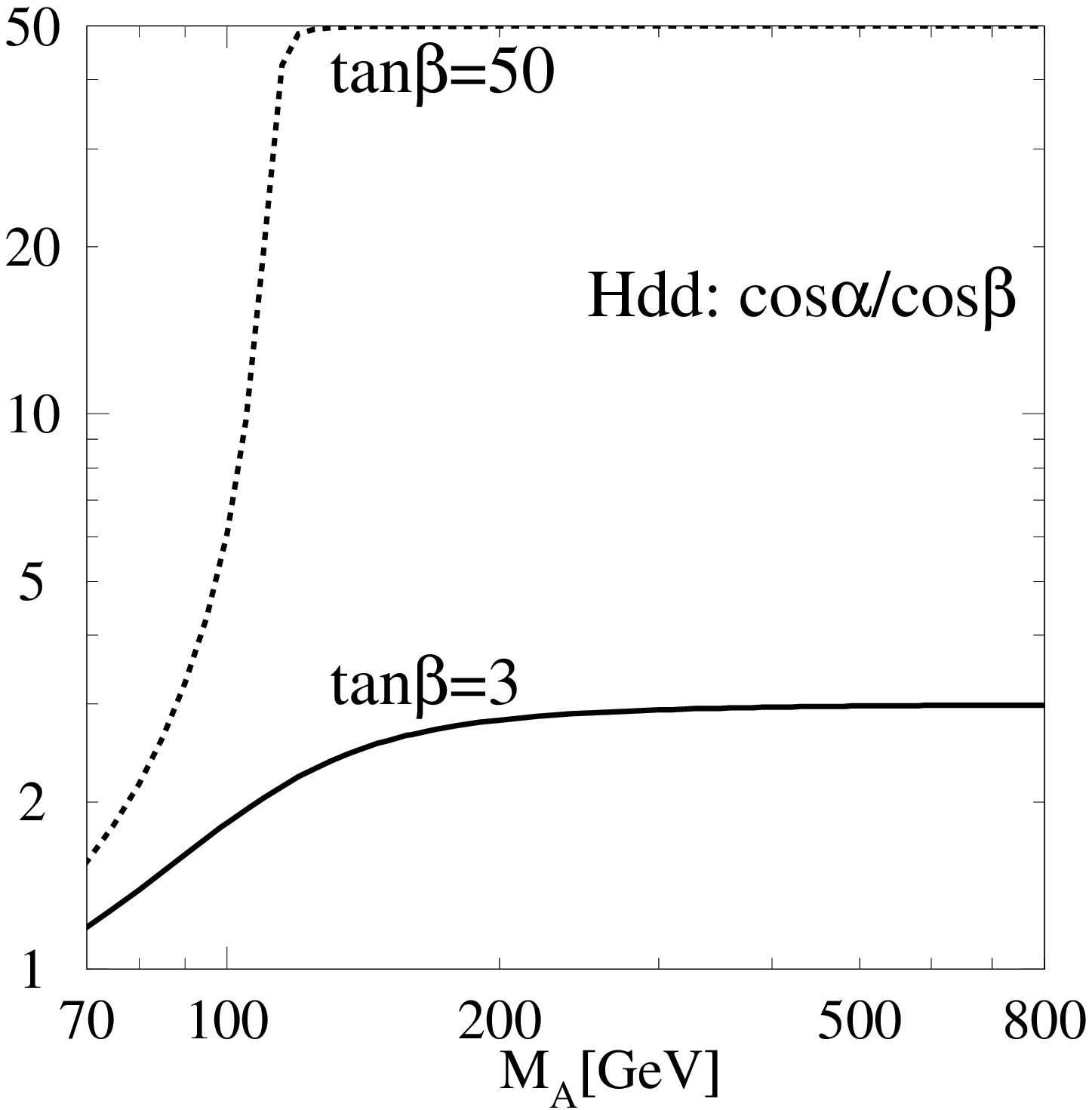,width=7.5cm}}
\put(-2.8,-4){\epsfig{figure=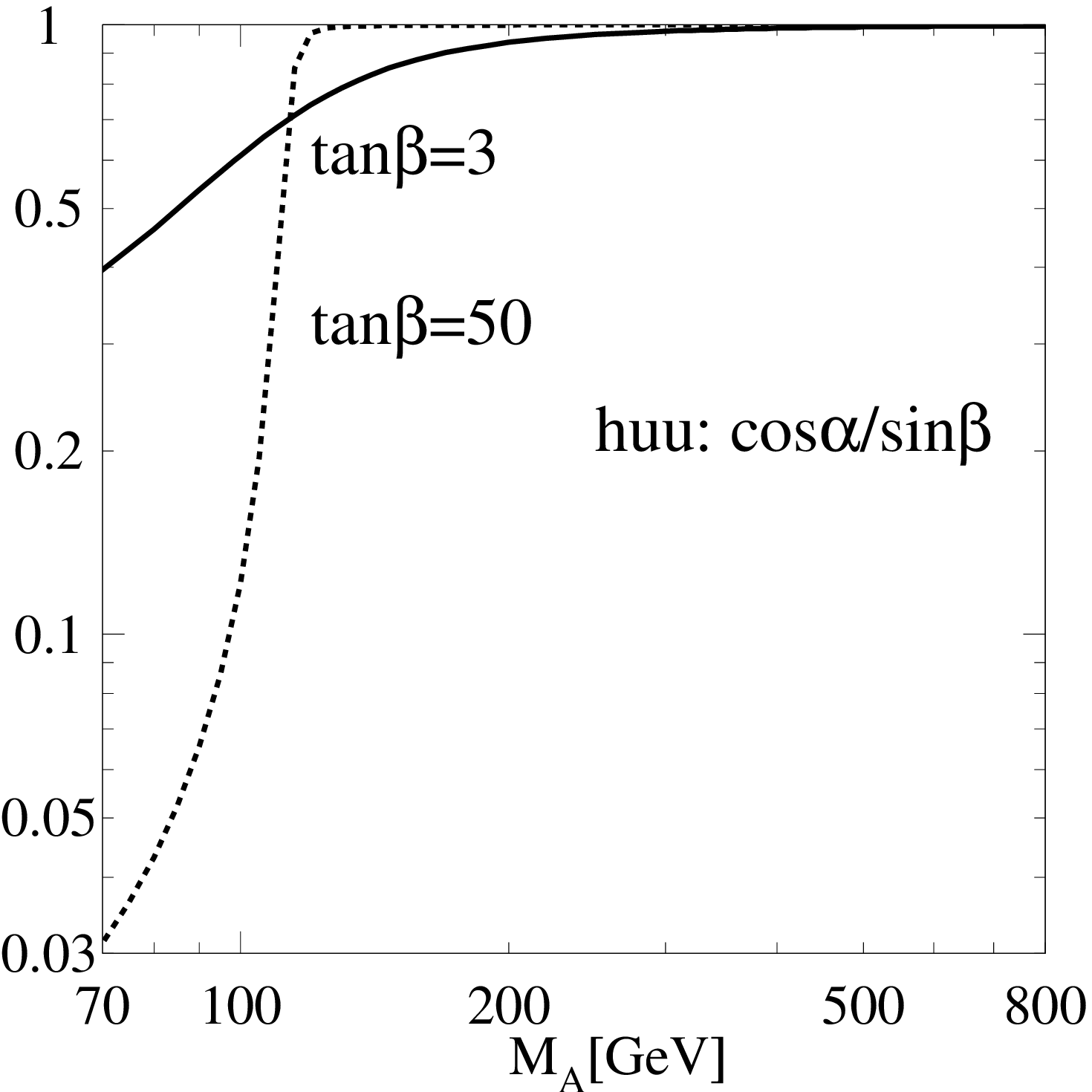,width=7.5cm}}
\put(5.3,-4){\epsfig{figure=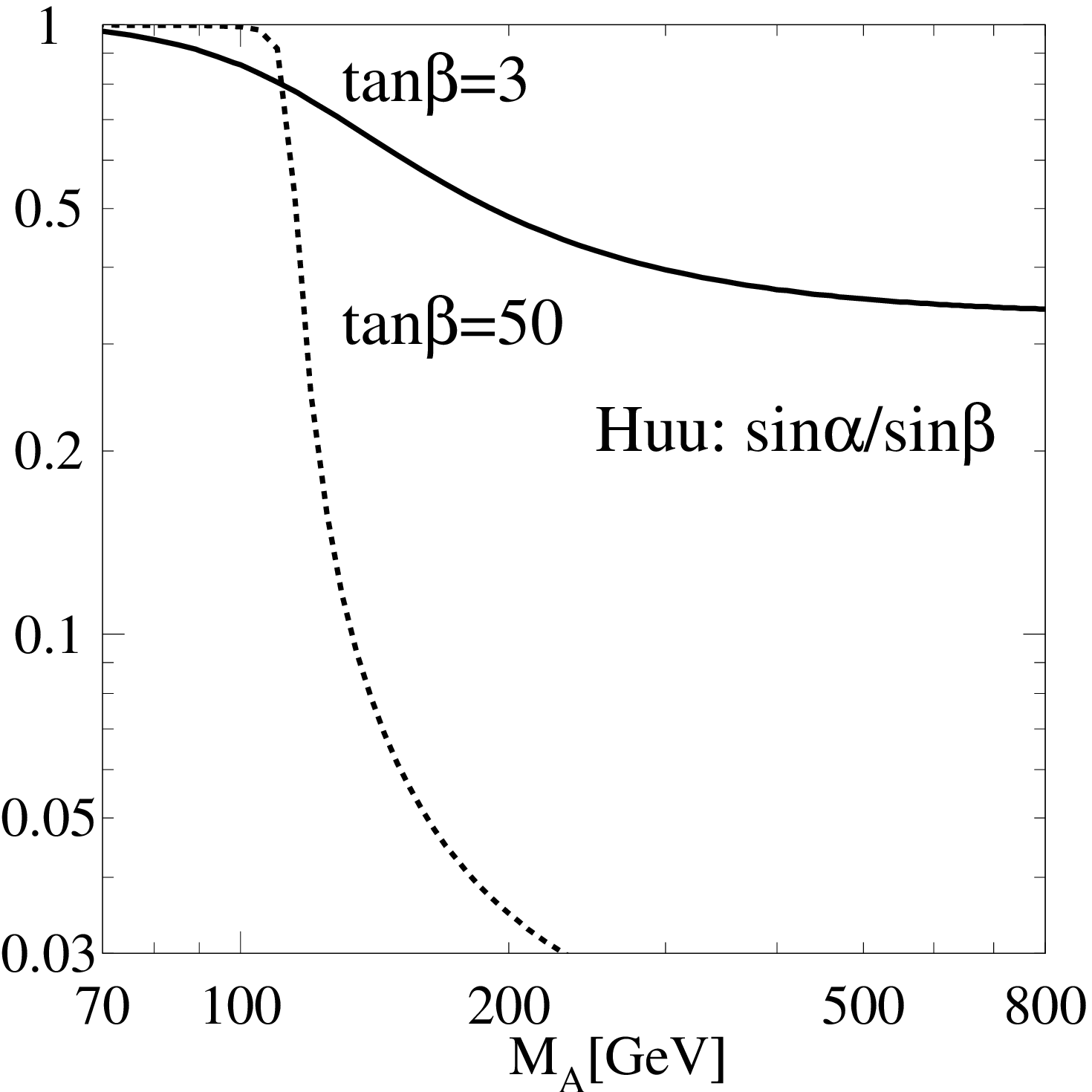,width=7.5cm}}
\end{picture}
\end{center}\vspace{3.8cm}
\caption{\textit{The Yukawa couplings of the neutral CP-even Higgs bosons to up- and down-type quarks, respectively, in units of the SM couplings as a function of $M_A$ for two values of $\tan\beta =3,50$ and vanishing mixing.}}
\label{yukmssm}
\end{figure}
The couplings of the MSSM Higgs particles to the fermions and gauge
bosons can be derived in the framework of the superspace formalism
from the Yukawa Lagrangean and the kinetic term for the two Higgs
doublets, respectively \cite{gunion,susy,susyprod}. The procedure is
analogous to the SM bearing in mind that two $SU(2)_L$ complex scalar
doublets are involved in the MSSM. Expanding around the VEVs and
transforming to the Higgs mass eigenstates, the couplings given in
Table~\ref{hcoup} will be found normalized to the SM
\cite{gunion,susy,susyprod}. They depend on the ratio of the vacuum
expectation values, $\tan\beta$, and the mixing angle $\alpha$ which
was introduced in order to diagonalize the mass matrix in the neutral
CP-even Higgs sector. \s

The Higgs gauge couplings $g_{hVV}$ and $g_{HVV}$ are suppressed with
respect to the corresponding SM couplings by the factors
$\sin(\beta-\alpha)$ and $\cos(\beta-\alpha)$, respectively. Only in
the decoupling limit the light CP-even Higgs boson gauge coupling
approaches the SM value whereas the coupling for $H$ becomes zero. For
the pseudoscalar Higgs particle $A$ there exists no coupling to the
gauge bosons at tree level due to CP-invariance. \s

The Yukawa gauge couplings of the CP-even neutral MSSM Higgs bosons to
up- (down-) type quarks are inversely proportional to $\sin\beta$
($\cos\beta$). Their variation with $M_A$ is shown for two values of
$\tan\beta = 3,50$ in Fig.~\ref{yukmssm}. Evidently, for large values
of $M_A$, {\it i.e.} in the decoupling limit, the couplings involving
$h$ approach the SM values whereas for small values of $M_A$ this
r\^{o}le is taken over by the heavy CP-even Higgs boson $H$. The
Yukawa couplings of the CP-odd Higgs boson $A$ to up- (down-)type quarks
are suppressed (enhanced) for large values of $\tan\beta$.\s

The couplings in Table~\ref{hcoup} determine the decay modes of the
MSSM Higgs bosons and therefore their experimental signatures. Thus it
will be difficult in the decoupling limit to distinguish the light scalar
MSSM Higgs boson from the SM Higgs particle. \s


%% file: chapter3.tex
\chapter{Higgs boson search \label{higgsmass}}
The mass of the Higgs boson cannot be predicted within the framework
of the SM. Its value can be constrained, however, by the assumption
that the model is valid up to an energy scale $\Lambda$. Demanding
that the SM remains perturbative up to the GUT scale ${\cal
  O}(10^{16}$~GeV$)$, an upper bound of the Higgs mass is given by
$\sim 200$~GeV. For $\Lambda \sim 1$~TeV and the constraint $M_H \le
\Lambda$ lattice simulations \cite{lat} and renormalization group
analyses \cite{ren1,ren2} predict an upper bound of $\sim 700$~GeV.\s

A lower bound on the Higgs mass is given by the requirement of vacuum
stability. The quantum corrections to the quartic self-coupling
$\lambda$ due to the Higgs self-interaction are positive whereas those
from the top-Yukawa coupling are negative. Depending on the value of
$\lambda$ at the scale of the Higgs mass the quartic coupling
increases with the energy or is driven to negative values so that the
vacuum becomes unstable depending on the top-quark mass. Since the
strength of the Higgs self-interaction is determined by the Higgs mass
at the scale $M_H$ the negative contribution can be compensated if
$M_H$ is large enough.  Assuming a top mass of 175~GeV the SM remains
weakly interacting up to $\sim 1$~TeV if the lower bound of the Higgs
mass is given by $\sim 55$~GeV. For $\Lambda \sim M_{{\rm GUT}}$ the
lower bound increases to 130~GeV. If the vacuum is metastable with a
lifetime exceeding the age of the universe, the lower Higgs mass bound
decreases. While the bound changes only slightly for $\Lambda \sim
M_{{\rm GUT}}$, the effect is significant for $\Lambda \sim 1$~TeV
\cite{ren1}.\s

The direct Higgs boson search in the Higgs-strahlung process at LEP2 ,
finally, constrains the Higgs mass from below \cite{exp}. The search
will possibly be extended up to Higgs masses of $\sim 115$~GeV. \s

In the MSSM, the Higgs sector at tree level is completely described by
the electroweak parameters and two additional parameters $M_A$ and
$\tan\beta$, so that an upper bound for the light scalar Higgs mass
can be derived, cf.~(\ref{massconstr}). \s

The crucial test for the existence of the Higgs particles will be
their experimental discovery. In this chapter an overview will be
presented over the SM and the MSSM Higgs boson search strategies at
$pp$ colliders and at $e^+e^-$ linear colliders. Subsequently, the
heavy MSSM Higgs particle production in the Compton mode of an
$e^+e^-$ linear collider will be examined in detail taking into
account the background reactions and interference effects.  These
processes are important for the search of heavy MSSM Higgs particles
in parameter regions that are difficult to exploit at the LHC and in
$e^+e^-$ collisions. \s

\section{Higgs boson search at $pp$ colliders}

\subsection{Standard Model Higgs boson}
\begin{figure}[ht]
\vspace{-1cm}
\begin{center}
\psfig{figure=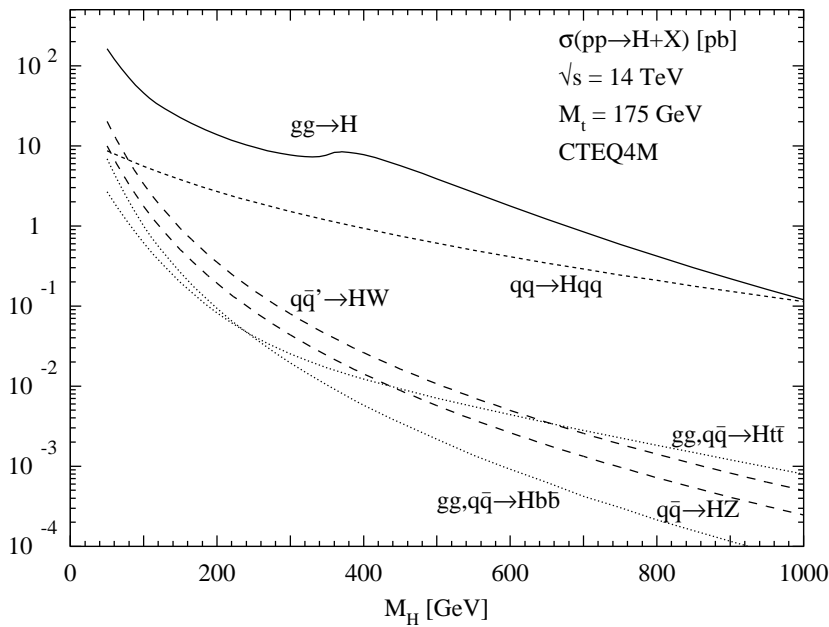,width=11cm,angle=-90}
\end{center}
\vspace{-2cm}
\caption{\textit{
  Higgs production cross sections at the LHC for $\sqrt{s} = 14$~TeV
  as a function of the Higgs mass. The full QCD corrections have been
  included apart from the processes $Hb\bar{b}$, $Ht\bar{t}$ where the
  QCD-corrected results are unknown \cite{spirahabil}.}}
\label{hsmprod}
\end{figure}

A primary goal of the Large Hadron Collider LHC will be the search for
Higgs particles. The main production mechanisms for the SM Higgs boson
are \cite{gunion,abdhabil,spirahabil}\s
\beq
\begin{array}{l@:l}
\mbox{gluon fusion} &\quad pp \to gg\to H \\
\mbox{vector-boson fusion} &\quad qq \to qqV^*V^* \to qqH \\
\mbox{Higgs-strahlung} &\quad q\bar{q}\to V^* \to VH\\
\mbox{associated production with $t\bar{t}$/$b\bar{b}$} &
\quad q\bar{q},gg \to Ht\bar{t}/b\bar{b}
\end{array}
\eeq
where $V=W,Z$. The cross sections are shown in Fig.~\ref{hsmprod} for
a center of mass energy of $\sqrt{s}=14$~TeV. Evidently, the gluon
fusion process is the dominant process in the entire mass range up to
1~TeV apart from vector boson fusion which is of the same order of
magnitude above 800~GeV. For Higgs masses less than about 100~GeV
Higgs-strahlung and associated production with $t\bar{t}$ become
competitive with vector boson fusion and provide additional
production mechanisms for the Higgs boson.\s

\begin{figure}[ht]
\begin{center}
\psfig{figure=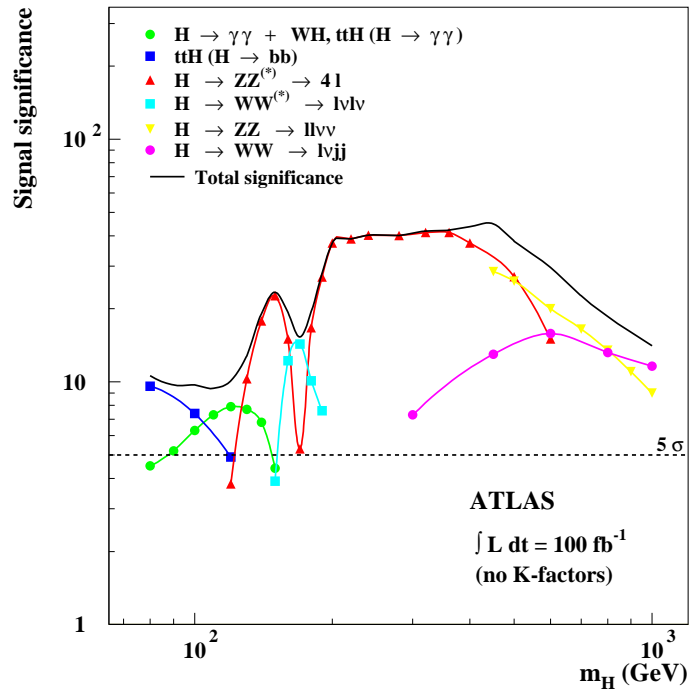,width=9cm}
\end{center}
\vspace{-0.5cm}
\caption{\textit{Expected significances in the various SM Higgs search channels at ATLAS as a function of the Higgs mass for an integrated luminosity of 100~fb$^{-1}$ \cite{atlascms}.}}
\label{atlas}
\end{figure}

\begin{figure}
\begin{center}
\psfig{figure=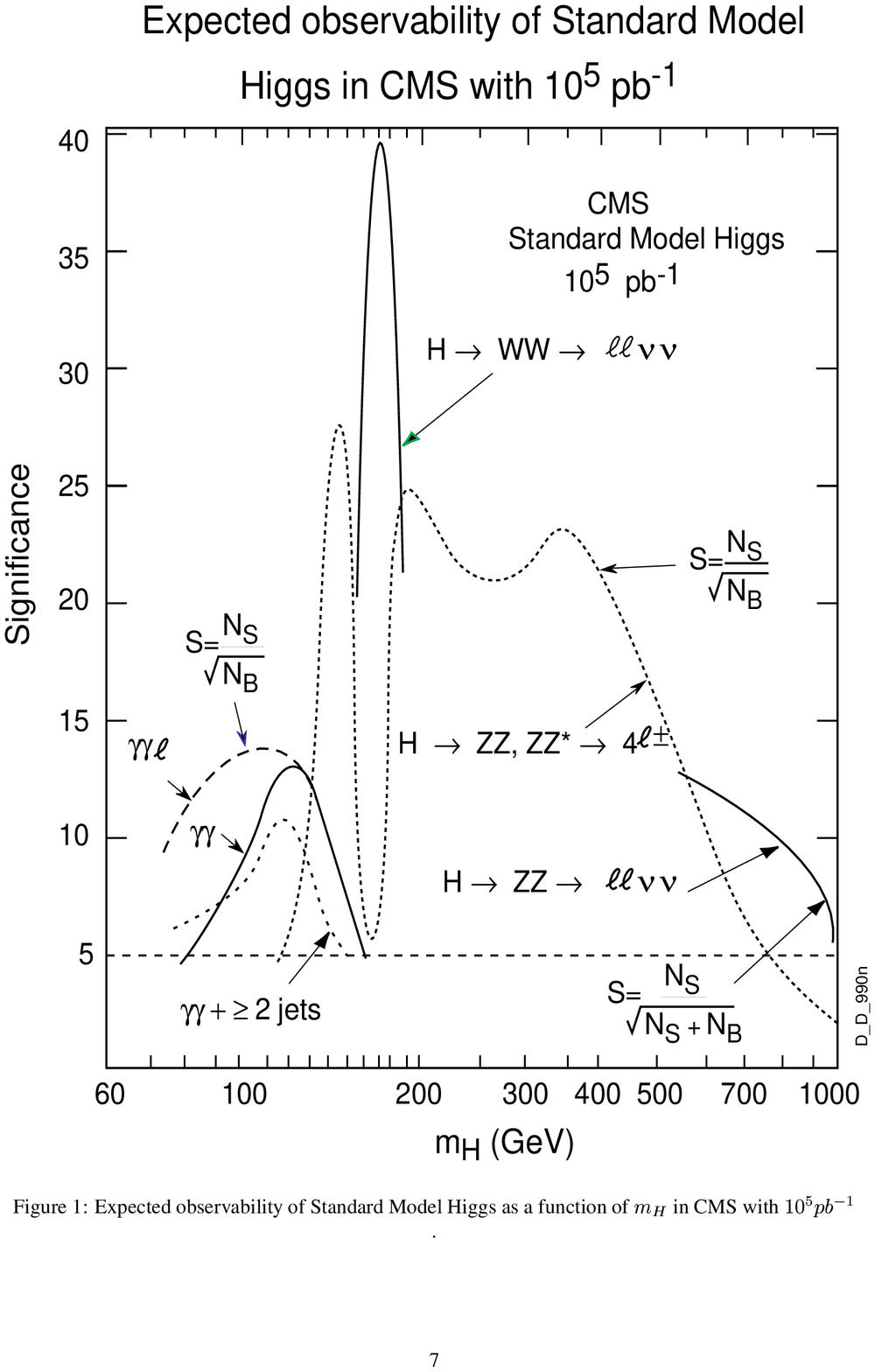,bbllx=83pt,bblly=170,bburx=513,bbury=681,width=8cm,clip=}
\end{center}
\vspace{-0.5cm}
\caption{\textit{Significances for the SM Higgs boson search as a function of the Higgs mass at CMS for an integrated luminosity of 100~fb$^{-1}$ \cite{kinnunen}.}}
\label{cms}
\end{figure}

The Higgs search strategies will depend on the mass of the Higgs
boson. Figs.~\ref{atlas} and \ref{cms} show the significances in the
various search channels of the SM Higgs boson in the ATLAS and the CMS
experiment, respectively, as a function of the Higgs mass
\cite{atlascms,kinnunen}. Evidently, at the LHC several channels are
available for the detection of a Higgs boson with a mass between
95~GeV and 1~TeV. \s

\begin{figure}
\begin{center}
\psfig{figure=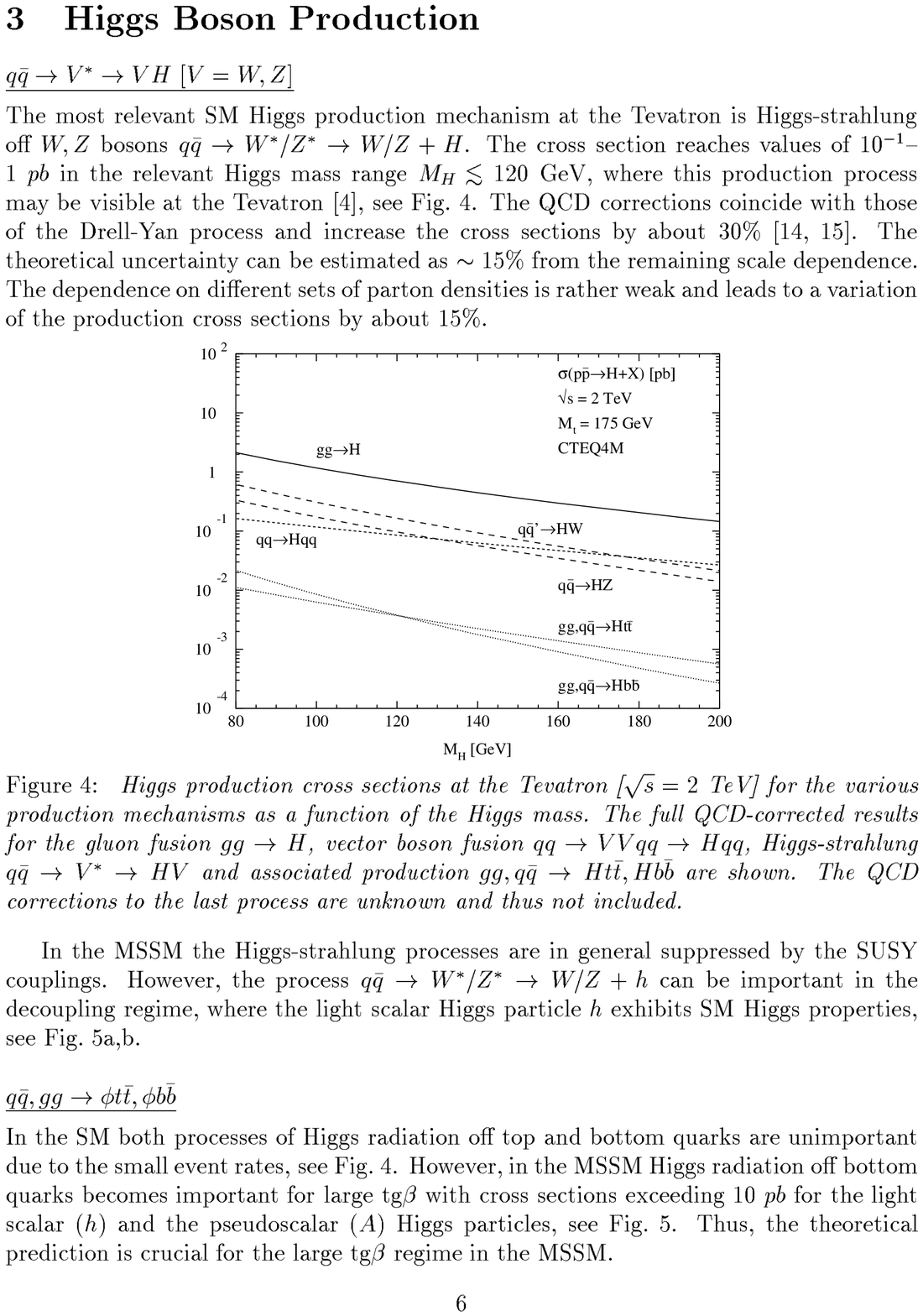,bbllx=171pt,bblly=344,bburx=440,bbury=553,width=11cm,clip=}
\end{center}
\vspace{-0.5cm}
\caption{\textit{The cross sections of the various Higgs production mechanisms at the Tevatron as a function of the Higgs mass \cite{tevspir}. The plot shows the full QCD corrected results for gluon fusion $gg\to H$, vector boson fusion $qq \to qqVV \to qqH$ and Higgs-strahlung $q\bar{q} \to V^* \to VH$. The QCD corrections to the associated production processes $gg,q\bar{q} \to Ht\bar{t}, Hb\bar{b}$ are unknown and therefore not included.}}
\label{tevplot}
\end{figure}
For masses below $\sim 110$~GeV the Higgs boson produced in
association with a $t\bar{t}$ pair may be detected in its $b\bar{b}$
decay channel. Since the process is faced with a huge QCD background,
however, the most promising channel in the mass region $80\,\lesim\,
M_H \,\lesim\, 140$~GeV is the rare decay $H\to \gamma\gamma$ with a
branching ratio of ${\cal O}(10^{-3})$. The Higgs boson is dominantly
produced in gluon fusion. Additional production mechanisms are provided
by Higgs radiation off a $W$ boson and associated production of a
Higgs boson with $t\bar{t}$. \s

In the mass range $120\,\lesim\, M_H\,\lesim\, 800$~GeV, the
'gold-plated' channel $H\to ZZ\to 4l$ is considered to be the most
reliable channel for the SM Higgs boson discovery at the LHC. For
masses $160\,\lesim\, M_H\,\lesim\, 180$~GeV where the $ZZ$ branching
ratio is only about $\sim 2$\% the Higgs boson search can be
supplemented by the dominant decay $H\to WW$ with $l\nu l\nu$ final
states. \s

If the Higgs mass is large the Higgs boson becomes very broad with a
width of up to 600~GeV for $M_H=1$~TeV. This implies a broad resonance
peak in the final state and due to the reduced phase space the event
rates decrease for a heavy Higgs boson. Since the channels $H\to ZZ\to
l\nu l\nu$ and $H\to WW \to l\nu jj$ involve larger branching ratios
than the 'gold-plated' decay they provide a means of detecting the
Higgs particle in the range $M_H \,\gesim\, 800$~GeV.\s

For an integrated luminosity of 300~fb$^{-1}$, the expected precision
for the mass measurement will be 0.1\% in the Higgs mass range
80-400~GeV \cite{atlascms,kintalk}.\s

Fig.~\ref{tevplot} shows the cross sections for the various Higgs
production mechanisms at the Tevatron as a function of the Higgs mass
for a c.m.~energy of 2~TeV. The most important process for the Higgs
search is Higgs-strahlung off $W,Z$ bosons, $q\bar{q} \to W^*/Z^* \to
W/Z+H$. Due to the lower luminosity and the large backgrounds at the
Tevatron several search channels have to be combined and the
statistical power of both experiments CDF and D0 has to be exploited
in order to reach significances above three for the detection of the
Higgs boson in certain mass windows. For an integrated luminosity of
20~fb$^{-1}$ as might be expected in RUNIIb a Higgs boson with mass up
to $\sim 180$~GeV can be detected at the 3$\sigma$ level using
neural network selection for the investigated channels. A 5$\sigma$
discovery will only be possible below $\sim 120$~GeV.  The Higgs can
be excluded up to $\sim 190$~GeV with 95\% CL \cite{conway}.

\subsection{Supersymmetric extension}
Except for vector boson fusion and Higgs-strahlung the production of
the MSSM Higgs bosons proceeds via the same mechanisms as in the SM
case taking into account that contributions due to $b$-quarks may also
play a r\^{o}le for large $\tan\beta$ where the Yukawa couplings to
down-type quarks are enhanced, cf.~Table~\ref{hcoup} and
Fig.~\ref{yukmssm}. The production via vector boson fusion and
Higgs-strahlung is only possible for CP-even Higgs bosons since the
CP-odd boson $A$ does not interact with the the vector particles at
tree level. \s

In the MSSM the gluon fusion process proceeds dominantly via top- and
bottom-quark loops with the $b$-quark contribution becoming of the
same size as that due to the top-quark for large $\tan\beta$. Squarks
are also involved. However, for large squark masses they decouple.
Since the scalar Higgs vector boson couplings are suppressed with
$\sin(\beta-\alpha),\cos(\beta-\alpha)$ compared to the corresponding
SM couplings the production cross sections for the MSSM scalar Higgs
bosons in vector boson fusion are smaller than the SM fusion cross
sections. Analogously, the associated production with $W/Z$ is smaller
than in the SM case and has only to be considered for the light
CP-even Higgs boson and the heavier one provided its mass is small. As
can be inferred from Table~\ref{hcoup} and Fig.~\ref{yukmssm} the MSSM
Higgs $t\bar{t}$ couplings are below the SM counterpart for
$\tan\beta>1$ so that associated production with $t\bar{t}$ will
always yield smaller cross sections than in the SM.  Yet, the same
process involving a $b\bar{b}$ pair may become dominant in the large
$\tan\beta$ region.\s

\begin{figure}[ht]
\unitlength 1cm
\begin{center}
\begin{picture}(13,6)
\put(-1.2,-1.65){\epsfig{figure=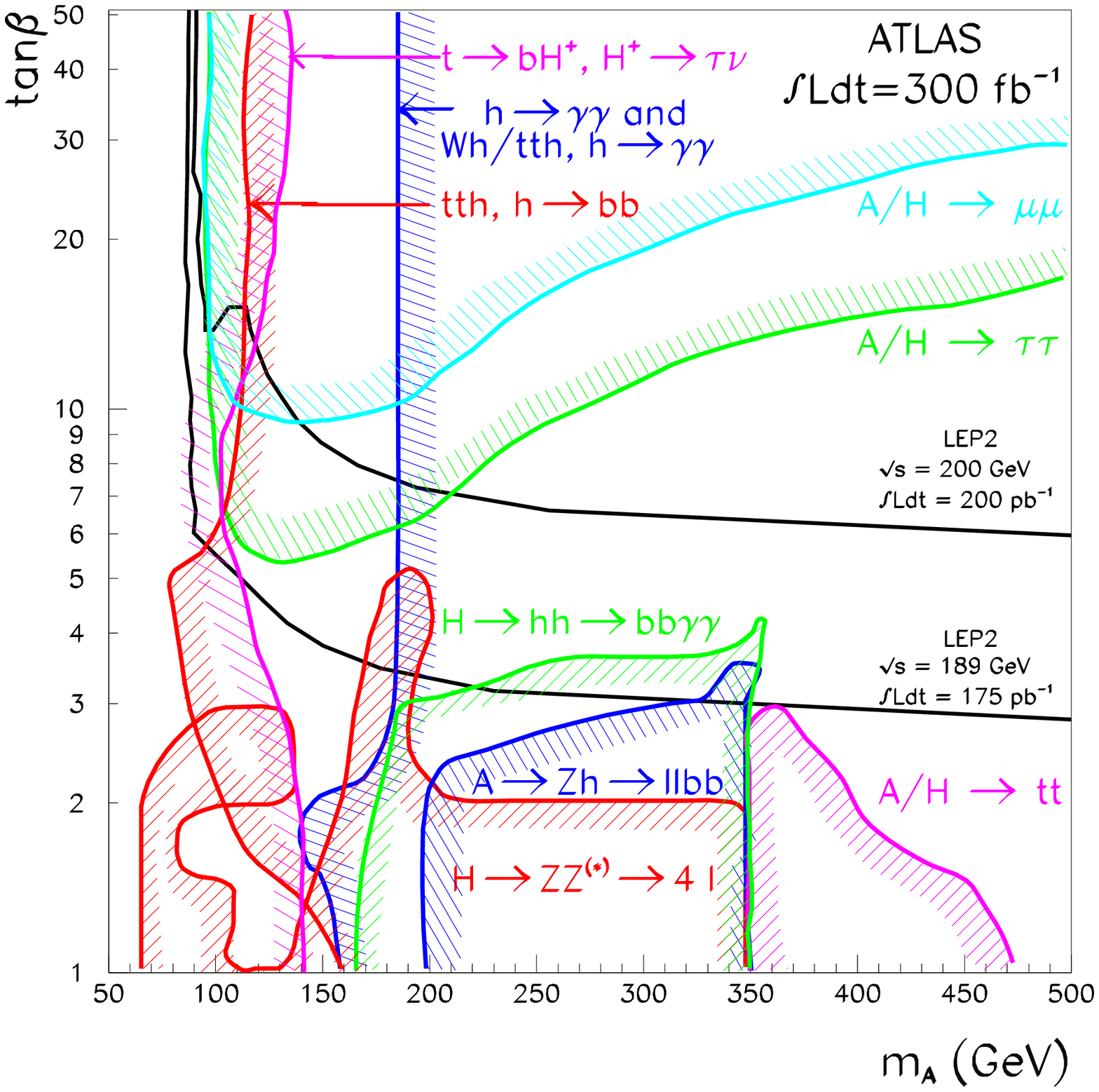,width=7.5cm}}
\put(6.2,-1.8){\epsfig{figure=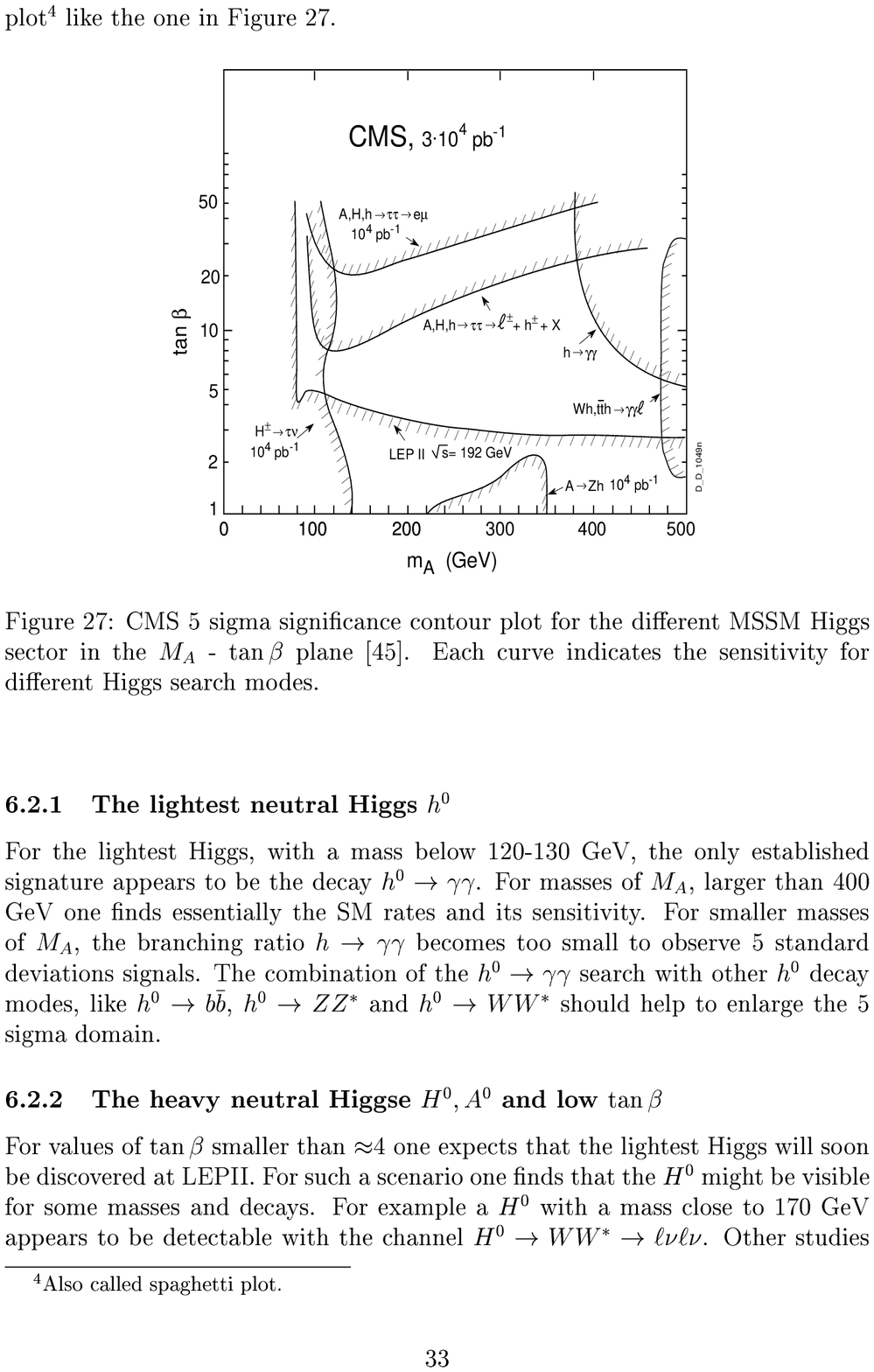,bbllx=162pt,bblly=409,bburx=457,bbury=652,width=9.2cm,clip=}}
\end{picture}
\end{center}\vspace{1.5cm}
\caption{\textit{Left: Expected discovery contours for MSSM Higgs bosons at ATLAS with $\int{\cal L} = 300$~fb$^{-1}$. Each curve indicates the sensitivity for different Higgs search modes \cite{atlascms}. Right: CMS 5 $\sigma$ significance contour plot in the $M_A-\tan\beta$ plane \cite{kinnunen}.}}
\label{lhc}
\end{figure}
The search for MSSM Higgs bosons is more complex than in the SM.
Fig.~\ref{lhc} shows the discovery contours for SUSY Higgs particles
at the ATLAS and CMS experiment, respectively. Referring to the ATLAS
experiment [the overall picture is essentially the same for the CMS
experiment, cf.~Fig.~\ref{lhc}], the basic features can be summarized
as follows: Taking into account the exclusion limits set by LEP2 at a
c.m.~energy of 200~GeV, the light Higgs state $h$ produced in
association with $t\bar{t}$ may be found in its $b\bar{b}$ decay
channel for pseudoscalar Higgs masses $M_A\,\gesim\, 110$~GeV. In the
mass range $M_A\,\gesim \,200$~GeV the CP-even Higgs boson $h$ can be
found via the $\gamma\gamma$ decay channel. The heavy Higgs particles
$H$ and $A$ can be found in the channel $H/A \to \tau \tau$ for
$M_A\,\gesim\, 100$~GeV and moderate and large values of $\tan\beta$.
Since the MSSM Higgs couplings to charged leptons grow with
$\tan\beta$, for larger values $\tan\beta\,\gesim\, 10$ the search
channel $H/A\to \mu\mu$ opens in the same $M_A$ region. As the masses
for the heavy CP-even and the CP-odd Higgs boson are almost the same
in this parameter space it will be very difficult to disentangle the
$H$ signal from the pseudoscalar signal. For $M_A\,\lesim\, 140$~GeV,
the charged Higgs boson produced in top decays can be searched for in
the channel $t\to bH^+$.\s

The mass of the light CP-even Higgs boson $h$ is expected be
determined with a precision of 200~MeV at the LHC \cite{atlascms}.
However, the final picture exhibits two difficult regions. For
$90\,\lesim\, M_A\,\lesim$\, 110~GeV, the light Higgs boson might be
seen neither by LEP2 nor by LHC. And for $M_A\,\gesim\, 200$~GeV and
$\tan\beta \,\gesim\, 6$ up to 15, no heavy neutral Higgs bosons can
be discovered so that the distinction between the SM and the MSSM
based solely on light scalar Higgs measurements will be difficult.
Note also that if neutralinos and charginos are light enough, Higgs
decays into these particles are possible and will change the search
strategies at the LHC \cite{baer}.  \s

By appropriate rescaling of the SM results an analysis for the MSSM
Higgs boson search has been carried out at the Tevatron \cite{conway}.
The analysed production channels for the scalar Higgs particles are
Higgs-strahlung with subsequent decay into $b\bar{b}$ and associated
production with $b\bar{b}$ for all neutral MSSM Higgs particles. The
results for the exclusion regions at 95\% CL and the 5$\sigma$
discovery regions depend on the chosen set of SUSY parameters. For
example, in the case of maximal mixing a light scalar Higgs with mass
$\lesim\, 130$~GeV can be discovered at the 5$\sigma$ level in almost
the whole $M_A-\tan\beta$ plane if an integrated luminosity of
30~fb$^{-1}$ can be achieved. Difficult regions are the low $M_A$
region for moderate and large values of $\tan\beta$. A detailed
prescription of the analysis can be found in \cite{conway,mrenna}.

\begin{figure}[ht]
\begin{center}
\psfig{figure=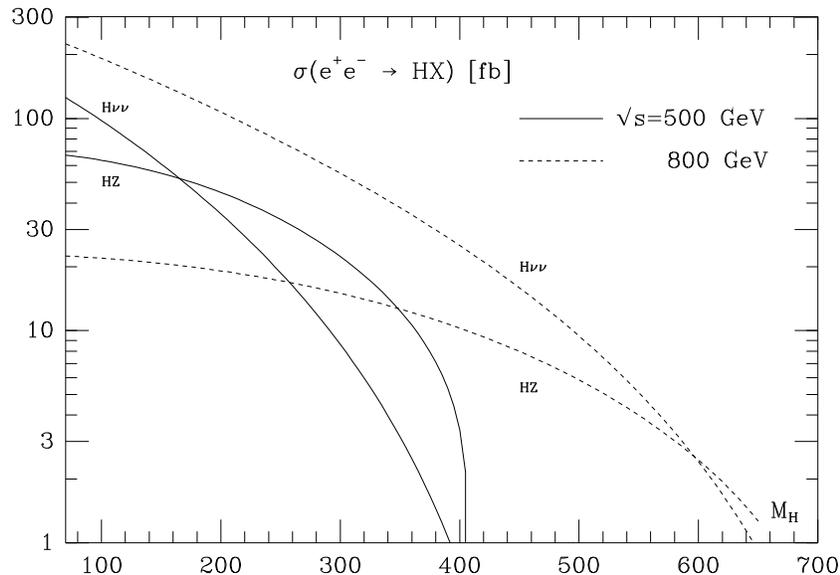,angle=-90,width=11cm,clip=}
\end{center}
\vspace{-0.5cm}
\caption{\textit{The cross sections of Higgs-strahlung $e^+e^-\to ZH$ and $WW/ZZ$ fusion $e^+e^-\to \bar{\nu}\nu/e^+e^- H$ for two collider energies $\sqrt{s}=500$~GeV (solid) and $\sqrt{s}=800$~GeV (dashed), \cite{accomando}.}}
\label{eeprod}
\end{figure}
\section{Higgs boson search at $e^+e^-$ colliders}
$e^+e^-$ linear colliders operating in the c.m. energy range between
500 GeV and 1 TeV are ideal instruments for the search of Higgs bosons
with masses in the intermediate mass range. In this section the Higgs
boson search at these future colliders will be reviewed.

\subsection{Standard Model}
The main production mechanisms for a SM Higgs boson at an $e^+e^-$
collider in the energy range $500$~GeV$\le \sqrt{s} \le 1$~TeV are the
processes
\beq
\begin{array}{l@:l}
\mbox{Higgs-strahlung} &\quad e^+e^- \to (Z) \to Z + H \\
\mbox{WW fusion} &\quad e^+e^- \to \bar{\nu}\nu(WW) \to \bar{\nu}{\nu} + H 
\end{array}
\eeq 
For moderate values of the Higgs mass and in the low energy range
Higgs-strahlung is the main production process \cite{hrad1,hrad2}.
Beyond the threshold region the cross section decreases proportional
$s^{-1}$ with the maximum being reached rather close to the threshold.
For larger energies $WW$ fusion starts playing a r\^{o}le
\citer{wfus1,wfus3} since in the high-energy limit, $M_H^2/s$ and
$M_W^2/s\ll 1$, the cross section increases with $\log s$. Another
production mechanism is provided by $ZZ$ fusion. Due to the small
$Ze^+e^-$ couplings this process is suppressed by an order of
magnitude compared to $WW$ fusion. Yet, the $Z$ decay yields two
leptons in the final state so that by reconstructing the recoil mass a
sufficiently large signal to background ratio can be achieved. \s

In Fig.~\ref{eeprod} the cross sections of $WW$ boson fusion and
Higgs-strahlung are shown for two typical collider energies,
$\sqrt{s}=500$~GeV and 800~GeV \cite{accomando}. Apparently, $WW$
fusion is the dominant process for higher energies. The cross sections
are ${\cal O}(1$-$100)$~fb yielding up to $10^5$ events in the
intermediate mass range for the foreseen integrated luminosity of
$500$~fb$^{-1}$ in the low-energy phase. \s

Due to the definite energy $E_Z$ of the recoiling $Z$ boson in the
Higgs-strahlung process the Higgs mass can be determined via the
relation $M_H^2 = s - 2\sqrt{s} E_Z + M_Z^2$. The Higgs boson search
will be independent from the decay channel and the signal will only
slightly be modified when detector properties are taken into account
\cite{schreiber}. Also in $WW$ fusion it will be no problem to find
the Higgs boson since the main background $e^+e^- \to (e^+) \nu_e W^-$
with the positron in the final state escaping detection is small.
The mass of the SM Higgs particle can be determined at $e^+e^-$
colliders very precisely by exploiting the kinematic constraints in
the leptonic channels of Higgs-strahlung events. For an integrated
luminosity of 500~fb$^{-1}$ and a c.m.~energy of 350~GeV, a precision
of $\sim 150$~MeV can be reached for Higgs masses between 120 and
160~GeV independent of the Higgs decay mode \cite{lohmann}.  The
discovery limit is given by $M_H \,\lesim\, 0.7 \sqrt{s}$. $e^+e^-$
linear colliders are therefore optimal machines for the detection of
Higgs bosons in the intermediate mass range $M_H \,\lesim\, 2M_Z$.

\subsection{SUSY Higgs particles}
Higgs-strahlung, Higgs pair production and vector boson fusion are the
dominant production mechanisms for neutral MSSM Higgs bosons at
$e^+e^-$ colliders \cite{susyprod,djkalzer,ohmann}:
\beq
\begin{array}{lll}
\mbox{Higgs-strahlung}&: &\quad e^+e^- \to (Z) \to Z + h/H \\
\mbox{Pair production}&: &\quad e^+e^- \to (Z) \to A +h/H \\
\mbox{$WW$ fusion}&:  &\quad e^+e^- \to \bar{\nu}\nu (WW) \to \bar{\nu}\nu
+ h/H \\
\mbox{$ZZ$ fusion}&: & \quad e^+e^- \to e^+e^- (ZZ) \to e^+e^- + h/H 
\end{array}
\eeq 
The cross sections for Higgs-strahlung and pair production can be cast 
into the form
\beq
\sigma (e^+e^-\to Zh/H) &=& \sin^2 / \cos^2(\beta-\alpha) \sigma_{SM} 
\non\\
\sigma (e^+e^-\to Ah/H) &=& \cos^2 / \sin^2(\beta-\alpha) \sigma_{SM} 
\bar{\lambda}
\eeq
where $\sigma_{SM}$ denotes the SM cross section for Higgs-strahlung and
$\bar{\lambda}\sim \lambda_{Aj}^{3/2}/\lambda_{Zj}^{1/2}$
($\lambda_{ij}$ is the K\"allen function for particles with masses
$M_i$ and $M_j$) results from the P-wave suppression near the
threshold. \s

The cross sections for $h,H$ production via Higgs-strahlung and pair
production are among themselves and mutually complementary to each
other being proportional either to $\sin^2(\beta-\alpha)$ or
$\cos^2(\beta-\alpha)$. Since $\sigma_{SM}$ is large, at least the
light CP-even Higgs boson should be found: If $\sin^2(\beta-\alpha)$ is
small, also $M_A$ is small, and $h$ can then be detected in associated
production with a light $A$. \s

\begin{figure}[ht]
\begin{center}
\psfig{figure=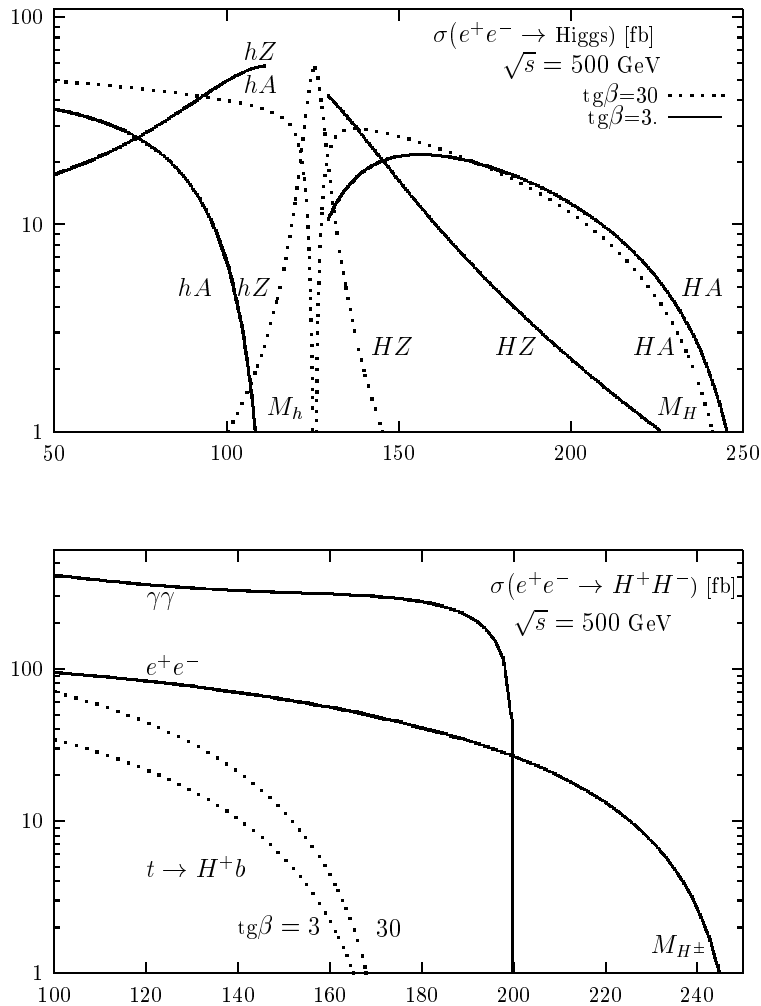,bbllx=140pt,bblly=244,bburx=480,bbury=698,width=10cm,clip=}
\end{center}
\vspace{-0.5cm}
\caption{\textit{Upper plot: Production cross sections of MSSM Higgs bosons at $\sqrt{s}=500$~GeV. Lower plot: Charged Higgs boson production for $\sqrt{s}=500$~GeV -- courtesy of A.~Djouadi.}}
\label{repprod}
\end{figure}
The upper plot in Fig.~\ref{repprod} shows representative examples of
the cross sections for the production of neutral Higgs bosons as a
function of the Higgs masses for $\tan\beta=3$ and 30. In the
decoupling limit where $\sin(\beta-\alpha)\to 1$ the $hZ$ cross
section becomes maximal being of ${\cal O}(60$~fb$)$. In contrast, the
$HZ$ cross section is large at the lower end of the mass range. The
signal signature, also for $HZ$ in most of the parameter space,
involves a $Z$ boson and a $b\bar{b}$ or $\tau^+\tau^-$ pair. Since
the process $Ah$ grows with $\cos^2(\beta-\alpha)$ it increases
towards lower $M_h$ masses where $\cos(\beta-\alpha)$ approaches its
maximum absolute value. The complementary process $AH$ is preferred
for large $M_H$. The signature will in both cases include four $b$
quarks in most of the parameter space so that a good $b$-tagging
performance will be required in order to separate the signal from the
background due to QCD jets and $Z$ boson pair production.\s

For small Higgs masses (below $\sim 160$~GeV for $\sqrt{s}=500$~GeV)
$WW$ fusion dominates over Higgs-strahlung \cite{abdhabil}. Yet, it
involves neutrinos in the final state thus rendering the extraction
of the signal more difficult. On the other hand the leptonic final
states of $ZZ$ fusion allow for the full signal reconstruction
although it is an order of magnitude smaller than the $WW$ fusion
process.\s

The lower plot in Fig.~\ref{repprod} shows the production cross
sections for charged Higgs bosons as a function of the Higgs mass for
$\tan\beta =3$ and 30. If kinematically allowed they are produced via
top-decays, $t\to b + H^+$. For $\tan\beta >1$, charged Higgs bosons
decay into $\tau\nu_\tau$ so that lepton universality will be broken
in the final state, since $\tau$ states dominate over $e,\mu$ final
states in $t$ decays. For large masses $M_{H^\pm}$ the bosons have to
be pair-produced in $e^+e^-$ collisions, $e^+e^-\to H^+H^-$.  The
cross section which only depends on $M_{H^\pm}$ decreases fast due to
$P$-wave suppression $\sim \beta^3$ near the threshold.
$\gamma\gamma$ collisions, however, yield larger cross sections.\s

The preceding discussion has shown that the light CP-even Higgs boson
will not escape detection at linear colliders. Its mass may be
determined with an accuracy of 50~MeV \cite{janot}. Furthermore, all MSSM
Higgs particles can be discovered in the mass range $M_H,M_A \,\lesim\,
1/2 \sqrt{s}$, independent of the value of $\tan\beta$ \cite{accomando}.

\section{Heavy MSSM Higgs production in $\gamma\gamma$ collisions}
\begin{figure}[p]
\unitlength 1cm
\begin{center}
\begin{picture}(9,9)
\put(-2.2,2.8){\epsfig{figure=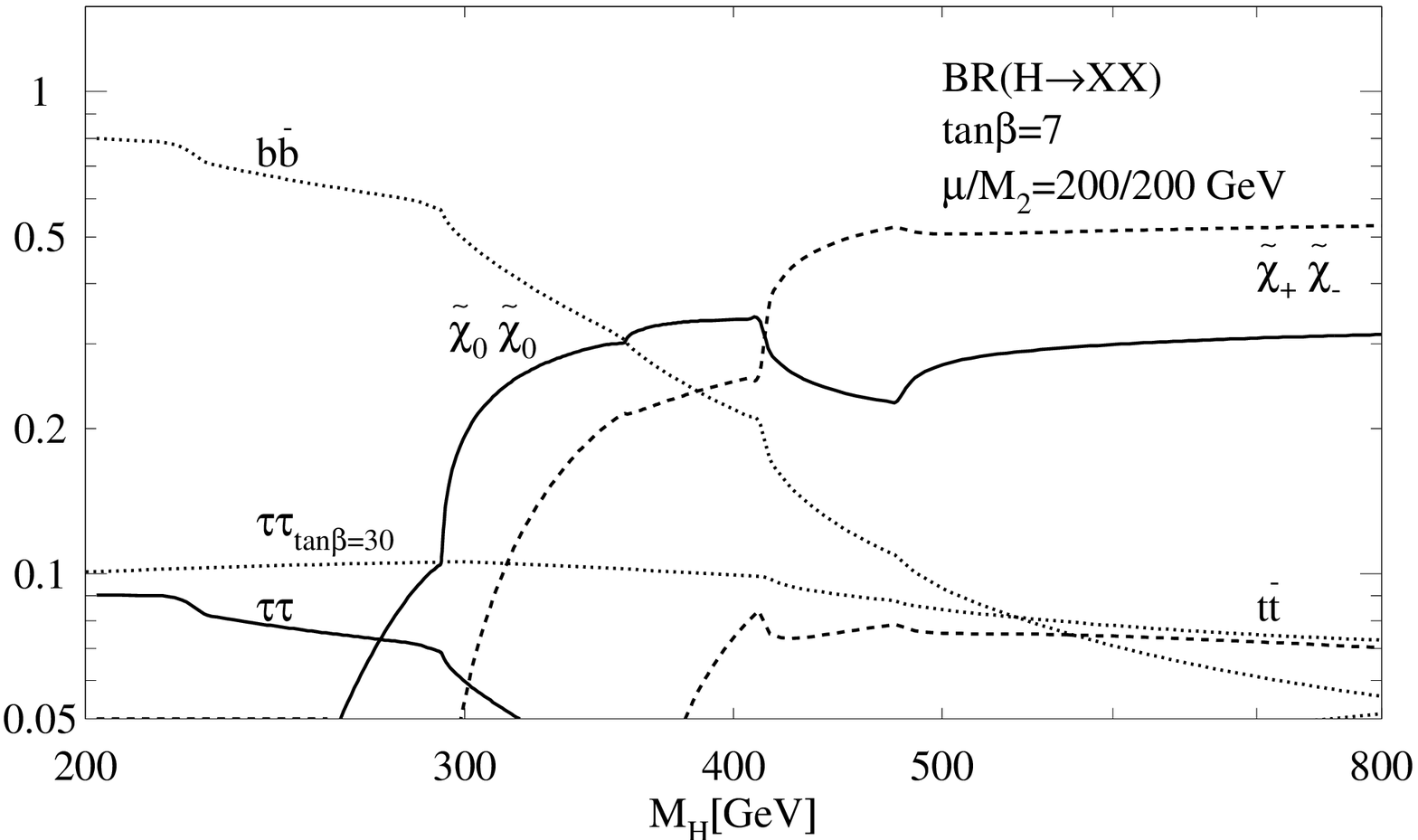,width=13cm}}
\put(-2.2,-6){\epsfig{figure=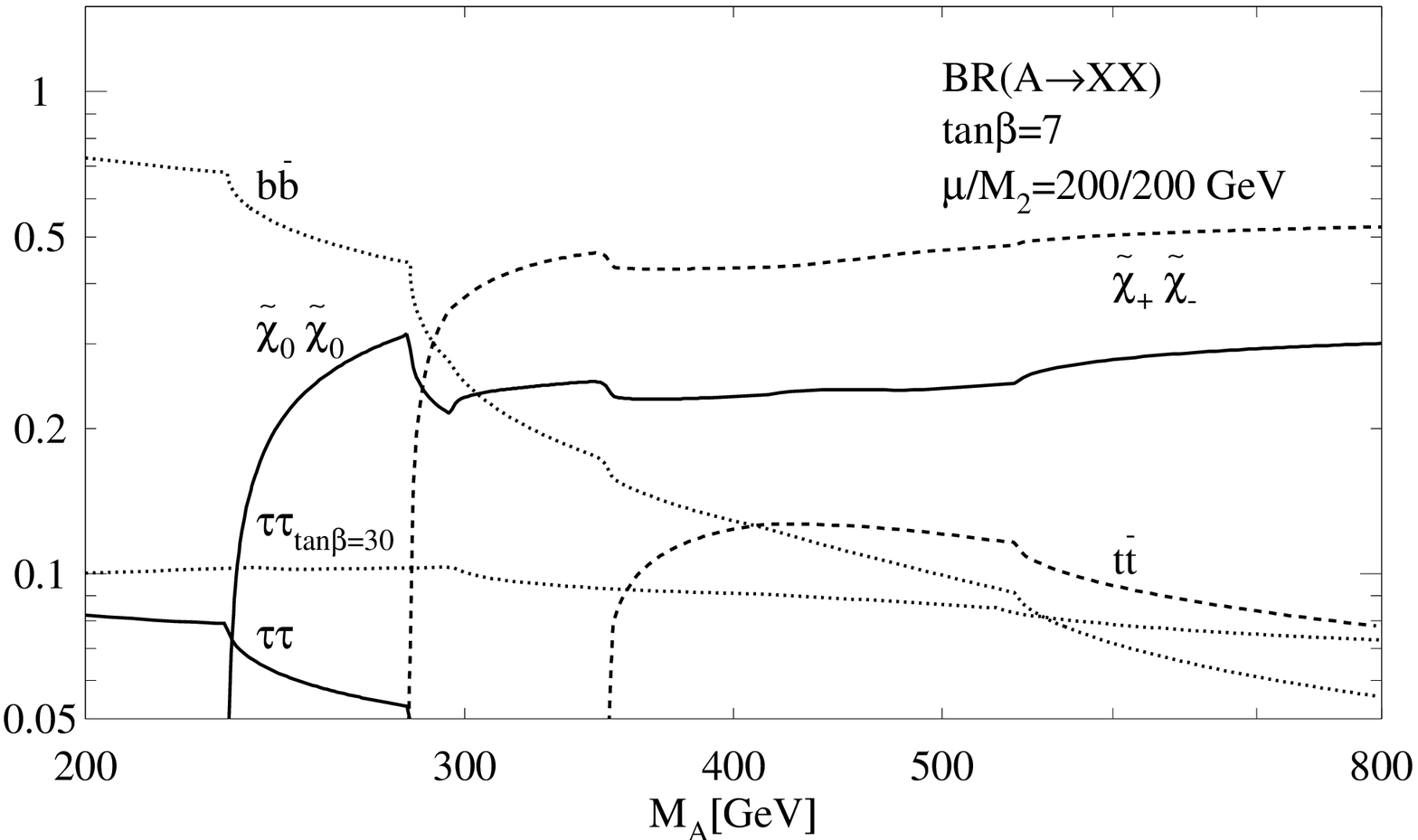,width=13cm}}
\end{picture}
\end{center}\vspace{6.5cm}
\caption{\textit{The branching ratios of the heavy MSSM Higgs bosons $H$ and $A$ into charginos and neutralinos, bottom and top quarks and $\tau^+\tau^-$. The visible SUSY final states have been summed up, {\it i.e.} $\tilde{\chi}^+\tilde{\chi}^-=\sum_{i,j=1,2} \tilde{\chi}^+_i\tilde{\chi}^-_j$ and $\tilde{\chi}^0\tilde{\chi}^0=\sum_{i,j=1..4} \tilde{\chi}^0_i\tilde{\chi}^0_j - \tilde{\chi}^0_1\tilde{\chi}^0_1$, assuming that the lightest neutralino $\tilde{\chi}^0_1$ is the LSP which cannot be observed. The SUSY parameters $\mu$ and $M_2$ have been chosen equal to 200~GeV and $\tan\beta=7$. (The branching fraction $BR(H/A\to \tau^+\tau^-)$ is also shown for $\tan\beta=30$.) The SUSY breaking squark and slepton masses have been set equal to 1~TeV and the SUSY breaking trilinear couplings are zero.}}
\label{fig:bra22}
\end{figure}
According to the present analyses neutral heavy MSSM Higgs bosons, $A$
and $H$, with masses $M_{A/H}\,\gesim\, 200$~GeV may escape detection
at the LHC for moderate values of $\tan\beta$. In addition, in
$e^+e^-$ collisions at linear colliders the discovery of all MSSM
Higgs particles with mass $M$ independently of $\tan\beta$ requires a
c.m.~energy $\sqrt{s}\,\gesim\, 2M$ \cite{accomando}, hence
$\sqrt{s}\,\gesim\, 1.6$~TeV for $M_{A/H}=800$~GeV. Since the photons
produced from Compton scattered laser light off energetic
electron/positron beams have approximately the same energy as the
initial beams and high luminosity \cite{ggvgl}, Higgs production in
$\gamma\gamma$ collisions provides an additional powerful mechanism
that can be used for the Higgs boson search.\s

In this section the heavy MSSM Higgs production via $\gamma\gamma$
fusion at a high-luminosity $e^+e^-$ linear collider will be analysed.
The branching ratios of the $A$ and $H$ decay modes that will be
exploited for this task are illustrated in Fig.~\ref{fig:bra22} as a
function of the corresponding Higgs mass in the range 200...800~GeV.
The higgsino mixing parameter is chosen $\mu= 200$~GeV, the universal
gaugino mass parameter $M_2 = 200$~GeV and $\tan\beta=7$. The SUSY
breaking sfermion masses are chosen as 1~TeV and the SUSY breaking
trilinear couplings are set equal to 0. This will be the SUSY
parameter set used in the following if not stated otherwise. In the
lower mass range $H$ and $A$ preferentially decay into $b\bar{b}$
whereas in the upper mass range, beyond the corresponding kinematic
thresholds, the chargino decays dominate followed by the neutralino
channels. Above the $t\bar{t}$ threshold the branching ratio into top
quarks amounts up to $\sim 13\%$. At the lower end of the mass range the
$\tau^+\tau^-$ branching ratio may reach $\sim 10$\%. \s

\noindent {\bf $b\bar{b}$ production} \newline
\noindent
The signal process $\gamma\gamma \to \mathrm{Higgs} \to b\bar{b}$ is
confronted with the background reaction $\gamma\gamma \to b\bar{b}$.
In order to suppress the background with respect to the signal it is
helpful to use polarized photons. The spin-0 Higgs particles are only
produced from an initial state with the third component of the angular
momentum $J_z=0$. The $J_z=0$ background process, however, is
suppressed by $m_b^2/s$ in leading order \citer{supp0,largel2}. The
channel $\gamma\gamma \to b\bar{b}$ hence provides an outstanding
signature for the heavy MSSM Higgs boson search.  The signal process
proceeding via massive $s$-channel Higgs bosons dominantly develops a
2--jet topology in the final state whereas the next-to-leading order
(NLO) background process favours the 3-jet topology due to the leading
order (LO) suppression $m_b^2/s$ for $J_z=0$. In the case of the light
$b$-quarks the subsequent analysis will therefore investigate the
two-jet final state. Next-to-leading order calculations have shown
that the $m_b^2/s$ suppression is removed by gluon brems-strahlung
\citer{supp1,largel2}. The significance of the 2-jet signal topology
can therefore be spoiled by the radiatively corrected background
process if partons are collinear or one of the partons is rather soft
thus faking a 2-jet final state.  This background has to be suppressed
by applying stringent cuts.  The two-jet configuration due to soft or
collinear partons involves large double logarithms
\citer{supp2,largel2}. In order to properly take into account higher
order corrections the logarithms have to be resummed leading to the
well-known Sudakov form factor \cite{sudak} denoted by ${\cal F}_g$
which corresponds to the emission of soft real and the exchange of
soft virtual gluons and a non-Sudakov form factor ${\cal F}_q$ related
to soft virtual quark contributions \cite{largel1,largel2}.\s

The subsequent analysis will investigate heavy MSSM Higgs production in
\beq
e^+ e^- \to \gamma\gamma \to A/H \to b\bar{b}
\eeq  
for polarized electron/positron and photon beams. The discovery reach
in the Higgs mass range $M_{A/H} = 200...800$~GeV for moderate
$\tan\beta=7$ is analysed by including the background and the
interference process in the two-jet topology. Radiative corrections
are tho\-rough\-ly incorporated considering the QCD corrections to
signal \citer{graudenz,spirdjou}, background \cite{supp2,rem} and also
to the interference process, calculated in this thesis for the first
time. The leading higher-order QCD corrections which are sizeable in
the two-jet configuration are considered through the resummation of
the Sudakov and non-Sudakov double logarithms
\cite{largel1,largel2}.\s

\noindent {\bf The $t\bar{t}$ and $\tau^+\tau^-$ channel}
\newline \noindent
As further options Higgs detection in 
\beq
e^+ e^- \to \gamma\gamma \to A/H \to t\bar{t} \quad \mathrm{and} 
\quad \tau^+\tau^-
\eeq
is investigated. In contrast to the top-quark final state the
$\tau^+\tau^-$ production does not acquire final state QCD
corrections. The $\tau^+\tau^-$ process will be analysed for the two
$\tan\beta$ values 7 and 30. Since the Yukawa couplings of $A$ and $H$ to
$\tau^+\tau^-$ increase with rising $\tan\beta$, thus enhancing the
corresponding branching fractions, the signal cross section will grow
with $\tan\beta$ while the background process is independent of
$\tan\beta$. \s

\noindent {\bf Chargino and neutralino production}
\newline \noindent
Since the branching ratios of the chargino and neutralino channels
are rather large the processes
\beq
e^+ e^- \to \gamma\gamma &\to& A/H \to  
\tilde{\chi}^+_i \tilde{\chi}^-_j \;, \qquad (i,j=1,2) \non\\
e^+ e^- \to \gamma\gamma &\to& A/H \to  
\tilde{\chi}^0_i \tilde{\chi}^0_j \;, \,\,\qquad (i,j=1,...,4; \; 
\mathrm{not} \; i=j=1)
\eeq
may provide additional interesting detection channels. The SUSY
particle pair production cross sections are summed over all possible
final states excluding however the production of $\tilde{\chi}_1^0
\tilde{\chi}_1^0$ where $\tilde{\chi}_1^0$ denotes the lightest
neutralino. This is due to the assumption that $\tilde{\chi}_1^0$ is
the lightest supersymmetric particle (LSP). Because of R-parity
conservation it cannot decay and the $\tilde{\chi}_1^0
\tilde{\chi}_1^0$ state will escape detection. As in the case of
lepton production the SUSY processes do not involve final state QCD
corrections.

\subsection{The quark final states \label{quarkfinal}}

\begin{figure}[t]
\begin{center}
\begin{picture}(360,100)(0,0)
\Text(-20,90)[]{$\gamma$}
\Text(-20,50)[]{$\gamma$}
\Photon(-15,90)(25,90){2}{5}
\Photon(-15,50)(25,50){2}{5}
\Line(25,50)(25,90)
\Line(25,90)(45,70)
\Line(45,70)(25,50)
\Text(62.5,75)[]{$h,H,A$}
\DashLine(45,70)(80,70){2}
\ArrowLine(100,50)(80,70)
\ArrowLine(80,70)(100,90)
\Text(105,90)[]{$q$}
\Text(105,50)[]{$\bar{q}$}
\Photon(-20,30)(10,30){2}{5}
\Photon(-20,-10)(10,-10){2}{5}
\Line(10,-10)(10,30)
\Line(10,30)(30,10)
\Line(30,10)(10,-10)
\DashLine(30,10)(50,10){2}
\ArrowLine(70,-10)(50,10)
\ArrowLine(50,10)(70,30)
\GlueArc(10,10)(9,270,450){2}{4}
\Photon(80,30)(110,30){2}{5}
\Photon(80,-10)(110,-10){2}{5}
\Line(110,-10)(110,30)
\Line(110,30)(130,10)
\Line(130,10)(110,-10)
\DashLine(130,10)(150,10){2}
\ArrowLine(170,-10)(150,10)
\ArrowLine(150,10)(170,30)
\Gluon(120,20)(110,10){2}{2}
\Photon(180,30)(210,30){2}{5}
\Photon(180,-10)(210,-10){2}{5}
\Line(210,-10)(210,30)
\Line(210,30)(230,10)
\Line(230,10)(210,-10)
\DashLine(230,10)(250,10){2}
\ArrowLine(270,-10)(250,10)
\ArrowLine(250,10)(270,30)
\Gluon(260,20)(260,0){2}{2}
\Photon(280,30)(310,30){2}{5}
\Photon(280,-10)(310,-10){2}{5}
\Line(310,-10)(310,30)
\Line(310,30)(330,10)
\Line(330,10)(310,-10)
\DashLine(330,10)(350,10){2}
\ArrowLine(370,-10)(350,10)
\ArrowLine(350,10)(370,30)
\Gluon(360,20)(380,20){2}{2}
\end{picture}
\end{center}
\caption{{\it Generic diagrams contributing to the signal process $\gamma\gamma \to h,H,A \to q\bar{q}$ at leading and next-to-leading order.}}
\label{diagsig}
\end{figure}
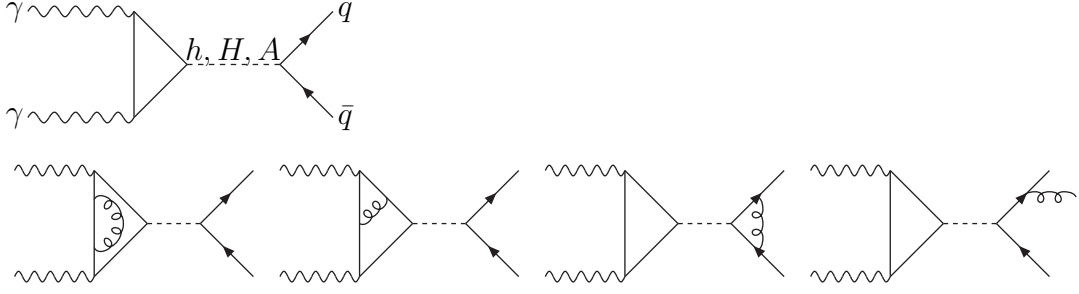

\noindent {\bf 1.) Signal process} \newline
\noindent 
\hspace{-0.15cm} The signal process $\gamma\gamma \to q\bar{q}$
proceeds via the MSSM Higgs bosons $h,H$ and $A$. Fig.~\ref{diagsig}
shows the generic diagrams contributing to the process at LO and at NLO. 
The two photons are coupled to the Higgs bosons
through triangle loops including heavy charged fermions, charginos and
in the case of CP-even Higgs bosons also $W$ bosons and charged Higgs
bosons as well as sfermions. $A$ does neither couple to gauge bosons
at tree level, cf.~Tab.~\ref{hcoup}, nor to charged Higgs particles,
cf.~Tab.~\ref{mssmcoup}. Furthermore $A$ flips the sfermion helicity
whereas it is conserved in the coupling to the photon so that sfermion
loops do not contribute to the $\gamma\gamma A$ coupling. The
effective $\gamma\gamma h/H$ coupling for incoming photons $k_1^\mu$,
$k_2^\nu$ is given by \cite{effcoup,dawson}
\beq
\frac{i\alpha\sqrt{\sqrt{2}G_F}}{2\pi} {\cal M}_\varphi [k_1^\nu k_2^\mu -k_1
k_2 g_{\mu\nu}] \;, \quad \varphi=h,H
\eeq
with the form factor \cite{spirahabil}
\beq
{\cal M}_\varphi &=& 
\sum_f N_c^f Q_f^2 g_{\varphi ff} A_f^\varphi(\tau_f) +  
g_{\varphi WW} A_W^\varphi(\tau_W) 
+ g_{\varphi H^+H^-} A_{H\pm}^\varphi 
(\tau_{H^\pm}) \non\\ 
&+& \sum_{\tilde{\chi}^\pm} 
\frac{2M_W}{M_{\tilde{\chi}_i^\pm}}
g_{\varphi \tilde{\chi}_i^+ \tilde{\chi}_i^-} 
A_{\tilde{\chi}^\pm}^\varphi(\tau_{\tilde{\chi}^\pm}) 
+ \sum_{\tilde{f}} N_c^f Q_{\tilde{f}}^2 g_{\varphi\tilde{f} 
\tilde{f}} A_{\tilde{f}}^\varphi (\tau_{\tilde{f}})
\eeq
and
\beq
\begin{array}{lll}
A_{f/\tilde{\chi}^\pm}^\varphi (\tau) &=& 2\tau [1+(1-\tau)f(\tau)] \\
A_{H^\pm/\tilde{f}}^\varphi (\tau) &=& -\tau[1-\tau f(\tau)] \\
A_W^\varphi (\tau) &=& -[2+3\tau+3\tau (2-\tau ) f(\tau)] 
\end{array}
\eeq
The parameter $\tau_x$ is defined as 
$\tau_x = 4M_x^2/M^2$ and the function $f(\tau)$ is given by 
\beq
f(\tau) = \left\{
\begin{array}{cc}
\arcsin^2 \frac{1}{\sqrt{\tau}} & \tau \ge 1 \\
-\displaystyle{\frac{1}{4}} 
\left[ \log \displaystyle{\frac{1+\sqrt{1-\tau}}{1-\sqrt{1-\tau}}}
 -i \pi 
\right]^2 & \tau < 1 
\end{array}
\right.
\eeq
In the cross sections given below, $M^2$ in the parameter $\tau$ is
equal to the c.m.~energy of the process. The couplings $g_{\varphi
  xx}$ are given in Tabs.~\ref{hcoup} and \ref{mssmcoup}. $N_c^f$
denotes the colour factor of the corresponding fermion $f$ and
$Q_{f(\tilde{f})}$ the electric fermion (sfermion) charges in units of
the positron charge. The effective coupling $\gamma\gamma A$ can be
cast into the form \cite{effcoup,dawson}
\beq
\frac{i\alpha\sqrt{\sqrt{2}G_F}}{2\pi} {\cal M}_A \epsilon_{\mu\nu\alpha\beta} 
k_1^\alpha k_2^\beta 
\eeq
where the form factor ${\cal M}_A$ is given by \cite{spirahabil}
\beq
{\cal M}_A = \sum_f N_c^f Q_f^2 g_{Aff} A_f^A (\tau_f) + 
\sum_{\tilde{\chi}^\pm} 
\frac{2M_W}{M_{\tilde{\chi}_i^\pm}} g_{A\tilde{\chi}_i^+ \tilde{\chi}_i^-} 
A^A_{\tilde{\chi}^\pm} (\tau_{\tilde{\chi}^\pm}) 
\eeq
with
\beq
\begin{array}{lll}
A_{f/\tilde{\chi}^\pm}^A (\tau) &=& 2\tau f(\tau) \\
\end{array}
\eeq
\noindent After integration over $|\cos\theta|<c$, where $\theta$ denotes the
scattering angle between the quark and the direction of the incoming
photons in the c.m.~frame, the polarized LO signal cross sections,
labeled by the index $S$, are given by
\beq
\sigma^{++/--}_{S,q\bar{q}} &=& 
\frac{N_c G_F^2 \alpha^2\beta m_q^2 c}{64\pi^3} \non\\
&\times& \left[
g_{hqq}^2\beta^2|{\cal G}_h|^2 + 
g_{Hqq}^2\beta^2|{\cal G}_H|^2 +
g_{Aqq}^2|{\cal G}_A|^2 +
2 g_{hqq}g_{Hqq}\beta^2{\mathrm Re}({\cal G}_h {\cal G}_H^*)
\right] \non\\
\sigma^{+-/-+}_{S,q\bar{q}} &=& 0 
\label{signsig}
\eeq
The generalized form factor ${\cal G}_\Phi$ ($\Phi=h,H,A$) is defined as
\beq
{\cal G}_\Phi \equiv \frac{{\cal M}_\Phi}{1-\mu_\Phi+i\gamma_\Phi}
\eeq
with the reduced mass $\mu_\Phi =M_\Phi^2/s$ and the reduced width
$\gamma_\Phi = [M_\Phi \Gamma_\Phi]/s$ where $\Gamma_\Phi$ denotes the
total width. The velocity $\beta$ is given by $\beta = (1-4
m_q^2/s)^{1/2}$.  Obviously, since the Higgs bosons carry spin 0, the
$J_z=\pm 2$ cross sections vanish. \s
\begin{table}[t]
\begin{center}$
\begin{array}{llll}\hline
\Phi & & g_{\Phi H^+ H^-} & g_{\Phi \tilde{\chi}_i^+ \tilde{\chi}_j^-} 
\\[0.1cm] \hline \\[-0.2cm]
SM & H & 0 & 0 \\[0.1cm] \hline \\[-0.2cm] 
MSSM &h& \frac{M_W^2}{M_{H^\pm}^2}\left[\sin(\beta-\alpha)+
\frac{\cos 2\beta \sin(\beta+\alpha)}{2\cos^2\theta_W}\right] & 
S_{ij}\cos\alpha-Q_{ij}\sin\alpha \\
&H& \frac{M_W^2}{M_{H^\pm}^2}\left[\cos(\beta-\alpha)-
\frac{\cos 2\beta \cos(\beta+\alpha)}{2\cos^2\theta_W}\right]  & 
S_{ij}\sin\alpha+Q_{ij}\cos\alpha \\
&A& 0 &  
-S_{ij}\cos\beta-Q_{ij}\sin\beta
\\[0.1cm] \hline
\end{array}
$\end{center}
\begin{center}$
\newline
\begin{array}{llll}\hline
\Phi & & g_{\Phi \tilde{f}_{L/R} \tilde{f}_{L/R}} &  
g_{\Phi \tilde{\chi}_i^0 \tilde{\chi}_j^0}\\[0.1cm] \hline \\[-0.2cm]
SM & H & 0 & 0 \\[0.1cm] \hline \\[-0.2cm] 
MSSM &h& \frac{M_f^2}{M_{\tilde{f}}^2}g_{hff}\mp 
\frac{M_Z^2}{M_{\tilde{f}}^2}(I_3^f-Q_f\sin^2\theta_W)\sin(\beta+\alpha) 
& -Q^{''}_{ij}\sin\alpha - S^{''}_{ij}\cos\alpha \\
&H&  \frac{M_f^2}{M_{\tilde{f}}^2}g_{Hff}\pm 
\frac{M_Z^2}{M_{\tilde{f}}^2}(I_3^f-Q_f\sin^2\theta_W)\cos(\beta+\alpha) 
& Q^{''}_{ij}\cos\alpha - S^{''}_{ij}\sin\alpha \\
&A& 0 & -Q^{''}_{ij}\sin\beta + S^{''}_{ij}\cos\beta \\[0.1cm] \hline
\end{array}
$\end{center}
\caption{{\it MSSM Higgs boson couplings to charged Higgs bosons, charginos, sfermions and neutralinos. $Q_{ij}$ and $S_{ij}$ ($i/j=1,2$) are related to the mixing angles in the chargino sector, $Q_{ij}^{''}$ and $S_{ij}^{''}$ ($i/j=1,..,4$) to those in the neutralino sector, cf.~Refs.~\cite{gunion,susy,susyprod}. $I_3^f$ denotes the third isospin component.}}
\label{mssmcoup}
\end{table}

The generic diagrams contributing to the NLO processes,
cf.~Fig.~\ref{diagsig}, involve the QCD corrections to the quark and
squark loops in the $\gamma\gamma\Phi$ form factor due to virtual gluon
exchange.  For the $t$ and $b$ loops the QCD corrections are known for
finite quark and Higgs masses \cite{nvoll,vollqcd} whereas for squark
loops they have only been calculated in the large squark mass limit
\cite{squark}. In order to improve the perturbative behaviour of the
quark loop contributions, the QCD corrections are expressed in terms
of the running on-shell masses $m_{on}(\mu)$. The scale $\mu$ has been
identified with $\mu = \sqrt{s}/2$. The running mass is normalized to
the pole mass $m_q$ via
\beq
m_{on}(\mu=m_q) = m_q
\eeq
This definition of the running mass allows to properly take into
account threshold effects at $\sqrt{s} = 2m_q$. In contrast, the
definition of the running $\overline{\mathrm{MS}}$ mass
\beq
m_q = \overline{m}_q (\mu) \left[
1+\frac{\alpha_s}{\pi}\left(\frac{4}{3}+\ln\frac{\mu^2}{m_q^2}
\right)\right] 
\eeq
leads to an artificial displacement between the running mass and the
pole mass at $\mu = m_q$:
\beq
m_{on}(\mu) = \overline{m}_q (\mu) \left[1+\frac{4}{3}\frac{\alpha_s(m_q)}{\pi}
\right]
\eeq
The QCD corrections to the Higgs decay into $q\bar{q}$ involving
virtual gluon exchange and real gluon radiation have been calculated,
too, \cite{qcdcorr,spirdjou}.  The arising large logarithmic
contributions are absorbed in the running $\overline{\mathrm{MS}}$
mass $\overline{m}_q(\mu)$ at the scale $\mu=\sqrt{s}$. The full
MSSM electroweak and SUSY-QCD corrections are known
\cite{susycorrec}. Since they are moderate they have not been
implemented.  Only the gluino corrections can become large if
$\tan\beta$ and the higgsino mixing parameter $\mu$ are large
\cite{susycorrec}. \s

\begin{figure}
\begin{center}
\begin{picture}(360,100)(0,0)
\Photon(0,90)(40,90){2}{5}
\Photon(0,50)(40,50){2}{5}
\ArrowLine(40,50)(40,90)
\ArrowLine(40,90)(80,90)
\ArrowLine(80,50)(40,50)
\Photon(100,90)(140,50){2}{5}
\Photon(100,50)(140,90){2}{5}
\ArrowLine(140,50)(140,90)
\ArrowLine(140,90)(180,90)
\ArrowLine(180,50)(140,50)
\Photon(0,30)(40,30){2}{5}
\Photon(0,-10)(40,-10){2}{5}
\GlueArc(40,10)(10,270,450){2}{4}
\ArrowLine(40,-10)(40,30)
\ArrowLine(40,30)(80,30)
\ArrowLine(80,-10)(40,-10)
\Photon(100,30)(140,30){2}{5}
\Photon(100,-10)(140,-10){2}{5}
\GlueArc(140,30)(10,270,360){2}{2}
\ArrowLine(140,-10)(140,30)
\ArrowLine(140,30)(180,30)
\ArrowLine(180,-10)(140,-10)
\Photon(190,30)(230,30){2}{5}
\Photon(190,-10)(230,-10){2}{5}
\Gluon(255,30)(255,-10){2}{4}
\ArrowLine(230,-10)(230,30)
\ArrowLine(230,30)(270,30)
\ArrowLine(270,-10)(230,-10)
\Photon(280,30)(320,30){2}{5}
\Photon(280,-10)(320,-10){2}{5}
\Gluon(320,10)(360,10){2}{4}
\ArrowLine(320,-10)(320,30)
\ArrowLine(320,30)(360,30)
\ArrowLine(360,-10)(320,-10)
\end{picture}
\end{center}
\caption{{\it Generic Diagrams contributing to the background process $\gamma\gamma \to q\bar{q}$ at leading and next-to-leading order.}}
\label{diagback}
\end{figure}
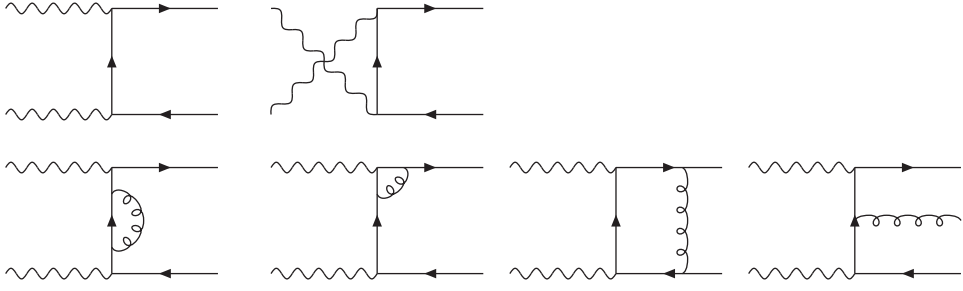

\noindent {\bf 2.) Background process}\newline
\noindent The generic diagrams contributing to the background process 
$\gamma\gamma \to q\bar{q}$ at Born level and next-to-leading order
are depicted in Fig.~\ref{diagback}. At leading order the polarized 
background cross sections after integration over $|\cos\theta|<c$ read
\beq
\sigma^{++/--}_{B,q\bar{q}} &=& 
\frac{N_c\alpha^2 Q_q^4 2\pi}{s(1-c^2\beta^2)} (1-\beta^4) \left[
2\beta c + (1-c^2\beta^2)
\ln \frac{1+\beta c}{1-\beta c} \right] \label{backsig}
\\
\sigma^{+-/-+}_{B,q\bar{q}} &=& 
\frac{N_c\alpha^2 Q_q^4 2\pi}{s(1-c^2\beta^2)} \left[
-2c\beta^5 + 8c\beta^3 + 4c^3 \beta^3 - 10\beta c 
+ (5-\beta^4)(1-c^2\beta^2) \ln \frac{1+\beta c}{1-\beta c}\right] \non
\eeq
$Q_q$ denotes the electric charge of the quark in units of the positron charge and $\alpha$ the
electromagnetic coupling. The NLO 2+3-jet cross section is built up by the
virtual corrections involving self-energy contributions, vertex
corrections and box diagrams and by the brems-strahlung processes,
cf.~Fig.~\ref{diagback}. The virtual cross sections involve UV and IR
divergences. The UV singularities are absorbed by the renormalization
of the quark mass and the $\gamma q \bar{q}$ vertex.  The
brems-strahlung process encounters IR singularities which cancel
those arising in the virtual cross section. The results for polarized
photons are given in Refs.~\cite{supp2,rem}. \s

\noindent {\bf 3.) Interference contribution}\newline
\noindent The polarized interference cross sections between signal and 
background process in leading order
\beq
\gamma(k_1) + \gamma(k_2) \to q(p_1) + \bar{q} (p_2)
\eeq
are given by 
\beq
\sigma^{++/--}_{I,q\bar{q}} &=& -\frac{N_c G_F \alpha^2 Q_q^2\beta(1-\beta^2)}{4\sqrt{2}\pi} \ln \frac{1+\beta c}{1-\beta c} \non\\
&\times & \left[ g_{hqq}\beta\, {\mathrm Re}
\left( {\cal G}_h \right)
+ g_{Hqq}\beta\, {\mathrm Re}
\left( {\cal G}_H \right) 
 - \frac{g_{Aqq}}{\beta}\, {\mathrm Re}
\left( {\cal G}_A \right) 
\right] \non\\
\sigma^{+-/-+}_{I,q\bar{q}} &=& 0
\label{intsig}
\eeq
after integrating out $\cos\theta$ in the interval [$-c,c$].\s

\noindent {\bf Virtual corrections}\newline
\noindent
The virtual interference cross section in next-to-leading order is
obtained from the interference of the virtual NLO background and the
LO signal diagrams and vice versa. The UV and IR singularities which
turn up in the loop integrals are treated in dimensional
regularization, {\it i.e.} all cross sections are calculated in
$n=4-2\epsilon$ space-time dimensions. The matrix $\gamma_5$ included
in the pseudoscalar Higgs-Yukawa coupling has been treated in the
't~Hooft/Veltman prescription \cite{hove} in $n$ dimensions. After
averaging over the photon spins and evaluating the $n$-dimensional
two-particle phase space the unpolarized virtual NLO cross section
reads
\beq
\left( s^2 \frac{d^2\sigma^{(1)}}{dt_1du_1} \right)^V &=& 
\frac{1}{4(1-\epsilon)^2}\frac{\pi S_\epsilon}{\Gamma(1-\epsilon)}
\left( \frac{t_1 u_1 -sm^2}{\mu^2 s} \right)^{-\epsilon} \delta(s+t_1+u_1)
\non \\
& & \sum 2{\mathrm Re} [{\cal M}^{LO}_{S}{\cal M}_{B}^{V*}+ 
{\cal M}^V_{S}{\cal M}_{B}^{LO*} ]
\eeq
where
\beq
s &\equiv&  (k_1 + k_2)^2 \non\\
t_1 &\equiv& t-m_q^2 = (k_1- p_1)^2 - m_q^2 \\
u_1 &\equiv& u-m_q^2 = (k_1- p_2)^2 - m_q^2 \non
\eeq
and ${\cal M}^{LO}$ denotes the Born amplitude, ${\cal M}^V$ the
virtual amplitude. The number of photon spin degrees of freedom in $n$
dimensions is $n-2$ so that averaging over the spin leads to
$(n-2)^{-2}=(1-\epsilon)^{-2}/4$. The mass parameter $\mu$ has been
introduced because the coupling constant is not dimensionless any more
in $n$ dimensions. The spacial angle constant \\ $S_\epsilon =
(4\pi)^{-2+\epsilon}$ and the remaining factors result from the
calculation of the $n$-dimensional 2-particle phase space. Since only
the initial photon state $J_z= 0$ couples to the Higgs boson the
polarized interference cross sections for $J_z = \pm 2$ are zero. Due
to CP-invariance the polarized $J_z=0$ interference cross section is
thus given by the unpolarized result:
\beq
\sigma^{++}_{I} = \sigma^{--}_{I} = 2 \sigma^{unpol}_{I}
\eeq
It is therefore sufficient to calculate the unpolarized interference
cross section in order to get the polarized result. The virtual loop
integrals are reduced to scalar integrals by means of the reduction
procedure outlined in Ref.~\cite{passa}. The scalar integrals are
listed in the Appendix. The virtual corrections encounter poles
$\epsilon^{-1}$ due to UV and IR singularities but no collinear
divergences since the quarks are massive. The UV divergences are
absorbed by the renormalization of the quark mass, the quark wave
function, the Yukawa coupling and the quark-photon coupling. The
renormalization conditions for the quark self-energy $\Sigma(p)$ in
the on-shell scheme
\beq
\Sigma_R (\slash{p}=m) = 0 \quad \mathrm{and} \quad
\left. \frac{\partial\Sigma_R (p)}{\partial\slash{p}} 
\right|_{\slash{p}=m} = 0 
\eeq 
result in the renormalization constants of the quark mass and the
quark field
\beq
Z_m = Z_2 = 1-\frac{3}{4}\frac{\alpha_s}{\pi} C_F C_\epsilon
 \left(\frac{1}{\epsilon} + 
\frac{4}{3} \right)
\eeq
The colour factor is given by $C_F= (N_c^2-1)/(2N_c)$. The constant
$C_\epsilon$ summarizes the terms that typically arise in dimensional
regularization
\beq
C_\epsilon = \Gamma(1+\epsilon) \left(
\frac{4\pi\mu^2}{m_b^2}\right)^\epsilon 
\eeq
The Ward identity
fixes the renormalization constant for the quark-photon vertex:
$Z_1=Z_2$. Since in NLO QCD processes the Higgs fields and the vacuum
expectation values are not renormalized the renormalization constant
for the scalar Higgs-Yukawa couplings is given by
\beq
Z_{\varphi f\bar{f}} = Z_m Z_2\;, \quad \varphi = h,H
\eeq
The pseudoscalar Higgs-Yukawa coupling includes the matrix $\gamma_5$.
The 't~Hooft/Veltman prescription, applied for the treatment of
$\gamma_5$ in $n$ dimensions, breaks Lorentz and chiral invariance.
This requires the introduction of a counterterm in order to restore
the symmetries. The renormalization constant for the A coupling to
fermions reads
\beq
Z_{Af\bar{f}} = Z_m Z_2 Z_5
\eeq
with the constant 
\beq
Z_5 = 1 - 2 C_F \frac{\alpha_s}{\pi}
\eeq
In order to absorb large logarithms in the QCD corrections the Yukawa
coupling is defined in terms of the running $\overline{\mathrm{MS}}$
quark mass. The remaining IR singularities are canceled by the
contribution from the soft gluons.\s

\noindent {\bf Gluon brems-strahlung}\newline
\noindent
The next-to-leading order correction due to gluon brems-strahlung
\beq
\gamma(k_1) + \gamma(k_2) \to q(p_1) + \bar{q}(p_2) + g(k)
\label{twotothree}
\eeq
is built up by the interference of the gluon radiation diagrams of the
signal and background process. It encounters IR singularities
resulting from soft gluon radiation. For the calculation of the ${\cal
  O}(\alpha_S)$ real correction to the interference part, the cross
section $\sigma_{I}^{(1)R}$ is divided into an IR finite and IR singular part
\beq
\sigma_{I}^{(1)R} = [\sigma_{I}^{(1)R}-\sigma_E] + \sigma_E
\eeq
The cross section $\sigma_E$ contains all IR singularities so that the
sum in brackets is manifestly IR finite. $\sigma_E$ can be obtained in
the Eikonal approximation where the soft gluon momentum is neglected
with respect to the quark momenta. In this limit the real singular
amplitude factorizes from the Born amplitude,
\beq
{\cal M}_{sing}^R = {\cal M}_{LO} T^a_{ij} g_s \left\{
\frac{p_1\epsilon^*(k)}{p_1 k} -
\frac{p_2\epsilon^*(k)}{p_2 k} \right\}
\eeq
$T^a_{ij}$ denote the generators of the colour group $SU(3)$ and
$\epsilon(k)$ the gluon polarization vector. Summing over the gluon
polarizations the matrix element squared reads
\beq
\sum | {\cal M}_{sing}^R |^2 = 4 C_F \pi \alpha_s \left\{
\frac{2p_1 p_2}{(p_1 k)(p_2 k)} - \frac{m_q^2}{(p_1 k)^2} - 
\frac{m_q^2}{(p_2 k)^2} \right\} \sum |{\cal M}_{LO}|^2
\label{eik}
\eeq
The differential three particle phase space $dPS_3(P\equiv k_1+k_2;
p_1,p_2,k)$ which turns up in the calculation of the $2\to3$ process
(\ref{twotothree}) can be decomposed into two two-particle phase
spaces:
\beq
dPS_3 (P;p_1,p_2,k) &=&  \frac{dQ^2}{2\pi} dPS_2(P;k,Q) dPS_2 (Q;p_1,p_2)
\label{decomp}
\eeq
The two differential phase space factors $dPS_2$ correspond to the
processes where the two photons with total momentum $P$ decay into the
gluon and an intermediate state $X$ with momentum $Q$ which
subsequently decays into the quark anti-quark pair. The invariant mass
of the state $X$ is given by $Q^2 = z s$ ($4m_q^2/s< z< 1$). The upper
limit of $z$ corresponds to the IR region where the gluon momentum $k$
approaches 0. The lower limit is given by the fact that the invariant
mass of $X$ must be large enough to produce two quarks. \s

In the calculation of $\sigma_E$ the decomposition (\ref{decomp})
allows to factorize the soft gluon kinematics from the remaining
kinematics which become equal to the kinematics of the Born process
$\gamma \gamma \to q\bar{q}$ in the IR region. Since the singularities
originate only from the soft gluon momentum, the variable $z$ is set
equal to 1 in the quark momenta in the IR limit so that the quark
velocity is given by $\beta = (1-4m_q^2/s)^{1/2}$. The differential
Eikonal cross section can then be written as
\beq
d\sigma_E = C_E d\sigma_{LO}
\eeq
where $C_E$ contains the IR singularities
\beq
C_E = \frac{dQ^2 dPS_2 (P;k,Q)}{2\pi} 
\;4 C_F \pi \alpha_s \left\{
\frac{2p_1 p_2}{(p_1 k)(p_2 k)} - \frac{m_q^2}{(p_1 k)^2} - 
\frac{m_q^2}{(p_2 k)^2} \right\}
\eeq
and $d\sigma_{LO}$ is the LO cross section of the process
$\gamma\gamma \to q\bar{q}$
\beq
d\sigma_{LO} = \frac{1}{8s(1-\epsilon)^2} dPS_2 (P;p_1,p_2) \sum |{\cal M}_{LO} |^2
\eeq
The factor $1/[8s(1-\epsilon)^2]$ results from the flux factor and the
average over the photon spin degrees of freedom. The differential two
particle phase space is given by
\beq
dPS_2 (P;p_1,p_2) = \frac{\beta^{1-2\epsilon}}{32\pi^2}
\left( \frac{16\pi\mu^2}{s} \right)^\epsilon 
\frac{(1-\cos^2\theta)^{-\epsilon}}{\Gamma(1-\epsilon)}\; d\cos\theta\; d\phi
\;, \quad \beta = \sqrt{1-4m_q^2/s}
\eeq 
The calculation of $dPS_2(P;k,Q)$ in $n$ dimensions leads to
\beq
dPS_2(P;k,Q) &=& \frac{dv}{8\pi} \left( \frac{4\pi\mu^2}{s} \right)^\epsilon 
\frac{(1-z)^{1-2\epsilon}}{\Gamma(1-\epsilon)} v^{-\epsilon} (1-v)^{-\epsilon}
\non \\
v &=& \frac{1}{2} (1-\cos\theta_0) \;, \quad z = \frac{Q^2}{s} 
\eeq
where $\theta_0$ denotes the angle between the gluon and
the incoming photons in the c.m.~system. With $dQ^2 = sdz$ we have
\beq
dPS_2(P;k,Q) \frac{dQ^2}{2\pi} 
= \frac{s(1-z)}{16\pi^2} F_\epsilon \;dz \;dv
\eeq
with $F_\epsilon$ containing the $\epsilon$ dependence
\beq
F_\epsilon = (1-z)^{-2\epsilon} \frac{v^{-\epsilon} (1-v)^{-\epsilon}}
{\Gamma (1-\epsilon)} \left( \frac{4\pi\mu^2}{s} \right)^\epsilon 
\eeq
The integration boundaries are given by
\beq
0 \le v \le 1\;, \quad \tau_0 \le z \le 1\;,
\quad \tau_0 = 4 \frac{m_q^2}{s} 
\eeq 
Keeping in mind that the quark momenta are parameterized according to
the Born kinematics the straightforward calculation of the Eikonal
factor $C_E$ results in
\beq
C_E = \bar{C}_\epsilon C_F 
\frac{\alpha_s}{\pi} \left\{ \frac{1}{\epsilon} [1-(1+\beta^2) f(\beta) ]
+ (1+\beta^2) g(\beta) + 2f(\beta) \right\} 
\eeq
where
\beq
\bar{C}_\epsilon = \Gamma(1+\epsilon) 
\left(\frac{4\pi\mu^2}{s(1-\tau_0)^2}\right)^\epsilon
\eeq
and 
\beq
f(\beta) = \frac{1}{2\beta}\ln \frac{1+\beta}{1-\beta} \;,
\quad
g(\beta) = \frac{1}{2\beta}\left\{ Li_2\left(\frac{-2\beta}{1-\beta}\right) 
- Li_2\left(\frac{2\beta}{1+\beta}\right) \right\} 
\eeq
The infrared singularities of $d\sigma_E = C_E d\sigma_{LO}$
cancel those of the virtual corrections. The finite contribution
to the interference radiation cross section
\beq
[d\sigma_{I}^{(1)R} - d\sigma_E] = \frac{1}{8s} dPS_3(P;k_1,k_2,k) \sum 
[|{\cal M}_{I}|^2-|{\cal M}_{sing}|^2]
\eeq
is calculated numerically in 4 dimensions, {\it i.e.}~$\epsilon = 0$.
The momenta in the interference matrix element ${\cal M}_I$ are
parameterized according to the three jet kinematics where $z \neq 1$.
In contrast, in the singular matrix element ${\cal M}_{sing}$,
calculated from Eq.~(\ref{eik}), only the gluon momentum is given in
the three jet kinematics and the quark momenta are parameterized in
the Born kinematics since they have to be treated in the same way as in the
integration of $d\sigma_E$.\s

Adding the ${\cal O}(\alpha_s)$ virtual and brems-strahlung
contributions provides the NLO interference cross section. The total
interference cross section without cuts is also given by the imaginary
part of the Higgs-$\gamma\gamma$ form factor via the optical theorem.
Since the form factor at NLO can be extracted from the literature
\cite{nvoll,vollqcd} it serves as a quantitative cross check for the
obtained result.\s

\noindent {\bf 4.) Two-jet final state} \\
\noindent
For reasons outlined above the photons will be taken polarized and the
bottom-quark production will be analysed in the two-jet final state
topology. A final state will be regarded as two-jet event if the gluon
energy $E_g$ is less than a minimal gluon energy $E_g^{min}$ or if two
of the three jets are collinear. In this analysis $E_g^{min}$ will be
chosen as
\beq
E_g^{min} = 0.1 \sqrt{s_{\gamma\gamma}}
\eeq
where $\sqrt{s_{\gamma\gamma}}$ denotes the c.m.~energy in the
$\gamma\gamma$ fusion process. Two jets will be combined to a single
jet if the angle $\alpha$ between them is less than
\beq
\alpha_{min} = 10^\circ
\eeq
The 2-jet result can be obtained from the NLO result and the 3-jet
cross section $\sigma_{3j}(\gamma\gamma\to b\bar{b} + g)$ which
has been calculated for polarized photons in the signal, background
and interference process. The two-jet cross section is then given by
\beq
\sigma}_{2j (\gamma\gamma \to b\bar{b}) &=& 
\sigma_{NLO} \\
&-&
\int_{E_g^{min}} dE_g \int_{\alpha_{min}} d\alpha(b\bar{b})\; d\alpha(bg)\;
d\alpha(\bar{b} g)\; \frac{\partial^4 \sigma_{3j}}{\partial E_g\, \partial \alpha(b\bar{b})\, \partial \alpha(bg)\, \partial \alpha(\bar{b} g)}
\non
\eeq
Leading higher order contributions due to large double logarithms,
which become important in the 2-jet topology for $J_z=0$, have been
resummed \cite{largel1,largel2}. The resummed cross section is given by
\beq
\sigma^{J_z=0}_{resum} &=& {\cal F} \;
\sigma^{J_z=0}_{LO} 
\label{theresum}
\eeq
For signal (S), background (B) and interference (I) process, respectively, 
the resummation form factor ${\cal F}$ reads
\beq
{\cal F}_S &=& {\cal F}_g \non\\
{\cal F}_B &=& {\cal F}_g {\cal F}_q^2 \label{resumf}\\
{\cal F}_I &=& {\cal F}_g {\cal F}_q \non
\eeq
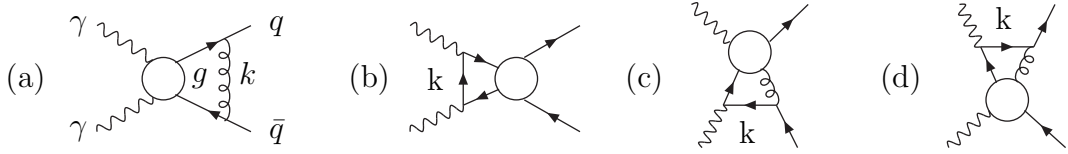
\begin{figure}
\begin{center}
\begin{picture}(360,60)(0,0)
\Text(-30,15)[]{(a)}
\Text(-10,35)[]{$\gamma$}
\Text(-10,-5)[]{$\gamma$}
\Photon(-3,35)(18,22.2){2}{4}
\Photon(-3,-5)(18,8){2}{4}
\CArc(22,15)(8,0,360)
\ArrowLine(26.5,21)(55,35)
\ArrowLine(55,-5)(26.5,9)
\Gluon(45,30)(45,-0.9){2}{4}
\Text(36,15)[]{$g$}
\Text(54,17)[]{$k$}
\Text(65,35)[]{$q$}
\Text(65,-5)[]{$\bar{q}$}
\Text(100,15)[]{(b)}
\Photon(115,35)(135,25){2}{4}
\Photon(115,-5)(135,5){2}{4}
\ArrowLine(135,5)(135,25)
\ArrowLine(135,25)(148.5,19)
\ArrowLine(148.5,11)(135,5)
\CArc(155,15)(8,0,360)
\ArrowLine(158,23)(179,35)
\ArrowLine(179,-5)(158,7)
\Text(125,15)[]{k}
\Text(204,15)[]{(c)}
\Photon(221,43)(235.5,29){2}{4}
\Photon(225,-11)(233,5){2}{4}
\ArrowLine(233,5)(239,18.5)
\CArc(243,25)(8,0,360)
\Gluon(247,18.5)(253,5){2}{2}
\ArrowLine(261,-11)(253,5)
\ArrowLine(253,5)(233,5)
\ArrowLine(250.5,29)(265,43)
\Text(243,-5)[]{k}
\Text(301,15)[]{(d)}
\Photon(322,43)(330,27){2}{4}
\Photon(318,-11)(333,3){2}{4}
\ArrowLine(336,13.5)(330,27)
\CArc(340,7)(8,0,360)
\Gluon(344,13.7)(350,27){2}{2}
\ArrowLine(330,27)(350,27)
\ArrowLine(350,27)(358,43)
\ArrowLine(362,-11)(346.5,3)
\Text(340,38)[]{k}
\end{picture}
\end{center}
\caption{{\it Schematic one-loop topologies contributing to the form factor ${\cal F}_g$ [diagram (a)] and ${\cal F}_q$ [diagrams (b)--(d)]. The blob denotes a hard $2\to 2$ subprocess compared to the soft momentum $k$.}}
\label{topologies}
\end{figure}
${\cal F}_g$ denotes the Sudakov resummation factor and ${\cal F}_q$
the non-Sudakov factor.  The dynamical origin of the signal,
background and interference form factors is explained in detail in
\cite{largel1}. A brief understanding at ${\cal O}(\alpha_s)$ shall be
given here. At NLO the Sudakov and non-Sudakov double logarithms in
the $J_z=0$ state arise in four non-overlapping kinematic regions.
They are depicted in Fig.~\ref{topologies}. One is related to the
emission of a soft real gluon and the exchange of a soft virtual gluon
that connects two helicity conserving vertices. This configuration
[diagram (a) in Fig.~\ref{topologies}] occurs both in the signal and
background process. The soft gluon contribution can be factorized from
the Born amplitude resulting in the NLO term of the
expansion of $\sqrt{{\cal F}_g}$. For the background there are three
more kinematic regimes [diagrams (b)--(d) in Fig.~\ref{topologies}].
They involve a soft virtual quark connecting a helicity conserving and
a helicity violating vertex. Factorizing these contributions from the
Born amplitude one is left with the NLO term in the
expansion of the non-Sudakov form factor ${\cal F}_q$.
Since the Sudakov and non-Sudakov factors factorize at the amplitude
level,
\beq
{\cal M}_S^{resum} &=& \sqrt{{\cal F}_g} {\cal M}_S^{LO} \non\\
{\cal M}_B^{resum} &=& \sqrt{{\cal F}_g} {\cal F}_q {\cal M}_B^{LO}
\eeq
the results of Eq.~(\ref{resumf}) are obtained.\s

The form factors ${\cal F}_g$ and ${\cal F}_q$, including the
leading and next-to-leading order running coupling $\alpha_s$, are
taken from \cite{largel2}. 
The cross section for $b\bar{b}$ production with initial state 
helicity $J_z$ in the 2-jet topology including resummation then reads
\beq
\sigma^{J_z}(\gamma\gamma\to b\bar{b}) = 
\sigma^{J_z}_{2j} + 
\sigma^{J_z}_{LO} \hat{{\cal F}}^{J_z}_{resum}
\label{bbsigma}
\eeq
For $J_z \neq 0$ a resummation of higher orders is not necessary. For
$J_z=0$ $\hat{{\cal F}}_{resum}$ reads
\beq
\hat{{\cal F}}^{J_z=0}_{resum} = 
{\cal F}-1-{\cal F}(\alpha_s)
\label{finalresum}
\eeq
${\cal F}(\alpha_s)$ denotes the ${\cal O} (\alpha_s)$ contribution of
the form factor ${\cal F}$, cf.~Eqs.~(\ref{theresum}), (\ref{resumf}).
The definition (\ref{finalresum}) takes into account that the LO cross
section and the full next-to-leading order result is already included
in $\sigma_{2j}^{J_z=0}$.  In the notation of \cite{largel2} the
${\cal O}(\alpha_s)$ part of the form factors ${\cal F}$ reads
\beq
{\cal F}_{S}(\alpha_s) &=& \frac{\alpha_s C_F}{\pi} \Big(
\ln \frac{s}{m_b^2} \left[ \frac{1}{2} - \ln \frac{s}{4l_c^2} \right]
+ \ln \frac{s}{4l_c^2} - 1 + \frac{\pi^2}{3} \Big) \non\\
{\cal F}_{B}(\alpha_s) &=& 6 \tilde{{\cal F}} + 
{\cal F}_{S}(\alpha_s) \\
{\cal F}_{I}(\alpha_s) &=& 3 \tilde{{\cal F}} + 
{\cal F}_{S}(\alpha_s) \non
\eeq
where the one-loop hard form factor $\tilde{{\cal F}}$ is given by
\beq
\tilde{{\cal F}} = -C_F \frac{\alpha_s}{4\pi} \ln^2 \frac{s}{m_b^2}
\eeq
The cut $l_c$ was introduced in the calculation of the resummation
factors in order to separate soft gluons from hard gluons.  In the
soft gluon region the Eikonal-approximation, exploited in the
calculation of the resummation factors, is valid. The cut $l_c \le
E_g^{min}$ further subdivides the 2-jet region that is characterized
by gluons with energies less than $E_g^{min}$. The dependence on this
phase space slicing parameter cancels at one-loop level since the
full NLO calculation is included. As suggested as best guess in
Ref.~\cite{largel3} the cut $l_c$ will be chosen equal to $E_g^{min}$
in the following. The strong coupling constant $\alpha_s$ is taken at
the scale $\sqrt{s}$.

\subsection{The $\tau^+\tau^-$ channel}
For the LO process $\gamma\gamma\to \tau^+\tau^-$ the signal,
background and interference cross section can be extracted from the
corresponding LO quark cross sections (\ref{signsig}),(\ref{backsig})
and (\ref{intsig}) by replacing the Yukawa coupling in the Higgs decay
appropriately, the quark mass and charge with the analogous $\tau$
lepton values and setting the colour factor $N_c=1$. At NLO there are
no QCD corrections in the final state but only in the
$\gamma\gamma\Phi$ form factor.

\subsection{Chargino and neutralino production}
Likewise the $\tau^+\tau^-$ channel the chargino and neutralino
production does not involve final state QCD corrections. Due to the
different masses in the final state the expression for chargino
production via intermediate Higgs states is more cumbersome than in
the case of quark anti-quark production. With the Higgs to chargino
couplings listed in Table~\ref{mssmcoup} and with the same notation as
in subsection \ref{quarkfinal} the chargino
$\tilde{\chi}_i^+\tilde{\chi}_j^-$ production cross sections for
photons with initial polarization $J_z=0$ after integration over
$|\cos\theta| <c$ can be cast into the form
\beq
\sigma^{++/--}_{S,ij} &=& 
\frac{G_F^2\alpha^2 \beta_{ij} M_W^2 c}{32\pi^3} \non \\
&\times& \left\{
\sum_{\varphi=h,H} |{\cal G}_\varphi|^2
\Big[ (1-\mu_i-\mu_j) \left(
|g_{\varphi \tilde{\chi}_i \tilde{\chi}_j}|^2 + 
|g_{\varphi \tilde{\chi}_j\tilde{\chi}_i}|^2\right) \right.\non \\
&& - \frac{2 M_{\tilde{\chi}_i} M_{\tilde{\chi}_j}}{s}
(g_{\varphi \tilde{\chi}_i\tilde{\chi}_j} 
g_{\varphi \tilde{\chi}_j\tilde{\chi}_i} + 
g_{\varphi \tilde{\chi}_i\tilde{\chi}_j}^* 
g_{\varphi \tilde{\chi}_j\tilde{\chi}_i}^*) \Big] \non\\
&+& |{\cal G}_A|^2 \Big[
(1-\mu_i-\mu_j)\left(|g_{A \tilde{\chi}_i\tilde{\chi}_j}|^2 +
|g_{A \tilde{\chi}_j\tilde{\chi}_i}|^2\right) \non\\
&& + \frac{2M_{\tilde{\chi}_i} M_{\tilde{\chi}_j}}{s} 
(g_{A \tilde{\chi}_i\tilde{\chi}_j} 
g_{A \tilde{\chi}_j\tilde{\chi}_i} + 
g_{A \tilde{\chi}_i\tilde{\chi}_j}^* g_{A \tilde{\chi}_j\tilde{\chi}_i}^*) 
\Big] \non \\
&+& 2 (1-\mu_i-\mu_j) \,\mathrm{Re}
\Big[ {\cal G}_h {\cal G}_H^* (
g_{h\tilde{\chi}_i\tilde{\chi}_j}g_{H\tilde{\chi}_i\tilde{\chi}_j}^* 
+ g_{h\tilde{\chi}_j\tilde{\chi}_i}^* g_{H\tilde{\chi}_j\tilde{\chi}_i}) \Big]
\non \\
& & - \frac{4M_{\tilde{\chi}_i} M_{\tilde{\chi}_j}}{s} 
\,\mathrm{Re} \Big[ {\cal G}_h {\cal G}_H^*
(g_{h\tilde{\chi}_i\tilde{\chi}_j}g_{H\tilde{\chi}_j\tilde{\chi}_i} 
+ g_{h\tilde{\chi}_j\tilde{\chi}_i}^* g_{H\tilde{\chi}_i\tilde{\chi}_j}^*) 
\Big] \non\\
&\mp& \sum_{\varphi=h,H} 
\Big( 2(1-\mu_i-\mu_j) \,\mathrm{Re} \Big[ {\cal G}_\varphi {\cal G}_A^* 
(g_{\varphi \tilde{\chi}_j\tilde{\chi}_i}^* g_{A\tilde{\chi}_j\tilde{\chi}_i} 
- g_{\varphi \tilde{\chi}_i\tilde{\chi}_j} 
g_{A\tilde{\chi}_i\tilde{\chi}_j}^*) \Big] \non\\
& & \left. +\frac{4 M_{\tilde{\chi}_i} M_{\tilde{\chi}_j}}{s} 
\,\mathrm{Re} \Big[{\cal G}_\varphi {\cal G}_A^*  
(g_{\varphi \tilde{\chi}_j\tilde{\chi}_i}^* 
g_{A\tilde{\chi}_i\tilde{\chi}_j}^* - 
g_{\varphi \tilde{\chi}_i\tilde{\chi}_j} g_{A\tilde{\chi}_j\tilde{\chi}_i}) 
\Big] \Big) \right\}
\label{charsig}
\eeq
The interference between the scalar and pseudoscalar Higgs channel is
proportional to the minus (plus) sign for $\sigma^{++}$
($\sigma^{--}$). $\mu_{i,j}$ denote the scaled chargino masses
$M_{\tilde{\chi}_{i,j}}^2/s$ and $\beta_{ij}$ the velocity
for non-degenerate chargino final states.  The signal cross sections for $J_z
=\pm 2$ are zero. The LO cross section for the background process
$\gamma\gamma \to \tilde{\chi}_i^+ \tilde{\chi}_i^-$ can be obtained
from the corresponding quark process Eq.~(\ref{backsig}) by replacing
the quark mass (charge) with the chargino mass (charge) and setting
$N_c=1$. After $\cos\theta$ has been integrated out in the range
[$-c,c$] the polarized cross sections of the interference process 
$\gamma\gamma \to \tilde{\chi}_i^+ \tilde{\chi}_i^-$ reads
\beq
\sigma^{++/--}_{I,ii} &=& -
\frac{\sqrt{2} G_F \alpha^2 M_W M_{\tilde{\chi}_i}}{4\pi s}
\ln\frac{1+\beta_{ii} c}{1-\beta_{ii} c} \non\\
&\times\Big\{& \sum_\varphi \Big[
\beta_{ii}^2 ({\cal G}_\varphi + 
{\cal G}_\varphi^* ) 
(g_{\varphi\tilde{\chi}_i\tilde{\chi}_i} + 
g_{\varphi\tilde{\chi}_i\tilde{\chi}_i}^*)  \mp 
({\cal G}_\varphi - 
{\cal G}_\varphi^* )
(g_{\varphi\tilde{\chi}_i\tilde{\chi}_i} - 
g_{\varphi\tilde{\chi}_i\tilde{\chi}_i}^*)  \Big] \non\\
&-& \Big[ ({\cal G}_A + 
{\cal G}_A^*)
(g_{A\tilde{\chi}_i\tilde{\chi}_i} + 
g_{A\tilde{\chi}_i\tilde{\chi}_i}^*) 
\mp \beta_{ii}^2 ({\cal G}_A - 
{\cal G}_A^*)
(g_{A\tilde{\chi}_i\tilde{\chi}_i} - 
g_{A\tilde{\chi}_i\tilde{\chi}_i}^*)  \Big]\Big\} \non\\
\sigma^{+-/-+}_{I,ii} &=& 0 
\label{charint}
\eeq
The minus (plus) sign in the contributions proportional to $({\cal
  G}_\Phi - {\cal G}_\Phi^* )$ ($\Phi = h,H,A$) corresponds to
$\sigma^{++}$ ($\sigma^{--}$). Assuming CP-invariance, the couplings
$g_{\Phi\tilde{\chi}_i\tilde{\chi}_j}$ are real and orthogonal so that
the formulae (\ref{charsig}) and (\ref{charint}) adopt a much simpler
form. \s

In the neutralino channel there are no background and interference
contributions at LO. The neutralino signal cross section is given by
the chargino result (\ref{charsig}) after replacing the masses
appropriately and the chargino couplings
$g_{\Phi\tilde{\chi}_i\tilde{\chi}_j}$ with the neutralino couplings
$g_{\Phi\tilde{\chi}^0_i\tilde{\chi}^0_j}$, cf.~Table~\ref{mssmcoup}.
For equal neutralinos in the final state Bose symmetry requires an
additional factor $1/2$.

\subsection{Results}
The cross section of the linear collider process
\beq
e^+e^- \to \gamma\gamma \to X_1 X_2 \;, \quad X_{1,2} = t,b,\tau,
\tilde{\chi}_i^\pm, \tilde{\chi}_i^0
\label{eeproc}
\eeq
is obtained by folding the cross section of the subprocess
$\gamma\gamma \to X_1 X_2$ denoted by $\hat{\sigma}$ with the
$\gamma\gamma$ luminosity. The $\gamma\gamma$ luminosity for Compton
scattered photons off (polarized) electron/positron beams has been
calculated \cite{kuehn}. It has been given as a function of the photon
energy, of the initial electron/positron and laser polarization as
well as of the final state photon helicities. \s

The differential photon luminosity ${\cal L}^{\gamma\gamma}$ is
maximal and peaked towards high energies if the electron/positron and
laser beams are polarized and have opposite helicity. The
electron/positron helicities $P_{e^{\pm}}$ will therefore be chosen
equal to 1 in the following and the incoming photon helicities
$P_{\gamma_{1,2}}$ are set equal to $-1$ each. For illustration,
Fig.~\ref{glum} shows the differential luminosity for this helicity
choice and $\sqrt{s_{ee}}=500$~GeV. The curves correspond to different
helicity combinations between the outgoing photons. As can be inferred
from the figure, the luminosity is maximal and peaked at the upper end
of the energy range of the scattered photons if their helicities are
$P_{\gamma^*_1}=P_{\gamma^*_2}=1$. \s
\begin{figure}
\begin{center}
\epsfig{figure=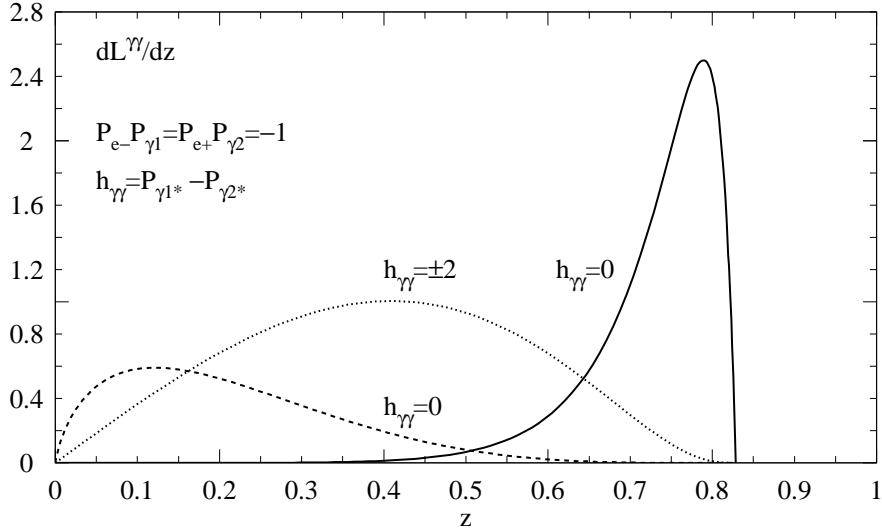,width=12cm}
\\
\caption{\textit{The differential $\gamma\gamma$ luminosity as a function of $z=\sqrt{s_{\gamma\gamma}}/\sqrt{s_{ee}}$ for the $e^+e^-$ c.m.~energy $\sqrt{s_{ee}}=500$~GeV. $\sqrt{s_{\gamma\gamma}}$ denotes the $\gamma\gamma$ c.m.~energy. The curves correspond to different polarizations of the final state photons. The full curve is the result for the outgoing photon helicities $P_{\gamma^*_1} = P_{\gamma^*_2} = 1$ and the dashed one for $P_{\gamma^*_1} = P_{\gamma^*_2} = -1$. The curves for $P_{\gamma^*_1} =1$, $P_{\gamma^*_2} = -1$ and $P_{\gamma^*_1} =-1$, $P_{\gamma^*_2} = 1$ are identical and added in one curve.}}
\label{glum}
\end{center}
\end{figure}

The polarized fermion, respectively SUSY particle production cross
section is given by a superposition of the different helicity
combinations of the scattered photons:
\beq
\sigma^{pol}(e^+e^-\to X_1 X_2) = \int_{\tau_0}^{\frac{x}{x+1}}
dz \sum_{h_i,h_j=+,-} 
\frac{d{\cal L}^{\gamma\gamma}(h_i h_j)}{dz}\;  
\hat{\sigma}^{h_i h_j}(\gamma\gamma 
\to X_1 X_2; \hat{s}_{\gamma\gamma}=z^2s_{ee})
\eeq
where $\tau_0 = (M_{X_1}+M_{X_2})/\sqrt{s_{ee}}$ with $\sqrt{s_{ee}}$
denoting the c.m.~energy of the incoming electron and positron.
$\sqrt{\hat{s}_{\gamma\gamma}}$ stands for the c.m.~energy of the
$\gamma\gamma$ fusion subprocess. The upper integration limit depends
on $x=(2\sqrt{s_{ee}}\omega_0)/M_e^2$ where $M_e$ denotes the electron
mass.  The laser energy $\omega_0$ will be chosen equal to 1.26~eV in
the subsequent analysis. \s

The formulae of the polarized subprocess cross sections $\hat{\sigma}$
for the signal, background and interference reaction have been given
in the previous subsections. In the case of $b\bar{b}$ production the
resummed cross section for the 2-jet final state,
cf.~Eq.~(\ref{bbsigma}), will be used. \s

\noindent {\bf $b\bar{b}$ final state} \newline
\noindent 
Fig.~\ref{bbchan} shows the total cross section and the result for the
signal and background part of $b\bar{b}$ production in the two-jet
topology as a function of the pseudoscalar Higgs mass $M_A$. The
c.m.~energy of the incoming electron and positron has been chosen such
that the energy squared of the subprocess
$\hat{s}_{\gamma\gamma}=z_0^2 s_{ee}$ is equal to $M_A^2$. $z_0$ is
the position where the differential luminosity $[d{\cal L}^{++} +
{\cal L}^{--}]/dz$ becomes maximal for $\sqrt{s}_{ee}$. The fusion
process folded with the $\gamma\gamma$ luminosity is then integrated
over the range
\beq
[z_{min},z_{max}] &=& 
[z_0-5\,\mathrm{ GeV}/\sqrt{s}_{ee}, z_0+5\,\mathrm{ GeV}/\sqrt{s}_{ee}] 
\non \\
&=& [(M_A - 5\,\mathrm{ GeV})/\sqrt{s}_{ee},(M_A + 5\,\mathrm{
  GeV})/\sqrt{s}_{ee}]
\eeq
This choice of $z_0$ and $\sqrt{s}_{ee}$ guarantees that the
pseudoscalar Higgs boson $A$ is nearly produced on-shell in the signal
process
\beq
\sigma_S^{pol} (e^+e^-\to b\bar{b}) = \int_{z_{min}}^{z_{max}} dz\; 
\hat{\sigma}^{++}_S (\gamma\gamma\to b\bar{b}; \hat{s}_{\gamma\gamma})
\left[\frac{d{\cal L}^{++}}{dz}+ \frac{d{\cal L}^{--}}{dz}\right]
\eeq
particularly since $[d{\cal L}^{++} + {\cal L}^{--}]/dz$ drops rapidly
around the maximum, cf.~Fig.~\ref{glum}. Furthermore, the signal
process is folded with the maximum luminosity for a given
electron/positron c.m.~energy. The integration interval $(2\times
5)$~GeV$/\sqrt{s}_{ee}$ accounts for the limited energy resolution of
about 10~GeV which may be achieved at a high-luminosity $e^+e^-$
linear collider.  \s

\begin{figure}
\begin{center}
\epsfig{figure=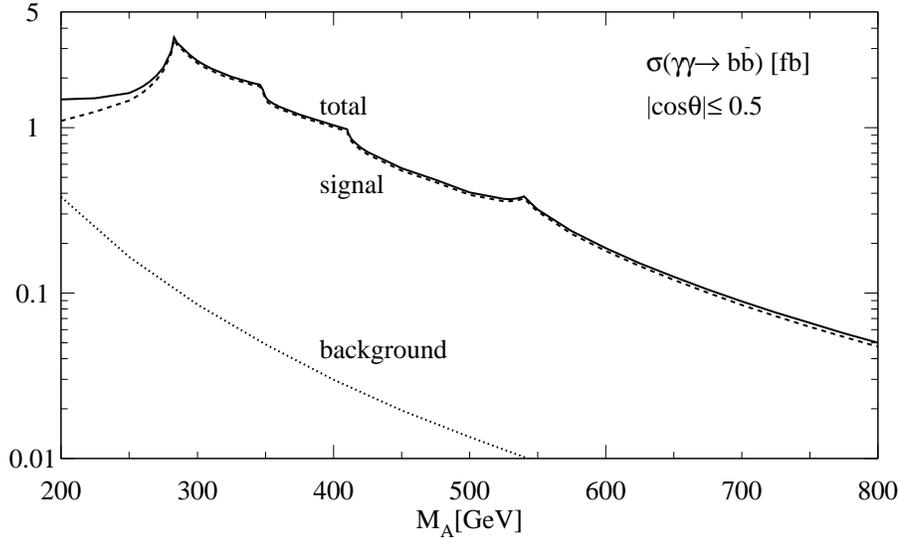,width=12cm}
\\
\caption{\textit{The total, signal and background cross sections of the process $\gamma\gamma \to b\bar{b}$ for $\mu=M_2=200$~GeV and $\tan\beta=7$. $\cos\theta$ has been integrated in the range $[-0.5...0.5]$.}}
\label{bbchan}
\end{center}
\end{figure}
The interference cross section here and in the following plots is not
shown separately since it is fairly small in this specific numerical
example, only of ${\cal O}(10^{-3}$~fb$)$. The heavy CP-even Higgs
mass $M_H$ and $M_A$ differ by at most $\sim 2.5$~GeV for the chosen
SUSY parameter set. The interference cross section, which is
proportional to the real part of the MSSM Higgs propagators, is
therefore almost vanishing since $A$ and $H$ are produced nearly
on-shell and the contribution of the light scalar Higgs is very much
suppressed as it is far off-shell. \s

The SUSY parameters are chosen as $\mu=M_2=200$~GeV, $\tan\beta=7$,
common sfermion mass $M_S=1$~TeV and the SUSY breaking trilinear
couplings are set equal to 0. The NLO corrections and leading higher
order contributions have been included in the way outlined above. The
${\cal O}(\alpha_s)$ real and virtual corrections to the signal part
amount up to $\sim 20$\% compared to the LO cross section. The NLO
background cross section is about a factor 10 to 60 larger than the LO
result in the considered mass range because at next-to-leading order
the $J_z=0$ background suppression by $m_b^2/s$ is removed and becomes
of ${\cal O}(1)$. In the case of the interference part the NLO
corrections are large.\s

The background contribution can be reduced enormously by cutting away
the 3-jet events. With the chosen cuts defining the 2-jet topology the
background is suppressed by 2 orders of magnitude compared to the
complete 3-jet NLO cross section. The reason is that the 2-jet NLO
cross section is dominated by the helicity-flipped $J_z=0$ states
which are suppressed.  [It is the non helicity-flipped $J_z=0$ states
which remove the $m^2_b/s$ suppression at NLO.] In contrast the signal
is reduced by $\sim 40$\% while the interference part is reduced by
an order of magnitude.\s

The background can further be suppressed by cutting the integration
region in $\cos\theta$ where $\theta$ denotes the angle between the
$b$-quark and the beam axis. The integration range has been chosen
$[-0.5,0.5]$ in Fig.~\ref{bbchan}. More than 90\% of the background
proceeding via $t$- and $u$-channel diagrams is cut away whereas the
isotropic signal is reduced by a factor 1/2 only.\s

In the 2-jet topology the resummation turns out to be important.  It
increases the 2-jet signal by up to $\sim 30$\% and changes the
interference 2-jet cross section by ${\cal O}(100\%)$. For the
background reaction the impact is even larger. At the upper mass range
of $M_A$ it is the main contribution to the complete 2-jet background
result.\s

The signal process shows a rich structure. The two peaks at $M_A
\approx 280$ and 540~GeV are due to the behaviour of the $\gamma\gamma
A$ form factor which peaks when the chargino thresholds
$\tilde{\chi}_1^+ \tilde{\chi}_1^-$ and $\tilde{\chi}_2^+
\tilde{\chi}_2^-$ are reached. The masses of the charginos for the
chosen parameter set are
\beq
M_{\tilde{\chi}_1^\pm} = 141\;{\mathrm GeV} \;, \qquad
M_{\tilde{\chi}_2^\pm} = 270\;{\mathrm GeV} 
\label{charmass}
\eeq
The kink around 410~GeV can be understood from the $H\to b\bar{b}$
branching fraction, c.f.~Fig.~\ref{fig:bra22}. It drops when the
$\tilde{\chi}_1^+ \tilde{\chi}_2^-$ channel opens. The peak at $\sim
350$~GeV, finally, is a mixture of the behaviour of the branching
fraction and $\gamma\gamma A$ form factor at the $t\bar{t}$
threshold.\s

As can be inferred from Fig.~\ref{bbchan} the resummed 2-jet signal
cross section amounts up to a few fb whereas the background is less
than 0.5~fb. By focusing on the 2-jet final states and cutting the
integration range in $\cos\theta$ the signal to background ratio can
be enhanced to significances $S/\sqrt{B}$ above 5 in the whole $M_A$
range 200...800~GeV for an integrated luminosity of $\int{\cal L} =
100$~fb$^{-1}$ in three years. Demanding the total cross section to
exceed 1~fb $\gamma\gamma$ fusion allows to find the pseudoscalar and
scalar Higgs bosons for masses less than about 400~GeV. Higgs
production via Compton back-scattered photons therefore provides a
powerful mechanism for the heavy MSSM Higgs boson search in the
$b\bar{b}$ channel.\s

\noindent {\bf $t\bar{t}$ and $\tau^+\tau^-$ final states}
\noindent \newline
Fig.~\ref{ttchan} illustrates the results for the $t\bar{t}$ and
$\tau^+\tau^-$ final states at next-to-leading order. If not stated
otherwise the SUSY parameters and the smearing procedure with the
$\gamma\gamma$ luminosity are the same as in the $b\bar{b}$ case in
the following. \s

For $t\bar{t}$ the (2+3)-jet NLO cross section has been plotted. Since
the electric top-quark charge is $2/3$ and enters in the fourth power in the
background process, the background is rather large. Furthermore,
the 2-jet topology is not suppressed very much because of the large
top-quark mass. Therefore the restriction on the 2-jet final state is
not efficient. Also due to the large top mass the cut in $\cos\theta$
does not lead to a significant enhancement in the signal to background
ratio though it may be improved if the cut is chosen equal to 0.7.
Nevertheless, the significance does not achieve the $5\sigma$ level.\s
\begin{figure}
\begin{center}
\epsfig{figure=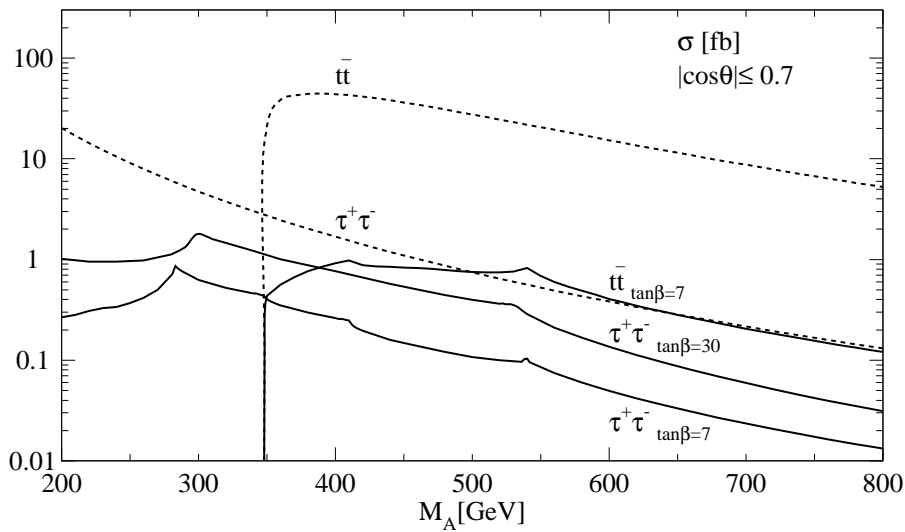,width=12cm}
\\
\caption{\textit{Full lines: The signal cross sections of the processes $\gamma\gamma \to t\bar{t}$ $(\tan\beta=7)$ and $\gamma\gamma \to \tau^+\tau^-$ $(\tan\beta=7,30)$ for $\mu=M_2=200$~GeV. Dashed lines: Background processes only. $\cos\theta$ has been integrated in the range $[-0.7...0.7]$.}}
\label{ttchan}
\end{center}
\end{figure}

Below the $t\bar{t}$ threshold the complete $\tau^+\tau^-$ production
cross section amounts up to ${\cal O}(10$~fb$)$. The $\tau^\pm$ charge
is unity so that the background is large. Since the $\tau$ mass is
rather small a cut in $\cos\theta$ is worthwhile. The best signal to
background ratio can be achieved for the integration region
$|\cos\theta|<0.7$. Nevertheless for moderate values of $\tan\beta=7$
the signal is too small compared to the background cross section to
reach significances above 5. The situation changes for larger values
of $\tan\beta$ since the branching fraction of $A\to \tau^+\tau^-$
increases with $\tan\beta$ as can be inferred from
Fig.~\ref{fig:bra22}. The signal cross section is therefore larger for
$\tan\beta=30$ whereas the background cross section remains the same
so that in the mass range $M_A \approx 300...480$~GeV significances
exceeding 5 with $\tau^+\tau^-$ production cross sections larger than
1~fb can be achieved. \s

In contrast to the $\tau^+\tau^-$ channel, increasing the value of
$\tan\beta$ will not enhance the signal to background ratio in
$t\bar{t}$ production since the $A$ Yukawa coupling to the top quarks
is proportional to $\cot\beta$ so that the $t\bar{t}$ signal cross
section decreases with rising $\tan\beta$.\s

The peaks and kinks in the $t\bar{t}$ and $\tau^+ \tau^-$ signal cross
sections can readily be explained by the behaviour of the $H$ and $A$
branching fractions in these channels and the $\gamma\gamma A$ form
factor when the various chargino and the $t\bar{t}$ production
thresholds are reached. For $\tan\beta = 30$ the chargino masses are
\beq
M_{\tilde{\chi}_1^\pm} = 149\;{\mathrm GeV} \;, \qquad
M_{\tilde{\chi}_2^\pm} = 266\;{\mathrm GeV} 
\eeq 

\noindent {\bf Chargino and neutralino production}
\noindent \newline
Fig.~\ref{susychan} presents the NLO cross sections for chargino and
neutralino production as a function of the pseudoscalar Higgs mass.
Each curve shows the sum of all possible chargino and neutralino final
states, respectively, except for $\tilde{\chi}^0_1 \tilde{\chi}^0_1$.
The chargino masses for the chosen parameter set have been given in
(\ref{charmass}) and the neutralino masses are
\beq
\begin{array}{rllrll}
M_{\tilde{\chi}_1^0} &=& 85\;{\mathrm GeV} &
M_{\tilde{\chi}_2^0} &=& 148\;{\mathrm GeV} \\
M_{\tilde{\chi}_3^0} &=& 208\;{\mathrm GeV} &
M_{\tilde{\chi}_4^0} &=& 271\;{\mathrm GeV}
\end{array}
\eeq
\begin{figure}
\begin{center}
\epsfig{figure=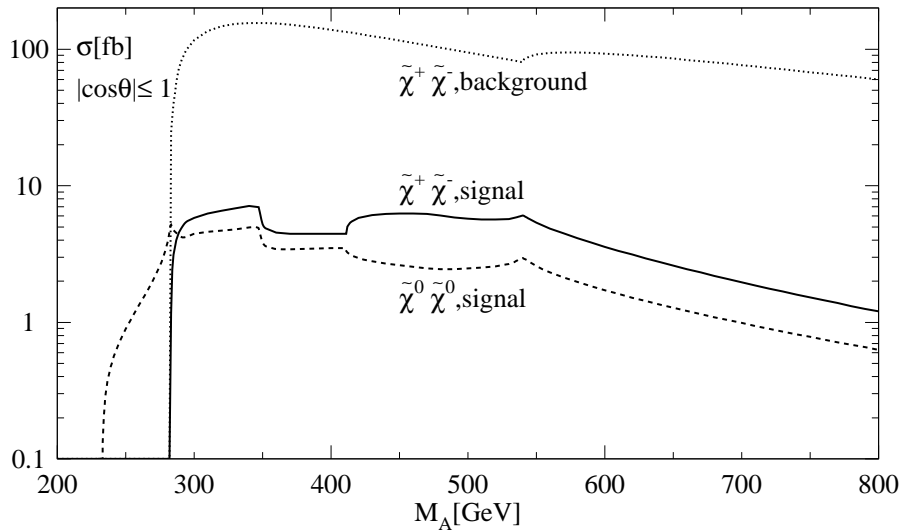,width=12cm}
\\
\caption{\textit{Full (dashed) line: Signal cross section of the process $\gamma\gamma \to \tilde{\chi}^+ \tilde{\chi}^-$ $(\gamma\gamma \to \tilde{\chi}^0 \tilde{\chi}^0)$. Dotted line: Background cross section of the process $\gamma\gamma \to \tilde{\chi}^+ \tilde{\chi}^-$. All possible chargino and neutralino final states except for $\tilde{\chi}^0_1 \tilde{\chi}^0_1$ are added up. The SUSY parameters are chosen $\mu=M_2=200$~GeV and $\tan\beta=7$. $\cos\theta$ has been integrated in the range $[-1,...,1]$.}}
\label{susychan}
\end{center}
\end{figure}
Since the chargino and neutralino masses are rather large a cut in
$\theta$ does not improve the signal to background ratio. Therefore
$\cos\theta$ has been integrated in the limits $-1$ and 1. \s

Above the $\tilde{\chi}_1^+\tilde{\chi}_1^-$ production threshold,
{\it i.e.} $M_A\approx 280$~GeV, the chargino channel opens. Since the
chargino charge is three times larger than the $b$-quark charge, the
background is large and amounts up to ${\cal O}(100$~fb$)$. The kink
at $\sim 540$~GeV is related to the $\tilde{\chi}_2^+
\tilde{\chi}_2^-$ threshold.\s

The signal cross section develops a rather complex structure due to
the behaviour of the $H$ and $A$ chargino branching fractions at the
$t\bar{t}$, $\tilde{\chi}_1^+\tilde{\chi}_2^-$ and the
$\tilde{\chi}_2^+\tilde{\chi}_2^-$ thresholds. The form of the signal
at the $t\bar{t}$ and $\tilde{\chi}_2^+\tilde{\chi}_2^-$ thresholds is
in addition influenced by the behaviour of the $\gamma\gamma A$ form
factor. The cross section varies between $\sim 1$ to 6~fb. It is too
small to allow for the Higgs detection in the chargino channel. \s

Above $M_A\approx 230$~GeV, the threshold for the
$\tilde{\chi}_1^0\tilde{\chi}_2^0$ production is reached and the
neutralino channel opens. It does not acquire any LO background and
interference contributions due to neutralinos. The kinks and peaks in
the signal curve correspond to the various chargino thresholds and the
$t\bar{t}$ threshold. The neutralino branching fraction drops when the
chargino channels open up. On the other hand the $\gamma\gamma A$ form
factor peaks when the $\tilde{\chi}_1^+\tilde{\chi}_1^-$ and
$\tilde{\chi}_2^+\tilde{\chi}_2^-$ thresholds are reached. The signal
cross section amounts up to a few fb. Due to the lower branching ratio
compared to the chargino channel the neutralino signal is smaller.
Above $M_A \approx 280$~GeV the neutralino channel has to fight
against the background stemming from the chargino decays: Neutralinos
can decay into a lighter neutralino and a $Z$ boson, for example:
\beq
\tilde{\chi}_2^0 \to \tilde{\chi}_1^0 + Z
\eeq
Charginos may decay into a neutralino and a $W^\pm$ boson:
\beq
\tilde{\chi}_1^\pm \to \tilde{\chi}_1^0 + W^\pm
\eeq
Due to the escaping LSPs and neutrinos the final states in the
chargino and neutralino cascade decays are rather similar. Since the
chargino background is very large it will be challenging to separate the
neutralino signal from the chargino background. Exploiting the
different decay topologies, however, the neutralino channel may serve
for the heavy MSSM Higgs boson search also in the mass range above the
chargino production threshold.

\subsection{Discovery reach}
The c.m.~energy in all production channels has been integrated in a
narrow energy range around $M_A$. Since the mass of the heavy scalar
Higgs $H$ differs hardly from $M_A$ it is also produced almost
on-shell. Though it will be very difficult if not impossible to
separate the pseudoscalar and scalar heavy Higgs particle, the
detection of additional heavy Higgs particles will help to disentangle
the MSSM from the SM if only one light Higgs boson has been discovered
at the LHC as it might be the case in the parameter region $M_A
\,\gesim\, 200$~GeV and $\tan\beta\,\gesim\, 6$ up to 15.\s

The analysis has demonstrated that heavy MSSM Higgs production in
polarized $\gamma\gamma$ fusion at a Compton collider provides an
excellent mechanism to detect the pseudoscalar and scalar Higgs bosons
in the $b\bar{b}$ channel for moderate values of $\tan\beta$. By this
way the discovery region which is not covered by the LHC can be
closed. In addition, the mass range up to which the Higgs bosons can
be detected in $\gamma\gamma$ fusion extends up to $\sim
0.8\sqrt{s_{ee}}$ in contrast to $e^+e^-$ collisions with a discovery
range of $\sim 0.5\sqrt{s_{ee}}$. For the detection in the $b\bar{b}$
channel it is crucial, however, to investigate 2-jet final states.
Including NLO and resummation of higher orders the analysis has been
completed to the currently most accurate level. \s

While the $t\bar{t}$ mode is not promising, an additional channel is
provided by $\tau^+\tau^-$ production for large values of $\tan\beta$.
The neutralino channel, finally, can be exploited in mass regions
below the chargino threshold, if it is kinematically allowed, where it
is not plagued by the large chargino background. Above the threshold
the different topologies of the final states may help to extract the
neutralino signal from the chargino background.

%% file: chapter4.tex
\chapter{The Lifetime of Higgs particles} 
The two basic properties characterizing particles are the mass and the
lifetime. As outlined in the previous chapter the masses of the SM and
MSSM Higgs particles can be determined with high accuracy at the LHC
and high-energy $e^+e^-$ linear colliders. The determination of the
lifetimes, {\it i.e.} the total widths of the Higgs bosons is a more
difficult task, especially for Higgs states with narrow widths.\s

Starting with a theoretical overview over the total widths and the
main decay channels of the SM Higgs boson and the MSSM Higgs scalars,
several mechanisms will be presented, subsequently, which can be
exploited at $e^+e^-$ linear colliders in order to determine the SM
Higgs lifetime for masses in the intermediate range where it is
narrow. The MSSM case will be discussed qualitatively.

\section{The SM Higgs boson decays}
The Higgs couplings to the $W$, $Z$ gauge bosons and to the fermions are
proportional to the masses of these particles,
cf.~Eqs.~(\ref{higgsgauge}), (\ref{higgsf}). Consequently, the Higgs
boson will predominantly decay into the heaviest particles of the SM,
i.e.  $W$, $Z$ gauge bosons, top and bottom quark, if this is
kinematically allowed. \s

The Higgs couplings to $\gamma\gamma$, $\gamma Z$ and $gg$ are
mediated by heavy particle loops. The Higgs decays into these
particles only play a r\^{o}le for Higgs masses below $\sim$ 140 GeV,
where the total Higgs decay width is rather small. They are
interesting since they are sensitive to scales far beyond the
Higgs mass and thus to new particles whose masses are generated by the
Higgs mechanism.\s

The main contributions to the $H\gamma\gamma$ coupling come from
top-quark and $W$ boson loops. In the intermediate mass range, $M_W
\le M_H \le 2M_Z$, the $W$-loop contribution dominates and interferes
destructively with the fermion amplitude. The same holds true for the
$H\gamma Z$ coupling. The Higgs coupling to gluons is practically
almost built up by top-quark loops and the branching ratio is only
sizeable below the $WW$ threshold.  \s

\begin{figure}[ht]
\unitlength 1cm
\begin{center}
\begin{picture}(11,11)
\put(-0.9,3.2){\epsfig{figure=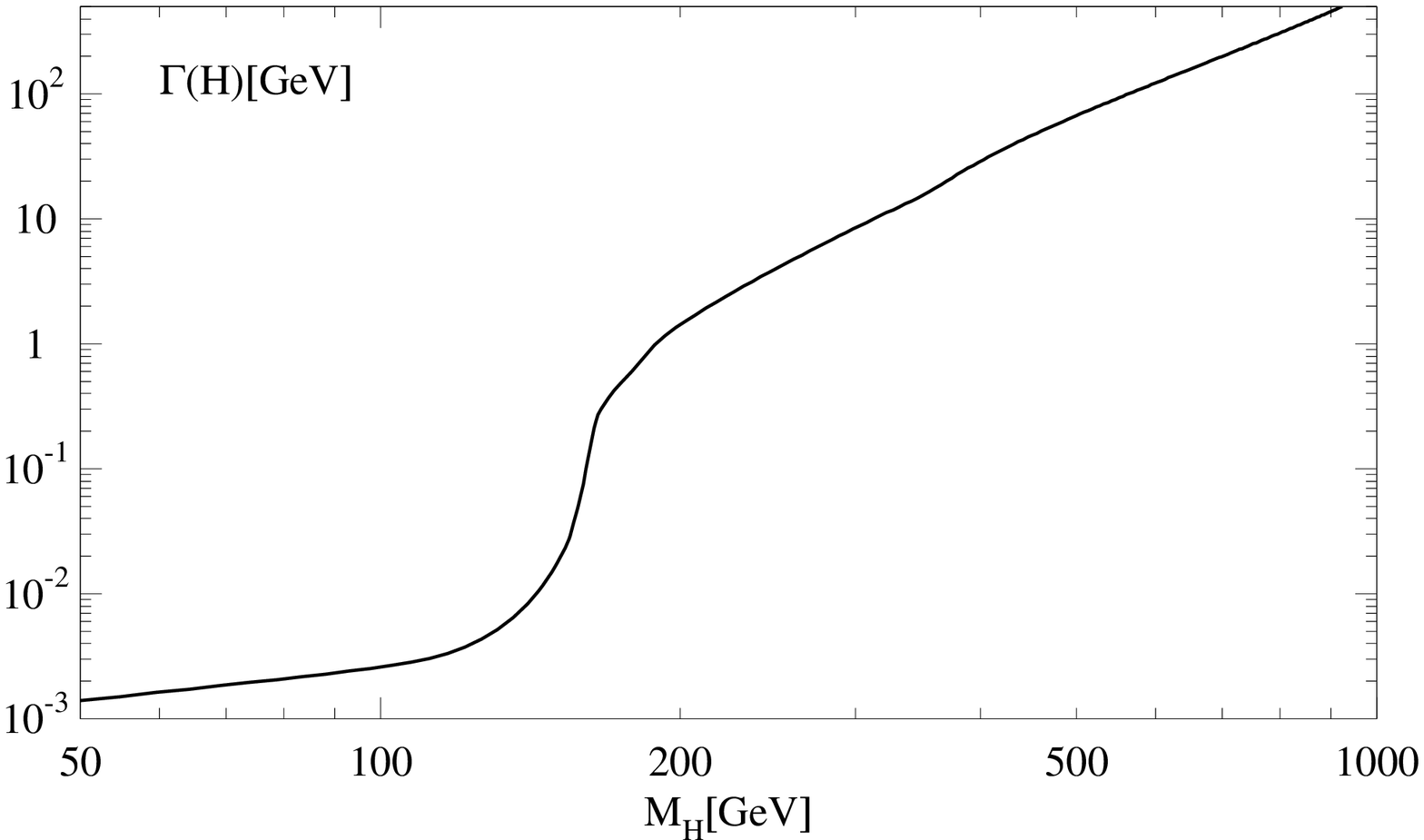,width=12.6cm}}
\put(-0.9,-4.5){\epsfig{figure=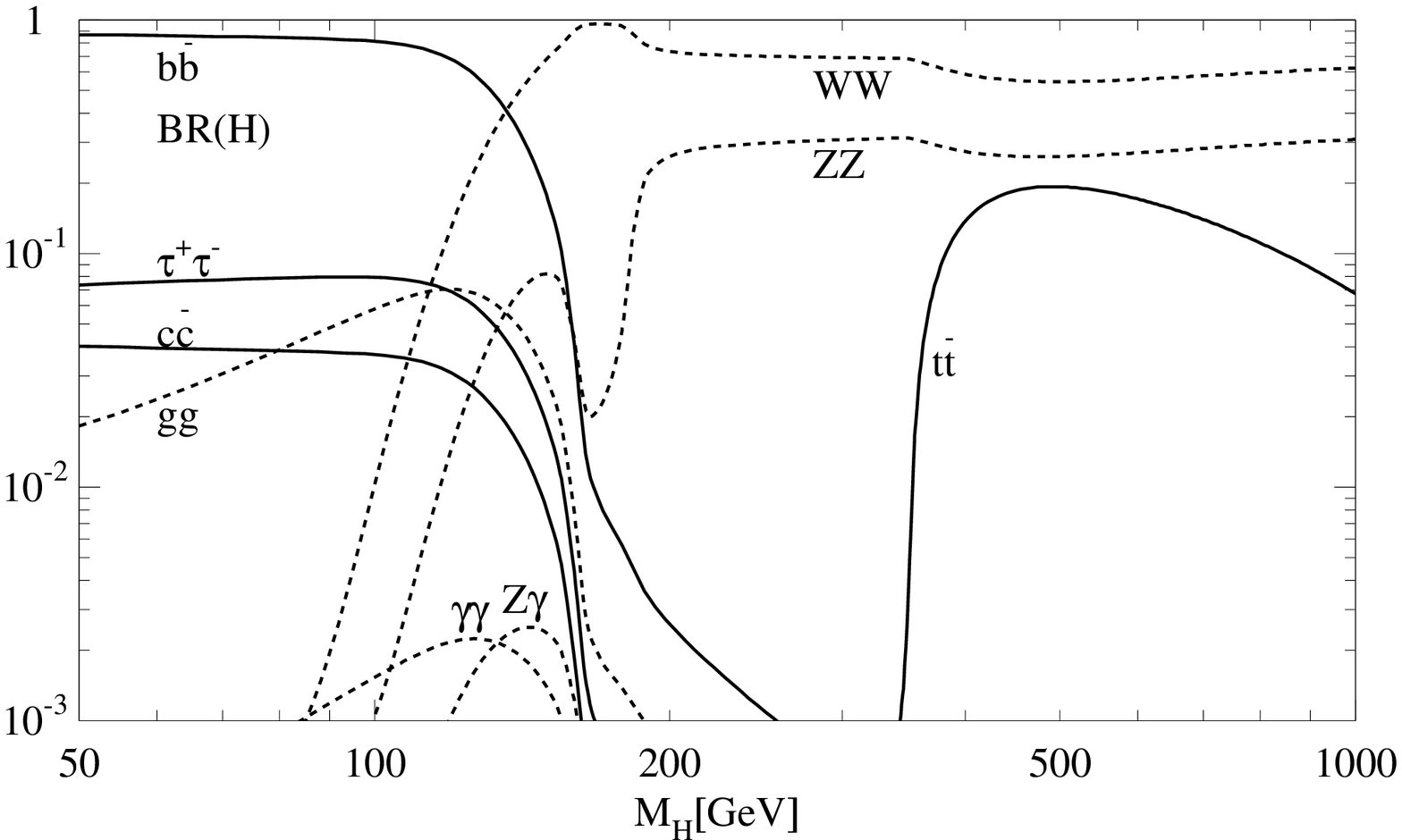,width=12.6cm}}
\end{picture}
\end{center}\vspace{3.8cm}
\caption{\textit{(a) Total decay width of the SM Higgs boson as a function of its mass. (b) Branching ratios of the dominant decay modes of the SM Higgs particle.}}
\label{smdecay}
\end{figure}
The decay modes discussed in the following are generated with the
FORTRAN program HDECAY \cite{hdecay}. They include electroweak
\cite{ewcorr} and QCD corrections \cite{qcdcorr,spirdjou}. Also the
decay into off-shell particles below the threshold is taken into
account \citer{zerwas,morettistir}.\s

In Fig.~\ref{smdecay} the total decay width and the branching ratios
of the SM Higgs particle are shown as a function of the Higgs mass.
Below about 140~GeV the total width is less than 10~MeV, {\it
  i.e.}~rather small. The Higgs preferentially decays into $b$-quarks
with a branching ratio of about 85\%. The next important decay
channels are provided by the decays into $\tau^+\tau^-$, charm-quarks
and gluons. Beyond $\sim 140$~GeV the decay into $W$ bosons becomes
dominant followed by the $ZZ$ decay mode. Above $M_H = 2m_t$ the decay
into top quarks is kinematically possible. The branching ratio,
however, always remains below 20\% since the decay width is
proportional to the Higgs mass whereas the leading $WW$ and $ZZ$ decay
widths grow with the third power of the Higgs mass. Thus the total
width strongly increases with the Higgs mass, up to 600~GeV for
$M_H=1$~TeV.
\newpage

\section{Decay modes of the MSSM Higgs particles}

\subsection{The total decay widths and branching ratios of non-SUSY decays}
The MSSM Higgs bosons mainly decay into heavy quarks and gauge bosons
if kinematically allowed. This is due to the fact that the
corresponding couplings are proportional to the masses of the
particles which are generated by the Higgs mechanism. In contrast to
the SM, however, the decay into bottom quarks becomes dominant for
large $\tan\beta$ since the couplings are enhanced in this case,
c.f.~Table~\ref{hcoup}.  Furthermore, the couplings to $W$ and $Z$
bosons are suppressed by $\cos(\beta-\alpha)$ or $\sin(\beta-\alpha)$
compared to the SM couplings. The decay widths and branching ratios
discussed below have been obtained by means of the FORTRAN program
HDECAY \cite{hdecay}. The Higgs sector contains the radiative
corrections due to top/bottom and squark loops calculated in the
effective potential approach, and the NLO QCD corrections as well as
the mixing in the stop and sbottom sectors are included \cite{carena}.
Off-shell decays into vector gauge bosons and top-quarks below
the threshold have also been taken into account \cite{zerwas1,morettistir}.\s

The leading order widths of $h$, $H$ into gluons are mediated by quark
and squark loops which are important for squark masses below about
400~GeV \cite{squarkl}. In contrast to the SM, the bottom quark
contribution can become important for large values of $\tan\beta$
where the relevant coupling is strongly enhanced. The pseudoscalar $A$
decay into gluons only involves quark loops and no squark loops since
$A$ flips the helicity of the squarks whereas the gluon coupling to
squarks conserves the squark-helicity. The decays of the CP-even
neutral Higgs particles $h,H$ into photons are generated by heavy
charged fermion, $W$ boson, charged Higgs boson, chargino and sfermion
loops whereas the corresponding $A$ decay is mediated by fermion and
chargino loops only.  \s

The heavy scalar particle $H$ can also decay into a pair of light
Higgs bosons or a pair of pseudoscalar bosons if kinematically
allowed. The radiatively corrected Higgs self-couplings in the
one-loop leading $m_t^4$ approximation are given by Eq.~(\ref{coup}).
Moreover, the heavy Higgs particles can decay into a gauge and a Higgs
boson. \s

\begin{figure}[ht]
\unitlength 1cm
\begin{center}
\begin{picture}(10,11)
\put(-0.8,3.8){\epsfig{figure=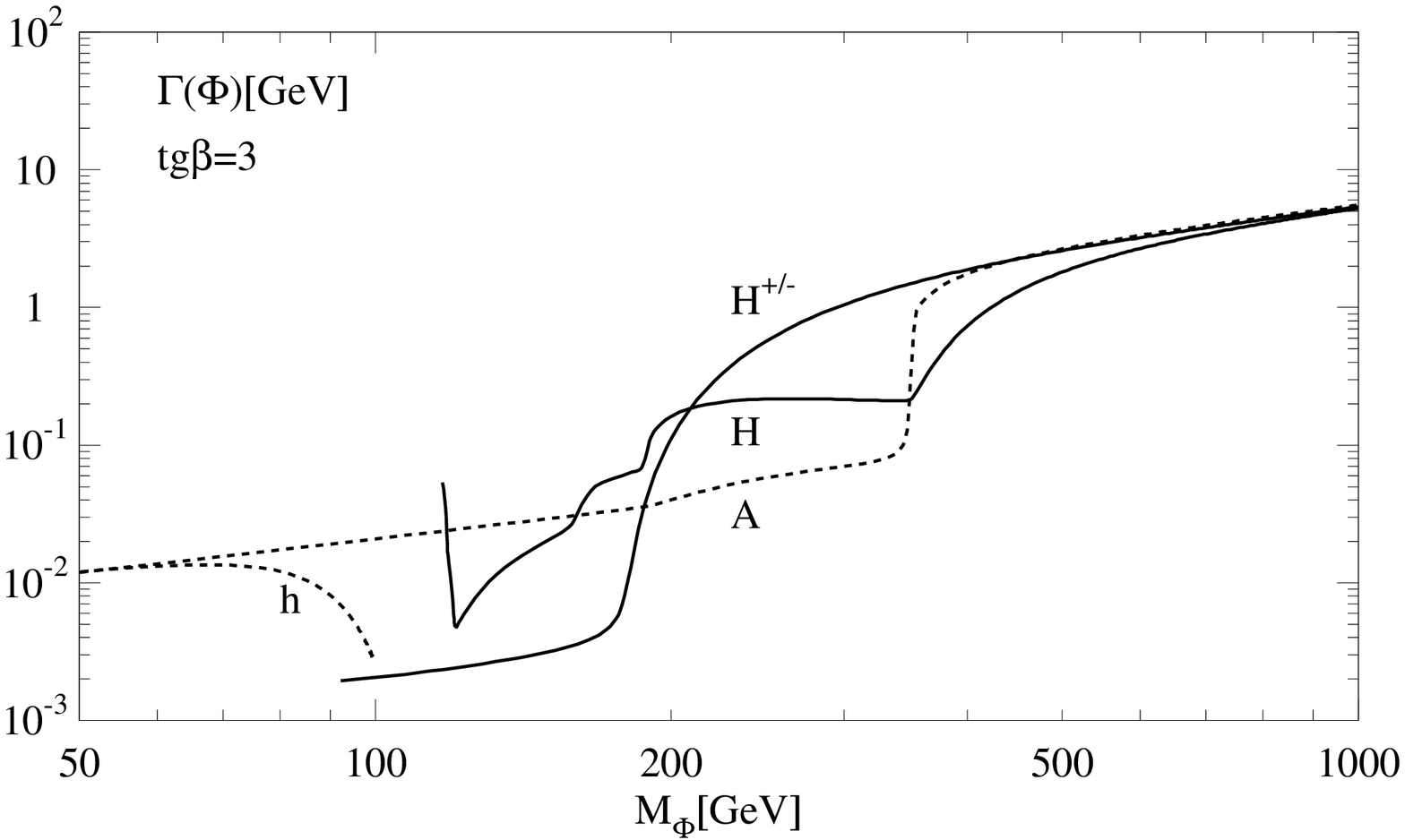,width=11.5cm}}
\put(-0.8,-3){\epsfig{figure=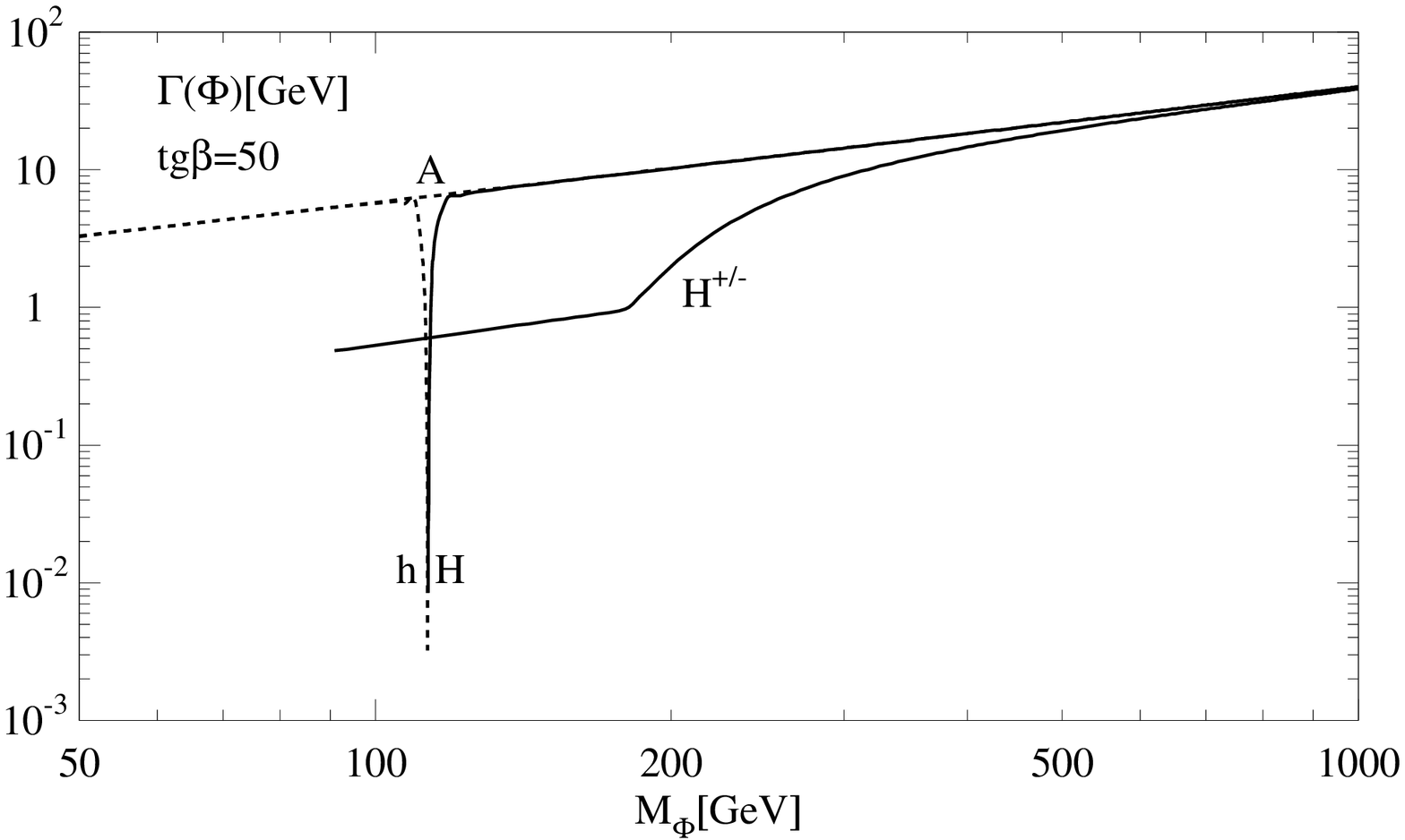,width=11.5cm}}
\end{picture}
\end{center}\vspace{2.4cm}
\caption{\textit{Total decay widths of the MSSM Higgs particles as a function of the corresponding Higgs mass for $\tan\beta=3,50$. Decays into SUSY particles have not been taken into account and the common squark mass has been set to $M_S=1$~TeV.}}
\label{mssmtot}
\end{figure}

\begin{figure}
\unitlength 1cm
\begin{center}
\begin{picture}(11,15)
\put(-0.4,9.6){\epsfig{figure=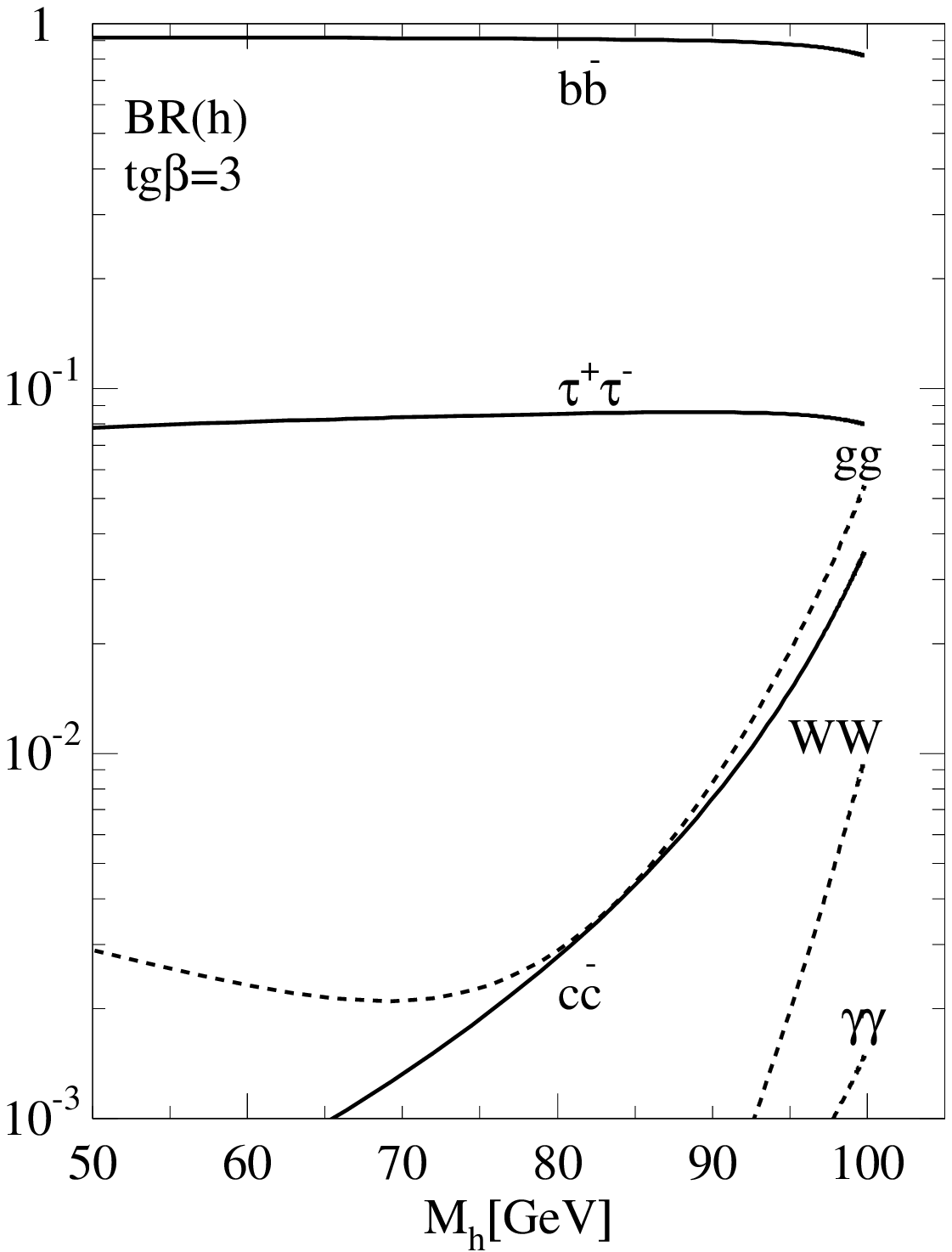,width=5.5cm}}
\put(5.9,9.6){\epsfig{figure=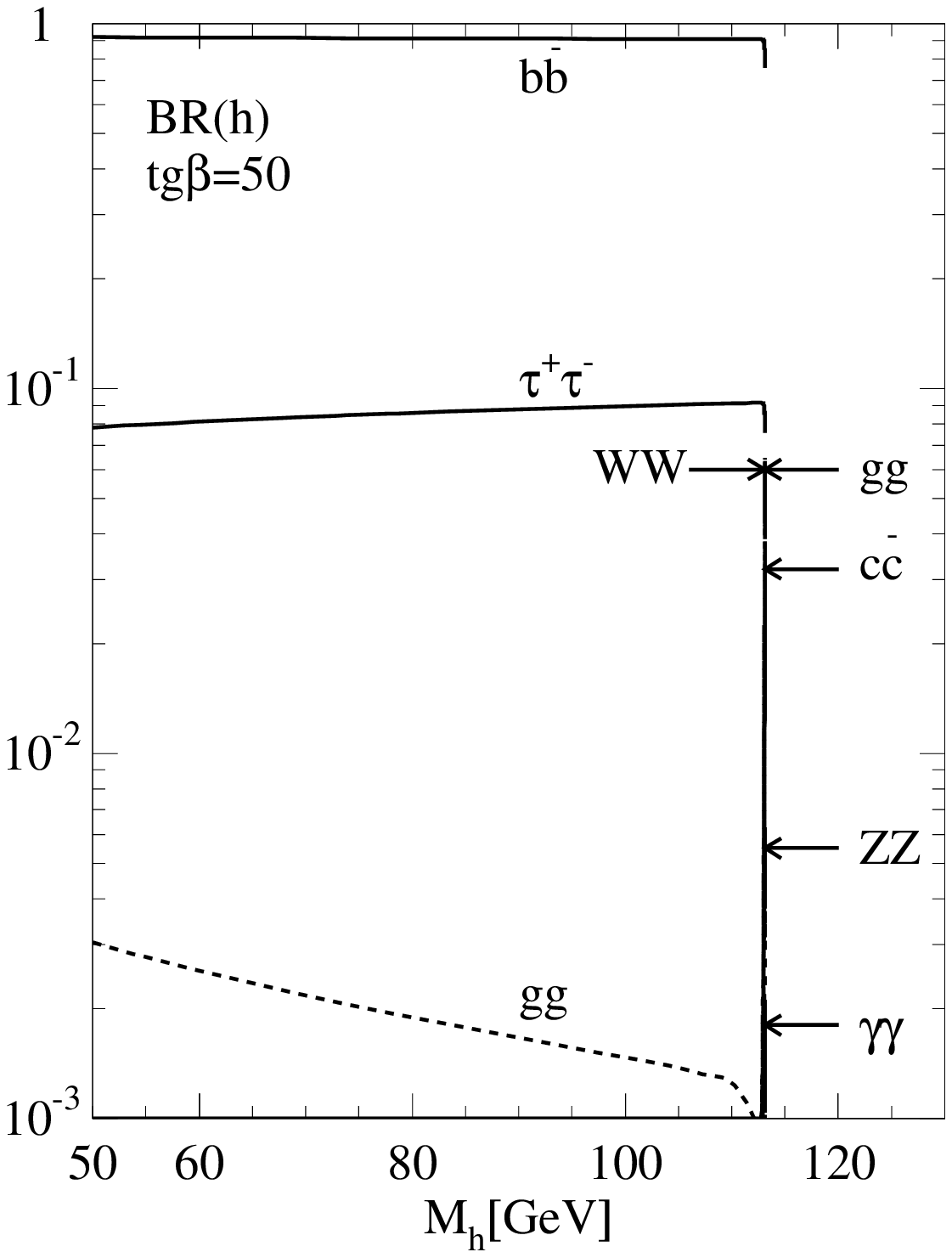,width=5.5cm}}
\put(-0.4,2.6){\epsfig{figure=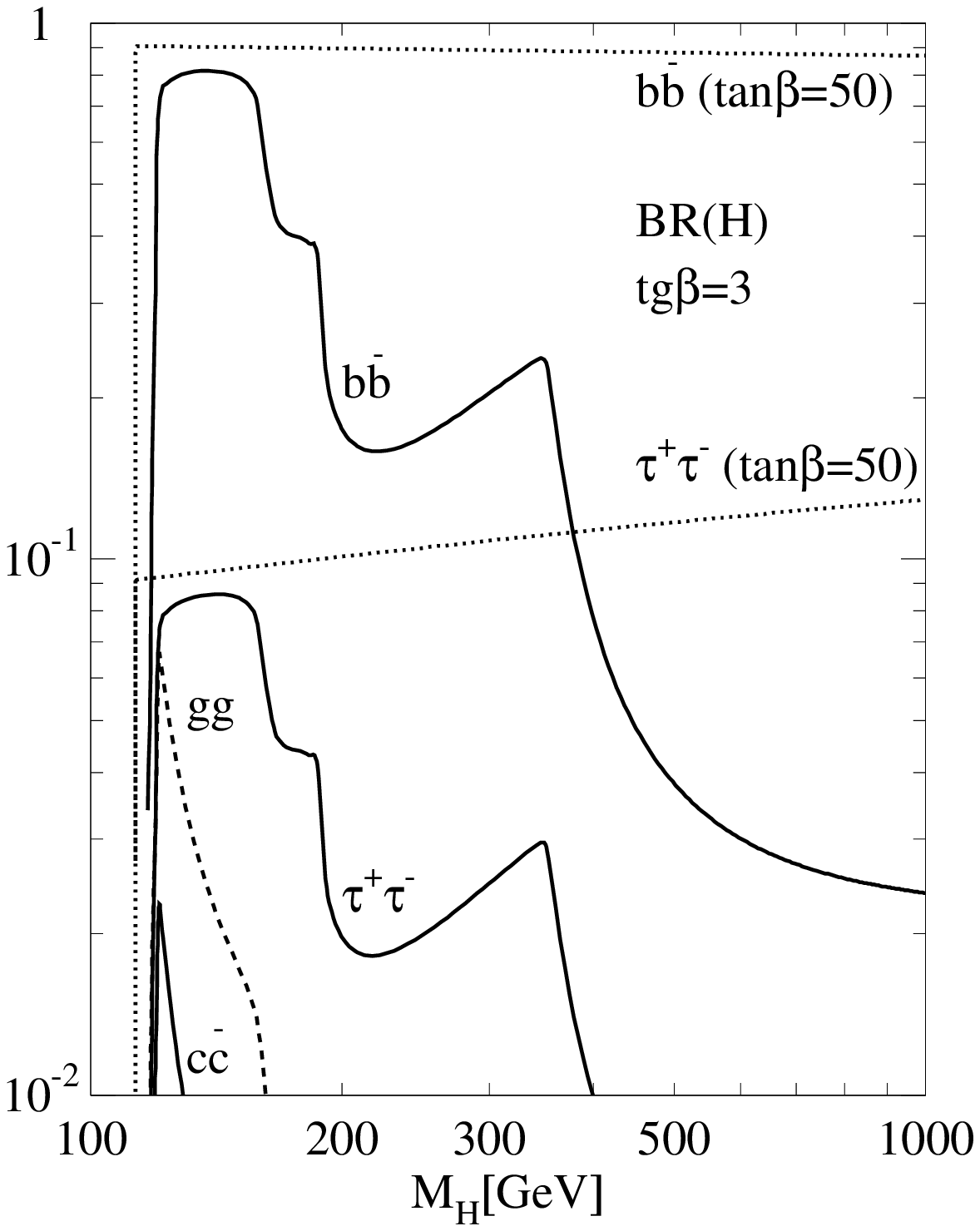,width=5.5cm}}
\put(5.9,2.6){\epsfig{figure=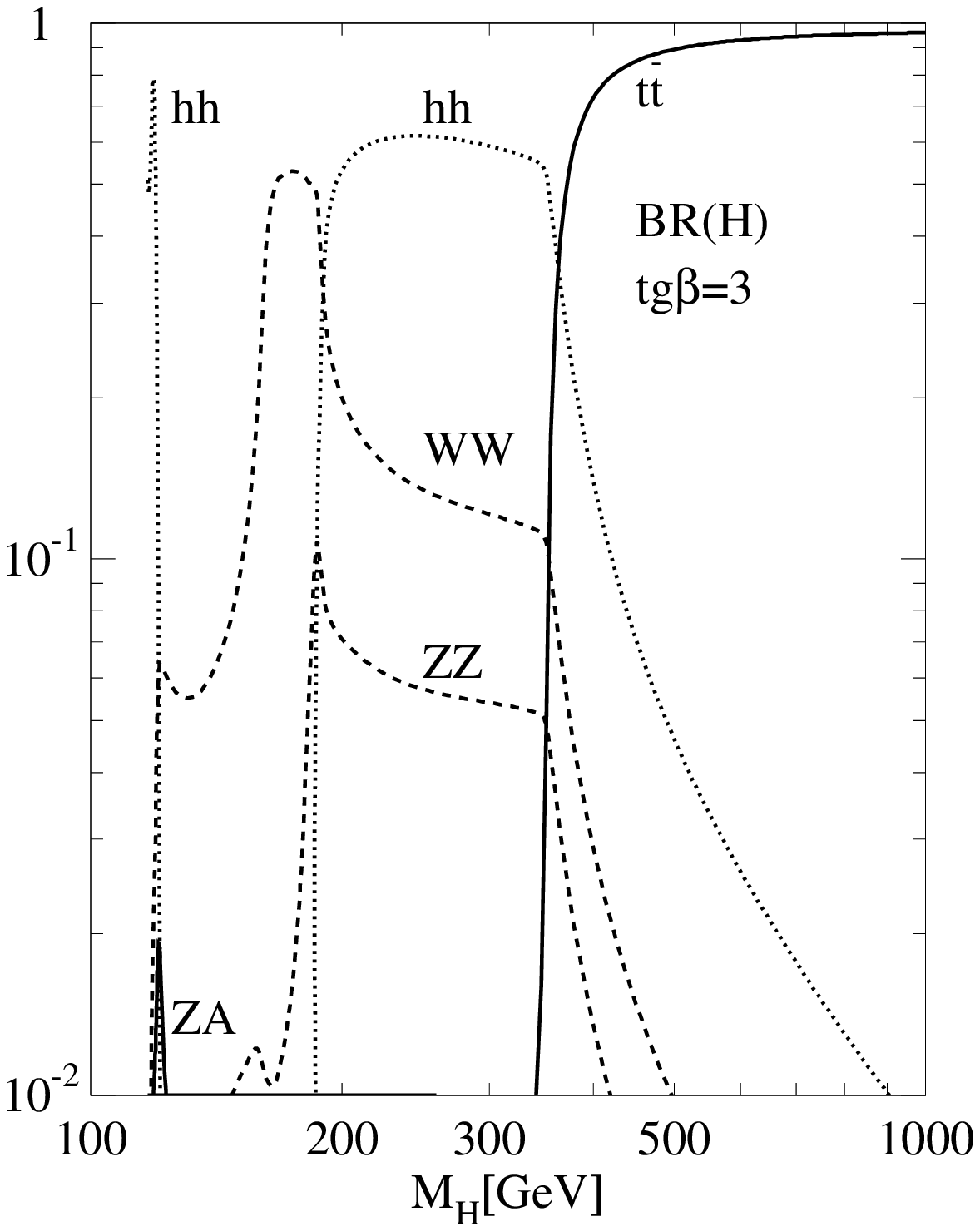,width=5.5cm}}
\put(-0.4,-4.4){\epsfig{figure=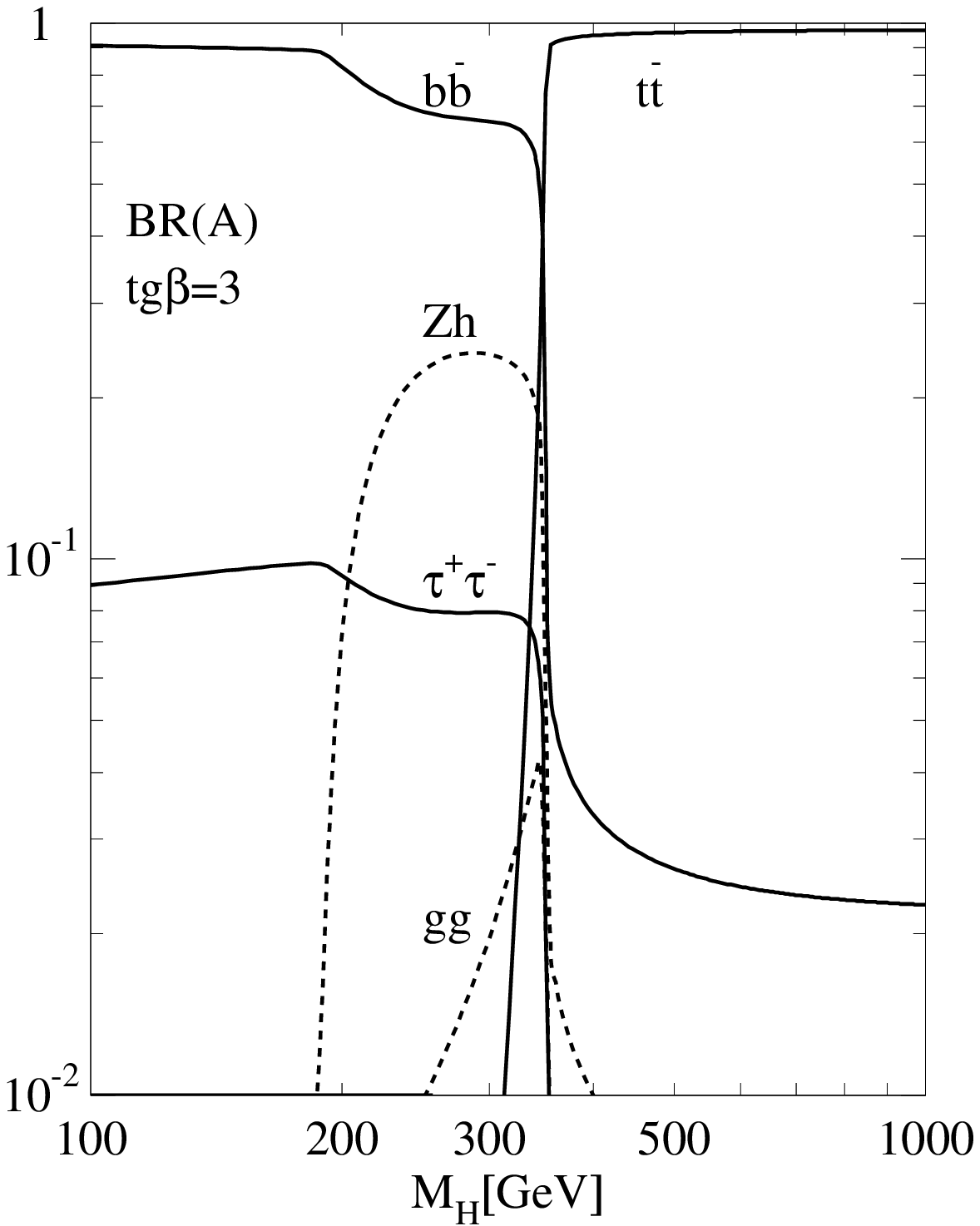,width=5.5cm}}
\put(5.9,-4.4){\epsfig{figure=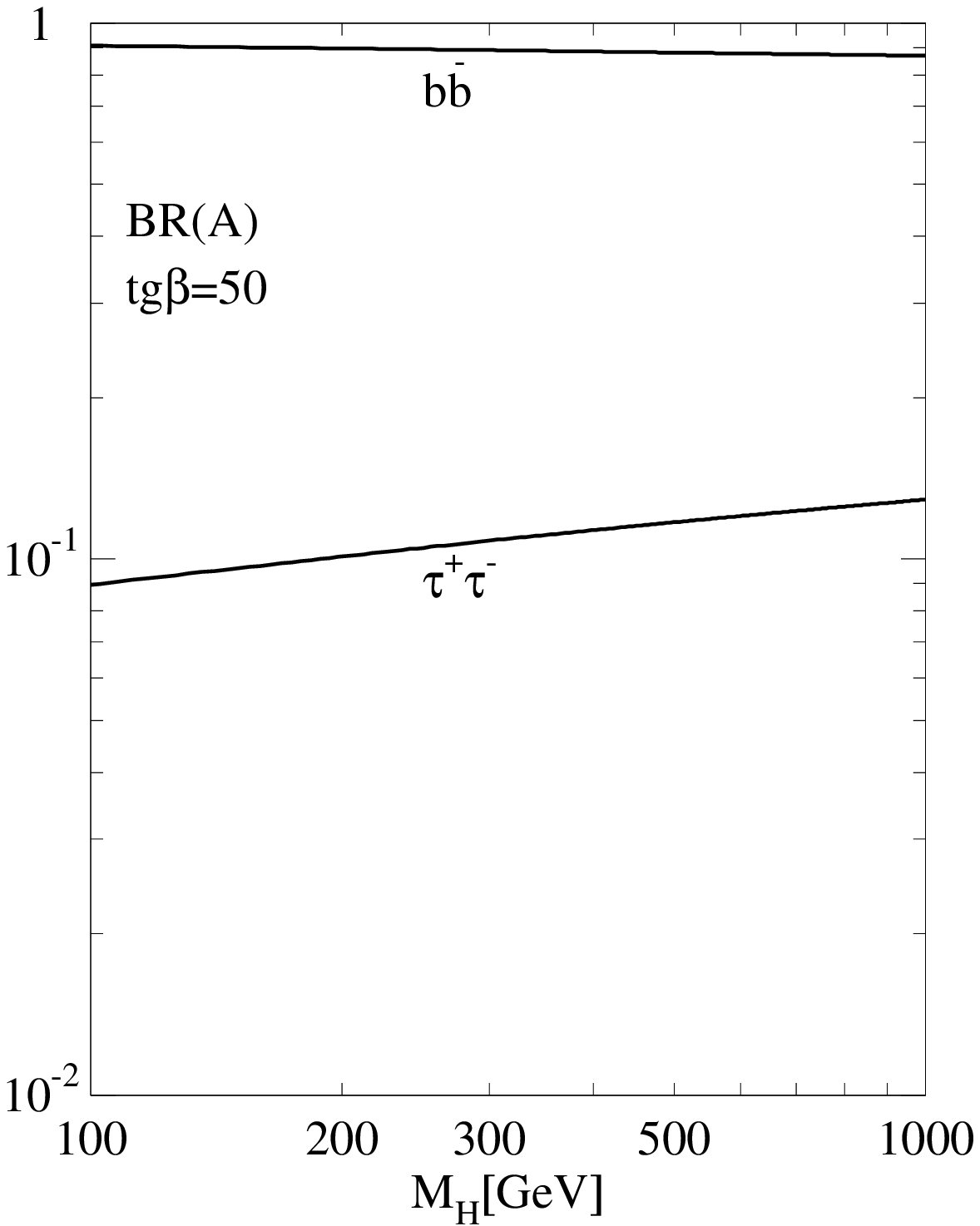,width=5.5cm}}
\end{picture}
\end{center}\vspace{3.8cm}
\caption{\textit{Branching ratios of the neutral MSSM Higgs bosons as a function of the corresponding Higgs mass for $\tan\beta=3,50$ and vanishing mixing. Decays into SUSY particles are not included and the common squark mass has been set to $M_S=1$~TeV.}}
\label{mssmbr}
\end{figure}

Figs.~\ref{mssmtot} and \ref{mssmbr} show the total decay widths of
the MSSM Higgs particles and the neutral Higgs boson branching ratios
into non-SUSY particles as a function of the corresponding Higgs mass for
two values of $\tan\beta = 3$, $50$. 
As can be inferred from Fig.~\ref{mssmbr} the light scalar Higgs boson
dominantly decays into bottom quarks, independent of $\tan\beta$, with
a branching ratio up to about 90\%.  The remaining 10\% are almost
solely supplemented by the decay into $\tau$'s.  At the maximum value
of the light Higgs mass, {\it i.e.} in the decoupling limit, the
branching ratios approach the SM values. \s

For large $\tan\beta$ the dominant decay mode of the heavy scalar
Higgs boson is the decay into bottom quarks (90\%), followed by the
decay into $\tau^+ \tau^-$ which amounts up to about 10\%. For small
values of $\tan\beta$, $H$ preferentially decays into a pair of light
Higgs bosons when kinematically allowed. Otherwise the $b$ decay takes
over the dominant r\^{o}le below the $t\bar{t}$ threshold. The next
important channels are provided by the decays into $W$ and $Z$ bosons,
respectively. Beyond the $t\bar{t}$ threshold the decay into top
quarks is by far dominant.\s

For large values of $\tan\beta$ the branching ratios of the
pseudoscalar Higgs show the same pattern as those for $H$.  For small
$\tan\beta$ values the bottom decay only dominates below the
$t\bar{t}$ threshold, the next important channels being the $Zh$ and
$\tau^+\tau^-$ decay. Above the $Zh$ and below the $t\bar{t}$
threshold the $gg$ channel turns up. As soon as the $t\bar{t}$
threshold is reached, the $t\bar{t}$ decay is overwhelming.

\subsection{Decays into SUSY particles}
If MSSM Higgs boson decays into charginos and neutralinos are
kinematically allowed they have to be taken into account. Their
branching ratios may yield 100\% below the various top-quark
thresholds so that the Higgs boson search at the LHC becomes difficult
because some of these decays develop invisible signatures
\cite{ohmann}.

The Higgs decays into sfermions of the first two generations are not
important. If decays into sfermions of the third generation are
kinematically allowed they can become sizeable, however, and reach
branching ratios of up to 80\%.

\section{Determination of the lifetime of Higgs bosons}
The lifetime of the Higgs bosons can be determined by exploiting the
relation
\beq
\Gamma_{tot} = \Gamma_i/BR_i
\eeq
where $\Gamma_i$ denotes the partial width of the decay $H\to
\mathrm{final \; state}\;i$ and $BR_i$ the corresponding branching
ratio. For this task several measurements have to be combined. The
branching ratio $BR_i$ is determined by measuring both the exclusive
process into the final state $i$ and the inclusive Higgs rate in the
same production channel $X\to H$:
\beq
\begin{array}{lll}
\sigma_{X\to H}^{incl} &\sim& \Gamma_{H\to X} \\
\sigma_{X\to i}^{excl} &\sim& \Gamma_{H\to X} BR_i
\end{array}
\eeq
Considering all possible production channels at $e^+e^-$ colliders the
Higgs-strahlung process $e^+e^-\to ZH$ is the only channel being
suitable for an inclusive measurement. The Higgs scalar is detected by
analysing the recoil spectrum of the $Z$ boson decay products that are
visible. In the case of the MSSM also the associated production
channels $Ah$ and $AH$ have to be taken into account.\s

The partial decay width $\Gamma_i$ is extracted from the inverse
process $i\to H$ or in associated production where the Higgs particle
is radiated off particle $i$. The Higgs boson can either decay into
the same state $i$ or another state $j$. In the latter case the
branching ratio $BR_j$ has also to be determined in the
Higgs-strahlung process.
\beq
\begin{array}{lrll}
{\phantom h} &\sigma_{i\to H\to i} & \sim & \Gamma_i BR_i \\
\mathrm{or} &\sigma_{i\to H\to j} & \sim & \Gamma_i BR_j 
\end{array}
\eeq
\indent If either the initial or the final state $i$ are not on-shell
the corresponding Higgs form factor will be probed at different scales
in production and decay, respectively, thus allowing new physics to
enter in the determination of the total width. These new effects are
associated with a scale $\Lambda$ at which they become sizeable. If
the scales $q^2$ at which the Higgs form factor is probed are small
compared to $\Lambda^2$, the error due to new physics in the extraction
of $\Gamma_{tot}$ is ${\cal O}(q^2/\Lambda^2)$.\s

In the following the mechanisms which can be exploited at high-energy
$e^+e^-$ colliders for the measurement of $\Gamma_{tot}$ both in the
SM and the MSSM in the intermediate mass range will be discussed from
a theoretical point of view and some illustrative numbers will be
given \cite{lpaper}. The analysis will be model-independent up to the
error originating from the scale dependence of the Higgs couplings.

\subsection{The total width of the SM Higgs boson}
In the intermediate mass range the relevant channels for the
extraction of the total width are given by $i=\gamma\gamma,WW$ and
$ZZ$. While the $\gamma$'s are on-shell, the $W$ and $Z$ bosons can be
off-shell. Demanding the $\gamma\gamma$ branching ratio
$BR_{\gamma\gamma}$ to be bigger than $10^{-3}$ and
$BR_{WW},BR_{ZZ}>10^{-2}$, the $\gamma\gamma$ channel can be
used for $M_H\,\lesim\, 155$~GeV, the $ZZ$ channel for $M_H\,\gesim\,
110$~GeV and the $WW$ channel in the whole intermediate range,
cf.~Fig.~\ref{smdecay}. For high-luminosity linear colliders this
corresponds to an event rate ${\cal O}(10)...{\cal O}(10^2)$ for
$\gamma\gamma$ and $WW/ZZ$ final states, respectively:
\beq
\begin{array}{lllr}
\sqrt{s} = 350\;\mathrm{GeV} & & & \non\\
M_h = 100...200\;\mathrm{GeV} & : & \sigma(ZH)=150...70\;\mathrm{fb} &
\mathrm{and} \; \int {\cal L} = 500\;\mathrm{fb}^{-1}
\end{array}
\eeq

\noindent {\bf a) $\gamma\gamma$ channel}\newline
From a theoretical point of view the $\gamma\gamma$ channel provides
the cleanest method for the de\-ter\-mi\-na\-tion of $\Gamma_{tot}$
since the photons are on-shell both in the production and the decay
channel. The $\gamma\gamma$ fusion Higgs production with the
$\gamma$'s resulting from the Compton scattering of laser light can be
used to extract the Higgs decay width into $\gamma\gamma$:
\beq
\sigma(\gamma\gamma\to H) = 8\pi^2 \frac{\Gamma(H\to \gamma\gamma)}{M_H}
\frac{d{\cal L}^{\gamma\gamma}}{ds_{\gamma\gamma}}
\eeq
$d{\cal L}^{\gamma\gamma}/ds_{\gamma\gamma}$ denotes the
$\gamma\gamma$ luminosity at $\sqrt{s_{\gamma\gamma}}=M_H$. The cross
section and hence the partial width can be determined with an accuracy
of about 2\% \cite{pwacc}. The on-shell $\gamma$'s ensure the
model-independence of the analysis. \s

Yet, the measurement of the branching ratio $BR_{\gamma\gamma}$ in the
Higgs-strahlung channel
\beq
e^+e^- \to ZH \to Z+\gamma\gamma
\eeq
is affected by a huge background rendering the error in
$BR_{\gamma\gamma}$ as big as about 15\% for $M_H\,\lesim\, 130$~GeV
\cite{brient}. This error translates into the maximum accuracy with which
$\Gamma_{tot}$ can be determined in the $\gamma\gamma$ channel.\s

\noindent {\bf b) $WW$ channel}\newline
The $WW$ channel is a valuable alternative to the $\gamma\gamma$ mode
\cite{bor}.  The branching ratio into $W$ bosons, with one $W$ being
virtual, in the lower intermediate Higgs mass range can be determined
with an error of ${\cal O}(5\%)$ down to Higgs masses $M_H\approx
120$~GeV \cite{bor,schrei} in the exclusive process
\beq
\sigma_{ZH\to ZWW} \equiv \sigma (e^+e^-\to ZH\to ZWW) 
\sim \Gamma_{ZZ} BR_{WW}
\eeq
[In the following the $W$ and $Z$ bosons may be understood as virtual
where appropriate.] The partial width in turn can be extracted from
the $WW$ fusion process $e^+e^-\to \bar{\nu}_e \nu_e H$
\citer{wfus2,wfus3} with
\beq
\sigma(e^+e^-\to \bar{\nu}_e\nu_eH) = \frac{2G_F^2M_W^4}{M_H^2\pi^2}
\ln\frac{s}{M_H^2} \frac{\Gamma(H\to WW)}{M_H}[1+\Delta_W]
\eeq
$\Delta_W$ parameterizes the non-zero mass effects for $H$ and $W$ in
the intermediate mass range. In the limit $M_W^2\ll M_H^2\ll s$ it
approaches zero, $\Delta_W\to 0$.\s

The $W$ bosons in the fusion process are off-shell with a maximum
virtuality of $\sim M_W/2$. Hence, the model dependence introduced by
anomalous $HWW$ couplings is negligible compared to the experimental
errors if this effect is associated with high scales $\Lambda =
1$~TeV.\s

The $H$ bosons produced in $WW$ fusion can be best detected in the
$b\bar{b}$ decay channel
\beq
\sigma_{WW\to b\bar{b}} \equiv \sigma(WW\to H\to b\bar{b}) \sim
\Gamma_{WW} BR_{b\bar{b}}
\eeq
Therefore in order to extract the partial width $\Gamma_{WW}$ the
branching ratio $BR(H\to b\bar{b})$ has to be measured, too. This can
be done in the exclusive process
\beq
\sigma_{ZH\to Zb\bar{b}} \equiv \sigma(e^+e^-\to ZH\to Zb\bar{b}) 
\sim \Gamma_{ZZ} BR_{b\bar{b}}
\eeq
Together with the inclusive process 
\beq
\sigma_{ZH} \equiv \sigma(e^+e^-\to ZH)\sim \Gamma_{ZZ}
\eeq
all ingredients necessary for the determination of $\Gamma_{tot}$ are
available
\beq
\Gamma_{tot} = \frac{\Gamma_{WW}}{BR_{WW}} \sim 
\frac{\sigma_{WW\to b\bar{b}}}{BR_{WW} BR_{b\bar{b}}}
\sim \frac{\sigma_{WW\to b\bar{b}}\sigma_{ZH}^2}{
\sigma_{ZH\to Zb\bar{b}}\sigma_{ZH\to ZWW}}
\eeq
For example, for $M_H=120$~GeV ($\sqrt{s}=350$~GeV, $\int {\cal
  L}=500$~fb$^{-1}$) the errors in the various cross sections are:
$(\delta\sigma/\sigma)_{WW \to b\bar{b}}\approx 2.6\%$ \cite{desch},
$(\delta\sigma/\sigma)_{ZH \to Zb\bar{b}}\approx 2.4\%$ \cite{bor} and
$(\delta\sigma/\sigma)_{ZH}\approx 2.1\%$ \cite{bor}. The error in
$\Gamma_{tot}$ is therefore dominated by the error in $BR_{WW}$ being
${\cal O}(5\%)$. This results in
$\delta\Gamma_{tot}/\Gamma_{tot}\approx 7.4\%$.\s

With rising Higgs mass the uncertainty in the determination
of $BR_{b\bar{b}}$ increases since the branching ratio decreases and
thus renders its measurement more difficult. For $M_H=140$~GeV an
error in $\Gamma_{tot}$ less than 10\% \cite{bor,desch,bpriv}
still seems feasible whereas for $M_H$ near and above the $H\to WW$
threshold the $b\bar{b}$ channel is not useful any more.\s

Instead the decay $H\to WW$ can be exploited in this mass range. The
Higgs mass has to be reconstructed from the four-jet final state
resulting from the decay of the $W$ bosons. In addition, for Higgs
masses where the decay $H\to ZZ$ is not negligible, the $ZZ$ and $WW$
modes have to be separated by reconstructing their invariant
2-jet-masses. For $M_H=160$~GeV ($\sqrt{s}=350$~GeV, $\int {\cal L} =
500$~fb$^{-1}$) the cross section $e^+e^-\to WW\to H \to WW$ can be
measured with an error of about 3.1\% if an efficiency of $\sim 30$\%
for the detection of the $WW$ mode is assumed. Hence, with
$(\delta\sigma/\sigma)_{ZH\to ZWW}=$3.5\% \cite{schrei} and
$(\delta\sigma/\sigma)_{ZH}=$2.8\% \cite{schrei} the error in
$\Gamma_{tot}$ will be 9.5\%.\s

\noindent {\bf c) $ZZ$ channel} \newline
Provided the Higgs mass is large enough the $ZZ$ channel can be
exploited for the measurement of the Higgs lifetime. The partial width
$\Gamma (H\to ZZ)$ is accessible in the Higgs-strahlung process
\cite{hrad2}
\beq
\sigma(e^+e^-\to ZH) = \frac{G_F(v_e^2+a_e^2)}{3\sqrt{2}} 
\frac{M_Z^4}{M_H^3 s}\Gamma(H\to ZZ)
[1+\Delta_Z]
\eeq
where $\Delta_Z$ accounts for the non-zero intermediate Higgs and $Z$
boson mass effects and $\Delta_Z\to 0$ for $M_Z^2\ll M_H^2 \ll s$. One
of the $Z$ bosons is virtual. The $HZZ$ coupling is measured at the
scale $q^2=s$ in this process whereas the exclusive process $e^+e^-\to
ZH \to Z+ZZ$ which provides the branching ratio $BR(H\to ZZ)$ probes
the $HZZ$ coupling at the scale $q^2=M_Z^2$. Hence, for
$\sqrt{s}=500$~GeV and $\Lambda=1$~TeV the error due to anomalous
effects in the $HZZ$ coupling may be $\sim 25\%$. By measuring
the Higgs-strahlung process at different c.m.~energies and
extrapolating down to $\sqrt{s}=M_Z$, the uncertainty due to new
physics effects can be reduced to ${\cal O}(M_Z^2/\Lambda^2)$. The
$ZZ$ fusion process $e^+e^-\to ZZ\to H$, though an order of magnitude
smaller than the $WW$ fusion process, may serve as a cross-check. With
a $Z$ boson virtuality of $\sim M_Z/2$ the error accounting for
anomalous $HZZ$ couplings is less than in Higgs-strahlung.\s

For the detection of the process $HZ\to ZZZ$ most of the $Z$ decay
channels can be used since the final state does not encounter any extra
neutrinos. Furthermore the $ZZZ$ continuum background is small so that
the error in the determination of the cross section is essentially
given by the total rate in this channel.

\subsection{Total widths of the MSSM Higgs particles}
Since the MSSM contains a quintet of Higgs bosons the extraction of
the lifetime of each Higgs scalar is more involved than in the SM. In
order to illustrate the strategy to be followed the lifetime of the
light CP-even Higgs boson $h$ will be discussed. The discussion may be
understood as synopsis of the qualitative features concerning
the determination of the lifetime. In order to make quantitative
statements, a detailed experimental analysis has to be performed. \s

Fig.~\ref{mssmbr} shows the dominant branching ratios of $h$ for
$\tan\beta=3$ and 50 (decays into SUSY particles are not included).
Outside the decoupling region the gauge boson channels play a minor
r\^{o}le whereas the $b\bar{b}$ channel is dominating in the whole
parameter range followed by the $\tau^+\tau^-$ channel. Especially for
large $\tan\beta$ values the Yukawa coupling $hbb\sim m_b\tan\beta$ is
strongly enhanced resulting in a total $h$ width much larger than in
the SM for the same Higgs mass. \s

\noindent {\bf a) Gauge boson channels} \newline
For the reasons described above the gauge boson channels are less
useful for the $h$ width measurement over most of the parameter range.
Only in the decoupling limit they approach the SM values and the same
methods as in the SM can be applied by properly taking into account
the changes in the couplings due to mixing effects in the Higgs
sector.\s

\noindent {\bf b) $b\bar{b}$ channel} \newline 
The branching ratio $BR(h\to b\bar{b})$ required for the
$\Gamma_{tot}$ determination in the $b\bar{b}$ channel can be
extracted by comparing the inclusive process $e^+e^-\to Zh$ with the
exclusive process $e^+e^-\to Zh\to Zb\bar{b}$. The measurement of the
partial width $\Gamma(h\to b\bar{b})$ or equivalently the coupling
$hb\bar{b}$ in the process $e^+e^-\to hb\bar{b}$ \citer{dkz,liao} is
much more involved. The Higgs-radiation cross section $e^+e^- \to
hb\bar{b}$ is small over a large part of the MSSM parameter range
including the decoupling region.  Being proportional to $\tan^2\beta$
it only becomes accessible for large values of $\tan\beta$. \s

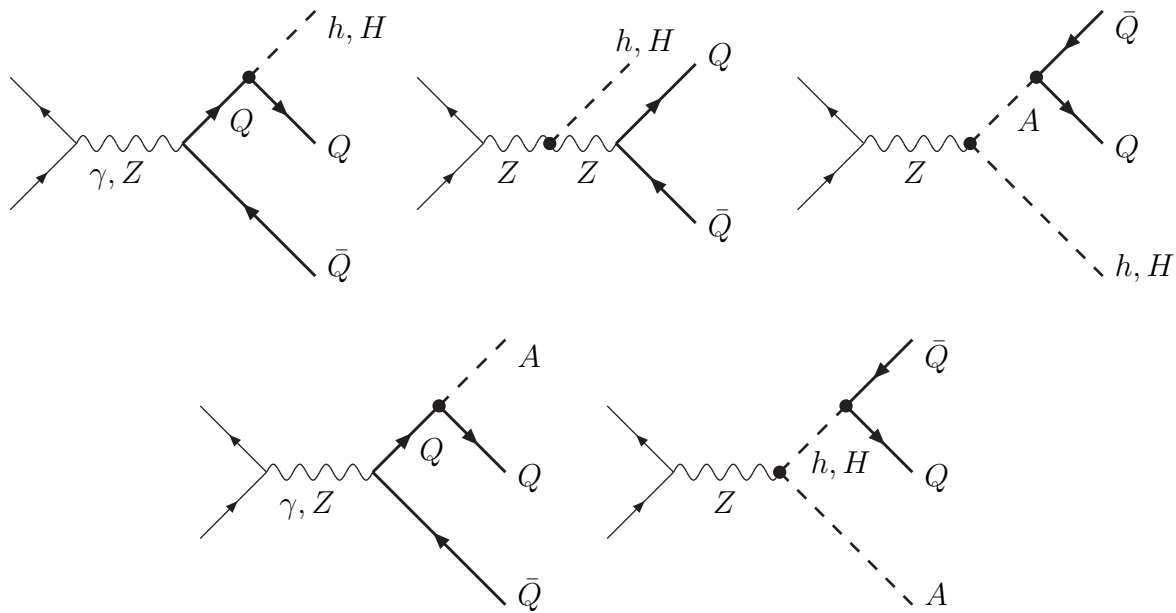
\begin{figure}[ht]
\begin{center}
\setlength{\unitlength}{1pt}
\noindent
\begin{picture}(140,100)(15,0)
\ArrowLine(15,25)(40,50)
\ArrowLine(40,50)(15,75)
\Photon(40,50)(80,50){3}{4}
\SetWidth{1.1}
\ArrowLine(80,50)(105,75)
\ArrowLine(105,75)(130,50)
\ArrowLine(130, 0)(80,50)
\SetWidth{1}
\DashLine(105,75)(130,100){5}
\Vertex(105,75){2.5}
\put(135,45){$Q$}
\put( 98,55){$Q$}
\put(135,90){$h,H$}
\put(135, 0){$\bar{Q}$}
\put(45,35){$\gamma,Z$}
\end{picture}
\begin{picture}(140,100)(5,0)
\ArrowLine(15,25)(40,50)
\ArrowLine(40,50)(15,75)
\Photon(40,50)(90,50){3}{5}
\SetWidth{1.1}
\ArrowLine(90,50)(120,80)
\ArrowLine(120,20)(90,50)
\SetWidth{1}
\DashLine(65,50)(95,80){5}
\Vertex(65,50){2.5}
\put(125,80){$Q$}
\put( 90,85){$h,H$}
\put(125,15){$\bar{Q}$}
\put(45,35){$Z$}
\put(75,35){$Z$}
\end{picture}
\begin{picture}(150,100)(5,0)
\ArrowLine(15,25)(40,50)
\ArrowLine(40,50)(15,75)
\Photon(40,50)(80,50){3}{4}
\SetWidth{1.1}
\ArrowLine(105,75)(130,50)
\ArrowLine(130,100)(105,75)
\SetWidth{1}
\DashLine(80,50)(105,75){5}
\DashLine(130, 0)(80,50){5}
\Vertex(105,75){2.5}
\Vertex(80,50){2.5}
\put(135,45){$Q$}
\put( 98,55){$A$}
\put(135,90){$\bar{Q}$}
\put(135, 0){$h,H$}
\put(55,35){$Z$}
\end{picture}
\\[2em]
\begin{picture}(140,100)(15,0)
\ArrowLine(15,25)(40,50)
\ArrowLine(40,50)(15,75)
\Photon(40,50)(80,50){3}{4}
\SetWidth{1.1}
\ArrowLine(80,50)(105,75)
\ArrowLine(105,75)(130,50)
\ArrowLine(130, 0)(80,50)
\SetWidth{1}
\DashLine(105,75)(130,100){5}
\Vertex(105,75){2.5}
\put(135,45){$Q$}
\put( 98,55){$Q$}
\put(135,90){$A$}
\put(135, 0){$\bar{Q}$}
\put(45,35){$\gamma,Z$}
\end{picture}
\begin{picture}(150,100)(5,0)
\ArrowLine(15,25)(40,50)
\ArrowLine(40,50)(15,75)
\Photon(40,50)(80,50){3}{4}
\SetWidth{1.1}
\ArrowLine(105,75)(130,50)
\ArrowLine(130,100)(105,75)
\SetWidth{1}
\DashLine(80,50)(105,75){5}
\DashLine(130, 0)(80,50){5}
\Vertex(105,75){2.5}
\Vertex(80,50){2.5}
\put(135,45){$Q$}
\put( 92,50){$h,H$}
\put(135,90){$\bar{Q}$}
\put(135, 0){$A$}
\put(55,35){$Z$}
\end{picture}
\end{center}
\caption[ ]{\it Diagrams contributing to the radiation of scalar and pseudoscalar MSSM Higgs bosons in $e^+e^-$ collisions; $Q=t,b$ \cite{liao}.}
\label{hraddia}
\end{figure}

The diagrams contributing to the processes $e^+e^-\to hb\bar{b}$ and
$e^+e^-\to H/A b\bar{b}$ are shown in Fig.~\ref{hraddia}. Since the
associated production of $h$ with $b\bar{b}$ also includes the coupling
$Ab\bar{b}$, the determination of the required Yukawa coupling is more
complicated than in the SM. Yet, each coupling $\Phi b\bar{b}$
($\Phi=h,H,A$) can be extracted by solving the equation system for all
three processes. Assuming polarized electron/positron beams and the
knowledge of the couplings $g_{ZZh/H}$ and $g_{ZAh/H}$ the cross
sections for the processes $e^+e^-\to H_i b\bar{b}$ read
\beq
\begin{array}{lll}
\sigma_{hb\bar{b}} &=& |g_{hb\bar{b}} A_1 + g_{Ab\bar{b}} A_2 + A_3 |^2 \\
\sigma_{Hb\bar{b}} &=& |g_{Hb\bar{b}} A'_1 + g_{Ab\bar{b}} A'_2 + A'_3 |^2 \\
\sigma_{Ab\bar{b}} &=& |g_{hb\bar{b}} A''_1 + g_{Hb\bar{b}} A''_2 +
 g_{Ab\bar{b}} A''_3|^2
\end{array}
\eeq
where $A_i,A_i',A_i''$ ($i=1,2,3$) denote the helicity amplitudes of
the corresponding subprocesses. Up to discrete ambiguities this system
is in principle solvable for all $b\bar{b}$ Higgs Yukawa couplings.\s

As is evident from the previous discussion and Fig.~\ref{hraddia} the
final state $b\bar{b}b\bar{b}$ includes several resonant channels. By
applying experimental cuts the various resonances have to be separated
so that the contribution due to $b\to bh$ can be extracted. The Yukawa
coupling $hb\bar{b}$ {\it e.g.}~could be determined in the region
where one of the possible $b$-pairings forms a resonance at the $h$
mass and the invariant mass of the remaining $b$-pair does not
reconstruct the pseudoscalar mass $M_A$.
In the decoupling limit $h$ behaves like a SM boson and is not
affected by heavy Higgs boson resonances. \s

\noindent {\bf c) $\tau^+\tau^-$ channel}\newline
Since the $\tau^+\tau^-$ Higgs Yukawa coupling is proportional
$m_{\tau}\tan\beta$ this channel can be exploited for large
$\tan\beta$, too. On the one hand no ambiguity in pairing the $b$-jets
arises here. On the other hand the neutrino in the $\tau$ decay
requires the reconstruction of the invariant $\tau^+\tau^-$ mass, which
is necessary for the separation of the $A$ resonance, from the mass
recoiling against the $b\bar{b}$ pair. 

%% file: chapter5.tex
\chapter{SM and MSSM Higgs self-couplings}
In the preceding chapters the discovery of Higgs particles in the SM
and the MSSM, the determination of their masses and lifetimes and the
Higgs couplings to gauge and matter particles have been described in
detail. In order to complete the profile of the Higgs bosons, the
Higgs self-couplings have to be probed.  This step is essential
for the clarification of the nature of the mechanism that creates
particle masses in the SM and its supersymmetric extensions. Only the
knowledge of the Higgs self-couplings will allow for the
reconstruction of the Higgs potential so that the Higgs mechanism can
be established experimentally. \s

In this chapter various double and triple Higgs production mechanisms
in the SM and the MSSM will be analysed \cite{ours}. [The relation to
general 2-Higgs doublet models has been discussed in
Ref.~\cite{dubinin}.] At hadron colliders and high-energy $e^+e^-$
linear colliders Higgs pair production via double Higgs-strahlung off
$W$ or $Z$ bosons \citer{gounaris,ilyin} and through $WW$ or $ZZ$
fusion \citer{ilyin,eboli} allows direct access to the trilinear Higgs
self-couplings. In addition at $pp$ colliders gluon-gluon fusion
\cite{dawson},\citer{eboli,dreeseal} can be exploited, and at photon
colliders high-energy $\gamma\gamma$ fusion
\cite{ilyin,boudjema,jikia} is sensitive to $\lambda_{HHH}$. For the
MSSM case all possible neutral multi-Higgs final states as defined in
Ref.~\cite{djouadi2} will be taken into account and thus a
theoretically complete picture for testing the trilinear Higgs
self-couplings will be elaborated. By including new results from
parton level background analyses \citer{moretti,prochouches} and
detector simulations \citer{houches,gay} the potential of measuring
the trilinear Higgs self-couplings at $e^+e^-$ linear colliders and at
LHC, respectively, will be discussed. A short remark will be made on
the present expectations of determining the quadrilinear Higgs
self-coupling(s).  The first section will be dedicated to the linear
collider, first examining the SM case before turning on to the MSSM.
The second section will focus on the LHC.\s

\section{Higgs self-couplings at $e^+e^-$ linear colliders}
The two main processes at $e^+e^-$ linear colliders which are
sensitive to $\lambda_{HHH}$ are double Higgs-strahlung and $WW$
double-Higgs fusion:
\beq
\begin{array}{l l l c l}
\mbox{double Higgs-strahlung}& \hspace{-0.3cm} : & e^+e^- & 
\hspace{-0.3cm} \longrightarrow &\hspace{-0.1cm}  ZHH \\
\\[-0.8cm]
& & & \hspace{-0.3cm} \scriptstyle{Z} & \\[0.1cm]
WW\ \mbox{double-Higgs fusion}& \hspace{-0.3cm} : & e^+e^- & 
\hspace{-0.3cm} \longrightarrow & \hspace{-0.1cm} \bar{\nu}_e \nu_e HH 
\\ \\[-0.8cm]
& & & \hspace{-0.3cm} \scriptstyle{WW} &
\end{array} 
\eeq
The cross section for $ZZ$ fusion is an order of magnitude smaller
than that for $WW$ fusion because of the small electron-$Z$ couplings.\s 
\begin{fmffile}{fd}
\begin{figure}
\begin{flushleft}
\underline{double Higgs-strahlung: $e^+e^-\to ZHH$}\\[1.5\baselineskip]
{\footnotesize
\unitlength1mm
\hspace{10mm}
\begin{fmfshrink}{0.7}
\begin{fmfgraph*}(24,12)
  \fmfstraight
  \fmfleftn{i}{3} \fmfrightn{o}{3}
  \fmf{fermion}{i1,v1,i3}
  \fmflabel{$e^-$}{i1} \fmflabel{$e^+$}{i3}
  \fmf{boson,lab=$Z$,lab.s=left,tens=3/2}{v1,v2}
  \fmf{boson}{v2,o3} \fmflabel{$Z$}{o3}
  \fmf{phantom}{v2,o1}
  \fmffreeze
  \fmf{dashes,lab=$H$,lab.s=right}{v2,v3} \fmf{dashes}{v3,o1}
  \fmffreeze
  \fmf{dashes}{v3,o2} 
  \fmflabel{$H$}{o2} \fmflabel{$H$}{o1}
  \fmfdot{v3}
\end{fmfgraph*}
\hspace{15mm}
\begin{fmfgraph*}(24,12)
  \fmfstraight
  \fmfleftn{i}{3} \fmfrightn{o}{3}
  \fmf{fermion}{i1,v1,i3}
  \fmf{boson,lab=$Z$,lab.s=left,tens=3/2}{v1,v2}
  \fmf{dashes}{v2,o1} \fmflabel{$H$}{o1}
  \fmf{phantom}{v2,o3}
  \fmffreeze
  \fmf{boson}{v2,v3,o3} \fmflabel{$Z$}{o3}
  \fmffreeze
  \fmf{dashes}{v3,o2} 
  \fmflabel{$H$}{o2} \fmflabel{$H$}{o1}
\end{fmfgraph*}
\hspace{15mm}
\begin{fmfgraph*}(24,12)
  \fmfstraight
  \fmfleftn{i}{3} \fmfrightn{o}{3}
  \fmf{fermion}{i1,v1,i3}
  \fmf{boson,lab=$Z$,lab.s=left,tens=3/2}{v1,v2}
  \fmf{dashes}{v2,o1} \fmflabel{$H$}{o1}
  \fmf{dashes}{v2,o2} \fmflabel{$H$}{o2}
  \fmf{boson}{v2,o3} \fmflabel{$Z$}{o3}
\end{fmfgraph*}
\end{fmfshrink}
}
\\[2\baselineskip]
\underline{$WW$ double-Higgs fusion: $e^+e^-\to \bar\nu_e\nu_e HH$}\\[1.5\baselineskip]
{\footnotesize
\unitlength1mm
\hspace{10mm}
\begin{fmfshrink}{0.7}
\begin{fmfgraph*}(24,20)
  \fmfstraight
  \fmfleftn{i}{8} \fmfrightn{o}{8}
  \fmf{fermion,tens=3/2}{i2,v1} \fmf{phantom}{v1,o2}
  \fmflabel{$e^-$}{i2}
  \fmf{phantom}{o7,v2} \fmf{fermion,tens=3/2}{v2,i7}
  \fmflabel{$e^+$}{i7}
  \fmffreeze
  \fmf{fermion}{v1,o1} \fmflabel{$\nu_e$}{o1}
  \fmf{fermion}{o8,v2} \fmflabel{$\bar\nu_e$}{o8}
  \fmf{boson}{v1,v3} 
  \fmf{boson}{v3,v2}
  \fmf{dashes,lab=$H$}{v3,v4}
  \fmf{dashes}{v4,o3} \fmf{dashes}{v4,o6}
  \fmflabel{$H$}{o3} \fmflabel{$H$}{o6}
  \fmffreeze
  \fmf{phantom,lab=$W$,lab.s=left}{v1,x1} \fmf{phantom}{x1,v3} 
  \fmf{phantom,lab=$W$,lab.s=left}{x2,v2} \fmf{phantom}{v3,x2}
  \fmfdot{v4}
\end{fmfgraph*}
\hspace{15mm}
\begin{fmfgraph*}(24,20)
  \fmfstraight
  \fmfleftn{i}{8} \fmfrightn{o}{8}
  \fmf{fermion,tens=3/2}{i2,v1} \fmf{phantom}{v1,o2}
  \fmf{phantom}{o7,v2} \fmf{fermion,tens=3/2}{v2,i7}
  \fmffreeze
  \fmf{fermion}{v1,o1}
  \fmf{fermion}{o8,v2}
  \fmf{boson}{v1,v3} 
  \fmf{boson}{v4,v2}
  \fmf{boson,lab=$W$,lab.s=left}{v3,v4}
  \fmf{dashes}{v3,o3} \fmf{dashes}{v4,o6}
  \fmflabel{$H$}{o3} \fmflabel{$H$}{o6}
\end{fmfgraph*}
\hspace{15mm}
\begin{fmfgraph*}(24,20)
  \fmfstraight
  \fmfleftn{i}{8} \fmfrightn{o}{8}
  \fmf{fermion,tens=3/2}{i2,v1} \fmf{phantom}{v1,o2}
  \fmf{phantom}{o7,v2} \fmf{fermion,tens=3/2}{v2,i7}
  \fmffreeze
  \fmf{fermion}{v1,o1}
  \fmf{fermion}{o8,v2}
  \fmf{boson}{v1,v3} 
  \fmf{boson}{v3,v2}
  \fmf{dashes}{v3,o3} \fmf{dashes}{v3,o6}
  \fmflabel{$H$}{o3} \fmflabel{$H$}{o6}
\end{fmfgraph*}
\end{fmfshrink}
}
\end{flushleft}
\caption{\textit{
Generic diagrams for the processes which contribute to Higgs-pair production 
in the Standard Model at $e^+e^-$ linear colliders: double Higgs-strahlung 
and $WW$ fusion.}}
\label{fig:diag}
\end{figure}
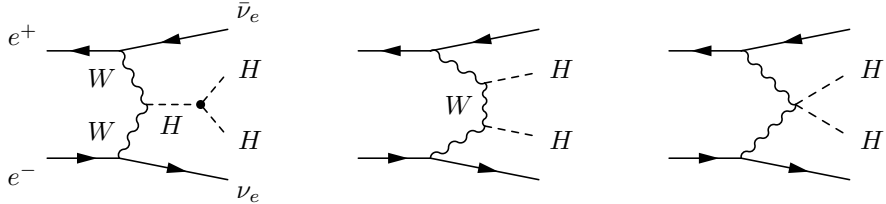

Fig.~\ref{fig:diag} shows the generic diagrams which contribute to the
two processes. Their cross sections are small, of the order of a
few fb and below, for Higgs masses in the intermediate range so that
high-luminosities as expected at $e^+e^-$ linear colliders are needed
in order to produce sufficiently high signal rates and to separate the
signal from the background.\s

\begin{figure}
\begin{center}
\epsfig{figure=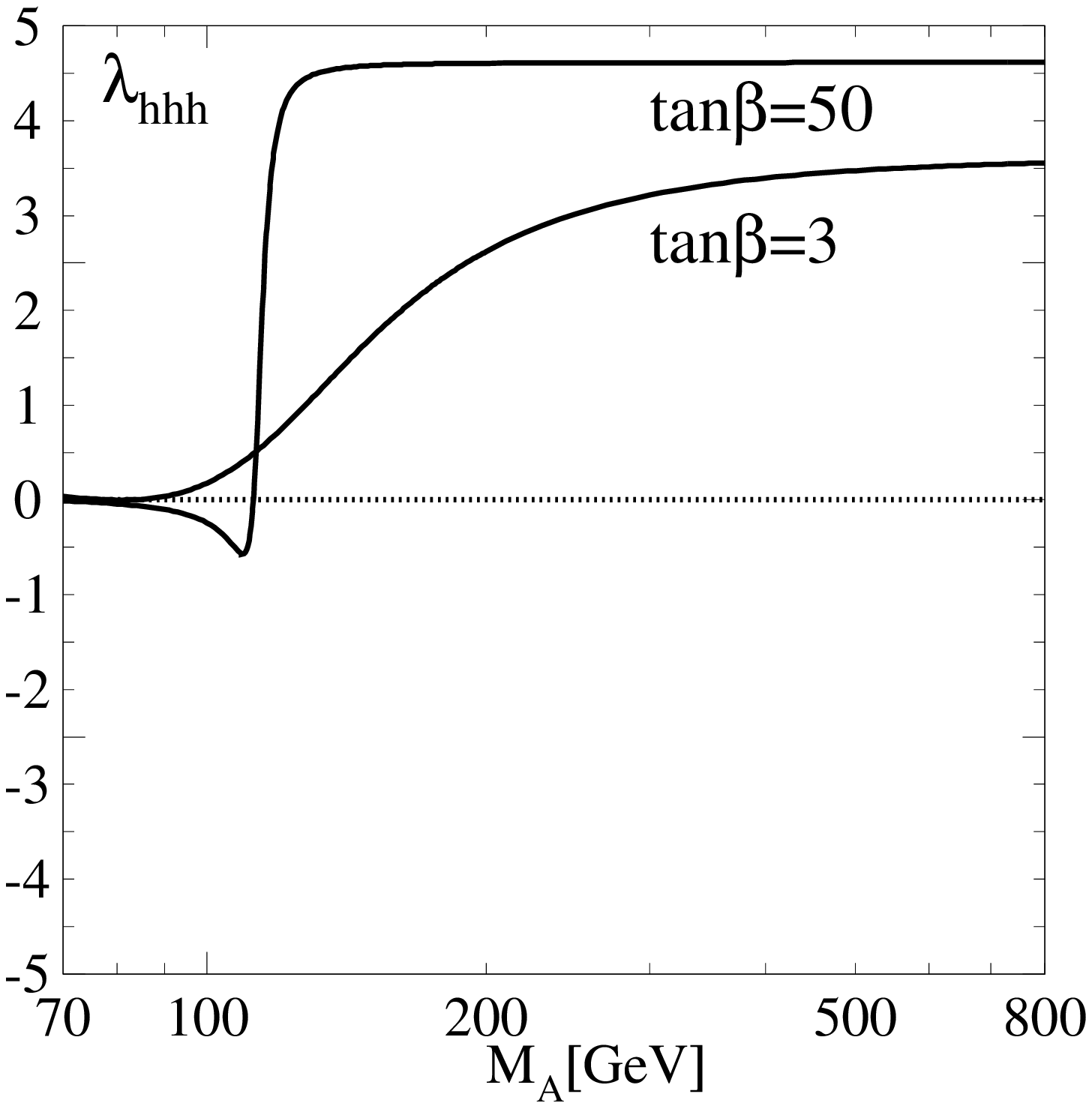,width=7.5cm}
\hspace{0.5cm}
\epsfig{figure=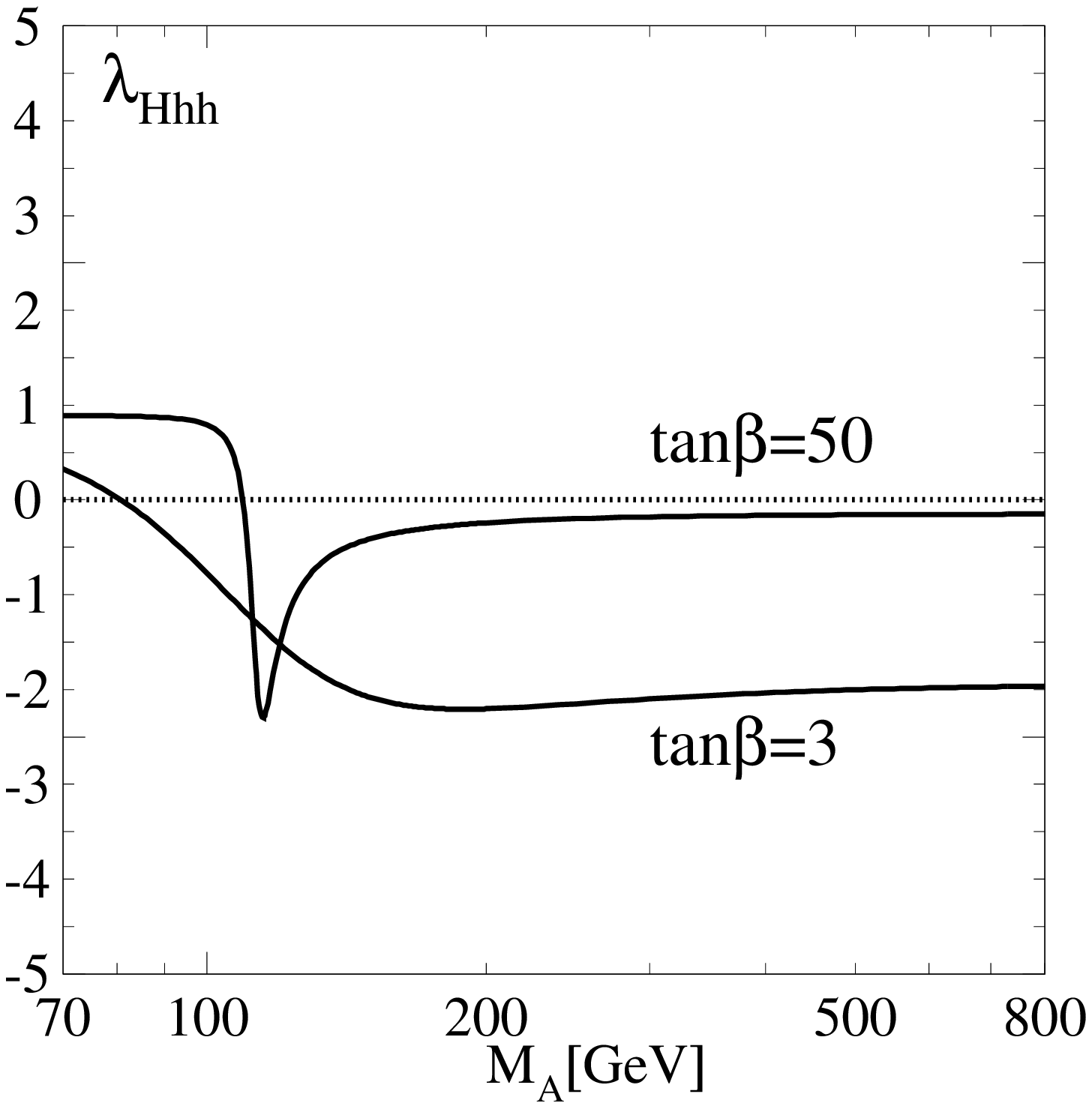,width=7.5cm}\\[2cm]
\end{center}
\begin{center}
\epsfig{figure=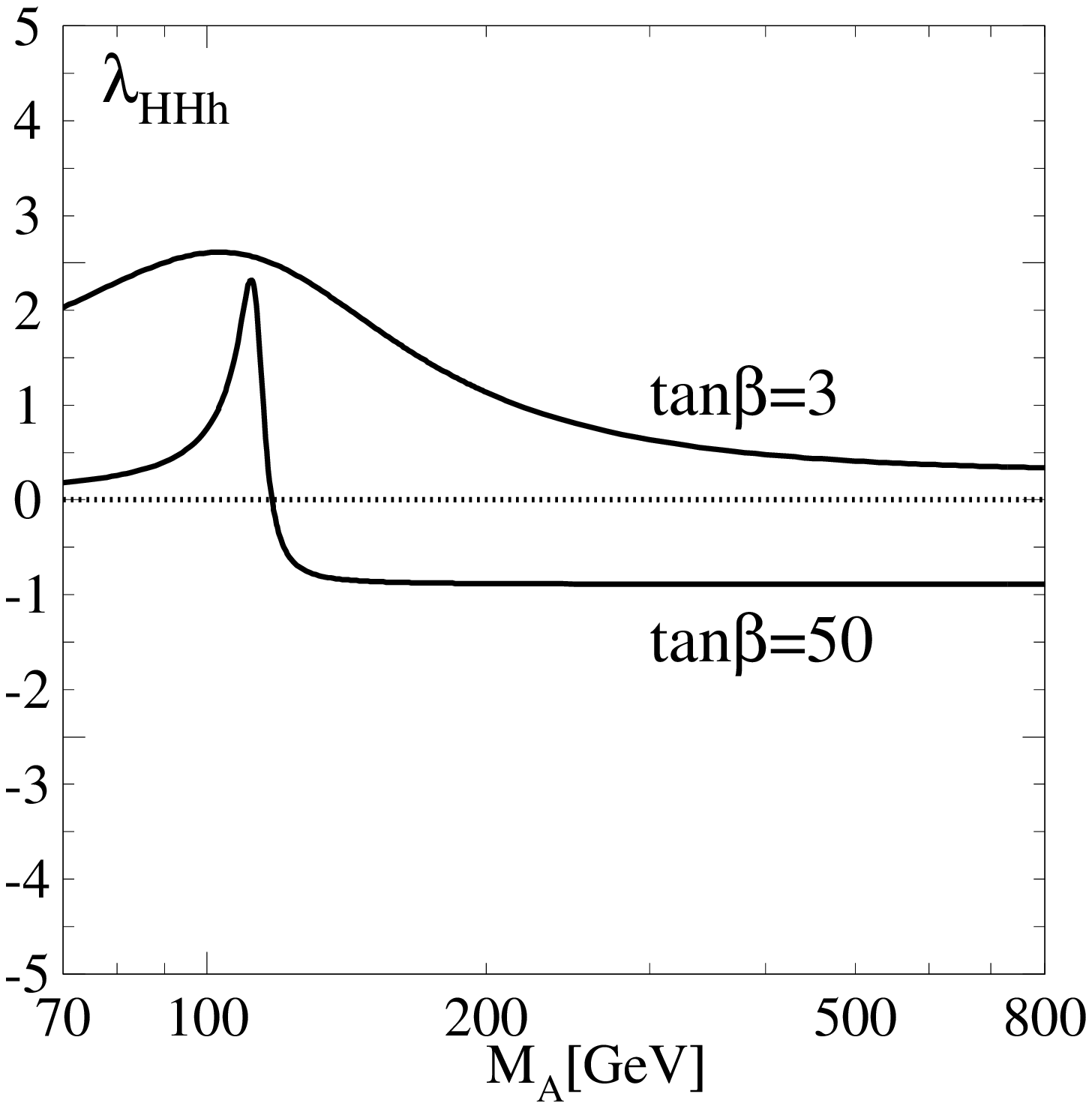,width=7.5cm}
\hspace{0.5cm}
\epsfig{figure=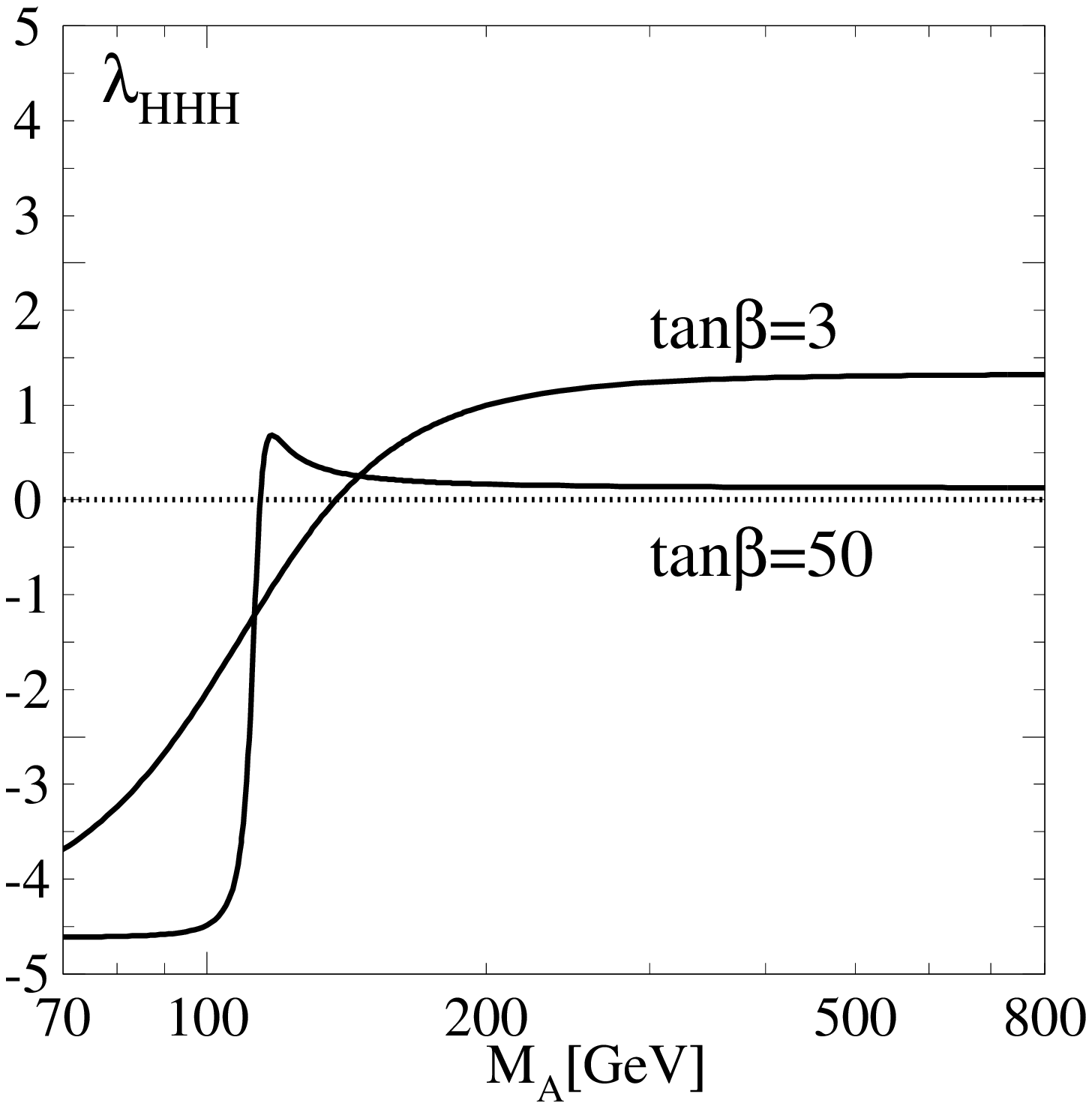,width=7.5cm}
\\[1cm]
\caption{\textit{Variation of the trilinear couplings between CP-even Higgs 
bosons with $M_A$ for tan$\beta = 3$ and $50$ in the MSSM; the region 
of rapid variations corresponds to the $h/H$ cross-over region in the 
neutral CP-even sector.}}
\label{fig:lambda1}
\end{center}
\end{figure}

\begin{figure}
\begin{center}
\epsfig{figure=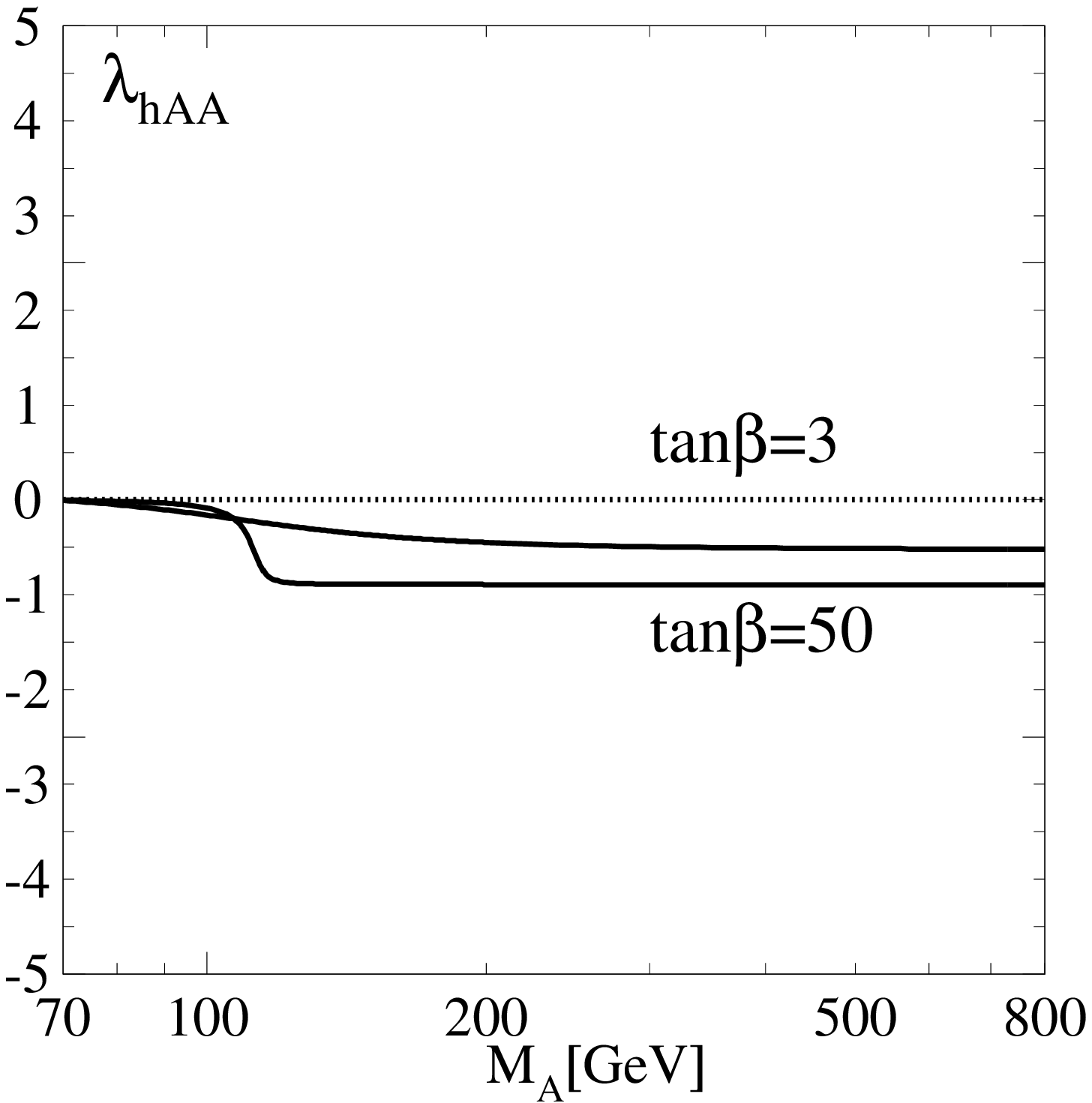,width=7.5cm}
\hspace{0.5cm}
\epsfig{figure=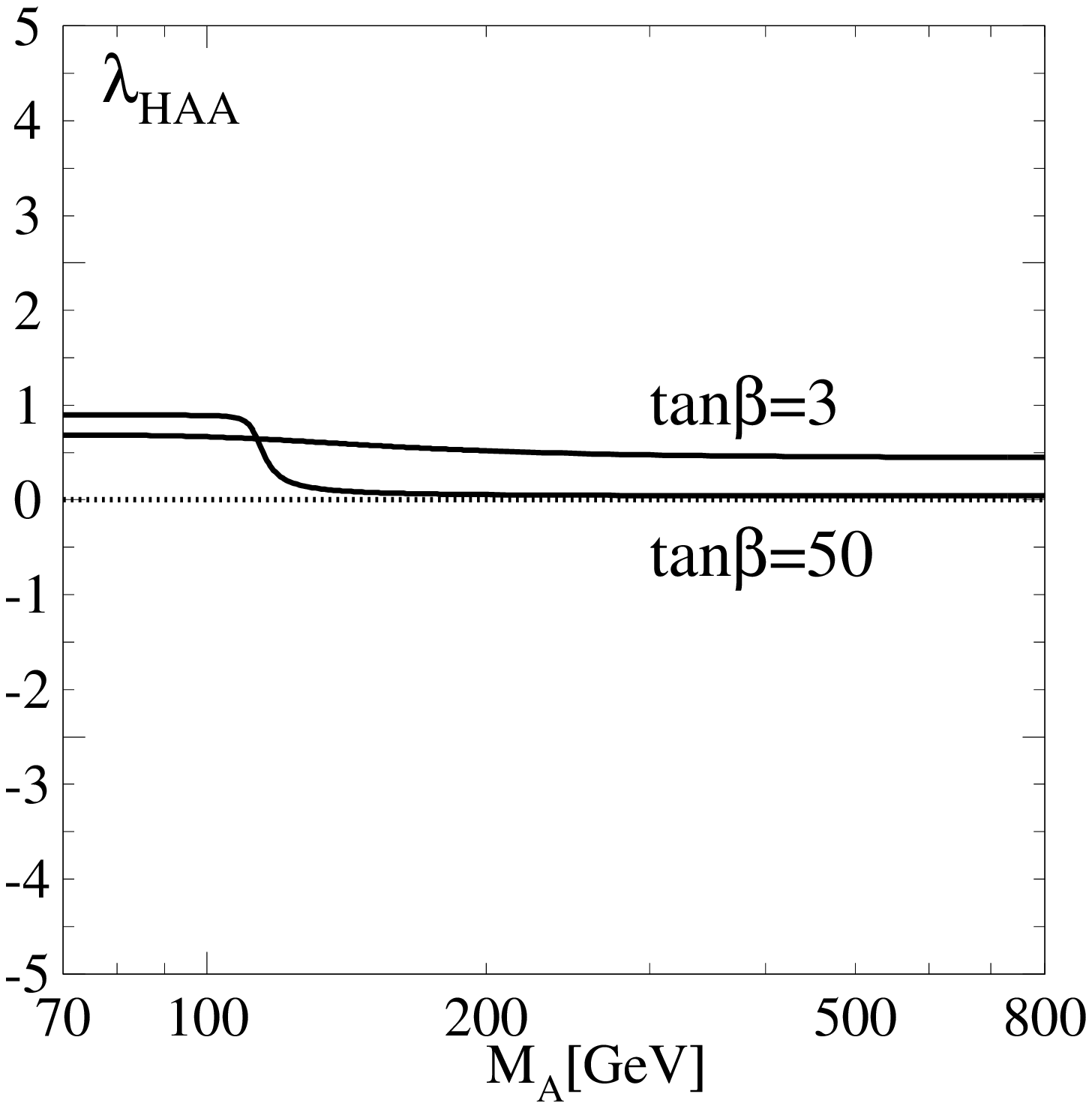,width=7.5cm}\\[2cm]
\end{center}
\begin{center}
\epsfig{figure=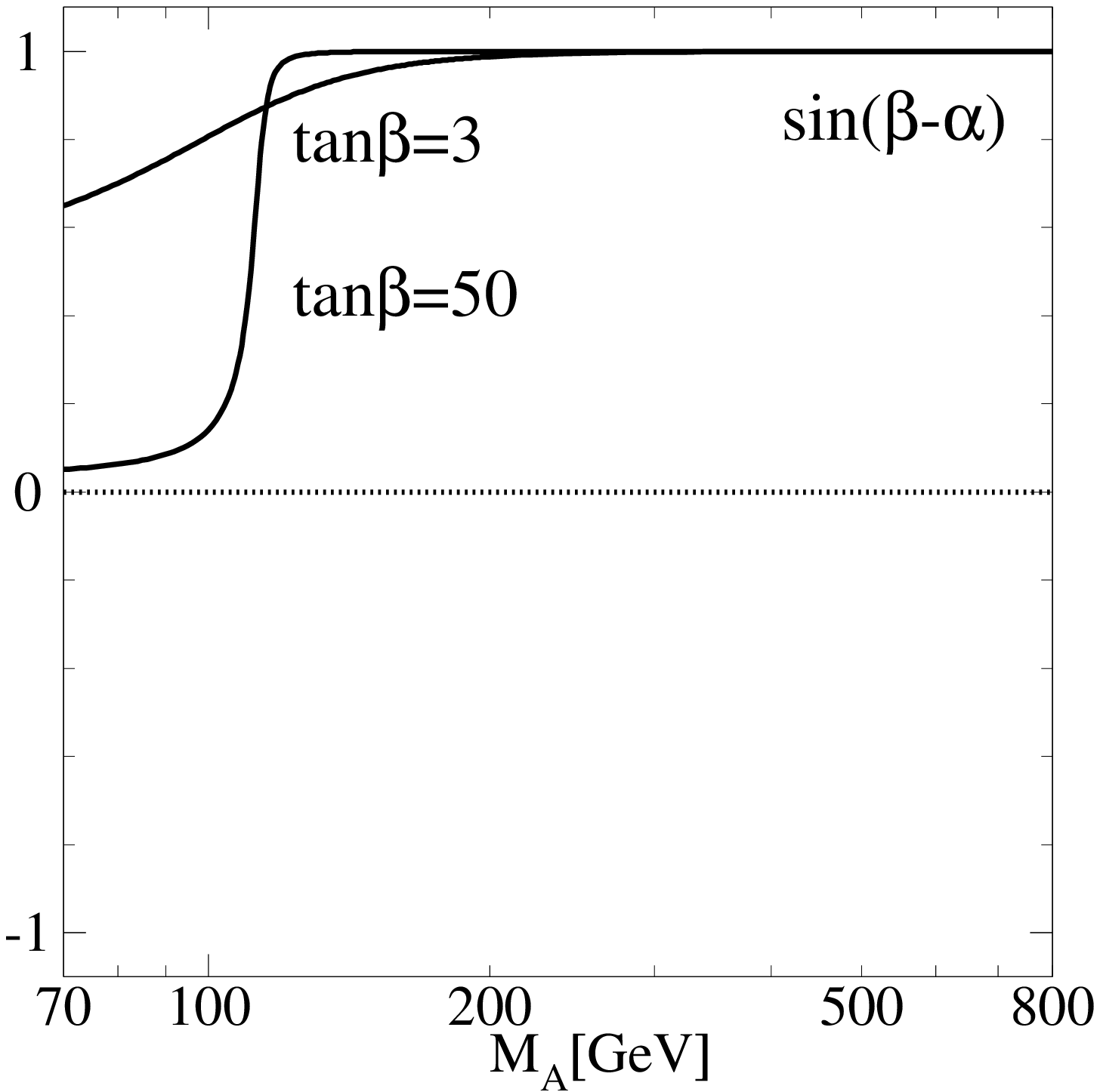,width=7.5cm}
\hspace{0.5cm}
\epsfig{figure=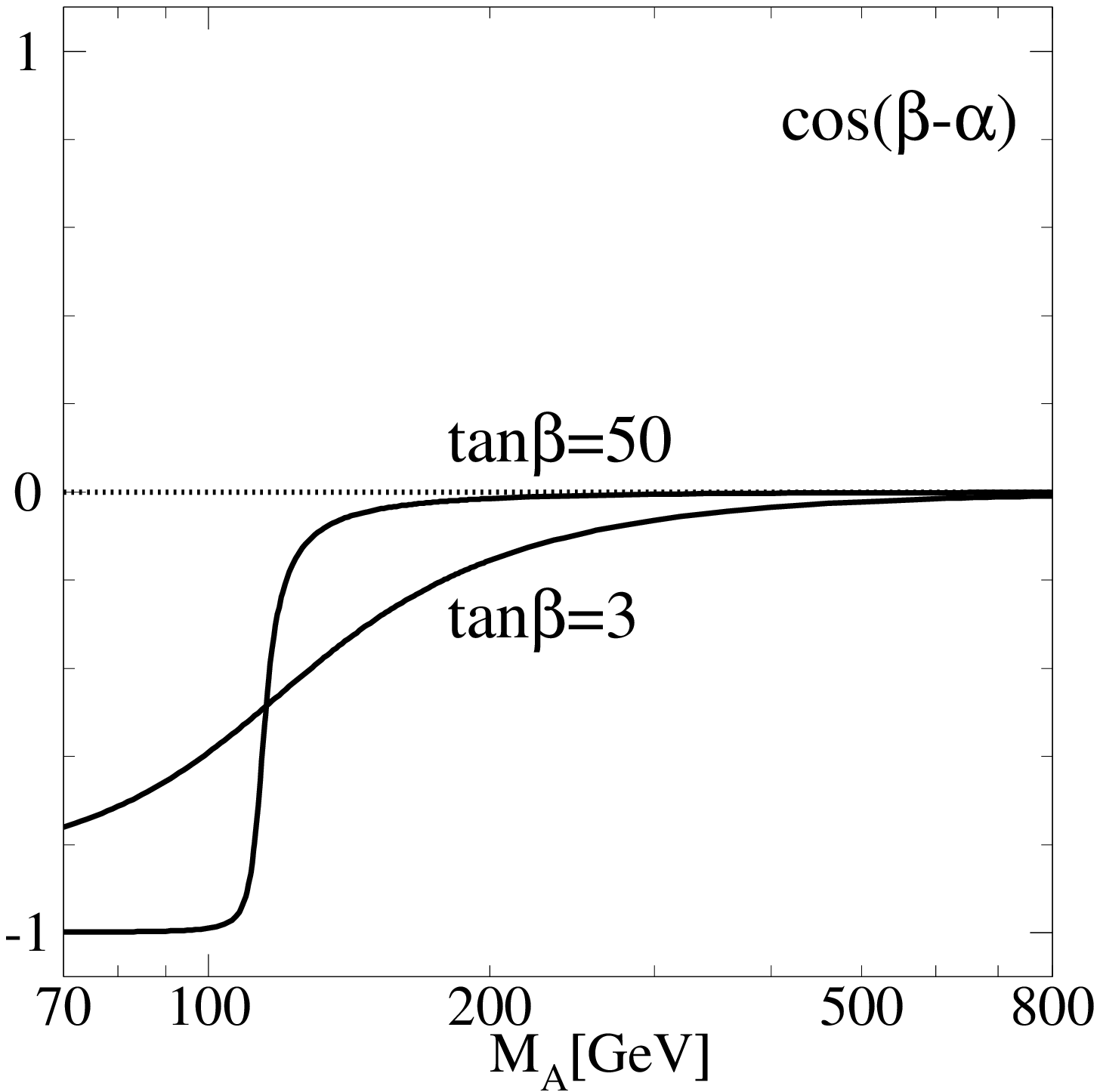,width=7.5cm}
\\[1cm]
\caption{\textit{Upper set: Variation of the trilinear couplings 
between CP-even and CP-odd Higgs bosons with $M_A$ for tan$\beta = 3$ 
and $50$ in the MSSM. Lower set: ZZh and ZZH gauge couplings in units 
of the SM coupling.}}
\label{fig:lambda2}
\end{center}
\end{figure}

\begin{figure}
\begin{center}
\epsfig{figure=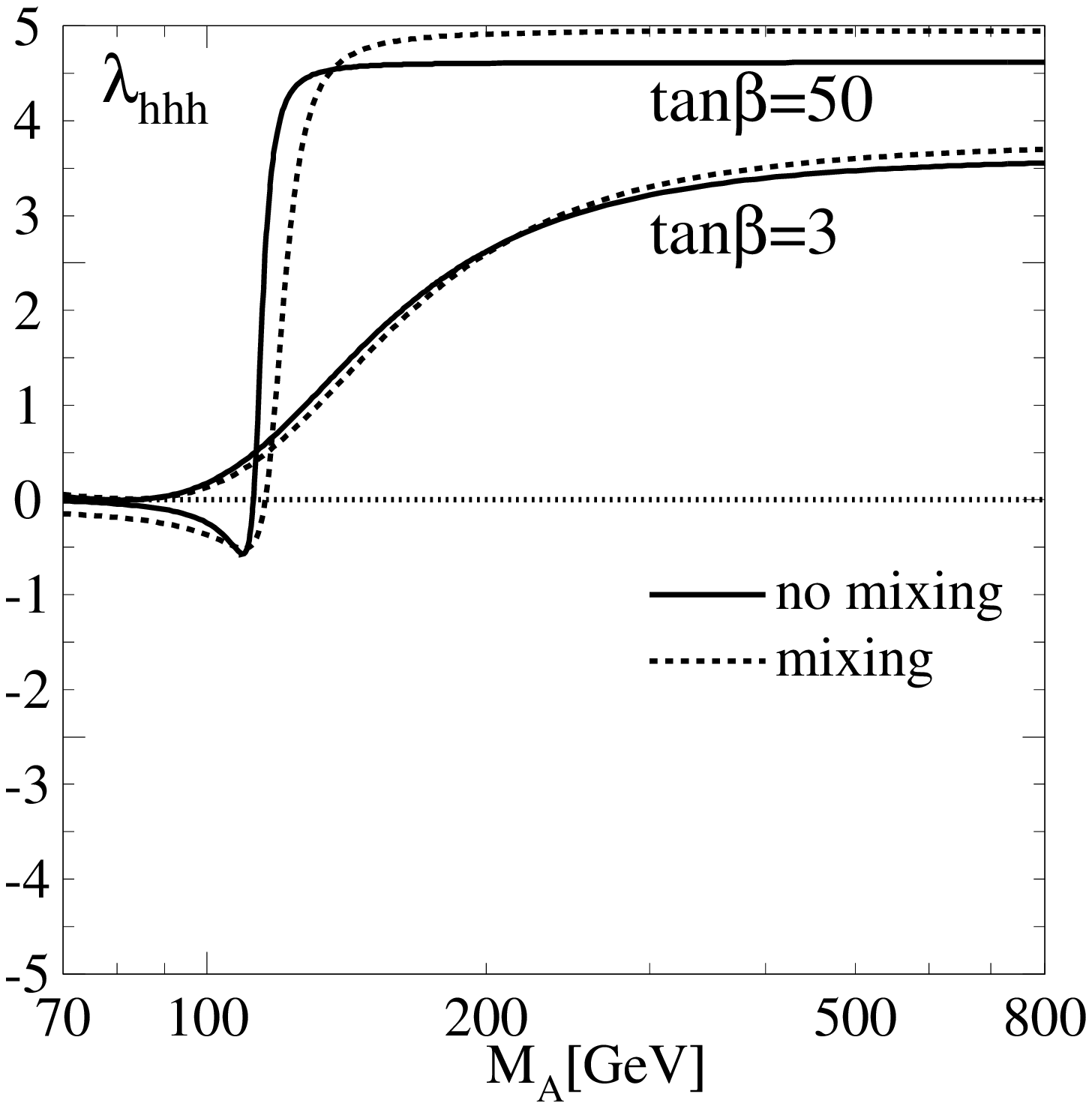,width=7.5cm}
\hspace{0.5cm}
\epsfig{figure=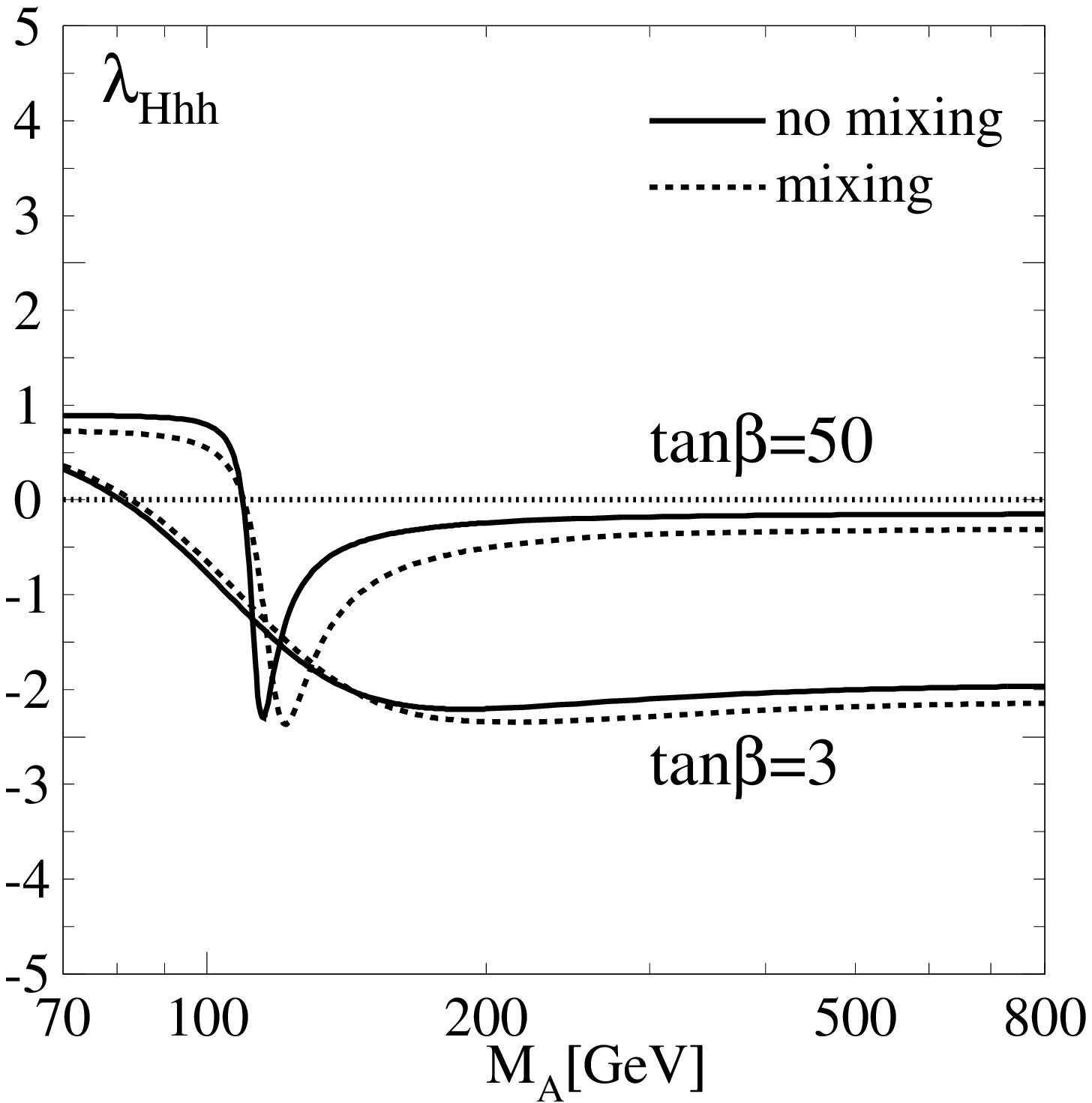,width=7.5cm}\\
\caption{\textit{Modification of the trilinear couplings $\lambda_{hhh}$ and 
$\lambda_{Hhh}$ due to mixing effects for $A=\mu=1$~TeV.}}
\label{fig:coupmix}
\end{center}
\end{figure}

In the minimal supersymmetric extension there are five physical Higgs
states as outlined in Chapter \ref{chap:higgssect}.  This leads to six
CP-invariant trilinear Higgs couplings among the neutral Higgs
bosons. The couplings including radiative corrections in the one-loop
leading $m_t^4$ approximation are given in Eq.~(\ref{coup}). In the
following analysis the dominant one-loop and the leading two-loop
corrections to the masses and couplings have been included,
cf.~\cite{carena,newhdecay}. The upper bound of the light Higgs mass
strongly depends on the radiative corrections and thus on the values
of the mixing parameters $A$ and $\mu$. In contrast, the trilinear
Higgs self-couplings show a weak dependence on the radiative
corrections when evaluated for the physical Higgs masses. In
Figs.~\ref{fig:lambda1} and \ref{fig:lambda2} the trilinear Higgs
self-couplings and the gauge-Higgs couplings are given as a function
of $M_A$ for two representative values of $\tan\beta$, {\it
  i.e.}~$\tan\beta=3,50$.  Around $M_A=120$~GeV the couplings vary
rapidly for large $\tan\beta$.  This region corresponds to the
cross-over of the mass branches in the neutral CP-even Higgs sector,
cf.~Eq.~(\ref{mass}). As can be inferred from Fig.~\ref{fig:lambda2}
the couplings involving a pair of pseudoscalar Higgs bosons are rather
small compared to the couplings involving pure CP-even Higgs states.
Note also that the trilinear Higgs couplings may become zero for some
parameter values.  Fig.~\ref{fig:coupmix} demonstrates for
$\lambda_{hhh}$ and $\lambda_{Hhh}$ the modification of the trilinear
couplings when mixing effects are included.  Evidently, the change is
rather small.  The mixing effects are also discussed in
Ref.~\cite{osland}.\s

At $e^+e^-$ linear colliders there are many processes which involve
the trilinear MSSM Higgs-boson couplings \cite{djouadi2}:
\beq
\begin{array}{l@{:\quad}l@{\;\to\;}l l l l}
\mbox{double Higgs-strahlung} & e^+e^-  & ZH_i H_j & \mathrm{and} 
& ZAA & [H_{i,j}=h,H] \\[0.2cm]
\mbox{triple Higgs production} & e^+e^- & AH_i H_j & \mathrm{and}
& AAA & \\[0.2cm]
WW\ \mbox{fusion} & e^+e^- & \bar{\nu}_e \nu_e H_i H_j & \mathrm{and}
& \bar{\nu}_e \nu_e AA & 
\end{array} 
\eeq
Table \ref{tab:coup} shows which trilinear couplings are involved in
the various processes, cf.~Fig.~\ref{fig:graphs}.  If all cross
sections would yield sufficiently high signal rates, the system could
be solved for all $\lambda$'s up to discrete ambiguities by using the
double Higgs-strahlung and the triple Higgs production processes $Ahh$
and $AAA$ ["bottom-up-approach"]. The processes $ZAA$ and $AAA$
can be used to solve for the couplings $\lambda (hAA)$ and $\lambda
(HAA)$.  The processes $Zhh$ and $Ahh$ provide the couplings $\lambda
(hhh)$ and $\lambda (Hhh)$. Subsequently the process $ZHh$ can be
exploited to extract the coupling $\lambda (HHh)$ and finally $ZHH$
for the determination of $\lambda (HHH)$ The remaining triple Higgs
production processes $AHh$ and $AHH$ provide additional redundant
information. \s
\begin{table}
\begin{center}$
\renewcommand{\arraystretch}{1.3}
\begin{array}{|l||cccc|c||ccc|}\hline
\phantom{\lambda} & 
\multicolumn{4}{|c|}{\mathrm{double\;Higgs\!-\!strahlung}} &
\multicolumn{4}{|c|}{\phantom{d} \mathrm{triple\;Higgs\!-\!production \phantom{d}}} \str \\
\phantom{\lambda i}\lambda & Zhh & ZHh & ZHH & ZAA & 
\multicolumn{2}{|c}{\phantom{d}Ahh \phantom{d}AHh} & \phantom{d}AHH & \!\! AAA \\ \hline\hline
hhh & \times & & & & \phantom{d}\times\phantom{d} & & &  \\
Hhh & \times & \times & & & \times & \times & & \\
HHh & & \times & \times & & & \times & \times & \\ 
HHH & & & \times & & & & \times & \\ 
\cline{1-6} & & & & & \multicolumn{2}{c}{\phantom{\times}} & & \\[-0.575cm]
\hline
hAA & & & & \times & \multicolumn{2}{c}{\,\,\,\times \quad\quad\, \times} & & \times \\ 
HAA & & & & \times & \multicolumn{2}{c}{\phantom{\times}\quad\quad\,\,\,\, \times}
 & \times & \times \\
\hline
\end{array}$
\end{center}
\caption{\textit{The trilinear couplings between neutral CP-even and CP-odd 
MSSM Higgs bosons which can generically be probed in double 
Higgs-strahlung and associated triple Higgs-production, are marked by 
a cross. [The matrix for WW fusion is isomorphic to the matrix for 
Higgs-strahlung.]}}
\label{tab:coup}
\end{table}
\begin{figure}
\begin{flushleft}
\underline{double Higgs-strahlung: $e^+e^-\to ZH_iH_j$, $ZAA$
[$H_{i,j}=h,H$]}\\[1.5\baselineskip]
{\footnotesize
\unitlength1mm
\hspace{5mm}
\begin{fmfshrink}{0.7}
\begin{fmfgraph*}(24,12)
  \fmfstraight
  \fmfleftn{i}{3} \fmfrightn{o}{3}
  \fmf{fermion}{i1,v1,i3}
  \fmf{boson,tens=3/2}{v1,v2}
  \fmf{boson}{v2,o3} \fmflabel{$Z$}{o3}
  \fmf{phantom}{v2,o1}
  \fmffreeze
  \fmf{dashes,lab=$H_{i,,j}$,lab.s=right}{v2,v3} \fmf{dashes}{v3,o1}
  \fmffreeze
  \fmf{dashes}{v3,o2} 
  \fmflabel{$H_{i,j}$}{o2} \fmflabel{$H_{i,j}$}{o1}
  \fmfdot{v3}
\end{fmfgraph*}
\hspace{15mm}
\begin{fmfgraph*}(24,12)
  \fmfstraight
  \fmfleftn{i}{3} \fmfrightn{o}{3}
  \fmf{fermion}{i1,v1,i3}
  \fmf{boson,tens=3/2}{v1,v2}
  \fmf{dashes}{v2,o1} \fmflabel{$H$}{o1}
  \fmf{phantom}{v2,o3}
  \fmffreeze
  \fmf{dashes,lab=$A$,lab.s=left}{v2,v3} 
  \fmf{boson}{v3,o3} \fmflabel{$Z$}{o3}
  \fmffreeze
  \fmf{dashes}{v3,o2} 
  \fmflabel{$H_{i,j}$}{o2} \fmflabel{$H_{i,j}$}{o1}
\end{fmfgraph*}
\hspace{15mm}
\begin{fmfgraph*}(24,12)
  \fmfstraight
  \fmfleftn{i}{3} \fmfrightn{o}{3}
  \fmf{fermion}{i1,v1,i3}
  \fmf{boson,tens=3/2}{v1,v2}
  \fmf{dashes}{v2,o1} \fmflabel{$H$}{o1}
  \fmf{phantom}{v2,o3}
  \fmffreeze
  \fmf{boson}{v2,v3,o3} \fmflabel{$Z$}{o3}
  \fmffreeze
  \fmf{dashes}{v3,o2} 
  \fmflabel{$H_{i,j}$}{o2} \fmflabel{$H_{i,j}$}{o1}
\end{fmfgraph*}
\hspace{15mm}
\begin{fmfgraph*}(24,12)
  \fmfstraight
  \fmfleftn{i}{3} \fmfrightn{o}{3}
  \fmf{fermion}{i1,v1,i3}
  \fmf{boson,tens=3/2}{v1,v2}
  \fmf{dashes}{v2,o1} \fmflabel{$H_{i,j}$}{o1}
  \fmf{dashes}{v2,o2} \fmflabel{$H_{i,j}$}{o2}
  \fmf{boson}{v2,o3} \fmflabel{$Z$}{o3}
\end{fmfgraph*}
\\[2\baselineskip]
\hspace{5mm}
\begin{fmfgraph*}(24,12)
  \fmfstraight
  \fmfleftn{i}{3} \fmfrightn{o}{3}
  \fmf{fermion}{i1,v1,i3}
  \fmf{boson,tens=3/2}{v1,v2}
  \fmf{boson}{v2,o3} \fmflabel{$Z$}{o3}
  \fmf{phantom}{v2,o1}
  \fmffreeze
  \fmf{dashes,lab=$H_{i,,j}$,lab.s=right}{v2,v3} \fmf{dashes}{v3,o1}
  \fmffreeze
  \fmf{dashes}{v3,o2} 
  \fmflabel{$A$}{o2} \fmflabel{$A$}{o1}
  \fmfdot{v3}
\end{fmfgraph*}
\hspace{15mm}
\begin{fmfgraph*}(24,12)
  \fmfstraight
  \fmfleftn{i}{3} \fmfrightn{o}{3}
  \fmf{fermion}{i1,v1,i3}
  \fmf{boson,tens=3/2}{v1,v2}
  \fmf{dashes}{v2,o1} \fmflabel{$H$}{o1}
  \fmf{phantom}{v2,o3}
  \fmffreeze
  \fmf{dashes,lab=$H_{i,,j}$,lab.s=left}{v2,v3} 
  \fmf{boson}{v3,o3} \fmflabel{$Z$}{o3}
  \fmffreeze
  \fmf{dashes}{v3,o2} 
  \fmflabel{$A$}{o2} \fmflabel{$A$}{o1}
\end{fmfgraph*}
\hspace{15mm}
\begin{fmfgraph*}(24,12)
  \fmfstraight
  \fmfleftn{i}{3} \fmfrightn{o}{3}
  \fmf{fermion}{i1,v1,i3}
  \fmf{boson,tens=3/2}{v1,v2}
  \fmf{dashes}{v2,o1} \fmflabel{$A$}{o1}
  \fmf{dashes}{v2,o2} \fmflabel{$A$}{o2}
  \fmf{boson}{v2,o3} \fmflabel{$Z$}{o3}
\end{fmfgraph*}
\end{fmfshrink}
}
\\[2\baselineskip]
\underline{triple Higgs production: $e^+e^-\to AH_iH_j$, $AAA$}
\\[1.5\baselineskip]
{\footnotesize
\unitlength1mm
\hspace{5mm}
\begin{fmfshrink}{0.7}
\begin{fmfgraph*}(24,12)
  \fmfstraight
  \fmfleftn{i}{3} \fmfrightn{o}{3}
  \fmf{fermion}{i1,v1,i3}
  \fmf{boson,tens=3/2}{v1,v2}
  \fmf{dashes}{v2,o3} \fmflabel{$A$}{o3}
  \fmf{phantom}{v2,o1}
  \fmffreeze
  \fmf{dashes,lab=$H_{i,,j}$,lab.s=right}{v2,v3} \fmf{dashes}{v3,o1}
  \fmffreeze
  \fmf{dashes}{v3,o2} 
  \fmflabel{$H_{i,j}$}{o2} \fmflabel{$H_{i,j}$}{o1}
  \fmfdot{v3}
\end{fmfgraph*}
\hspace{15mm}
\begin{fmfgraph*}(24,12)
  \fmfstraight
  \fmfleftn{i}{3} \fmfrightn{o}{3}
  \fmf{fermion}{i1,v1,i3}
  \fmf{boson,tens=3/2}{v1,v2}
  \fmf{dashes}{v2,o1} \fmflabel{$H_{i,j}$}{o1}
  \fmf{phantom}{v2,o3}
  \fmffreeze
  \fmf{dashes,lab=$A$,lab.s=left}{v2,v3} 
  \fmf{dashes}{v3,o3} \fmflabel{$A$}{o3}
  \fmffreeze
  \fmf{dashes}{v3,o2} 
  \fmflabel{$H_{i,j}$}{o2} \fmflabel{$A$}{o3}
\end{fmfgraph*}
\hspace{15mm}
\begin{fmfgraph*}(24,12)
  \fmfstraight
  \fmfleftn{i}{3} \fmfrightn{o}{3}
  \fmf{fermion}{i1,v1,i3}
  \fmf{boson,tens=3/2}{v1,v2}
  \fmf{dashes}{v2,o1}
  \fmf{phantom}{v2,o3}
  \fmffreeze
  \fmf{boson}{v2,v3} 
  \fmf{dashes}{v3,o3} \fmflabel{$A$}{o3}
  \fmffreeze
  \fmf{dashes}{v3,o2} 
  \fmflabel{$H_{i,j}$}{o2} \fmflabel{$H_{i,j}$}{o1}
\end{fmfgraph*}
\\[2\baselineskip]
\hspace{5mm}
\begin{fmfgraph*}(24,12)
  \fmfstraight
  \fmfleftn{i}{3} \fmfrightn{o}{3}
  \fmf{fermion}{i1,v1,i3}
  \fmf{boson,tens=3/2}{v1,v2}
  \fmf{dashes}{v2,o3} \fmflabel{$A$}{o3}
  \fmf{phantom}{v2,o1}
  \fmffreeze
  \fmf{dashes,lab=$H_{i,,j}$,lab.s=right}{v2,v3} \fmf{dashes}{v3,o1}
  \fmffreeze
  \fmf{dashes}{v3,o2} 
  \fmflabel{$A$}{o2} \fmflabel{$A$}{o1}
  \fmfdot{v3}
\end{fmfgraph*}
\end{fmfshrink}
}
\\[2\baselineskip]
\underline{$WW$ fusion: $e^+e^-\to\bar\nu_e\nu_e H_i H_j$, $AA$}
\\[1.5\baselineskip]
{\footnotesize
\unitlength1mm
\hspace{5mm}
\begin{fmfshrink}{0.7}
\begin{fmfgraph*}(24,20)
  \fmfstraight
  \fmfleftn{i}{8} \fmfrightn{o}{8}
  \fmf{fermion,tens=3/2}{i2,v1} \fmf{phantom}{v1,o2}
  \fmf{phantom}{o7,v2} \fmf{fermion,tens=3/2}{v2,i7}
  \fmffreeze
  \fmf{fermion}{v1,o1}
  \fmf{fermion}{o8,v2}
  \fmf{boson}{v1,v3} 
  \fmf{boson}{v3,v2}
  \fmf{dashes,lab=$H_{i,,j}$}{v3,v4}
  \fmf{dashes}{v4,o3} \fmf{dashes}{v4,o6}
  \fmflabel{$H_{i,j}$}{o3} \fmflabel{$H_{i,j}$}{o6}
  \fmffreeze
  \fmf{phantom,lab=$W$,lab.s=left}{v1,x1} \fmf{phantom}{x1,v3} 
  \fmf{phantom,lab=$W$,lab.s=left}{x2,v2} \fmf{phantom}{v3,x2}
  \fmfdot{v4}
\end{fmfgraph*}
\hspace{15mm}
\begin{fmfgraph*}(24,20)
  \fmfstraight
  \fmfleftn{i}{8} \fmfrightn{o}{8}
  \fmf{fermion,tens=3/2}{i2,v1} \fmf{phantom}{v1,o2}
  \fmf{phantom}{o7,v2} \fmf{fermion,tens=3/2}{v2,i7}
  \fmffreeze
  \fmf{fermion}{v1,o1}
  \fmf{fermion}{o8,v2}
  \fmf{boson}{v1,v3} 
  \fmf{boson}{v4,v2}
  \fmf{boson,lab=$W$,lab.s=left}{v3,v4}
  \fmf{dashes}{v3,o3} \fmf{dashes}{v4,o6}
  \fmflabel{$H_{i,j}$}{o3} \fmflabel{$H_{i,j}$}{o6}
\end{fmfgraph*}
\hspace{15mm}
\begin{fmfgraph*}(24,20)
  \fmfstraight
  \fmfleftn{i}{8} \fmfrightn{o}{8}
  \fmf{fermion,tens=3/2}{i2,v1} \fmf{phantom}{v1,o2}
  \fmf{phantom}{o7,v2} \fmf{fermion,tens=3/2}{v2,i7}
  \fmffreeze
  \fmf{fermion}{v1,o1}
  \fmf{fermion}{o8,v2}
  \fmf{boson}{v1,v3} 
  \fmf{boson}{v4,v2}
  \fmf{dashes,lab=$H^\pm$,lab.s=left}{v3,v4}
  \fmf{dashes}{v3,o3} \fmf{dashes}{v4,o6}
  \fmflabel{$H_{i,j}$}{o3} \fmflabel{$H_{i,j}$}{o6}
\end{fmfgraph*}
\hspace{15mm}
\begin{fmfgraph*}(24,20)
  \fmfstraight
  \fmfleftn{i}{8} \fmfrightn{o}{8}
  \fmf{fermion,tens=3/2}{i2,v1} \fmf{phantom}{v1,o2}
  \fmf{phantom}{o7,v2} \fmf{fermion,tens=3/2}{v2,i7}
  \fmffreeze
  \fmf{fermion}{v1,o1}
  \fmf{fermion}{o8,v2}
  \fmf{boson}{v1,v3} 
  \fmf{boson}{v3,v2}
  \fmf{dashes}{v3,o3} \fmf{dashes}{v3,o6}
  \fmflabel{$H_{i,j}$}{o3} \fmflabel{$H_{i,j}$}{o6}
\end{fmfgraph*}
\\[2\baselineskip]
\hspace{5mm}
\begin{fmfgraph*}(24,20)
  \fmfstraight
  \fmfleftn{i}{8} \fmfrightn{o}{8}
  \fmf{fermion,tens=3/2}{i2,v1} \fmf{phantom}{v1,o2}
  \fmf{phantom}{o7,v2} \fmf{fermion,tens=3/2}{v2,i7}
  \fmffreeze
  \fmf{fermion}{v1,o1}
  \fmf{fermion}{o8,v2}
  \fmf{boson}{v1,v3} 
  \fmf{boson}{v3,v2}
  \fmf{dashes,lab=$H_{i,,j}$}{v3,v4}
  \fmf{dashes}{v4,o3} \fmf{dashes}{v4,o6}
  \fmflabel{$A$}{o3} \fmflabel{$A$}{o6}
  \fmfdot{v4}
\end{fmfgraph*}
\hspace{15mm}
\begin{fmfgraph*}(24,20)
  \fmfstraight
  \fmfleftn{i}{8} \fmfrightn{o}{8}
  \fmf{fermion,tens=3/2}{i2,v1} \fmf{phantom}{v1,o2}
  \fmf{phantom}{o7,v2} \fmf{fermion,tens=3/2}{v2,i7}
  \fmffreeze
  \fmf{fermion}{v1,o1}
  \fmf{fermion}{o8,v2}
  \fmf{boson}{v1,v3} 
  \fmf{boson}{v4,v2}
  \fmf{dashes,lab=$H^\pm$,lab.s=left}{v3,v4}
  \fmf{dashes}{v3,o3} \fmf{dashes}{v4,o6}
  \fmflabel{$A$}{o3} \fmflabel{$A$}{o6}
\end{fmfgraph*}
\hspace{15mm}
\begin{fmfgraph*}(24,20)
  \fmfstraight
  \fmfleftn{i}{8} \fmfrightn{o}{8}
  \fmf{fermion,tens=3/2}{i2,v1} \fmf{phantom}{v1,o2}
  \fmf{phantom}{o7,v2} \fmf{fermion,tens=3/2}{v2,i7}
  \fmffreeze
  \fmf{fermion}{v1,o1}
  \fmf{fermion}{o8,v2}
  \fmf{boson}{v1,v3} 
  \fmf{boson}{v3,v2}
  \fmf{dashes}{v3,o3} \fmf{dashes}{v3,o6}
  \fmflabel{$A$}{o3} \fmflabel{$A$}{o6}
\end{fmfgraph*}
\end{fmfshrink}
}
\end{flushleft}
\caption{\textit{
Processes contributing to double and triple Higgs production involving
trilinear couplings in the MSSM.}}
\label{fig:graphs}
\end{figure}

In practice, not all the cross sections can be measured
experimentally since some are too small. In this case, however, one can
compare the theoretical predictions of the cross sections with the
experimental results for the accessible channels and by this means
test the trilinear Higgs couplings stringently ["top-down
approach"].\s

In the MSSM, the couplings involving a CP-odd Higgs boson pair, {\it
  i.e.}~$\lambda (hAA)$ and $\lambda (HAA)$, are small. The analysis
can then be carried out without making use of the processes $ZAA$ and
$AAA$. [The validity of this approach can be checked experimentally in
a model-independent way: Assuming the knowledge of the predetermined
couplings between gauge and Higgs bosons the measurement of
$\sigma(ZAA)$ and $\sigma(AAA)$ provides an upper bound on
$\lambda_{hAA}$ and $\lambda_{HAA}$.] The processes $ZH_i H_j$ and
$Ahh$ are then sufficient to solve for the couplings among the neutral
CP-even Higgs bosons in the manner described above, c.f.~the
double-line box of Table \ref{tab:coup}. \s

The processes $e^+e^- \to ZH_i A$ and $e^+e^- \to \bar{\nu}_e \nu_e
H_i A$ [$H_i=h,H$] with a CP even-odd Higgs pair in the final state
cannot be used to extract the trilinear Higgs self-couplings. They are
mediated by a virtual $Z$ boson which subsequently decays into a
parity-mixed final state $Z^* \to H_i A$. Hence, they only involve gauge
interactions and no Higgs boson self-interactions.\s

\subsection{SM double-Higgs production in $e^+e^-$ collisions 
\label{smzhhprod}}
The (unpolarized) differential cross section for the double
Higgs-strahlung process $e^+e^- \to ZHH$, cf.~Fig.~\ref{fig:diag}, can
be cast into the form \cite{djouadi2}
\beq 
\frac{d \sigma (e^+ e^- \to ZHH)}{d x_1 d x_2} = 
\frac{\sqrt{2} G_F^3 M_Z^6}{384 \pi^3 s}
\frac{v_e^2 + a_e^2}{(1- \mu_Z)^2}\, {\cal Z} 
\eeq 
after the angular dependence has been integrated out. The vector and
axial-vector $Z$ charges of the electron are defined by $v_e
= -1 + 4\sin^2 \theta_W$ and $a_e = -1$. $x_{1,2} = 2
E_{1,2}/\sqrt{s}$ are the scaled energies of the two Higgs particles,
$x_3 = 2 - x_1 -x_2$ is the scaled energy of the $Z$ boson, and $y_i =
1 - x_i$; the square of the reduced masses is denoted by $\mu_i =
M_i^2/s$, and $\mu_{ij}=\mu_i-\mu_j$. In terms of these variables, the
coefficient ${\cal Z}$ can be written as:
\beq 
{\cal Z} =  {\mathfrak a}^2 f_0 +
\frac{1}{4 \mu_Z (y_1+\mu_{HZ})} \left[ 
\frac{f_1}{y_1+\mu_{HZ}} + \frac{f_2}{y_2+\mu_{HZ}} 
+ 2\mu_Z {\mathfrak a} f_3 \right] 
+ \Bigg\{ y_1 \leftrightarrow y_2 \Bigg\} 
\eeq
with
\beq
{\mathfrak a} = \frac{\lambda_{HHH}}{y_3-\mu_{HZ}}
 + \frac{2}{y_1+\mu_{HZ}} + 
\frac{2}{y_2+\mu_{HZ}} + \frac{1}{\mu_Z}
\eeq
The coefficients $f_i$ are given by 
\beq
f_0 &=& \mu_Z[(y_1+y_2)^2 + 8\mu_Z]/8 \non\\
f_1 &=& (y_1-1)^2(\mu_Z-y_1)^2-4\mu_H y_1(y_1+y_1\mu_Z-4\mu_Z) \non\\
& & {} + \mu_Z(\mu_Z-4\mu_H)(1-4\mu_H)-\mu_Z^2  
\non\\
f_2 &=& [\mu_Z(y_3+\mu_Z - 8\mu_H)-(1+\mu_Z)y_1 y_2](1+y_3+2\mu_Z) 
\non\\
& & {}+ y_1 y_2[y_1 y_2 + 1 + \mu_Z^2+4\mu_H (1+\mu_Z)]
+ 4\mu_H \mu_Z(1+\mu_Z+4\mu_H)+ \mu_Z^2 
\non\\
f_3 &=& y_1(y_1-1)(\mu_Z-y_1)-y_2(y_1+1)(y_1+\mu_Z)+2\mu_Z
(1+\mu_Z-4\mu_H) 
\eeq
The first term in the coefficient ${\mathfrak a}$ involves the
trilinear Higgs self-coupling. The other terms are due to sequential
Higgs-strahlung and due to the 4-Higgs-gauge-coupling. The various
terms can be identified by analysing the propagators they are
associated with.\s

The double Higgs-strahlung process is mediated by an s-channel $Z$
boson. Since $Z$ has spin one it only couples to a left-handed
electron and a right-handed positron and vice versa. Due to
CP-invariance the total cross section for double Higgs-strahlung
therefore doubles if electron-positron beams with opposite
polarization are used.\s
\begin{figure}
\begin{center}
\epsfig{figure=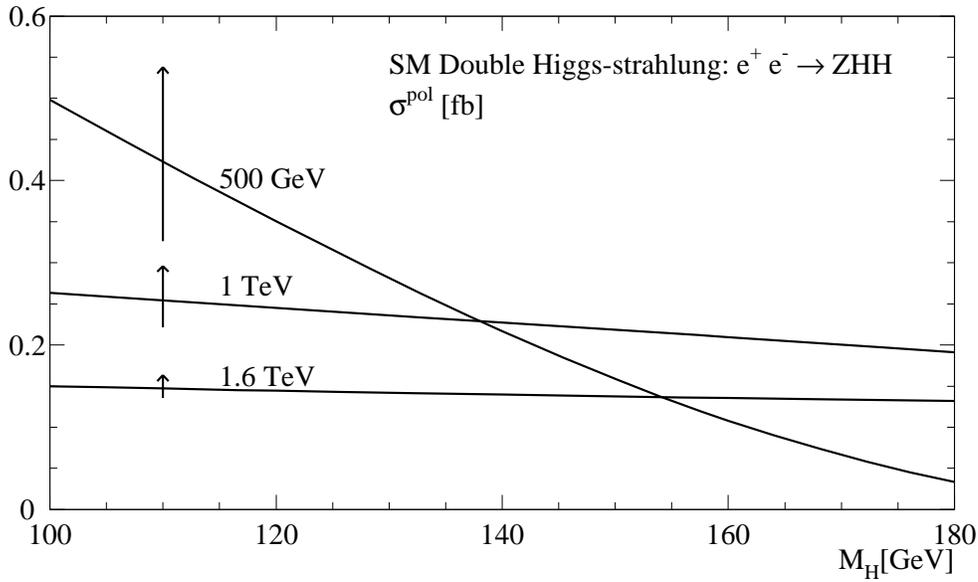,width=13cm}
\\
\end{center}
\caption{\textit{The cross section for double Higgs-strahlung in the SM 
at three collider energies: $500$~GeV, $1$~TeV and $1.6$~TeV. The 
electron/positron beams are taken oppositely polarized. The vertical 
arrows correspond to a variation of the trilinear Higgs coupling from 
$1/2$ to $3/2$ of the SM value.}}
\label{fig:SM1}
\end{figure}
\begin{figure}
\begin{center}
\epsfig{figure=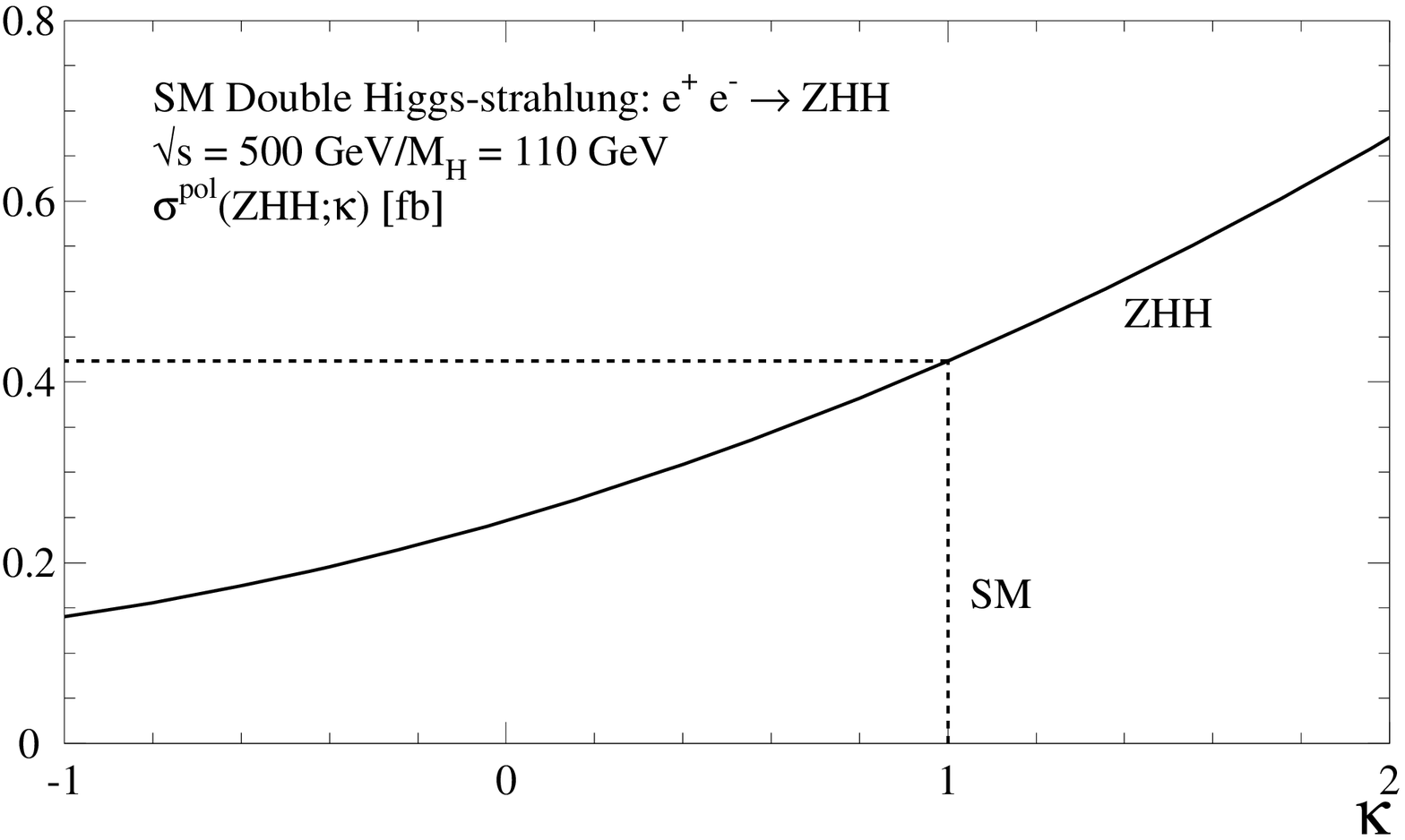,width=13cm}
\end{center}
\caption{\textit{Variation of the cross section $\sigma (ZHH)$ with the 
modified trilinear coupling $\kappa \lambda_{HHH}$ at a collider 
energy of $\sqrt{s}=500$~GeV and $M_H=110$~GeV.}}
\label{fig:smvar1}
\end{figure}
\begin{figure}
\begin{center}
\epsfig{figure=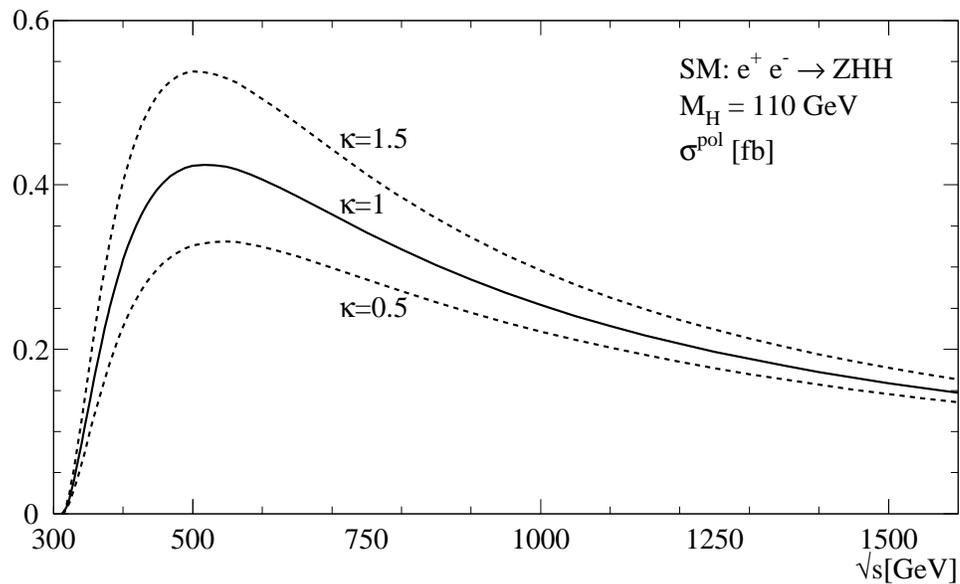,width=13cm}
\end{center}
\caption{\textit{The energy dependence of the cross section for double 
Higgs-strahlung for a fixed Higgs mass $M_H=110$~GeV. The variation of 
the cross section for modified trilinear couplings 
$\kappa\lambda_{HHH}$ is indicated by the dashed lines.}}
\label{fig:smvar2}
\end{figure}

Fig.~\ref{fig:SM1} shows the total cross section of double
Higgs-strahlung as a function of the Higgs mass for three typical
$e^+e^-$ collider energies, {\it i.e.}~$\sqrt{s}= 500$~GeV, 1~TeV and
1.6~TeV.  The electron-positron beams are taken oppositely polarized
[$\lambda_{e^-}\lambda_{e^+}=-1$] so that the cross section is
enhanced by a factor two. Due to the s-channel $Z$ boson propagator,
the cross sections show scaling behaviour beyond the threshold region.
The arrows indicate the modifications of the cross sections with a
variation of the trilinear Higgs coupling from 1/2 to 3/2 in units of
the SM coupling. They increase with rising coupling.
Fig.~\ref{fig:smvar1} illustrates the sensitivity to $\lambda_{HHH}$
if the coupling is varied in the range
[$-\lambda_{HHH},2\lambda_{HHH}$]. Evidently, the sensitivity of the
double Higgs-strahlung process to the value of the Higgs self-coupling
is not washed out by the irreducible background diagrams which do not
involve $\lambda_{HHH}$. The influence of these diagrams increases
with rising energy so that the sensitivity to the trilinear coupling
is largest near the kinematic threshold as can be inferred form
Fig.~\ref{fig:smvar2}. This behaviour results from the propagator of
the virtual Higgs boson connecting the two real Higgs bosons which is
maximal near the threshold. Furthermore Fig.~\ref{fig:smvar2} shows
that the maximum double Higgs-strahlung cross section is reached for
energies $\sqrt{s} \sim 2M_H + M_Z + 200$~GeV, {\it i.e.}~$\sqrt{s} =
500$~GeV for Higgs masses in the lower part of the intermediate range.
A c.m.~energy of 500 GeV is therefore a good choice in this mass range
since the cross section as well as the sensitivity to $\lambda_{HHH}$
is maximal.\s

\subsection{SM $WW$ double-Higgs fusion \label{wwsmfusion}}
For high energies the logarithmic behaviour of the $t$-channel diagram
contributing to the $WW$ fusion process, cf.~Fig.~\ref{fig:diag},
dominates over the scaling behaviour due to the $Z$ boson propagator
in the s-channel of the double Higgs-strahlung process. Therefore the
$WW$ double-Higgs fusion process provides the largest cross section at
high c.m.~energies for Higgs masses in the intermediate range,
especially when the initial state beams are taken oppositely
polarized.\s

In order to get a rough estimate of the cross section for $WW$
double-Higgs fusion the equivalent particle approximation
\cite{wfus2a,repko} can be applied. In this approximation the $W$
bosons are regarded as partons in the electron and positron,
respectively, with the $W$ bosons taken on-shell. The production
amplitude for the dominant longitudinal degrees of freedom is given by
\cite{kalli}
\beq
\hspace{-0.4cm}
{\cal M}_{LL} = \,\scriptstyle{\frac{G_F \hat{s}}{\sqrt{2}}} 
\, \left\{ \textstyle{(1 \!+ \beta_W^2)} \left[ 1 \! + \, 
\scriptstyle{\frac{\lambda_{HHH}}{(\hat{s}-M_H^2)/M_Z^2}} \right] 
\! + \,\scriptstyle{\frac{1}{\beta_W \beta_H}} \left[ 
\scriptstyle{\frac{(1-\beta_W^4)+ 
(\beta_W - \beta_H \cos\theta)^2}{\cos\theta - x_W}} \,\textstyle{-}\,
\scriptstyle{\frac{(1-\beta_W^4) + 
(\beta_W + \beta_H \cos\theta)^2}{\cos\theta + x_W}}  
\right] \right\}
\eeq
with $\beta_{W,H}$ denoting the $W$, $H$ velocities in the c.m.\ 
frame, and $x_W = (1- 2 M_H^2/\hat{s})/(\beta_W \beta_H)$.
$\hat{s}^{1/2}$ is the invariant energy of the $WW$ pair; $\theta$ is
the Higgs production angle in the c.m.\ frame of $WW$. After
integrating out the angular dependence the corresponding total cross
section reads \cite{djouadi2}
\beq
\hat{\sigma}_{LL} &=& \frac{G_F^2 M_W^4}{4\pi \hat{s}} \frac{\beta_H}
{\beta_W (1-\beta_W^2)^2} \Bigg\{ (1+\beta_W^2)^2 \left[1 + 
\frac{\lambda_{HHH}}
{(\hat{s} -M_H^2)/M_Z^2} \right]^2 \non\\
& + &\frac{16}{(1+\beta_H^2)^2-4\beta_H^2 \beta_W^2} 
\left[ \beta_H^2(
-\beta_H^2 x_W^2+4\beta_W \beta_H x_W -4\beta_W^2)
 + (1+\beta_W^2 -\beta_W^4)^2 \right] \non \\
& + &\frac{1}{\beta_W^2 \beta_H^2} \left( l_W +
 \frac{2x_W} {x_W^2-1} \right) 
 \left[ \beta_H ( \beta_H x_W-4\beta_W) (1+ \beta_W^2-
\beta_W^4 +3 x_W^2 \beta_H^2) \right. \non\\ 
&& \left. \hspace{4.2cm} + \beta_H^2 x_W 
(1-\beta_W^4 +13 \beta_W^2) 
- \frac{1}{x_W}(1+\beta_W^2-\beta_W^4)^2 \right] \non\\
& + & \frac{2(1+\beta_W^2)}{ \beta_W \beta_H} 
\left[1 + \frac{\lambda_{HHH}}{(\hat{s} -M_H^2)/M_Z^2} \right] 
\left[ l_W ( 1+\beta_W^2-\beta_W^4 -2\beta_W \beta_H x_W +
\beta_H^2 x_W^2) \right.
\non\\
& & \left. \hspace{5.75cm} +2\beta_H (x_W \beta_H  -2\beta_W ) \right] 
\Bigg\} 
\eeq
with $ l_W = \log [(x_W-1)/(x_W+1)]$. Folding $\hat{\sigma}_{LL}$ with the 
longitudinal $W_L$ spectra \cite{wfus2a,repko},
\beq
f_L(z) = \frac{G_F M_W^2}{2\sqrt{2} \pi^2} \frac{1-z}{z}
\qquad [z = E_W/E_e]
\eeq
a rough estimate of the cross section for the process $e^+e^- \to WW
\to HH \bar{\nu}_e \nu_e$ can be obtained. Though the value of the
exact calculation is overestimated by a factor 2 to 5, depending on
the collider energy, the approximated cross section is helpful for a
transparent interpretation of the exact result. \s
\begin{table}
\begin{center}$
\begin{array}{|rl|rl||c|c|}\hline
\multicolumn{4}{|c||}{\sigma\;[\mathrm{fb}]} & WW & ZZ \str \\
\hline \hline
\sqrt{s}\!=\!\!\!\!&1\;\mathrm{TeV}& 
M_H\!=\!\!\!\!&110\;\mathrm{GeV} &
0.104 & 0.013 \str \\ 
\phantom{val} & \phantom{val} & \phantom{val} & 150\;\mathrm{GeV} &
0.042 & 0.006 \str \\
\phantom{val} & \phantom{val} & \phantom{val} & 190\;\mathrm{GeV} &
0.017 & 0.002 \str \\ \hline
\sqrt{s}\!=\!\!\!\!&1.6\;\mathrm{TeV}& 
M_H\!=\!\!\!\!&110\;\mathrm{GeV} &
0.334 &  0.043\str \\ 
\phantom{val} & \phantom{val} & \phantom{val} & 150\;\mathrm{GeV} &
0.183 & 0.024 \str \\
\phantom{val} & \phantom{val} & \phantom{val} & 190\;\mathrm{GeV} &
0.103 & 0.013 \str \\ \hline
\end{array}$
\end{center}
\caption{\textit{Total cross sections for SM pair production in $WW$ and $ZZ$ fusion at $e^+e^-$ colliders for two characteristic energies and masses in the intermediate range (unpolarized beams).}}
\label{tab:SM}
\end{table}

For high energies the cross section is dominated by the $t$-channel
exchange which does not include the trilinear Higgs self-coupling.
The convoluted $WW$ fusion process $e^+e^- \to HH \bar{\nu}_e \nu_e$
nevertheless remains sensitive to $\lambda_{HHH}$ because the major
contribution to the cross section stems from the lower end of the
$WW$ spectrum so that also in the high-energy limit the sensitivity to
the Higgs self-coupling is maintained.\s

\begin{figure}
\begin{center}
\includegraphics{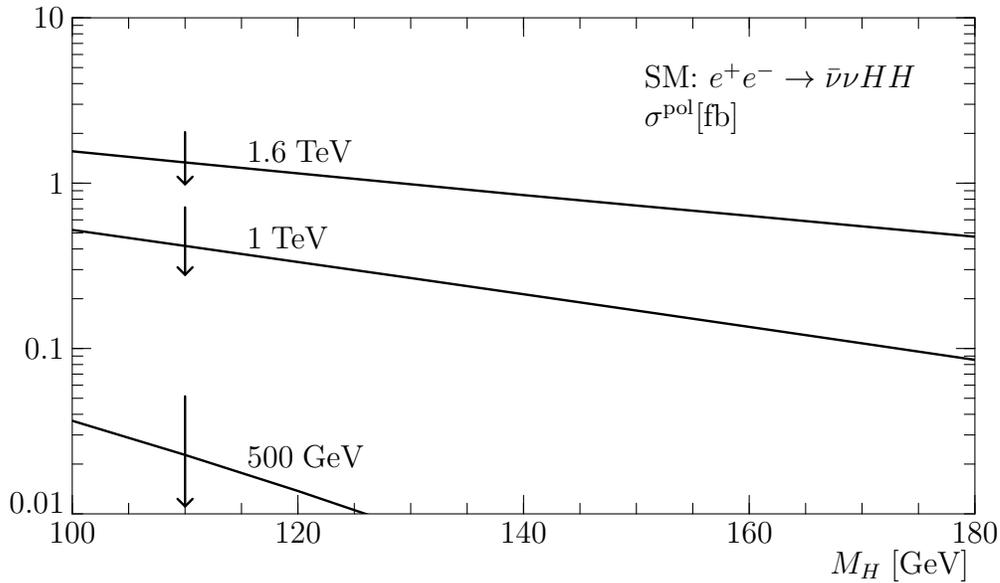} \\[1cm]
\end{center}
\caption{\textit{The total cross section for WW double-Higgs fusion in 
the SM at three collider energies: $500$~GeV, $1$~TeV and $1.6$~TeV. 
The vertical arrows correspond to a variation of the trilinear Higgs 
coupling from $1/2$ to $3/2$ of the SM value.}}
\label{fig:SM2}
\end{figure}

In the subsequent analysis the exact values for the $WW$ fusion cross
sections have been used with the $W$ bosons being off-shell and the
transverse degrees of freedom included. The cross sections have been
calculated numerically with the semi-analytical CompHEP program
\cite{boos}. The electron-positron beams are taken oppositely
polarized thus enhancing the cross section by a factor four since the
$W^-$ boson only couples to left-handed electrons. The results are
shown in Fig.~\ref{fig:SM2} for three collider energies,
$\sqrt{s}=500$~GeV, 1~TeV and 1.6~TeV, as a function of the Higgs
mass. As anticipated the $WW$ fusion cross section increases with
rising energy. The arrows indicate the modification of the cross
sections due to the variation of the Higgs self-coupling from
1/2$\lambda_{HHH}$ to 3/2 $\lambda_{HHH}$.  Fig.~\ref{fig:wwvar1}
shows the variation of the cross section with $\kappa\lambda$,
$\kappa= -1$ to 2. Due to destructive interference with the gauge
diagrams the cross section decreases with rising $\lambda_{HHH}$. As
expected the sensitivity to a variation of the Higgs self-coupling is
smaller for high c.m.\ energies.\s
\begin{figure}
\begin{center}
\includegraphics{sm-WW-HH-2.eps} \\[1cm]
\end{center}
\caption{\textit{Variation of the cross section 
$\sigma(e^+ e^- \lra \bar{\nu}_e \nu_e HH)$ with the modified 
trilinear coupling $\kappa \lambda_{HHH}$ at a collider energy of 
$\sqrt{s}=1$~TeV and $M_H=110$~GeV.}}
\label{fig:wwvar1}
\end{figure}

The cross sections for $ZZ$ fusion are not shown because they are an
order of magnitude below the $WW$ fusion values as can be inferred
from Table~\ref{tab:SM}. This is due to the smallness of the
$Ze^+e^-$ coupling.\s

The preceding discussion shows that for moderate energies double
Higgs-strahlung $e^+e^- \to ZHH$ is the preferred channel for
measurements of the trilinear Higgs-self couplings whereas for
energies in the TeV range $WW$ double Higgs fusion $e^+e^- \to WW \to
HH \bar{\nu}_e \nu_e$ is the more suitable channel. Since for both
processes the cross sections are nevertheless small high luminosities
as foreseen for $e^+e^-$ linear colliders are needed. A further
enhancement of a factor two for Higgs-strahlung and four for $WW$
fusion can be achieved by taking the electron-positron beams
oppositely polarized.  Clear multi-$b$ signatures like $e^+e^- \to
Z(b\bar{b})(b\bar{b})$ and $e^+e^- \to (b\bar{b})(b\bar{b}) +
\slash{E}$ will enable the separation of the signal from the
background. Experimental simulations taking into account detector
properties have demonstrated that for Higgs masses in the intermediate
range and an integrated luminosity of $\int {\cal L} = 500$~fb$^{-1}$
the trilinear Higgs coupling may be determined with 20\% accuracy
\cite{lutzcom}.\s

For the reconstruction of the Higgs potential also the quadrilinear
Higgs self-coupling has to be determined. This coupling can in
principle be measured directly in triple Higgs production: $e^+e^- \to
ZHHH$ and $e^+e^- \to HHH \bar{\nu}_e \nu_e$. Yet, these cross
sections are reduced by three orders of magnitude with respect to the
corresponding double Higgs production cross sections. This is due to
the suppression of the quadrilinear coupling compared to the trilinear
coupling and due to the additional particle in the final state so that
the total signal cross section is reduced by a factor
$[\lambda_{HHHH}^2\lambda_0^4/16/\pi^2]/[\lambda_{HHH}^2\lambda_0^2/M_Z^2]
\sim 10^{-3}$. Likewise, the irreducible background diagrams are
suppressed. A few exemplifying results are given in
Table~\ref{tab:quadri}.
\begin{table}
\begin{center}$
\begin{array}{|rlrl|lrl|}\hline
\multicolumn{4}{|c|}{\phantom{\sigma(HH)\;[\mathrm{ab}]}} &
\multicolumn{3}{|c|}{\sigma(e^+e^-\to ZHHH)[\mathrm{ab}]} 
\str \\
\hline 
\sqrt{s}\!=\!\!\!\!&1\;\mathrm{TeV}& 
M_H\!=\!\!\!\!&110\;\mathrm{GeV} &
\hspace{0.5cm} 0.44 & [0.41/\!\!\!\!& 0.46] \str \\ 
\phantom{val} & \phantom{val} & \phantom{val} & 150\;\mathrm{GeV} &
\hspace{0.5cm}0.34 & [0.32/\!\!\!\!& 0.36] \str \\ 
\phantom{val} & \phantom{val} & \phantom{val} & 190\;\mathrm{GeV} &
\hspace{0.5cm}0.19 & [0.18/\!\!\!\!& 0.20] \str \\ \hline
\sqrt{s}\!=\!\!\!\!&1.6\;\mathrm{TeV}& 
M_H\!=\!\!\!\!&110\;\mathrm{GeV} &
\hspace{0.5cm}0.30 & [0.29/\!\!\!\!& 0.32] \str \\ 
\phantom{val} & \phantom{val} & \phantom{val} & 150\;\mathrm{GeV} &
\hspace{0.5cm}0.36 & [0.34/\!\!\!\!& 0.39] \str \\ 
\phantom{val} & \phantom{val} & \phantom{val} & 190\;\mathrm{GeV} &
\hspace{0.5cm}0.39 & [0.36/\!\!\!\!& 0.43] \str \\ \hline
\end{array}$
\end{center}
\caption{\textit{Representative values for triple SM Higgs-strahlung 
(unpolarized beams). The sensitivity to the quadrilinear coupling is 
illustrated by the variation of the cross sections when 
$\lambda_{HHHH}$ is altered by factors $1/2$ and $3/2$, as indicated 
in the square brackets.}}
\label{tab:quadri}
\end{table}

\subsection{Double and triple Higgs production in the MSSM}
In the MSSM the following trilinear self-couplings arise among the
neutral Higgs bosons
\beq
\begin{array}{l l l l}
hhh, & Hhh, & HHh, & HHH \non \\
hAA, & HAA &     & \non
\end{array} 
\eeq
There are many more quadrilinear couplings \cite{dubinin}. The double
and triple Higgs production processes and the trilinear couplings
involved are listed in Table~\ref{tab:coup}.  Since in most of the
MSSM parameter space the Higgs bosons $H$ and $A$ are heavy the
emphasis of the subsequent analysis will be on the production of a
light Higgs boson pair with heavy Higgs boson final states taken into
account where necessary. \s

In contrast to the SM, the resonant production of a heavy Higgs boson
$H$ that subsequently decays into a light Higgs boson pair is
possible. The resonant production is important in part of the
parameter space with $H$ masses between 200 and 350~GeV and for
moderate values of $\tan\beta$ \cite{zerwas1}. In this case the
branching ratio obtained from the partial width
\beq
\Gamma (H \to hh) = 
\frac{\sqrt{2} G_F M_Z^4 \beta_h}{32 \pi M_H} \lambda_{Hhh}^2 
\eeq
is neither too small nor too close to unity to be accessible directly.
The resonant $H$ decay enhances the total production cross section for
$hh$ by an order of magnitude \cite{djouadi2} thus improving the
potential for the measurement of the Higgs self-coupling
$\lambda_{Hhh}$. Apart from the exceptional case of a light pseudoscalar
Higgs boson $A$ the coupling $\lambda_{Hhh}$ is the only trilinear
Higgs self-coupling which is accessible via resonant decay. All the
other couplings have to be extracted from the continuum pair
production. The generic diagrams contributing to double and triple Higgs
production and $WW$ double Higgs fusion have been shown in
Fig.~\ref{fig:graphs}.

\subsection{MSSM double Higgs-strahlung}
The (unpolarized) production cross section for a pair of light Higgs
bosons via Higgs-strahlung in the MSSM has the same structure as in
the SM yet involving $H$ and $A$ exchange diagrams
\cite{djouadi2,ours,osland}, cf.~Fig.~\ref{fig:graphs}:
\beq
\frac{d \sigma (e^+ e^- \to Zhh)}{d x_1 d x_2} &=& 
\frac{\sqrt{2} G_F^3 M_Z^6}{384 \pi^3 s} 
\frac{v_e^2 + a_e^2}{(1- \mu_Z)^2}\, {\cal Z}_{11}
\label{zhh1}
\eeq
with
\beq
{\cal Z}_{11} &=&  {\mathfrak a}^2 f_0 + 
\frac{{\mathfrak a}}{2} \left[ 
\frac{\sin^2 (\beta-\alpha) f_3}{y_1 + \mu_{1Z}} + 
\frac{\cos^2 (\beta-\alpha) f_3}{y_1 + \mu_{1A}} \right] \non \\ 
&& {} + \frac{\sin^4 (\beta-\alpha)}{4\mu_Z (y_1+\mu_{1Z})} \left[ 
\frac{f_1}{y_1+\mu_{1Z}} + \frac{f_2}{y_2+\mu_{1Z}} \right] \non 
\\
&& {} + \frac{\cos^4 (\beta-\alpha)}{4\mu_Z (y_1+\mu_{1A})} \left[ 
\frac{f_1}{y_1+\mu_{1A}} + \frac{f_2}{y_2+\mu_{1A}} \right] \non\\
&& {} + \frac{\sin^2 2(\beta-\alpha)}{8\mu_Z (y_1+\mu_{1A})} \left[ 
\frac{f_1}{y_1+\mu_{1Z}} + \frac{f_2}{y_2+\mu_{1Z}} \right]
+ \Bigg\{ y_1 \leftrightarrow y_2 \Bigg\}
\label{zhh2}
\eeq
and
\beq
{\mathfrak a} = \left[ 
\frac{\lambda_{hhh}\sin(\beta-\alpha)}{y_3-\mu_{1Z}}
+ \frac{\lambda_{Hhh}\cos(\beta-\alpha)}{y_3 - \mu_{2Z}} \right] 
+ \frac{2 \sin^2(\beta-\alpha)}{y_1+\mu_{1Z}} 
+ \frac{2 \sin^2(\beta-\alpha)}{y_2+\mu_{1Z}} 
+ \frac{1}{\mu_Z}
\label{zhh3}
\eeq
The notation is the same as in the Standard Model, with
$\mu_1=M_h^2/s$ and $\mu_2=M_H^2/s$. Taking into account that in part
of the parameter space the heavy neutral Higgs boson $H$ or the
pseudoscalar Higgs boson $A$ may become resonant, the decay widths are
included implicitly by shifting the masses to complex values $M \to M
- i\Gamma/2$, {\it i.e.}~$\mu_i \to \mu_i - i \gamma_i$ with the
reduced width $\gamma_i = M_i\Gamma_i/s$, and by replacing products of
propagators $\pi_1 \pi_2$ with Re$(\pi_1 \pi_2^*)$.\s

\begin{figure}
\begin{center}
\epsfig{figure=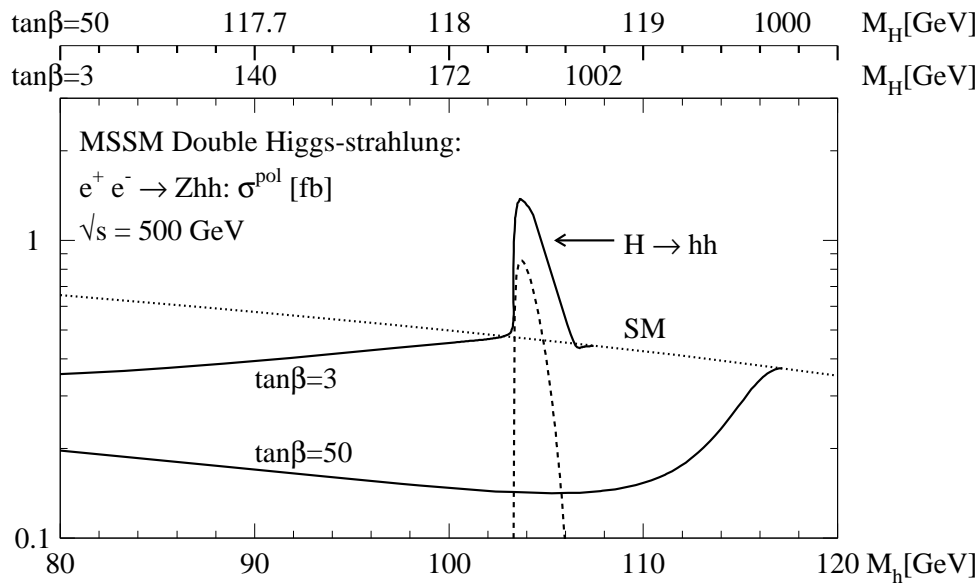,width=13cm}
\caption{\textit{
  Total cross sections for MSSM $hh$ production via double
  Higgs-strahlung at $e^+e^-$ linear colliders for $\tan\beta =3$, $50$ 
  and $\sqrt{s}=500\;\mathrm{GeV}$, including mixing effects 
  [$A = 1$~TeV, $\mu=-1/1$~TeV for $\tan\beta=3/50$]. 
  The dotted line indicates the SM cross section. The dashed line represents 
  the resonant contribution.}}
\label{fig:NLC/SUSY}
\end{center}
\end{figure}
Fig.~\ref{fig:NLC/SUSY} shows the total polarized cross section $Zhh$
for $\sqrt{s}=500$~GeV as a function of the Higgs masses. The mixing
parameters are chosen $A=1$~TeV and $\mu=-1$(1)~TeV for
$\tan\beta=3$(50). Choosing $\tan\beta$ and $M_h$ as
input parameters the masses of the heavy Higgs bosons are fixed in the
MSSM for given values of the mixing parameters \cite{zerwas1}. As can
be inferred from Fig.~\ref{fig:NLC/SUSY} the continuum cross section
is smaller compared to the SM result. This originates from the
suppression of the MSSM vertices by $\sin/\cos$ functions of the
mixing angles $\alpha$ and $\beta$. In contrast, the cross section is
enhanced by an order of magnitude when resonant $H$ production with
subsequent decay $H\to hh$ is kinematically possible.  Approaching the
maximum value of $M_h$ for $\tan\beta$ fixed, $H$ and $A$ become very
heavy and decouple so that the MSSM $Zhh$ cross section reaches the SM
value.  Due to the decoupling theorem resonant $H$ production does not
increase the MSSM cross section for large values of $\tan\beta$ since
the decay $H\to hh$ is kinematically not possible unless the
decoupling region is reached where the $ZZH$ coupling is too small to
produce a sizeable cross section.\s

\begin{figure}
\begin{center}
\epsfig{figure=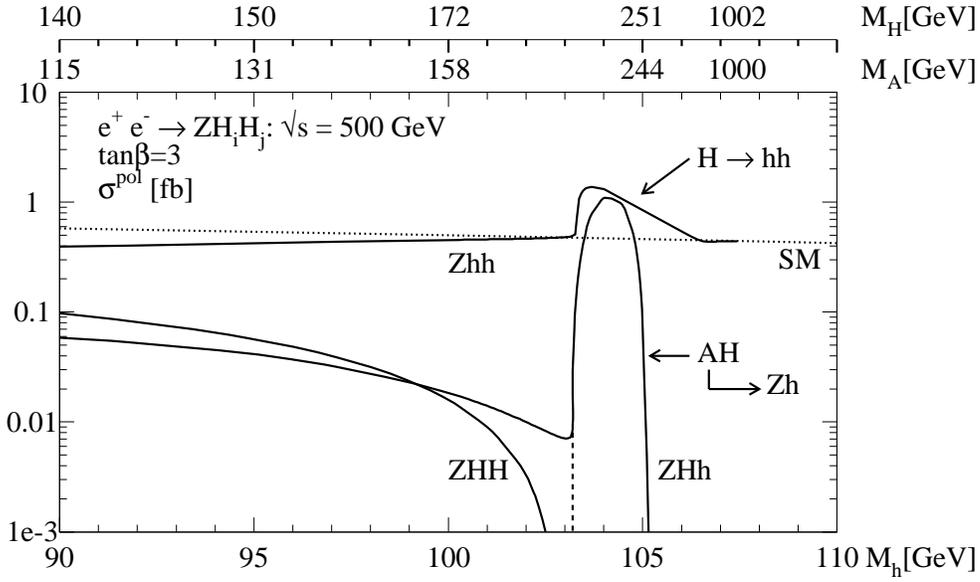,width=13cm}
\\
\end{center}
\caption{\textit{Cross sections for the processes $Zhh$, $ZHh$ and 
$ZHH$ for $\sqrt{s}=500$~GeV and tan$\beta=3$, including mixing 
effects ($A = 1$~TeV, $\mu=-1$~TeV).}}
\label{fig:add3}
\end{figure}

The formulae for the Higgs-strahlung processes $ZH_iH_j$
[$H_i,H_j=h,H$] are deferred to the Appendix. They are more
complicated due to the different masses of the final state Higgs
particles. The corresponding polarized cross sections are shown
together with the SM cross section as a function of $M_h$ in
Fig.~\ref{fig:add3}. The c.m.~energy is chosen equal to 500~GeV and
$\tan\beta =3$. Evidently, if kinematically possible, the cross
sections $Zhh,$ $ZHh$ and $ZHH$ add up approximately to the SM result.

\subsection{Triple Higgs production}
The processes $e^+e^- \to ZH_i$ and $e^+e^- \to AH_i$ are proportional
to $\sin^2(\beta-\alpha)/\cos^2(\beta-\alpha)$ and
$\cos^2(\beta-\alpha)/ \sin^2(\beta-\alpha)$ for $H_i =h,H$,
respectively, so that Higgs-strahlung and associated Higgs production
are complementary to each other. In addition the processes are
among themselves complementary \cite{djkalzer}. Since the double and
triple Higgs production processes $ZH_i H_j$, $ZAA$ and $AH_i H_j$,
$AAA$ proceed via virtual $h,H$ bosons this behaviour will be found
there, too, being of more complex matrix form, however, since the
combination of the different mechanisms is more involved,
cf.~Fig.~\ref{fig:graphs}. 
In this subsection the processes $e^+e^- \to Ahh$ and $e^+e^- \to AAA$
will be discussed in more detail. The more complicated results for
triple Higgs production involving heavy Higgs bosons are listed in the
Appendix. \s

The unpolarized cross section for $e^+e^- \to Ahh$ is given by 
\beq
\frac{d\sigma [e^+ e^- \to Ahh]}{dx_1 dx_2} = 
\frac{G_F^3 M_Z^6}{768 \sqrt{2} \pi^3 s} 
\frac{v_e^2 + a_e^2}{(1-\mu_Z)^2} {\mathfrak A}_{11}
\eeq
where the function ${\mathfrak A}_{11}$ can be cast into the form
\beq
{\mathfrak A}_{11} &=& \left[ 
\frac{c_1 \lambda_{hhh}} {y_3-\mu_{1A}}
+ \frac{c_2 \lambda_{H hh}} {y_3-\mu_{2A}} \right]^2 \frac{g_0}{2}
+ \frac{c_1^2 \lambda^2_{hAA}} {(y_1+\mu_{1A})^2} g_1
+ \frac{c_1^2 d_1^2 } {(y_1+\mu_{1Z})^2} g_2 
\non \\
&+& \left[ \frac{c_1 \lambda_{hhh} } {y_3-\mu_{1A}}
+ \frac{c_2 \lambda_{H hh}} {y_3-\mu_{2A}} \right]
\left[ \frac{c_1 \lambda_{hAA}} {y_1+\mu_{1A}} g_3
+ \frac{c_1 d_1 } {y_1+\mu_{1Z}} g_4 \right] \non \\
&+& 
\frac{c_1^2\lambda_{h AA}^2}{2(y_1+\mu_{1A})(y_2+\mu_{1A})}
g_5
+ \frac{c_1^2 d_1 \lambda_{h AA}}{(y_1+\mu_{1A})(y_1+\mu_{1Z})}g_6
\non \\
&+&\frac{c_1^2 d_1 \lambda_{h AA}}{(y_1+\mu_{1A})(y_2+\mu_{1Z})}g_7
+ \frac{c_1^2 d_1^2}{2(y_1+\mu_{1Z})(y_2+\mu_{1Z})}g_8
\non \\
&+& \Bigg\{ y_1 \leftrightarrow y_2 \Bigg\}
\eeq
with $\mu_{1,2} = M_{h,H}^2/s$ and the vertex coefficients
\beq
c_1/c_2 = \cos (\beta-\alpha)/-\sin(\beta-\alpha) \qquad \mathrm{and} 
\qquad d_1/d_2 = \sin (\beta - \alpha)/\cos(\beta-\alpha)
\eeq
The coefficients $g_k$ are given by
\beq
g_0 &=& \mu_Z[(y_1+y_2)^2-4\mu_A] \non \\
g_1 &=& \mu_Z[y_1^2-2y_1-4\mu_1+1] \non \\
g_2 &=& \mu_Z [y_1(y_1+2)+ 4y_2(y_2+y_1-1)+1- 4(\mu_1+2\mu_A)]
+ (\mu_1-\mu_A)^2 \non\\
&& [8+[(1-y_1)^2-4\mu_1]/\mu_Z] + (\mu_1-\mu_A) 
[4 y_2(1+y_1)+2(y_1^2-1)]\non\\
g_3 &=& 2\mu_Z(y_1^2- y_1+y_2+y_1 y_2-2\mu_A) \non \\
g_4 &=& 2\mu_Z(y_1^2+y_1+2y_2^2-y_2+3y_1 y_2 -6 
\mu_A) \non\\ 
&&+2(\mu_1-\mu_A)(y_1^2- y_1+ y_2  + y_1y_2 -2\mu_A) 
    \non \\
g_5 &=& 2\mu_Z(y_1+ y_2+y_1y_2+4\mu_1-2\mu_A-1) \non \\
g_6 &=& 2\mu_Z(y_1^2+ 2 y_1y_2+ 2y_2+4\mu_1-4\mu_A-1) \non \\
 && + 2(\mu_1-\mu_A)(y_1^2-2y_1-4\mu_1+1) \non \\
g_7 &=& 2[ \mu_Z(2y_1^2-3y_1+y_1y_2+y_2-4\mu_1-2\mu_A+1) 
\non \\
 && + (\mu_1-\mu_A)(y_1+y_1y_2+y_2 +4\mu_1-2\mu_A-1)] \non \\
g_8 &=& 2 \left\{ \mu_Z(y_1+y_2+2y_1^2+2y_2^2+5y_1 y_2 -1 + 4\mu_1 
 -10\mu_A)\right. \non \\
&& +4(\mu_1-\mu_A)(-2\mu_1-\mu_A-y_1-y_2+1) \non \\
&& + [2(\mu_1-\mu_A)((y_1+y_2+y_1y_2+y_1^2+y_2^2-1)\mu_Z
+2\mu_1^2+4\mu_A^2-\mu_1+\mu_A)
\non\\
&& + \left.
6\mu_A(\mu_A^2-\mu_1^2) + (\mu_1-\mu_A)^2 (1+y_1)(1+y_2)]/\mu_Z \right\}
\label{gfuncs}
\eeq
The notation of the kinematics is the same as for double Higgs-strahlung. \s

The cross section for triple $A$ production is much simpler since there are 
only a few diagrams involved in the process, cf.~Fig.~\ref{fig:graphs}:
\beq
\frac{d\sigma [e^+ e^- \to AAA]}{dx_1 dx_2} = 
\frac{G_F^3 M_Z^6}{768 \sqrt{2} \pi^3 s} 
\frac{v_e^2 + a_e^2}{(1-\mu_Z)^2} {\mathfrak A}_{33}
\eeq
where
\beq
{\mathfrak A}_{33} = D_3^2 g_0+ D_1^2 g_1 +D_2^2 g_1'
- D_3 D_1 g_3- D_3 D_2 g_3' + D_1 D_2 g_5
\eeq
and
\beq
D_k= \frac{\lambda_{hAA} c_1} {y_k-\mu_{1A}}
   + \frac{\lambda_{HAA} c_2} {y_k-\mu_{2A}} 
\eeq
The scaled mass parameter $\mu_1$ has to be replaced by $\mu_A$ in the
coefficients $g_i$ and $g_i'$ where the functions $g_i'$ are related
to $g_i$ via $g_i'(y_1,y_2)=g_i(y_2,y_1)$.\s

\begin{figure}
\begin{center}
\epsfig{figure=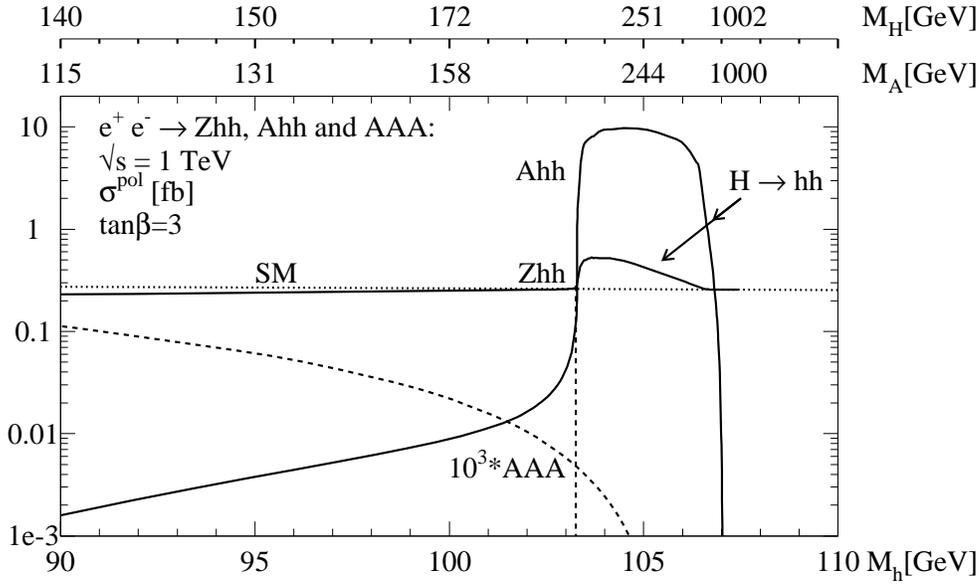,width=13cm}\\[0.9cm]
\end{center}
\caption{\textit{Cross sections of the processes Zhh, Ahh and AAA for 
$\tan\beta = 3$ and $\sqrt{s}=1$~TeV, including mixing effects 
($A=1$~TeV, $\mu=-1$~TeV.)}}
\label{fig:ahh}
\end{figure}
Fig.~\ref{fig:ahh} shows the results for the triple Higgs production
processes $e^+e^- \to Ahh$ and $e^+e^- \to AAA$ compared to double
Higgs-strahlung $e^+e^- \to Zhh$ and the SM counterpart, $e^+e^- \to
ZHH$. The triple Higgs processes are small in the
continuum. This behaviour can be explained by analysing the effective
couplings. The process $e^+e^- \to Ah_{virt} \to Ahh$ involves the
coupling $\cos (\beta-\alpha ) \lambda_{hhh}$ and $e^+e^- \to
AH_{virt} \to Ahh$ is proportional to
$\sin(\beta-\alpha)\lambda_{Hhh}$. In both cases the products of the
gauge and Higgs self-coupling are small as can be inferred from
Figs.~\ref{fig:lambda1} and \ref{fig:lambda2}. Only if resonant $H$
production with subsequent $H\to hh$ decay is kinematically allowed
the cross section increases by about three orders of magnitude. The
process $e^+e^- \to Ah_{virt} \to AAA$ exhibits the effective
coupling $\cos (\beta-\alpha ) \lambda_{hAA}$, one of the factors
always being small, and $e^+e^- \to AH_{virt} \to AAA$ is proportional
to $\sin (\beta-\alpha )\lambda_{HAA}$ with $\lambda_{HAA}$ being of
order $1/2$. Taking into account the reduction of the phase space due
to three heavy Higgs bosons $A$ in the final state the pseudoscalar
triple Higgs production process is always small.

\subsection{MSSM $WW$ double Higgs fusion}
The $WW$ double Higgs fusion process in the MSSM involves also $H$
and $H^\pm$ exchange diagrams in contrast to the SM. The dominant
longitudinal amplitude for on-shell $W$ bosons with a pair of light
Higgs bosons in the final state is given by:
\beq
{\cal M}_{LL} &=& \frac{G_F \hat{s}}{\sqrt{2}} \left\{ (1+\beta_W^2) 
\left[ 1 + 
\frac{\lambda_{hhh}\sin(\beta-\alpha)}{(\hat{s}-M_h^2)/M_Z^2} + 
\frac{\lambda_{Hhh}\cos(\beta-\alpha)}{(\hat{s}-M_H^2)/M_Z^2} \right] 
\right. \non\\
&+& \frac{\sin^2(\beta-\alpha)}{\beta_W \beta_h} \left[ 
\frac{(1-\beta_W^4) + 
(\beta_W - \beta_h \cos\theta)^2}{\cos\theta-x_W} - 
\frac{(1-\beta_W^4) + 
(\beta_W + \beta_h \cos\theta)^2}{\cos\theta+x_W} \right]
\non \\
&+& \left. \frac{\cos^2(\beta-\alpha)}{\beta_W \beta_h} \left[ 
\frac{(\beta_W - \beta_h \cos\theta)^2}{\cos\theta-x_+} - 
\frac{(\beta_W + \beta_h \cos\theta)^2}{\cos\theta+x_+} \right] 
\right\}
\eeq
$\hat{s}^{1/2}$ is the c.m.\ energy of the subprocess,
$\theta$ the scattering angle, $\beta_W$ and $\beta_h$ denote the
velocities of the $W$ and $h$ bosons, and
\begin{equation}
x_W = \frac{1-2\mu_h}{\beta_W \beta_h} \qquad {\rm and} \qquad
x_+ = \frac{1-2\mu_h+2\mu_{H^\pm}-2\mu_W}{\beta_W\beta_h} 
\end{equation}
After integrating out the angular dependence, the total cross section
of the fusion subprocess reads
\beq 
\hat{\sigma}_{LL} &=&
\frac{G_F^2 M_W^4}{4\pi \hat{s}} \frac{\beta_h}{ \beta_W
  (1-\beta_W^2)^2} \left\{ (1+\beta_W^2)^2 \left[ \frac{
      \lambda_{hhh} d_1}{(\hat{s} -M_h^2)/M_Z^2} + \frac{
      \lambda_{Hhh} d_2}{(\hat{s} -M_H^2)/M_Z^2} + 1\right]^2 \right. 
    \non \\
&& {}+ \frac{2(1+\beta_W^2)} {\beta_W \beta_h } \left[
  \frac{\lambda_{hhh}d_1}{(\hat{s} -M_h^2)/M_Z^2} + \frac{
    \lambda_{Hhh}d_2}{(\hat{s} -M_H^2)/M_Z^2} + 1 \right] \left[
  d_1^2 a_1^W  + c_1^2  a_1^+   \right] \non \\
&& {}+ \left. \left( \frac{d_1^2}{\beta_W \beta_h } \right)^2 a_2^W +
  \left( \frac{c_1^2}{\beta_W \beta_h } \right)^2 a_2^+ + 4
  \left(\frac{c_1^2 d_1^2}{\beta_W^2 \beta_h^2 } \right)
  [a_3^W + a_3^+ ] \right\} 
\eeq 
with 
\beq 
a_1^W &=& [ (x_W \beta_h
-\beta_W)^2 + r_W ] l_W +2 \beta_h (x_W \beta_h -2\beta_W ) \non \\
a_2^W &=& \left[ \frac{1}{x_W} l_W + \frac{2} {x_W^2-1} \right] \bigg[
x_W^2 \beta_h^2 (3 \beta_h^2 x_W^2
+2 r_W +14 \beta_W^2) \non \\
&& {}-(\beta_W^2+ r_W)^2 -4 \beta_h \beta_W x_W
(3 \beta_h^2 x_W^2 +\beta_W^2+r_W ) \bigg] \non \\
&& {}- \frac{4}{x_W^2-1} \left[ \beta_h^2 (\beta_h^2 x_W^2
  +4 \beta^2_W - 4 \beta_h x_W \beta_W) - (\beta_W^2 +r_W)^2 \right] 
\non \\
a_3^W&=& \frac{1}{x_+^2 - x_W^2} \, l_W \bigg[ 2 \beta_W \beta_h x_W
[(\beta_W^2+x_W^2 \beta_h^2)(x_W+x_+)
+ x_W r_W +x_+ r_+]  \non \\
&& {}-x_+( r_+ + r_W + \beta_h^2 x_W^2)(\beta_W^2 + \beta_h^2 x_W^2)
-\beta_W^2( x_+ \beta_W^2 +4\beta_h^2 x_W^3+
x_+ x_W^2 \beta_h^2 ) \non \\
&& {}- x_+ r_W r_+ \bigg] + \beta_h^2 \left[ 
\beta_h^2 x_+ x_W -2 \beta_W
  \beta_h (x_W+x_+) + 4\beta_W^2 \right] \non \\
a_i ^+ &\equiv& a_i^W \ 
(x_W \leftrightarrow x_+ \ , \ r_W \leftrightarrow r_+ )
\eeq 
and $r_W = 1- \beta^4_W$, $r_+ = 0$. \s

\begin{figure}
\begin{center}
\includegraphics{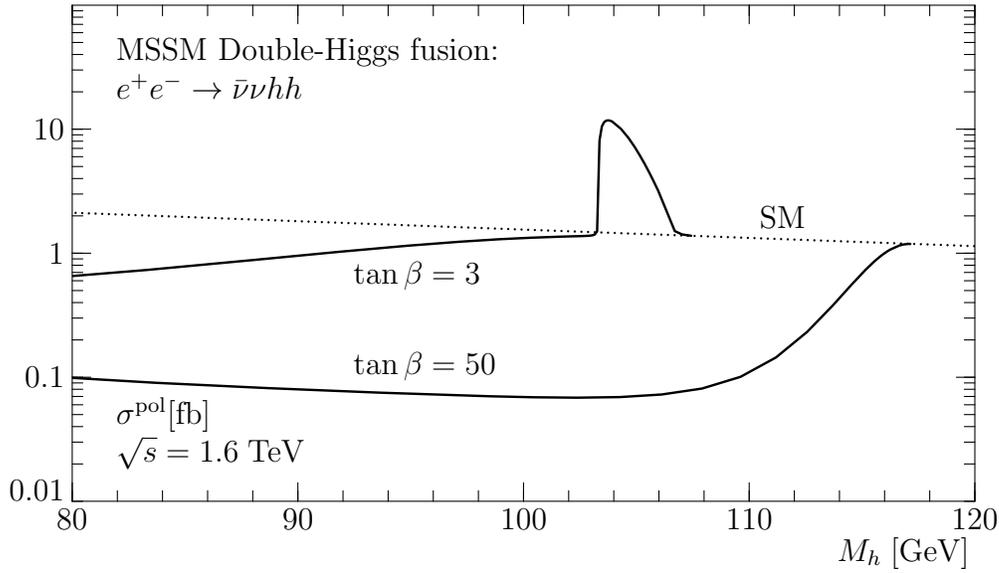}\\[1cm]
\end{center}
\caption{\textit{Total cross sections for MSSM $hh$ production via 
double WW double-Higgs fusion at $e^+e^-$ linear colliders for 
$\tan\beta = 3,$ $50$ and $\sqrt{s}=1.6$~TeV, including mixing effects 
($A = 1$~TeV, $\mu=-1/1$~TeV for $\tan\beta=3/50$).}}
\label{fig:WW1/SUSY}
\end{figure}

\begin{figure}
\begin{center}
\vspace{1.3cm}
\includegraphics{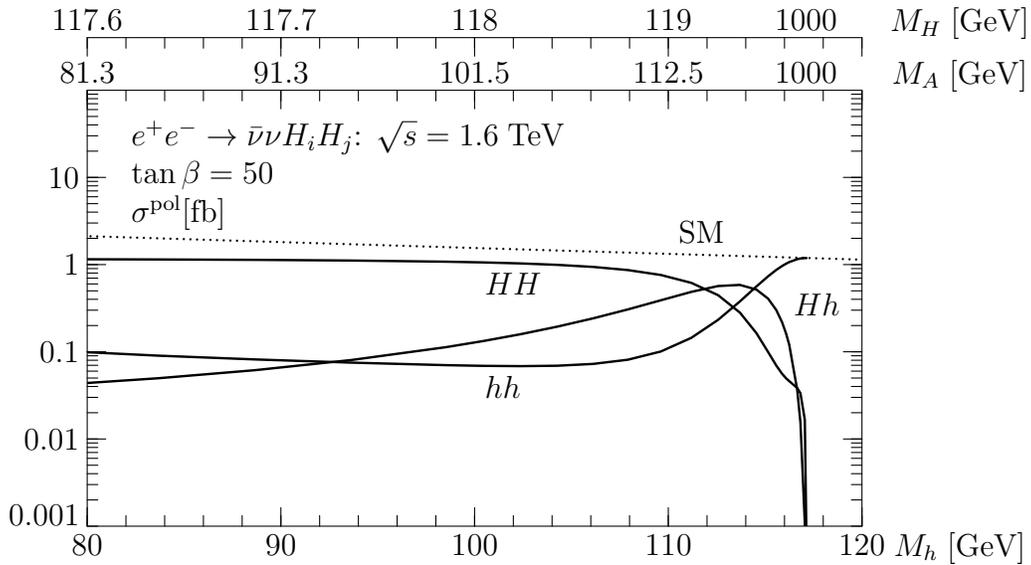} \\[1cm]
\end{center}
\caption{\textit{Total cross sections for WW double-Higgs fusion with 
$hh$, $Hh$ and $HH$ final states for $\sqrt{s}=1.6$~TeV and 
tan$\beta = 50$, including mixing effects ($A = 1$~TeV, $\mu=1$~TeV).}}
\label{fig:WW2/SUSY}
\end{figure}

The cross sections presented in Figs.~\ref{fig:WW1/SUSY} and
\ref{fig:WW2/SUSY} have been calculated exactly by means of the
program CompHEP~\cite{boos} without using any approximations. The
electron-positron beams are taken oppositely polarized. The process
involving a light Higgs bosons pair in the final state is suppressed
compared to the SM unless resonance $H\to hh$ decay is possible for
modest $\tan\beta$ values. In the case of large $\tan\beta$ the MSSM
fusion process is much more suppressed and resonance production does
not play any r\^{o}le since in this parameter range the gauge
couplings involved are very small. Yet, the $H$ mass is rather light
so that the processes $hh$, $Hh$ and $HH$ approximately add up to the
SM cross section. The formulae for the longitudinal amplitudes $W_L
W_L \to Hh, HH$ and $AA$ can be found in the Appendix.

\subsection{Sensitivity areas}
In this subsection the perspectives for the determination of the Higgs
self-couplings in the SM and the MSSM will be analysed. For the MSSM
case, the feasibility of measuring the trilinear Higgs self-couplings
shall be represented in compact form in sensitivity plots in the
[$M_A,\tan\beta$] plane \cite{djouadi2,ours}.\s

\noindent
{\bf 1.) SM WW Double Higgs fusion}\newline 
\noindent 
Since for SM Higgs masses below 140 GeV the main decay channel is
given by $H\to b\bar{b}$ the final state for the $WW$ fusion process
is characterized by two $b\bar{b}$ pairs and missing energy in this
mass region. Analogously, for $M_H \,\gesim\, 140$~GeV the final state
consists of two $WW^{(*)}$ pairs and missing energy since in this case
$H$ dominantly decays in $WW^{(*)}$, see Fig.~\ref{smdecay}. Possible
background processes yielding $b\bar{b} b\bar{b} + \slash{E}$ in the
final state may proceed via the pure EW $WW$ fusion process being
${\cal O}(\alpha^6)$ or $\gamma\gamma$ fusion with a gluon exchanged
between the $b$-quarks of ${\cal O}(\alpha^4\alpha_s^2)$ whereas the
signal process is ${\cal O}(\alpha^4)$. Hence the $WW$ fusion
background process is suppressed by ${\cal O}(\alpha^2)$ compared to
the signal. $\gamma\gamma$ fusion results in large event rates in
forward direction. In contrast the bottom quarks from the decay of the
massive Higgs bosons are back-to-back. Applying appropriate angular cuts
will hence help to reduce the background with respect to the signal.
For $M_H\,\gesim\, 140$~GeV a possible background process is generated by
$\gamma\gamma$ fusion being suppressed by ${\cal O}(\alpha^2)$ compared 
to the signal. \s

\noindent
{\bf 2.) SM Double Higgs-strahlung} \newline
\noindent
In analogy to $WW$ fusion the double Higgs-strahlung process being
${\cal O}(\alpha^3)$ consists of a $Z$ boson and two $b\bar{b}$ pairs
in the final state if $M_H\,\lesim\, 140$~GeV and of
$ZWW^{(*)}WW^{(*)}$ if $M_H\,\gesim\, 140$~GeV. For the low mass range
signal to background analyses have been performed on parton level
\cite{moretti} as well as on detector level \cite{gay}. Assuming
efficient $b$-tagging and high purity sampling of $b$-quarks the main
background is due to irreducible EW processes of ${\cal O}(\alpha^5)$
and QCD backgrounds of ${\cal O}(\alpha^3\alpha_s^2)$. Another source
is the intrinsic background resulting from the signal diagrams which
do not contain the trilinear Higgs self-coupling $\lambda_{HHH}$. The
parton level analysis demonstrates that after applying typical
selection cuts the signal to background ratio which may be reached for
$M_H=110$~GeV is $S/B=25/60/140$ in the case of
$\sqrt{s}=500/1000/1500$~GeV \cite{moretti}. These results are based
on the assumption of a very high luminosity (chosen equal to
$\int{\cal L} = 1$~ab$^{-1}$ in \cite{moretti}), excellent $b$-tagging
performances and high di-jet resolution. \s

Whereas the previous results have been obtained without taking into
account detector effects, the analyses of Ref.~\cite{gay} are based on
a detector simulation and include hadronization effects. The examined
final state consists of $b\bar{b}b\bar{b}q\bar{q}$, the $q\bar{q}$
pair resulting from the $Z$ boson decay. The study of the 6-jet
topology demonstrates that for $M_H\sim 100$~GeV and $\int{\cal L} =
500$~fb$^{-1}$ a significance of 5 might be achievable, high
$b$-tagging efficiency and purity provided ($\epsilon_b = 0.85$,
$\epsilon_{not\,b}=0.9$).\s

\noindent
{\bf 3.) MSSM Double Higgs-strahlung} \newline
\noindent
For the process $Zhh$, there exists a parton level analysis of the
signal to background ratio \cite{houches,prochouches}. The final state
is given by $Zb\bar{b}b\bar{b}$ since $h$ dominantly decays into
$b\bar{b}$ independent of the value of $\tan\beta$
\cite{spirdjou,zerwas1,morettistir}. A low and a high $\tan\beta$
scenario has been investigated, mixing effects included ($A=2.4$~TeV,
$\mu=1$~TeV). After applying appropriate acceptance cuts the main
background results from pure EW and EW/QCD mixed processes. Applying
selection cuts according to the kinematics of the signal process, it
yields 156 events for $\int{\cal L} = 500$~fb$^{-1}$ and
$\tan\beta=3$, $M_h=104$~GeV. For these parameters resonant $H$ production
with subsequent decay $H\to hh$ is possible. The background only
amounts to a $10$\% correction. In the case of $\tan\beta=50$ where no
resonant production is possible and the cross section is strongly
suppressed, only for Higgs masses in the decoupling limit a reasonable
event rate of 15 may be achieved. The analysis does not take into
account $b$-tagging efficiency and $Z$ boson decay.\s
\begin{figure}
\begin{center}
\epsfig{figure=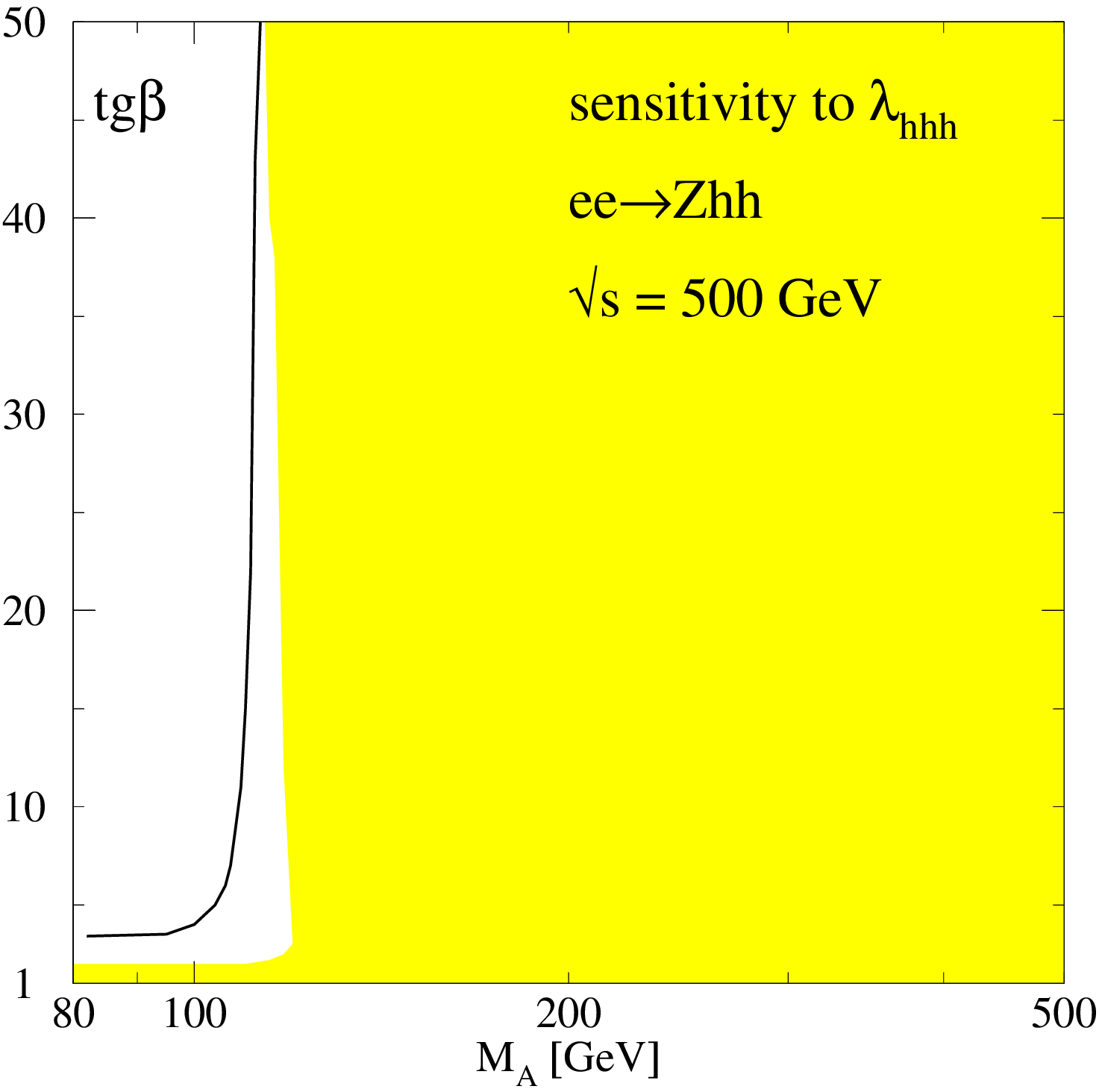,width=7cm}
\hspace{1cm}
\epsfig{figure=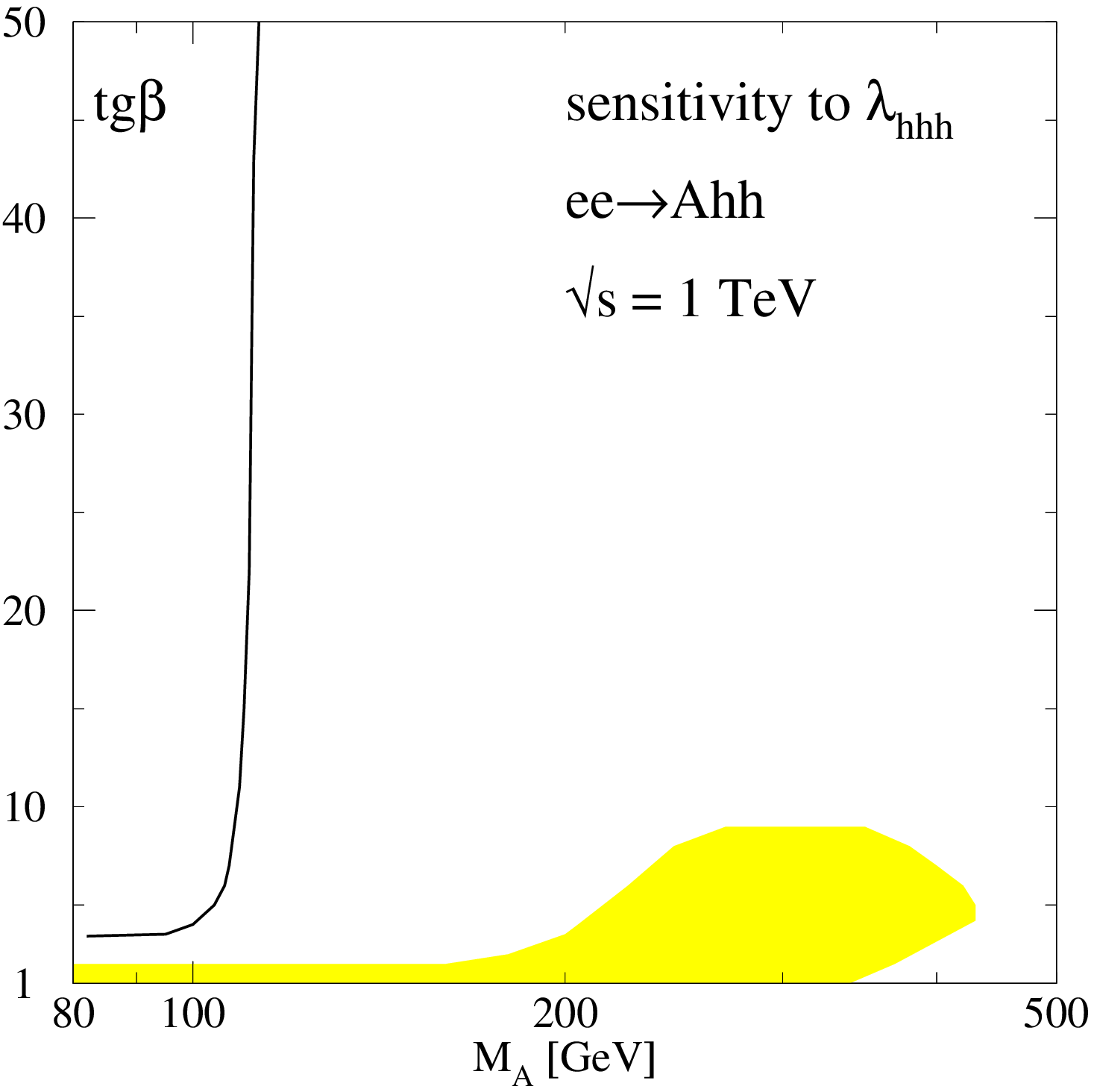,width=7cm}
\end{center}
\vspace{1.5cm}
\begin{center}
\epsfig{figure=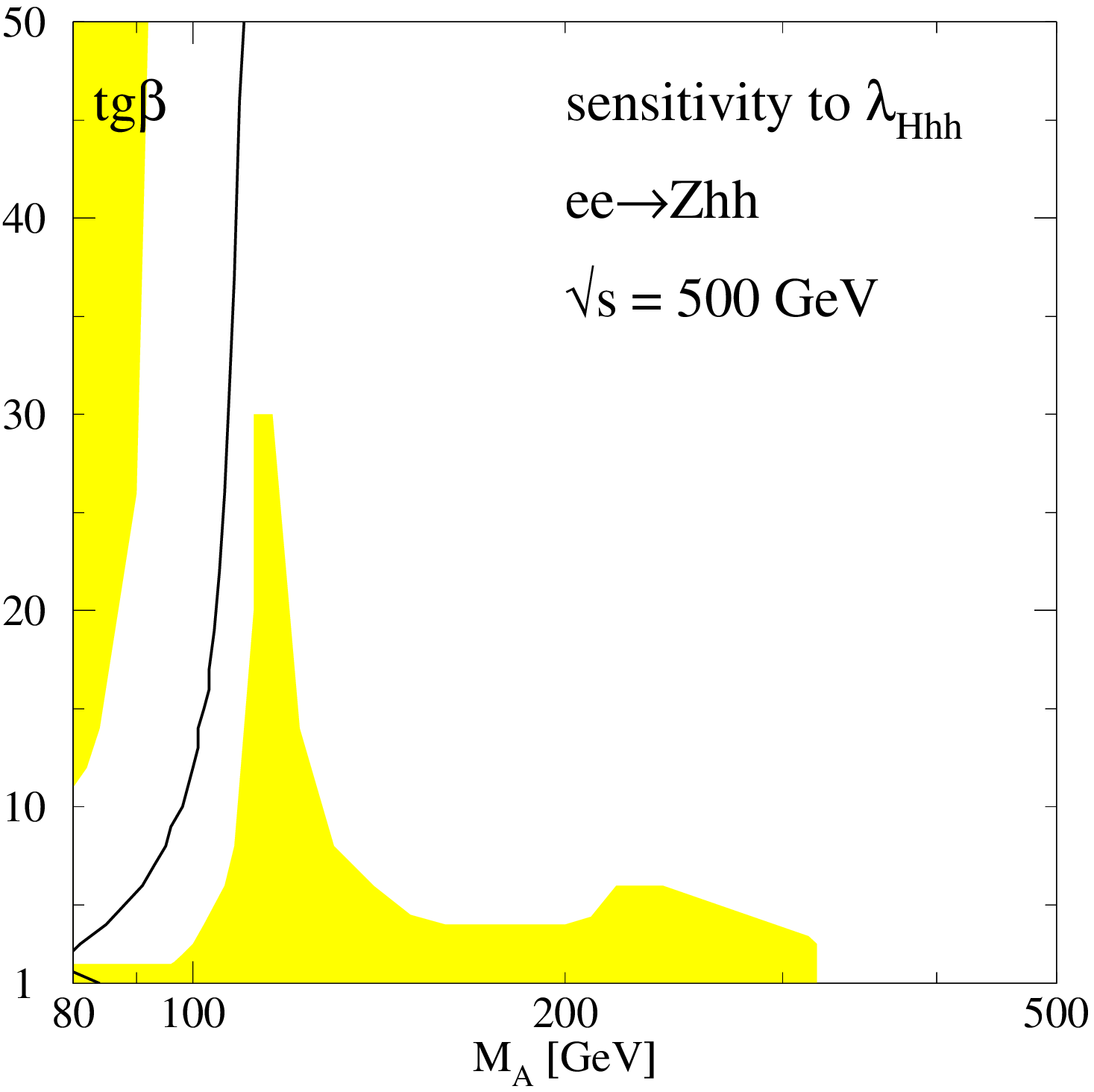,width=7cm}
\hspace{1cm}
\epsfig{figure=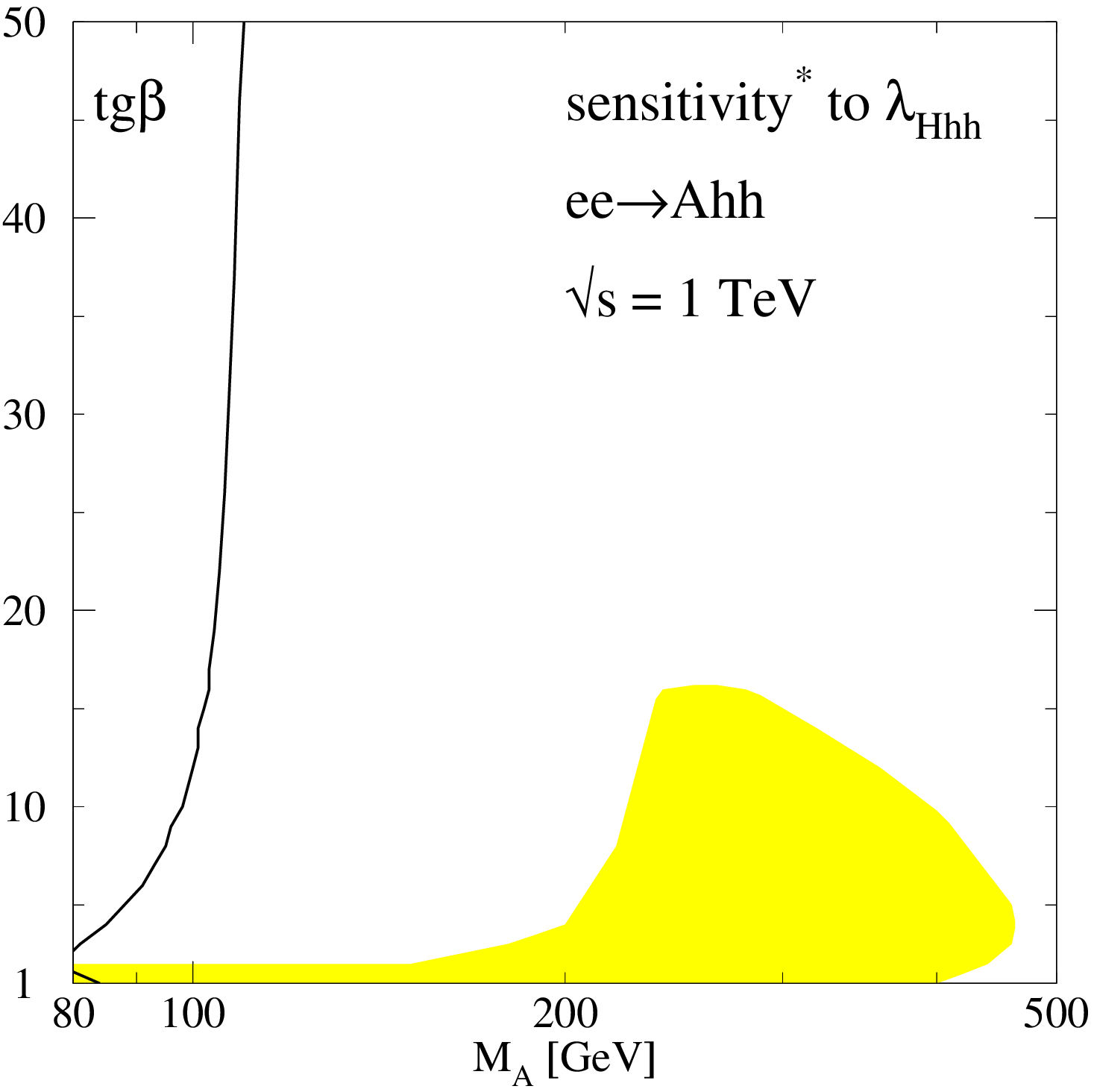,width=7cm}\\[0.5cm]
\end{center}
\caption{\textit{Sensitivity [* eff$\{\lambda\to (1\pm\frac{1}{2})\lambda)\}>2
$st.dev.] to the couplings 
$\lambda_{hhh}$ and $\lambda_{Hhh}$ in the 
processes $e^+e^-\to Zhh$ and $e^+e^-\to Ahh$ for collider energies 
$500$~GeV and $1$~TeV, respectively (no mixing). [Vanishing trilinear 
couplings are indicated by contour lines.]}}
\label{fig:sens1}
\end{figure}
\begin{figure}
\begin{center}
\epsfig{figure=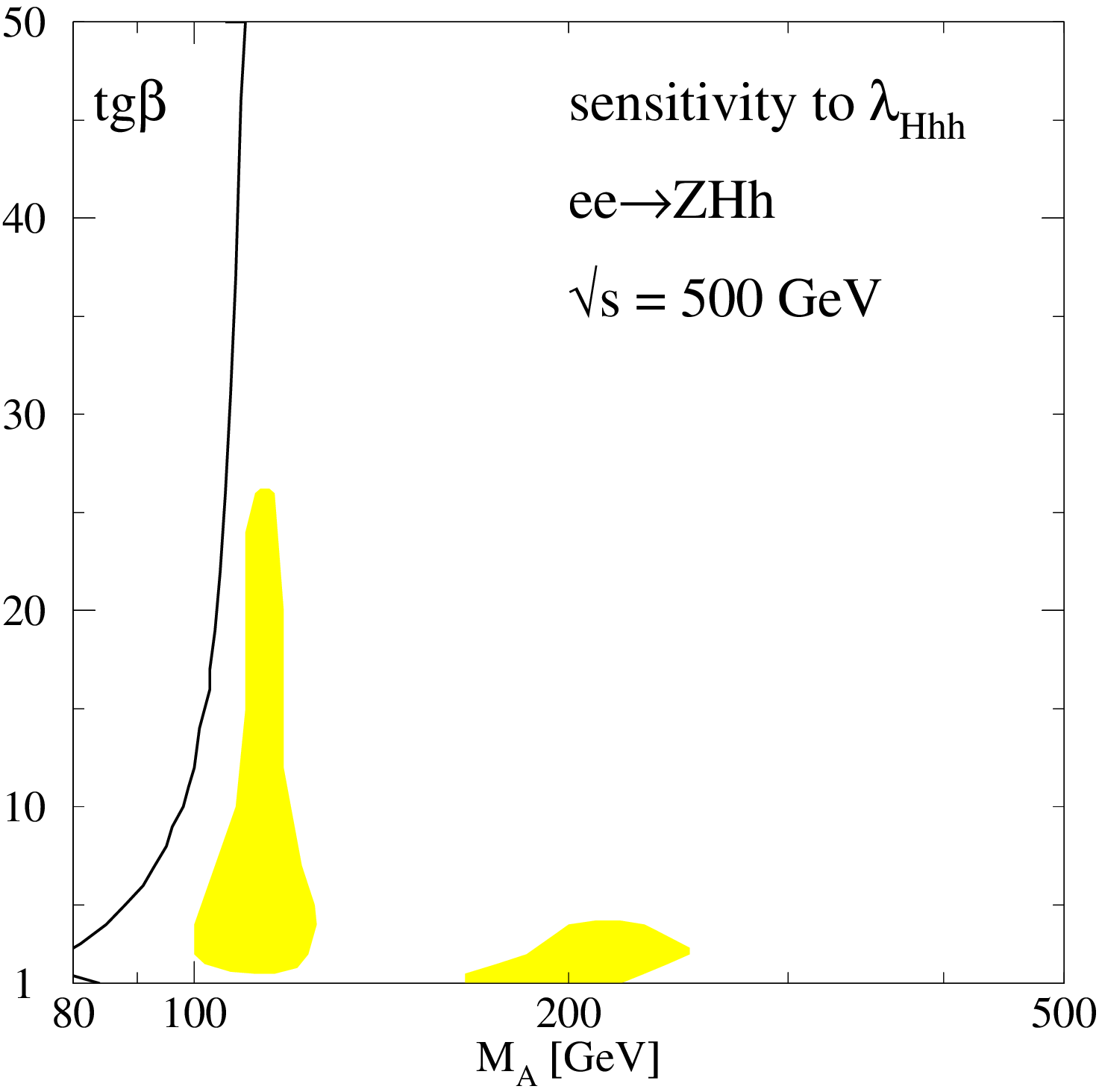,width=7cm}
\hspace{1cm}
\epsfig{figure=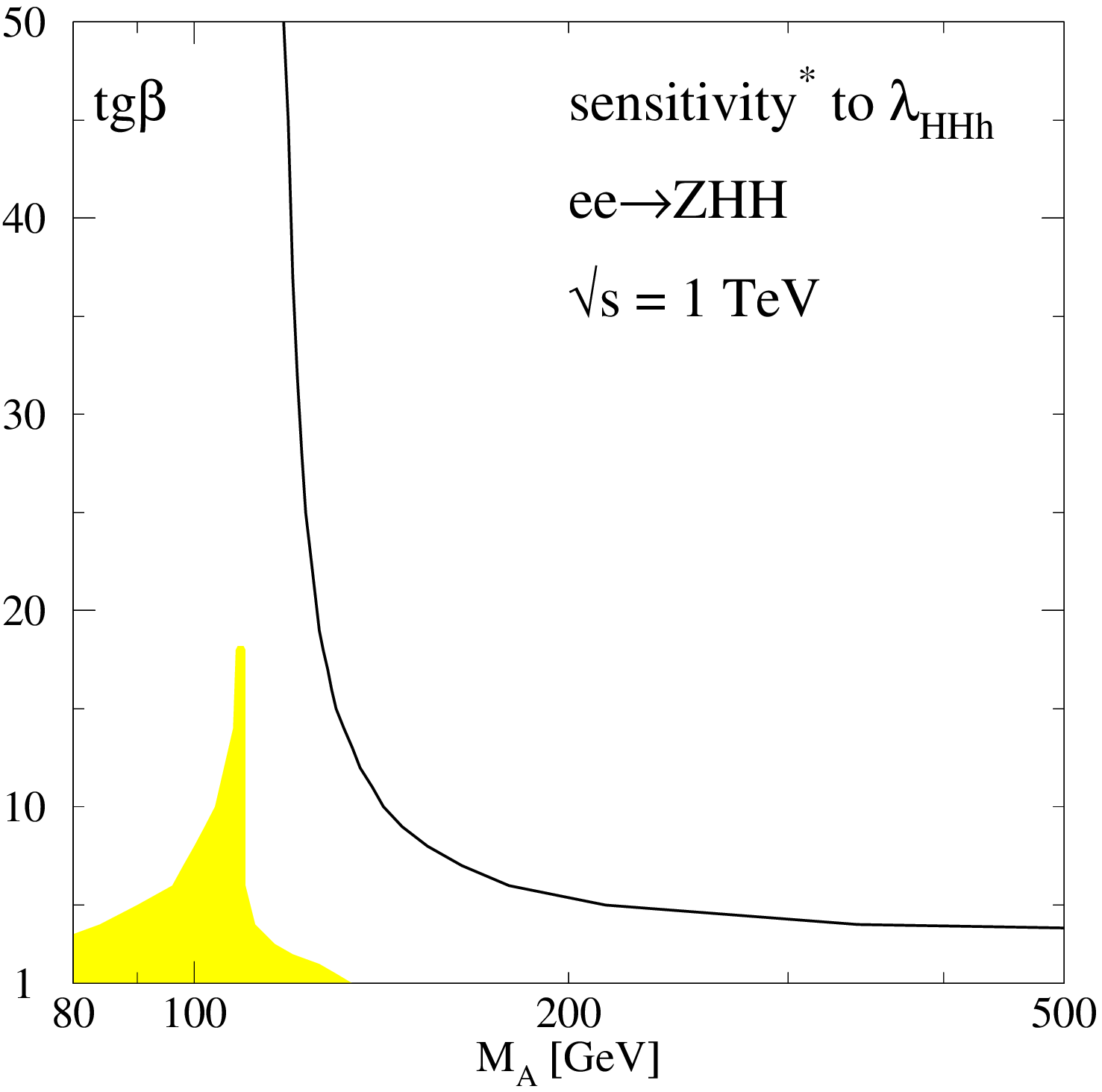,width=7cm}
\end{center}
\vspace{1.5cm}
\begin{center}
\epsfig{figure=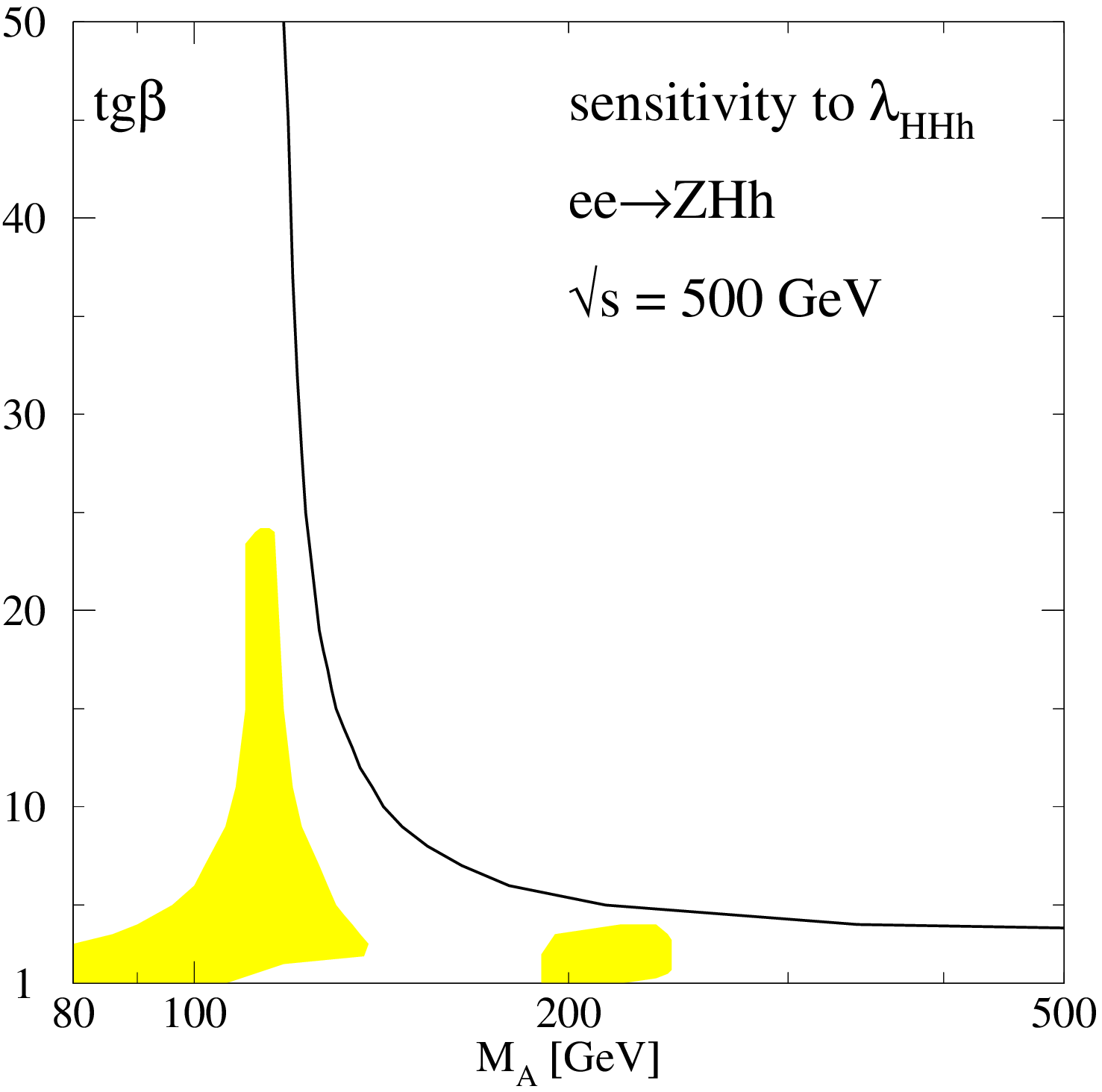,width=7cm}
\hspace{1cm}
\epsfig{figure=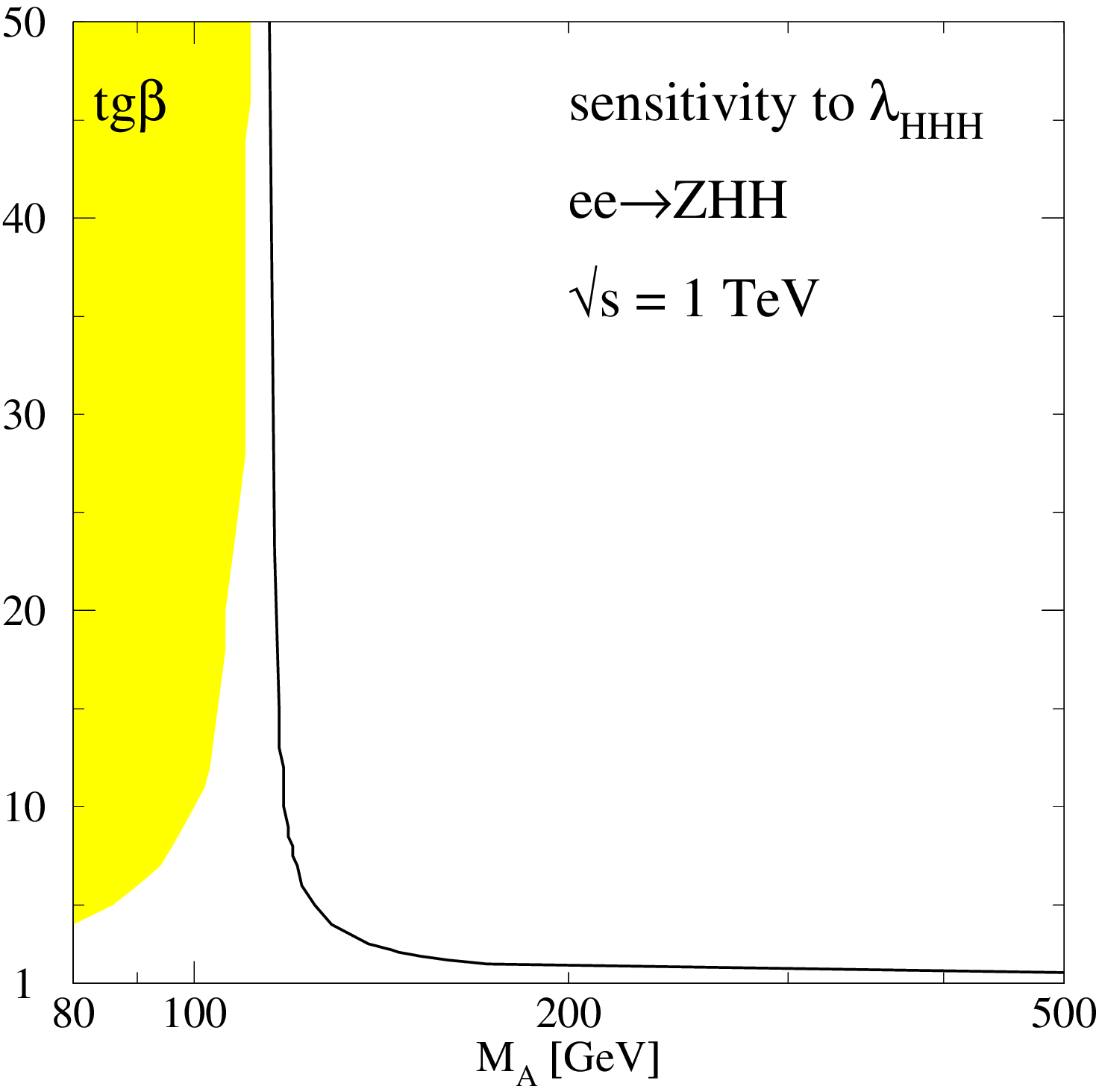,width=7cm}\\[0.5cm]
\end{center}
\caption{\textit{Sensitivity [* eff$\{\lambda\to 0\}>1$st.dev.] to 
the couplings $\lambda_{Hhh}$, $\lambda_{HHh}$ and $\lambda_{HHH}$ 
in the processes $e^+e^-\to ZHh$ and $e^+e^-\to ZHH$ for collider energies 
$500$~GeV and $1$~TeV, respectively (no mixing).}}
\label{fig:sens2}
\end{figure}

The feasibility of measuring the trilinear Higgs self-couplings
of the MSSM can be summarized in sensitivity plots in the
[$M_A,\tan\beta$] plane \cite{djouadi2,ours}. The sensitivity criteria for a
point in the plane to allow for the measurement of the trilinear
coupling in the corresponding channel have been chosen as follows:
\beq
\begin{array}{l l} 
(i) & \sigma [\lambda] > 0.01~{\rm fb} \\
(ii) & {\rm eff}\{ \lambda \to 0 \} > 
2~{\rm st.dev.} \quad {\rm for} \quad \int {\cal L} = 2~{\rm ab}^{-1}
\end{array} 
\label{crit}
\eeq
The first criterion demands the value of the cross section of the
examined process to be larger than 0.01~fb, corresponding to at least
20 events for an integrated luminosity of $\int {\cal L} = 2~{\rm
  ab}^{-1}$. The second condition is fulfilled if the effect of the
trilinear coupling on the cross section exceeds 2 standard-deviations.
(The second criterion is changed slightly where appropriate,
cf.~Figs.~\ref{fig:sens1} and \ref{fig:sens2}.) If more stringent cuts
are applied the results will not change dramatically,
cf.~Ref.~\cite{osland}. The sensitivity areas have been analysed based
on the double Higgs-strahlung and triple Higgs production processes.
$WW$ double Higgs fusion would provide additional information
especially for higher c.m.~energies. For the sake of simplicity no
mixing effects have been included.\s

Figs.~\ref{fig:sens1} and \ref{fig:sens2} show the sensitivity areas
for the trilinear couplings among the neutral CP-even Higgs bosons.
The c.m.~energy has been chosen equal to 500~GeV for double
Higgs-strahlung if at most one heavy Higgs boson is present in the
final state. Otherwise $\sqrt{s}=1$~TeV yields larger sensitivity
regions. Likewise the areas for triple Higgs production involving $A$
are larger for 1~TeV.  Unless interference effects become important
the size of the sensitivity areas in $\tan\beta$ can be explained by
analysing the values of the effective couplings
$\lambda\sin(\beta-\alpha)$ and $\lambda\cos(\beta-\alpha)$ involved
in the various channels. The couplings are shown separately in
Figs.~\ref{fig:lambda1} and \ref{fig:lambda2}. For large values of
$M_A$ the areas are limited by phase space effects and due to the
suppression of the $H$ and $A$ propagators for large masses.  The
figures show that the coupling $\lambda_{hhh}$ is accessible in most
of the parameter space whereas the parameter range for the other
couplings is much smaller.\s

The sensitivity areas have been constructed without taking into
account any experimental efficiencies and background effects so that
the given results must be regarded as best case. Considering
experimental boundary conditions the areas will shrink. This effect
may be reduced, however, if sophisticated cuts on signal and
background are applied.

\section{Higgs pair production at the LHC}
Higgs pair production at the LHC follows the same pattern as at the
linear collider with the exception that gluon-gluon fusion provides an
additional channel. The partonic cross sections can be derived from
the corresponding $e^+e^-$ processes by replacing the couplings
appropriately. The cross sections at hadron level are obtained from
the partonic results $\hat{\sigma} (qq'\to HH;\hat{s}= \tau s)$ of the
quark subprocess by folding with the appropriate luminosities $d{\cal
  L}^{qq'}/d\tau$:
\beq
\sigma (pp\to HH) = \int_{4M_H^2/s}^1 d\tau 
\frac{d{\cal L}^{qq'}}{d\tau} 
\hat{\sigma} (qq' \to HH; \hat{s} = \tau s)
\eeq
where
\beq
\frac{d{\cal L}^{qq'}}{d\tau} = \int_\tau^1 \frac{dx}{x}
q(x;Q^2) q'(\tau/x;Q^2)
\eeq
with $q$ and $q'$ denoting the quark densities in the proton
\cite{martin}, taken at a typical factorization scale $Q\sim M_H$.\s

Being of up to ${\cal O}(10$~fb$)$ in the SM case the results for the
double Higgs production cross sections at the LHC are larger than in
$e^+e^-$ collisions. Yet, the signal processes are plagued by an
overwhelming QCD background

\subsection{SM double Higgs production}
The processes for the production of a pair of Higgs bosons in the
final state at the LHC are given by double Higgs-strahlung off $W$ and
$Z$ bosons \cite{bargerhan}, $WW$ and $ZZ$ fusion \citer{dicus,eboli}
and gluon-gluon fusion \cite{dawson},\citer{eboli,plehn}, in generic notation:
\beq
\begin{array}{l l l c l c l}
\mbox{double Higgs-strahlung}& \hspace{-0.3cm} : & q\bar{q} & 
\hspace{-0.3cm} \to &\hspace{-0.1cm}  W^*/Z^* &
\hspace{-0.3cm} \to &\hspace{-0.1cm}  W/Z + HH 
\non \\[0.1cm]
WW/ZZ\ \mbox{double-Higgs fusion}& \hspace{-0.3cm} : & qq & 
\hspace{-0.3cm} \to & \hspace{-0.1cm}  qq + WW/ZZ &
\hspace{-0.3cm} \to & \hspace{-0.1cm}  HH 
\non \\[0.1cm]
\mbox{gluon} \; \mbox{fusion} &  \hspace{-0.3cm} : & gg &
\hspace{-0.3cm} \to & \hspace{-0.1cm}  HH & & \non
\end{array} 
\eeq
\begin{figure}[ht]
\begin{flushleft}
\underline{double Higgs-strahlung: $q\bar q\to ZHH/WHH$}\\[1.5\baselineskip]
{\footnotesize
\unitlength1mm
\hspace{10mm}
\begin{fmfshrink}{0.7}
\begin{fmfgraph*}(24,12)
  \fmfstraight
  \fmfleftn{i}{3} \fmfrightn{o}{3}
  \fmf{fermion}{i1,v1,i3}
  \fmflabel{$q$}{i1} \fmflabel{$\bar q$}{i3}
  \fmf{boson,lab=$W/Z$,lab.s=left,tens=3/2}{v1,v2}
  \fmf{boson}{v2,o3} \fmflabel{$W/Z$}{o3}
  \fmf{phantom}{v2,o1}
  \fmffreeze
  \fmf{dashes,lab=$H$,lab.s=right}{v2,v3} \fmf{dashes}{v3,o1}
  \fmffreeze
  \fmf{dashes}{v3,o2} 
  \fmflabel{$H$}{o2} \fmflabel{$H$}{o1}
  \fmfdot{v3}
\end{fmfgraph*}
\hspace{15mm}
\begin{fmfgraph*}(24,12)
  \fmfstraight
  \fmfleftn{i}{3} \fmfrightn{o}{3}
  \fmf{fermion}{i1,v1,i3}
  \fmf{boson,tens=3/2}{v1,v2}
  \fmf{dashes}{v2,o1} \fmflabel{$H$}{o1}
  \fmf{phantom}{v2,o3}
  \fmffreeze
  \fmf{boson}{v2,v3,o3} \fmflabel{$W/Z$}{o3}
  \fmffreeze
  \fmf{dashes}{v3,o2} 
  \fmflabel{$H$}{o2} \fmflabel{$H$}{o1}
\end{fmfgraph*}
\hspace{15mm}
\begin{fmfgraph*}(24,12)
  \fmfstraight
  \fmfleftn{i}{3} \fmfrightn{o}{3}
  \fmf{fermion}{i1,v1,i3}
  \fmf{boson,tens=3/2}{v1,v2}
  \fmf{dashes}{v2,o1} \fmflabel{$H$}{o1}
  \fmf{dashes}{v2,o2} \fmflabel{$H$}{o2}
  \fmf{boson}{v2,o3} \fmflabel{$W/Z$}{o3}
\end{fmfgraph*}
\end{fmfshrink}
}
\\[2\baselineskip]
\underline{$WW/ZZ$ double-Higgs fusion: $qq\to qqHH$}\\[1.5\baselineskip]
{\footnotesize
\unitlength1mm
\hspace{10mm}
\begin{fmfshrink}{0.7}
\begin{fmfgraph*}(24,20)
  \fmfstraight
  \fmfleftn{i}{8} \fmfrightn{o}{8}
  \fmf{fermion,tens=3/2}{i2,v1} \fmf{phantom}{v1,o2}
  \fmflabel{$q$}{i2}
  \fmf{fermion,tens=3/2}{i7,v2} \fmf{phantom}{v2,o7} 
  \fmflabel{$q$}{i7}
  \fmffreeze
  \fmf{fermion}{v1,o1} 
  \fmf{fermion}{v2,o8} 
  \fmf{boson}{v1,v3} 
  \fmf{boson}{v3,v2}
  \fmf{dashes,lab=$H$}{v3,v4}
  \fmf{dashes}{v4,o3} \fmf{dashes}{v4,o6}
  \fmflabel{$H$}{o3} \fmflabel{$H$}{o6}
  \fmffreeze
  \fmf{phantom,lab=$W/Z$,lab.s=left}{v1,x1} \fmf{phantom}{x1,v3} 
  \fmf{phantom,lab=$W/Z$,lab.s=left}{x2,v2} \fmf{phantom}{v3,x2}
  \fmfdot{v4}
\end{fmfgraph*}
\hspace{15mm}
\begin{fmfgraph*}(24,20)
  \fmfstraight
  \fmfleftn{i}{8} \fmfrightn{o}{8}
  \fmf{fermion,tens=3/2}{i2,v1} \fmf{phantom}{v1,o2}
  \fmf{fermion,tens=3/2}{i7,v2} \fmf{phantom}{v2,o7} 
  \fmffreeze
  \fmf{fermion}{v1,o1}
  \fmf{fermion}{v2,o8}
  \fmf{boson}{v1,v3} 
  \fmf{boson}{v4,v2}
  \fmf{boson,lab=$W/Z$,lab.s=left}{v3,v4}
  \fmf{dashes}{v3,o3} \fmf{dashes}{v4,o6}
  \fmflabel{$H$}{o3} \fmflabel{$H$}{o6}
\end{fmfgraph*}
\hspace{15mm}
\begin{fmfgraph*}(24,20)
  \fmfstraight
  \fmfleftn{i}{8} \fmfrightn{o}{8}
  \fmf{fermion,tens=3/2}{i2,v1} \fmf{phantom}{v1,o2}
  \fmf{fermion,tens=3/2}{i7,v2} \fmf{phantom}{v2,o7} 
  \fmffreeze
  \fmf{fermion}{v1,o1}
  \fmf{fermion}{v2,o8}
  \fmf{boson}{v1,v3} 
  \fmf{boson}{v3,v2}
  \fmf{dashes}{v3,o3} \fmf{dashes}{v3,o6}
  \fmflabel{$H$}{o3} \fmflabel{$H$}{o6}
  \fmffreeze
\end{fmfgraph*}
\end{fmfshrink}
}
\\[2\baselineskip]
\underline{$gg$ double-Higgs fusion: $gg\to HH$}\\[1.5\baselineskip]
{\footnotesize
\unitlength1mm
\hspace{10mm}
\begin{fmfshrink}{0.7}
\begin{fmfgraph*}(30,12)
  \fmfstraight
  \fmfleftn{i}{2} \fmfrightn{o}{2}
  \fmflabel{$g$}{i1}  \fmflabel{$g$}{i2}
  \fmf{gluon,tens=2/3}{i1,v1} \fmf{phantom}{v1,v2,v3,o1}
  \fmf{gluon,tens=2/3}{w1,i2} \fmf{phantom}{w1,w2,w3,o2}
  \fmffreeze
  \fmf{fermion}{w1,x2,v1}
  \fmf{dashes, lab=$H$}{x2,x3}
  \fmf{dashes}{o1,x3,o2}
  \fmffreeze
  \fmf{fermion,label=$t$,label.s=left}{v1,w1}
  \fmflabel{$H$}{o1}  \fmflabel{$H$}{o2}
  \fmfdot{x3}
\end{fmfgraph*}
\hspace{15mm}
\begin{fmfgraph*}(30,12)
  \fmfstraight
  \fmfleftn{i}{2} \fmfrightn{o}{2}
  \fmf{gluon}{i1,v1} \fmf{phantom}{v1,v3} \fmf{dashes}{v3,o1}
  \fmf{gluon}{w1,i2} \fmf{phantom}{w1,w3} \fmf{dashes}{w3,o2}
  \fmffreeze
  \fmf{fermion}{v1,w1,w3,v3,v1}
  \fmflabel{$H$}{o1}  \fmflabel{$H$}{o2}
\end{fmfgraph*}
\end{fmfshrink}
}
\end{flushleft}
\caption{\textit{
Processes contributing to Higgs-pair production in the Standard Model
at the LHC: double Higgs-strahlung, $WW/ZZ$ fusion, and $gg$ fusion
(generic diagrams).}}
\label{smdiag}
\end{figure}
\hspace{-0.225cm} Fig.~\ref{smdiag} shows generic diagrams
contributing to these processes.  Since high energetic protons contain
a large number of gluons, gluon-gluon fusion \cite{dawson},
\citer{eboli,plehn}, provides an important mechanism for Higgs pair
production at the LHC. The fusion proceeds via heavy top-quark
triangle and box diagrams, cf.~Fig.~\ref{smdiag}. As in the case of
single Higgs production \cite{singleh} QCD radiative corrections play
an important r\^{o}le.  In the limit $M_H^2 \ll 4M_t^2$ they yield a
$K$ factor of about 1.9 \cite{dawson}. For Higgs masses beyond the
top-quark threshold a similar $K$ factor is expected.\s

\begin{figure}
\begin{center}
\epsfig{figure=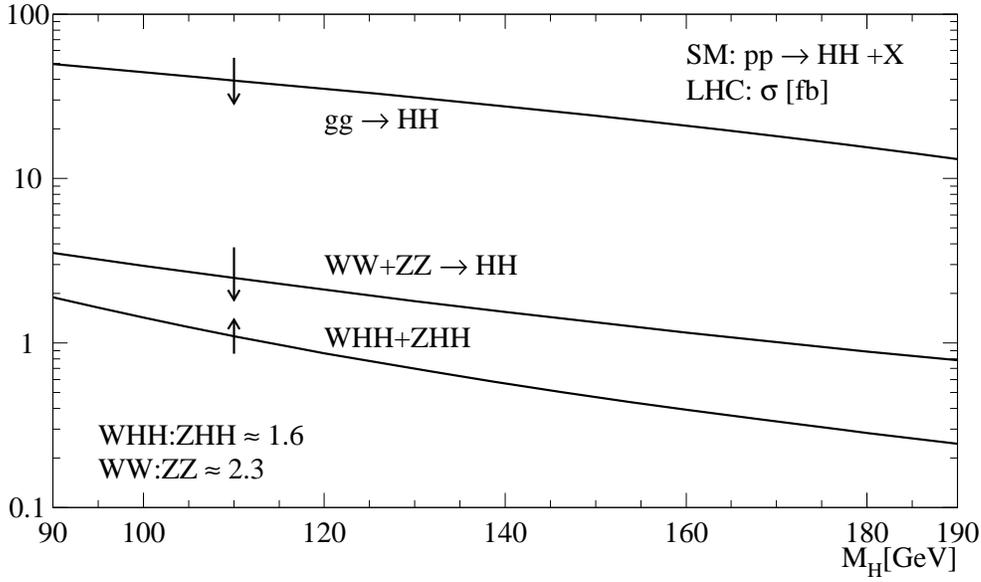,width=13cm}
\caption{\textit{The cross sections for gluon fusion, $WW/ZZ$ fusion and 
double Higgs-strahlung $WHH$, $ZHH$ in the SM. The vertical arrows 
correspond to a variation of the trilinear Higgs coupling from $1/2$ 
to $3/2$ of the SM value.}}
\label{fig:LHCSM}
\end{center}
\end{figure}
Fig.~\ref{fig:LHCSM} shows the results of the Higgs pair production
mechanisms as a function of the Higgs mass in the intermediate
mass range. The dominating process is gluon-gluon fusion followed by
$WW+ZZ$ fusion being about one order of magnitude smaller.  The ratio
of $WW$ to $ZZ$ fusion is $\sim 2.3$. For $M_H\,\lesim\, 140$~GeV, $WW/ZZ$
fusion yields two $b\bar{b}$ pairs resulting from the $H$ decay in the
final state. For $M_H \,\gesim\, 140$~GeV the final state is characterized
by $WW^{(*)}WW^{(*)}$. The additional light quark jets due to the
fragmentation $q\to W/Z+q$ with a transverse momentum of $p_T \sim
{\scriptstyle \frac{1}{2}} M_{W/Z}$ may be exploited to tag the fusion
process. The double Higgs-strahlung process $WHH+ZHH$ provides the
smallest cross section due to the scaling behaviour, {\it
  i.e.}~$\hat{\sigma}\sim 1/\hat{s}$.  Because of the smaller
$q\bar{q}Z$ coupling $WHH$ dominates over $ZHH$ by a factor of about
1.6. The arrows indicate the sensitivity of the double Higgs
production processes to a variation of the trilinear Higgs
self-coupling in the range [${\scriptstyle
  \frac{1}{2}}\lambda_{HHH},{\scriptstyle \frac{3}{2}}\lambda_{HHH}$].
The size of the cross sections does not exceed ${\cal O}(10$~fb$)$.
High luminosities are therefore needed to enable the extraction of the
signal from the large QCD background.

\subsection{Higgs pairs in the MSSM}
\begin{figure}
\begin{flushleft}
\underline{double Higgs-strahlung: $q\bar q\to Zhh/Whh$}\\[1.5\baselineskip]
{\footnotesize
\unitlength1mm
\hspace{5mm}
\begin{fmfshrink}{0.7}
\begin{fmfgraph*}(24,12)
  \fmfstraight
  \fmfleftn{i}{3} \fmfrightn{o}{3}
  \fmf{fermion}{i1,v1,i3}
  \fmflabel{$q$}{i1} \fmflabel{$\bar q$}{i3}
  \fmf{boson,tens=3/2,label=$W/Z$, label.s=left}{v1,v2}
  \fmf{boson}{v2,o3} \fmflabel{$W/Z$}{o3}
  \fmf{phantom}{v2,o1}
  \fmffreeze
  \fmf{dashes,lab=$h,,H$,lab.s=right}{v2,v3} \fmf{dashes}{v3,o1}
  \fmffreeze
  \fmf{dashes}{v3,o2} 
  \fmflabel{$h$}{o2} \fmflabel{$h$}{o1}
  \fmfdot{v3}
\end{fmfgraph*}
\hspace{15mm}
\begin{fmfgraph*}(24,12)
  \fmfstraight
  \fmfleftn{i}{3} \fmfrightn{o}{3}
  \fmf{fermion}{i1,v1,i3}
  \fmf{boson,tens=3/2}{v1,v2}
  \fmf{dashes}{v2,o1}
  \fmf{phantom}{v2,o3}
  \fmffreeze
  \fmf{dashes,lab=$H^\pm/A$,lab.s=left}{v2,v3} 
  \fmf{boson}{v3,o3} \fmflabel{$W/Z$}{o3}
  \fmffreeze
  \fmf{dashes}{v3,o2} 
  \fmflabel{$h$}{o2} \fmflabel{$h$}{o1}
\end{fmfgraph*}
\hspace{15mm}
\begin{fmfgraph*}(24,12)
  \fmfstraight
  \fmfleftn{i}{3} \fmfrightn{o}{3}
  \fmf{fermion}{i1,v1,i3}
  \fmf{boson,tens=3/2}{v1,v2}
  \fmf{dashes}{v2,o1}
  \fmf{phantom}{v2,o3}
  \fmffreeze
  \fmf{boson}{v2,v3,o3} \fmflabel{$W/Z$}{o3}
  \fmffreeze
  \fmf{dashes}{v3,o2} 
  \fmflabel{$h$}{o2} \fmflabel{$h$}{o1}
\end{fmfgraph*}
\hspace{15mm}
\begin{fmfgraph*}(24,12)
  \fmfstraight
  \fmfleftn{i}{3} \fmfrightn{o}{3}
  \fmf{fermion}{i1,v1,i3}
  \fmf{boson,tens=3/2}{v1,v2}
  \fmf{dashes}{v2,o1} \fmflabel{$h$}{o1}
  \fmf{dashes}{v2,o2} \fmflabel{$h$}{o2}
  \fmf{boson}{v2,o3} \fmflabel{$W/Z$}{o3}
\end{fmfgraph*}
\end{fmfshrink}
}
\\[2\baselineskip]
\underline{triple Higgs production: $q\bar q\to Ahh$}
\\[1.5\baselineskip]
{\footnotesize
\unitlength1mm
\hspace{5mm}
\begin{fmfshrink}{0.7}
\begin{fmfgraph*}(24,12)
  \fmfstraight
  \fmfleftn{i}{3} \fmfrightn{o}{3}
  \fmf{fermion}{i1,v1,i3}
  \fmflabel{$q$}{i1} \fmflabel{$\bar q$}{i3}
  \fmf{boson,tens=3/2,label=$Z$, label.s=left}{v1,v2}
  \fmf{dashes}{v2,o3} \fmflabel{$A$}{o3}
  \fmf{phantom}{v2,o1}
  \fmffreeze
  \fmf{dashes,lab=$H,,h$,lab.s=right}{v2,v3} \fmf{dashes}{v3,o1}
  \fmffreeze
  \fmf{dashes}{v3,o2} 
  \fmflabel{$h$}{o2} \fmflabel{$h$}{o1}
  \fmfdot{v3}
\end{fmfgraph*}
\hspace{15mm}
\begin{fmfgraph*}(24,12)
  \fmfstraight
  \fmfleftn{i}{3} \fmfrightn{o}{3}
  \fmf{fermion}{i1,v1,i3}
  \fmf{boson,tens=3/2}{v1,v2}
  \fmf{dashes}{v2,o1} \fmflabel{$h$}{o1}
  \fmf{phantom}{v2,o3}
  \fmffreeze
  \fmf{dashes,lab=$A$,lab.s=left}{v2,v3} 
  \fmf{dashes}{v3,o3}
  \fmffreeze
  \fmf{dashes}{v3,o2} 
  \fmflabel{$h$}{o2} \fmflabel{$A$}{o3}
  \fmfdot{v3}
\end{fmfgraph*}
\hspace{15mm}
\begin{fmfgraph*}(24,12)
  \fmfstraight
  \fmfleftn{i}{3} \fmfrightn{o}{3}
  \fmf{fermion}{i1,v1,i3}
  \fmf{boson,tens=3/2}{v1,v2}
  \fmf{dashes}{v2,o1}
  \fmf{phantom}{v2,o3}
  \fmffreeze
  \fmf{boson}{v2,v3} 
  \fmf{dashes}{v3,o3} \fmflabel{$A$}{o3}
  \fmffreeze
  \fmf{dashes}{v3,o2} 
  \fmflabel{$h$}{o2} \fmflabel{$h$}{o1}
\end{fmfgraph*}
\end{fmfshrink}
}
\\[2\baselineskip]
\underline{$WW/ZZ$ double-Higgs fusion: $qq\to qqhh$}
\\[1.5\baselineskip]
{\footnotesize
\unitlength1mm
\hspace{5mm}
\begin{fmfshrink}{0.7}
\begin{fmfgraph*}(24,20)
  \fmfstraight
  \fmfleftn{i}{8} \fmfrightn{o}{8}
  \fmf{fermion,tens=3/2}{i2,v1} \fmf{phantom}{v1,o2}
  \fmf{fermion,tens=3/2}{i7,v2} \fmf{phantom}{v2,o7} 
  \fmflabel{$q$}{i2} \fmflabel{$q$}{i7}
  \fmffreeze
  \fmf{fermion}{v1,o1}
  \fmf{fermion}{v2,o8}
  \fmf{boson}{v1,v3} 
  \fmf{boson}{v3,v2}
  \fmf{dashes,lab=$H,,h$}{v3,v4}
  \fmf{dashes}{v4,o3} \fmf{dashes}{v4,o6}
  \fmflabel{$h$}{o3} \fmflabel{$h$}{o6}
  \fmffreeze
  \fmf{phantom,lab=$W/Z$,lab.s=left}{v1,x1} \fmf{phantom}{x1,v3} 
  \fmf{phantom,lab=$W/Z$,lab.s=left}{x2,v2} \fmf{phantom}{v3,x2}
  \fmfdot{v4}
\end{fmfgraph*}
\hspace{15mm}
\begin{fmfgraph*}(24,20)
  \fmfstraight
  \fmfleftn{i}{8} \fmfrightn{o}{8}
  \fmf{fermion,tens=3/2}{i2,v1} \fmf{phantom}{v1,o2}
  \fmf{fermion,tens=3/2}{i7,v2} \fmf{phantom}{v2,o7}
  \fmffreeze
  \fmf{fermion}{v1,o1}
  \fmf{fermion}{v2,o8}
  \fmf{boson}{v1,v3} 
  \fmf{boson}{v4,v2}
  \fmf{boson,lab=$W/Z$,lab.s=left}{v3,v4}
  \fmf{dashes}{v3,o3} \fmf{dashes}{v4,o6}
  \fmflabel{$h$}{o3} \fmflabel{$h$}{o6}
\end{fmfgraph*}
\hspace{15mm}
\begin{fmfgraph*}(24,20)
  \fmfstraight
  \fmfleftn{i}{8} \fmfrightn{o}{8}
  \fmf{fermion,tens=3/2}{i2,v1} \fmf{phantom}{v1,o2}
  \fmf{fermion,tens=3/2}{i7,v2} \fmf{phantom}{v2,o7} 
  \fmffreeze
  \fmf{fermion}{v1,o1}
  \fmf{fermion}{v2,o8}
  \fmf{boson}{v1,v3} 
  \fmf{boson}{v4,v2}
  \fmf{dashes,lab=$H^\pm/A$,lab.s=left}{v3,v4}
  \fmf{dashes}{v3,o3} \fmf{dashes}{v4,o6}
  \fmflabel{$h$}{o3} \fmflabel{$h$}{o6}
\end{fmfgraph*}
\hspace{15mm}
\begin{fmfgraph*}(24,20)
  \fmfstraight
  \fmfleftn{i}{8} \fmfrightn{o}{8}
  \fmf{fermion,tens=3/2}{i2,v1} \fmf{phantom}{v1,o2}
  \fmf{fermion,tens=3/2}{i7,v2} \fmf{phantom}{v2,o7}
  \fmffreeze
  \fmf{fermion}{v1,o1}
  \fmf{fermion}{v2,o8}
  \fmf{boson}{v1,v3} 
  \fmf{boson}{v3,v2}
  \fmf{dashes}{v3,o3} \fmf{dashes}{v3,o6}
  \fmflabel{$h$}{o3} \fmflabel{$h$}{o6}
\end{fmfgraph*}
\end{fmfshrink}
}
\\[2\baselineskip]
\underline{$gg$ double-Higgs fusion: $gg\to hh$}\\[1.5\baselineskip]
{\footnotesize
\unitlength1mm
\hspace{5mm}
\begin{fmfshrink}{0.7}
\begin{fmfgraph*}(30,12)
  \fmfstraight
  \fmfleftn{i}{2} \fmfrightn{o}{2}
  \fmflabel{$g$}{i1}  \fmflabel{$g$}{i2}
  \fmf{gluon,tens=2/3}{i1,v1} \fmf{phantom}{v1,v2,v3,o1}
  \fmf{gluon,tens=2/3}{w1,i2} \fmf{phantom}{w1,w2,w3,o2}
  \fmffreeze
  \fmf{fermion}{w1,x2,v1}
  \fmf{dashes, lab=$h,,H$}{x2,x3}
  \fmf{dashes}{o1,x3,o2}
  \fmffreeze
  \fmf{fermion,label=$t,,b$,label.s=left}{v1,w1}
  \fmflabel{$h$}{o1}  \fmflabel{$h$}{o2}
  \fmfdot{x3}
\end{fmfgraph*}
\hspace{15mm}
\begin{fmfgraph*}(30,12)
  \fmfstraight
  \fmfleftn{i}{2} \fmfrightn{o}{2}
  \fmf{gluon}{i1,v1} \fmf{phantom}{v1,v3} \fmf{dashes}{v3,o1}
  \fmf{gluon}{w1,i2} \fmf{phantom}{w1,w3} \fmf{dashes}{w3,o2}
  \fmffreeze
  \fmf{fermion}{v1,w1,w3,v3,v1}
  \fmflabel{$h$}{o1}  \fmflabel{$h$}{o2}
\end{fmfgraph*}
\end{fmfshrink}
}
\end{flushleft}
\caption{\textit{
Processes contributing to double and triple Higgs production involving
trilinear couplings in the MSSM.}}
\label{fig:mssm}
\end{figure}
\begin{figure}
\begin{flushleft}
\underline{cascade decay: $q\bar q\to AH/HH^\pm \to ZHh/WHh$}\\[1.5\baselineskip
]
{\footnotesize
\unitlength1mm
\hspace{5mm}
\begin{fmfshrink}{0.7}
\begin{fmfgraph*}(24,12)
  \fmfstraight
  \fmfleftn{i}{3} \fmfrightn{o}{3}
  \fmf{fermion}{i1,v1,i3}
  \fmflabel{$q$}{i1} \fmflabel{$\bar q$}{i3}
  \fmf{boson,tens=3/2,lab=$W/Z$,lab.s=right}{v1,v2}
  \fmf{dashes}{v2,o1}
  \fmf{phantom}{v2,o3}
  \fmffreeze
  \fmf{dashes,lab=$H^\pm/A$,lab.s=left}{v2,v3} 
  \fmf{boson}{v3,o3} \fmflabel{$W/Z$}{o3}
  \fmffreeze
  \fmf{dashes}{v3,o2} 
  \fmflabel{$h$}{o2} \fmflabel{$H$}{o1}
\end{fmfgraph*}
\end{fmfshrink}
}
\end{flushleft}
\caption{\textit{
Processes which contribute to double light plus heavy Higgs production 
in the MSSM but do not involve trilinear couplings.}}
\label{nocoup}
\end{figure}
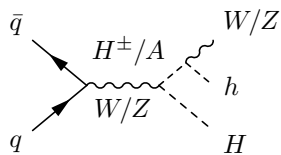
\end{fmffile}
There are many processes at the LHC which involve the trilinear
MSSM Higgs self-couplings \cite{dawson,ours,plehn,dreeseal}:
\beq
\begin{array}{l@{:\quad}l@{\;\to\;}l l l l}
\mbox{double Higgs-strahlung} & $\qq$  & W/Z + H_i H_j & \mathrm{and} 
& W/Z + AA & [H_{i,j}=h,H] \non\\[0.2cm]
\mbox{triple Higgs production} & $\qq$ & AH_i H_j & \mathrm{and}
& AAA & \non \\[0.2cm]
WW/ZZ\ \mbox{double-Higgs fusion} & qq & qq+ H_i H_j & \mathrm{and}
& qq+ AA & \non \\[0.2cm]
gg\ \mbox{fusion} & gg & H_i H_j,\quad H_iA & \mathrm{and}
& AA & \non
\end{array} 
\eeq
As in the SM case also gluon-gluon fusion plays a r\^{o}le in contrast
to the linear collider. Since in the major part of the parameter space
$H$ and $A$ are quite heavy the main focus of the subsequent analysis
will be on the production of a light Higgs boson pair. The diagrams
contributing to the individual production processes are shown in
Fig.~\ref{fig:mssm}. Some results will also be presented for heavy
Higgs boson final states. The cross sections are small unless they
include resonant decays. For example in the case of $WHh/ZHh$ they
proceed via 
\beq
q\bar{q} &\to& Z^* \to AH \to ZHh \\
q\bar{q} &\to& W^* \to H^{\pm}H \to WHh \non
\eeq
The corresponding diagrams are shown in Fig.~\ref{nocoup}. These
resonant heavy Higgs boson decays can largely enhance the cross
sections. Since they only involve gauge interactions they are useless
for the measurement of the Higgs self-couplings, however.\s

\begin{figure}
\begin{center}
\epsfig{figure=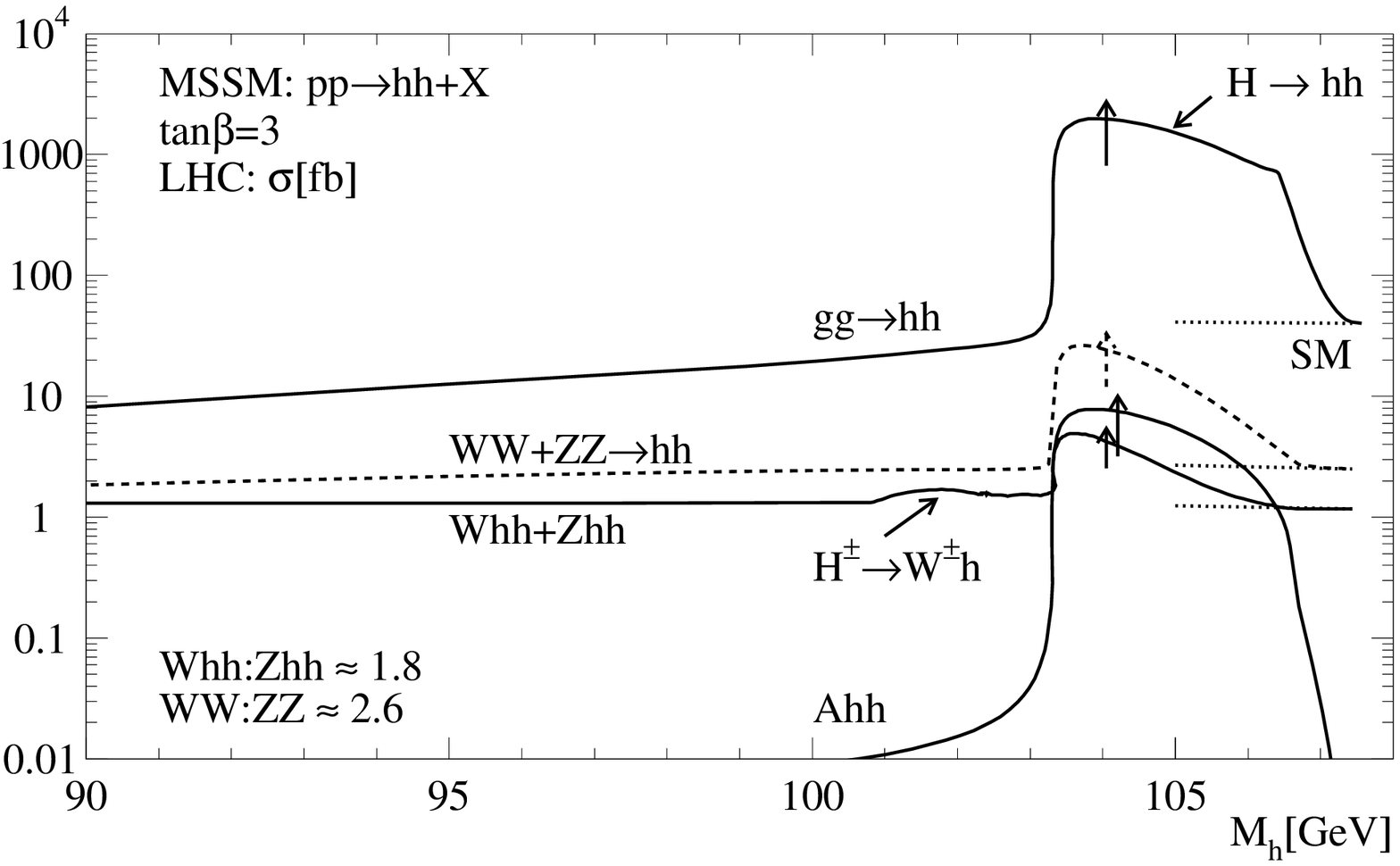,width=13cm}
\end{center}
\caption{\textit{Total cross sections for MSSM hh production via double 
Higgs-strahlung $Whh$ and $Zhh$, $WW/ZZ$ fusion and gluon fusion at the 
LHC for $\tan\beta=3$, including mixing effects ($A=1$~TeV, 
$\mu=-1$~TeV).}}
\label{fig:lhct3}
\end{figure} 
\begin{figure}
\begin{center}
\epsfig{figure=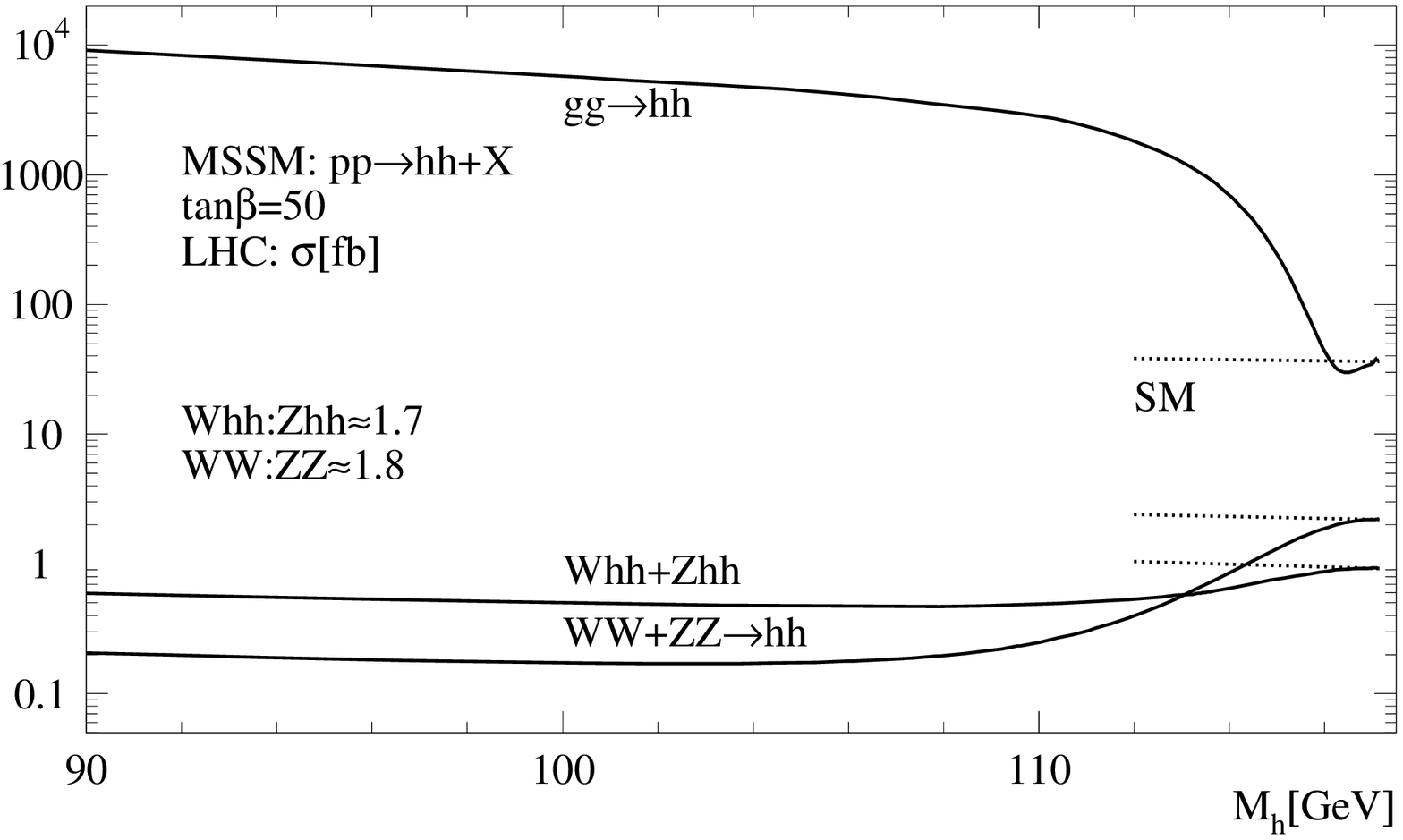,width=13cm}
\end{center}
\caption{\textit{Total cross sections for MSSM hh production via double 
Higgs-strahlung $Whh$, $Zhh$, $WW/ZZ$ fusion and gluon fusion at the 
LHC for $\tan\beta=50$, including mixing effects ($A=1$~TeV, 
$\mu=1$~TeV).}}
\label{fig:lhct50}
\end{figure}
The results for the $hh$ production processes, {\it i.e.}
\beq
pp &\to& gg \to hh \\
pp &\to& Z/W + hh \qquad {\rm and} \quad A+hh \non
\eeq
are demonstrated in Fig.~\ref{fig:lhct3}. For moderate values of
$\tan\beta$ the order of the cross sections is the same as in the SM.
Yet, the cross sections rise enormously if cascade decays proceeding
via intermediate resonant heavy Higgs bosons \cite{djkalzer}
\beq
H \to hh, \quad A \to Zh \quad {\rm and} \quad H^{\pm} \to W^{\pm} 
h
\eeq
are kinematically possible. The resonant $H$ decay increases the cross
sections by up to 2 orders of magnitude giving rise to about a million
events in gluon-gluon fusion. This channel therefore allows for the
light MSSM Higgs boson search at the LHC
\cite{atlascms,prochouches,froide,daigu}. The resonant decay regions
are indicated by arrows in Fig.~\ref{fig:lhct3}.  In the case of $Whh$
also the resonant decay $H^\pm \to W^\pm h$ leads to an enhancement of
the cross section. The vertical arrows illustrate the sensitivity to a
variation of $\lambda_{Hhh}$ in the range [${\scriptstyle
  \frac{1}{2}}\lambda_{Hhh},{\scriptstyle \frac{3}{2}}\lambda_{Hhh}$].
In the continuum region the same variation of $\lambda_{hhh}$ induces
a 10\% modification of the cross sections. \s

\begin{table}
\begin{center}$
\begin{array}{|rll||c|c|c|c||c|}\hline
\multicolumn{3}{|c||}{\sigma\;[\mathrm{fb}]} & Whh & WHh & WHH & 
\sum\limits_{i,j} WH_i H_j & 
WHH_{SM} \str \\
\hline \hline
M_A\!=\!\!\!\!&100\!\!\!\!\!\!\!\!&\mathrm{GeV}& 0.332 & 0.017 & 0.540 & 0.889 
& 0.908 \str \\
M_h\!=\!\!\!\!&98.53\!\!\!\!\!\!\!\!&\mathrm{GeV} & & & & & \str\\
\hline \hline
\multicolumn{3}{|c||}{\sigma\;[\mathrm{fb}]} & Zhh & ZHh & ZHH & 
\sum\limits_{i,j} ZH_i H_j & 
ZHH_{SM} \str \\
\hline \hline
M_A\!=\!\!\!\!&112\!\!\!\!&\mathrm{GeV}& 0.175 & 0.036 & 0.287 & 0.498 
& 0.431 \str \\
M_h\!=\!\!\!\!&109.61\!\!\!\!&\mathrm{GeV} & & & & & \str\\
\hline
\end{array}$
\end{center}
\caption{\textit{Total cross sections for $W/Z + H_i H_j$ compared to the corresponding SM cross section. The mixing parameters are chosen $A=\mu=1$~TeV and $\tan\beta=50$.}}
\label{tab:50rep}
\end{table}
For large values of $\tan\beta$ cascade decays do not play any
r\^{o}le since they are kinematically forbidden until the decoupling
region has been reached where they are not sizeable any more, see
Fig.~\ref{fig:lhct50}. The gluon fusion cross section is nevertheless very
large due to the enhancement of the $hbb$ Yukawa coupling
$\sim m_b\tan\beta$ growing with $\tan\beta$. Since this coupling
enters quadratically in the box diagram and only linearly in the
triangle loop connecting the gluons to the Higgs boson the
sensitivity to the trilinear Higgs self-coupling is small. Yet, the
huge cross section leading to the multi-$b$ final states $pp\to hh\to
(b\bar{b})(b\bar{b})$ with two resonance structures and large transverse
momenta allows for the $h$ Higgs boson search in the large $\tan\beta$
region at the LHC. The $WW/ZZ$ fusion and Higgs-strahlung cross
sections are suppressed in the continuum compared to their SM
counterparts until the decoupling limit is reached. The heavy Higgs
boson is fairly light in this parameter region so that the cross
sections with $hh$, $Hh$ and $HH$ final states approximately add up to
the SM cross section as can be inferred from Table~\ref{tab:50rep} for
some representative examples. \s

\begin{figure}[ht]
\begin{center}
\epsfig{figure=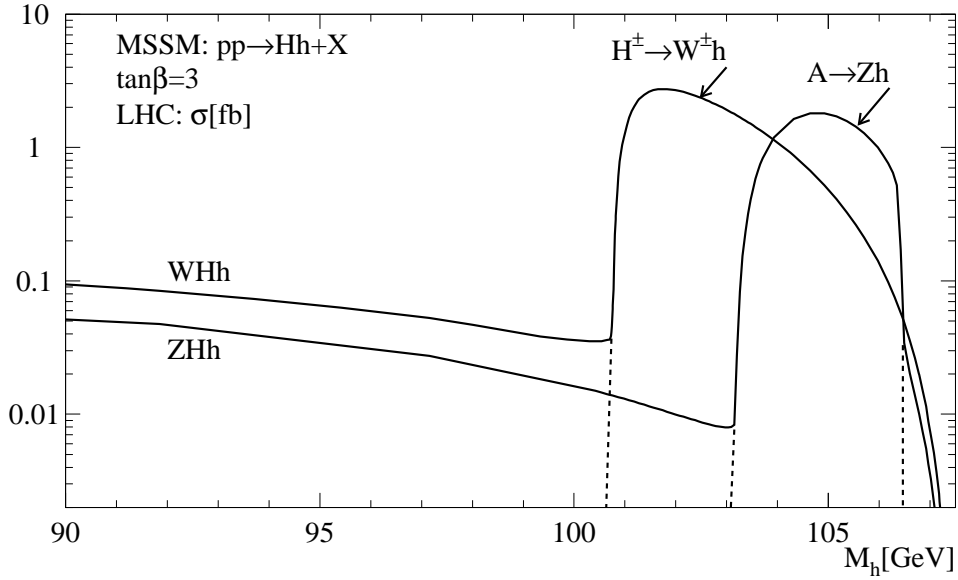,width=13cm}
\caption{\it Total cross sections for MSSM $Hh$ production in the 
processes $WHh$ and $ZHh$ for $\tan\beta=3$, including mixing effects 
($A=1$~TeV, $\mu=-1$~TeV).}
\label{heavy} 
\end{center} 
\end{figure}
Fig.~\ref{heavy} shows the cross sections of double Higgs-strahlung
involving a heavy and a light CP-even Higgs boson in the final state.
In the continuum they are below 0.1~fb. Cascade decays,
cf.~Fig.~\ref{nocoup},
\beq
\begin{array}{l l l l l}
pp & \hspace{-0.1cm} \to & AH &  
\hspace{-0.3cm} \to &\hspace{-0.1cm}  ZHh \\
\\[-0.8cm]
& \hspace{-0.1cm} \scriptstyle{Z}& & &  \\[0.1cm]
pp & \hspace{-0.1cm} \to & H^{\pm} H & 
\hspace{-0.3cm} \to & \hspace{-0.1cm} W^{\pm} Hh
\\ \\[-0.8cm]
& \hspace{-0.1cm} \scriptstyle{W}& & &
\end{array} 
\eeq
give rise to an enhancement by up to 2 orders of magnitude. Mediated
by pure gauge interactions they cannot be exploited for the
determination of the Higgs self-couplings.

\subsection{Extraction of the signal}
In Ref.~\cite{houches,prochouches} the extraction of the double Higgs
production signal in the channel
\beq
gg\to hh\to (b\bar{b})(b\bar{b})
\label{ggchannel}
\eeq
has been studied both at parton level and in a more realistic detector
simulation. As the previous analysis has shown, in the SM case the
gluon-gluon fusion cross section is of ${\cal O}(10$~fb$)$ only.
Bearing in mind the huge QCD background at the LHC the extraction of
the SM signal does not seem to be realistic. In contrast, in the case
of the MSSM light Higgs boson $h$ there are regions in the MSSM
parameter space where the signal is largely enhanced due to resonant
$H$ production with subsequent decay $H\to hh$. Independent of the
value of $\tan\beta$, $h$ dominantly decays into a $b\bar{b}$ pair
\cite{spirdjou,zerwas1,morettistir} yielding a four $b$ final state
in gluon-gluon fusion. \s

In the parton level analysis of Ref.~\cite{houches,prochouches} the
potential of extracting the trilinear Higgs coupling $\lambda_{Hhh}$
in the resonant region has been investigated for the parameter set
$\tan\beta=3$, $M_A = 210$~GeV, the mixing parameters $A=-\mu=1$~TeV
and the common squark mass set equal to 1~TeV. Assuming efficient
$b$-quark tagging and high purity the main background is due to
irreducible QCD modes \cite{froide} of ${\cal O}(\alpha_s^4)$, EW
processes of ${\cal O}(\alpha_{em}^4)$ and QCD and EW mixed
contributions of ${\cal O}(\alpha_s^2\alpha_{em}^2)$.  Applying
typical LHC detector cuts to the partons and selection cuts
appreciating the kinematics of the signal process, the signal and
background rates can be reduced to the same level. In a conservative
scenario the signal cross section reduces to 102~fb and the background
yields 453~fb. Assuming an integrated luminosity of $\int{\cal L} =
100~$fb$^{-1}$ a sufficiently high signal to background ratio should
be achievable. A more realistic analysis based on a detector
simulation predicts a signal rate of 38 events per year after all
efficiencies and selection cuts have been included
\cite{houches,prochouches}.  Though a background study on detector
level has not been performed yet the results from the parton level
analysis predict an encouraging level of background suppression thus
demanding for a detailed detector study. Nevertheless, according to
the present results the signal seems to be feasible in this parameter
region. \s

For large values of $\tan\beta$ the signal is strongly enhanced due to
the large Yukawa coupling $hb\bar{b}$ as has been pointed out
previously. Since the box diagram is more favored by this effect than
the triangle diagram the sensitivity to the Higgs self-couplings is
lost. The large $\tan\beta$ scenario may, however, provide an
additional channel for the $h$ detection at moderate $M_A$ values. The
detector simulation yields for the signal 1044 events for
$M_h=105$~GeV and $A=\mu=1$~TeV.\s

Though the QCD background is very large the determination of the
self-coupling $\lambda_{Hhh}$ seems feasible at the LHC in the cascade
decay $H\to hh$. In the large $\tan\beta$ regime the extraction of the
coupling will be very difficult since no resonance production takes
place and the signal sub-process is dominated by the box diagram
contributions. In this case, however, the gluon fusion process can be
exploited for the light scalar Higgs boson search. The experimental
determination of the remaining trilinear Higgs couplings at the LHC
will be a difficult task. Since $h$ cannot decay in a pair of
pseudoscalar Higgs states and $H \to AA$ is kinematically forbidden in
the considered MSSM parameter space, trilinear couplings involving
pseudoscalar Higgs bosons have to be extracted from the continuum. The
same holds true for $\lambda_{hhh}$, $\lambda_{HHh}$ and
$\lambda_{HHH}$ which are not involved in cascade decays. The
continuum signal might be swamped by the jetty QCD background if it is
not rejected sufficiently well.

%% file: chapter6.tex
\chapter{Conclusions} 
This thesis has presented a theoretical analysis of the properties of
the SM and MSSM Higgs bosons which can be investigated at the LHC and
$e^+e^-$ linear colliders. \s

The search for Higgs bosons is crucial to prove their existence.
Although there are several production mechanisms at the LHC that allow
to find the SM or the MSSM Higgs bosons, experimental studies exhibit
a region around moderate values of $\tan\beta$ where the heavy MSSM
Higgs particles $H$ and $A$ may escape detection. An alternative to
the search for $H$ and $A$ at the LHC is provided by Higgs boson
production at $\gamma\gamma$ colliders which may be realized by
Compton backscattering of laser light from high-energy
electron/positron beams.  The most promising search channels with
large branching ratios are given by the Higgs decays into $b\bar{b}$,
$t\bar{t}$, $\tau^+\tau^-$ and chargino or neutralino pairs. They have
been investigated by including the polarized NLO results for the
signal, background and interference processes and also the resummation
of higher orders where necessary.  Polarization of the
electron/positron and laser beams helps to increase the significances
by suppressing the helicity modes of the background processes, that
are not compatible with the signal.  \s

For the anticipated luminosities at future $\gamma\gamma$ colliders,
the $b\bar{b}$ mode develops sufficiently high significances for the $H$
and $A$ discovery in the whole analysed mass range from 200 to 800~GeV
for moderate and large values of $\tan\beta$. The restriction to
two-jet topologies in the final state and a cut in the scattering
angle of the $b$-quark with respect to the beam axis have been crucial
for the reduction of the background.  Since the Higgs Yukawa coupling
to $\tau^+\tau^-$ increases with $\tan\beta$, also the $\tau^+\tau^-$
channel provides a discovery potential in parts of the $A/H$ mass range
for large values of $\tan\beta$. For the $t\bar{t}$ and chargino decay
channels the extraction of the signal will be more challenging due to
larger backgrounds. The Higgs search in the neutralino channel is
confronted with decay products stemming from charginos beyond the
corresponding kinematic thresholds. This is due to the final states of
the neutralino and chargino cascade decays being rather similar
because of the escaping LSP and neutrinos. The different decay
topologies may be exploited, however, in order to extract the signal.
\s

Summarizing the results, $\gamma\gamma$ fusion at $e^+e^-$ linear
colliders provides a valuable alternative for the heavy Higgs boson
search in a variety of decay channels of which the $b\bar{b}$ mode is
outstanding. \s

In addition to the mass, the second basic feature characterizing Higgs
particles is their lifetime, or equivalently their total width.
Mechanisms that can be exploited at $e^+e^-$ linear colliders for the
determination of the small lifetime of a SM Higgs boson in the
intermediate mass range have been compared. They use the relation
$\Gamma_{tot} = \Gamma_i/BR_{H\to i}$ where $\Gamma_i$ denotes the
partial width and $BR_{H\to i}$ the branching ratio of the Higgs decay
into the final state $i$.  The $\gamma\gamma$ channel allows a
model-independent determination of $\Gamma_{tot}$. Alternatively, the
lifetime can be measured in the $WW$ channel. Due to possible
anomalous $HWW$ couplings involved in the $WW$ fusion process, the
analysis is model-dependent. However, the model dependence will be
reduced for small momentum transfer at the $HWW$ vertex.  The error of
the lifetime extracted from the $WW$ mode is smaller than in the
$\gamma\gamma$ channel. Taking into account the uncertainties in the
various cross sections the total width can be deduced with an accuracy
better than 10\% for Higgs bosons with $M_H \,\lesim \, 160$~GeV from
the $WW$ mode. \s

A qualitative discussion has demonstrated that the lifetime of a light
MSSM Higgs scalar $h$ can be determined from the $b\bar{b}$ channel.
The $b\bar{b}$ branching ratio occurs in the Higgs-strahlung process
and the Higgs Yukawa coupling to $b$-quarks is accessible in
associated production of $h$ with a $b\bar{b}$ pair. \s

To complete the profile of the Higgs bosons their self-couplings have
to be measured. The knowledge of the Higgs self-interactions allows
the reconstruction of the Higgs potential and thus the experimental
verification of the Higgs mechanism. In this thesis the theoretical
framework has been set up for the measurement of the trilinear Higgs
couplings.
%
%
They are accessible in double and triple
Higgs production. The processes for SM Higgs bosons in the
intermediate mass range and for MSSM Higgs particles turn out to be
small at $e^+e^-$ linear colliders. Although at the LHC the cross
sections are larger, the signal is confronted with a large QCD
background. The extraction of the Higgs self-couplings will therefore
be easier in an $e^+e^-$ environment where the background is
significantly smaller. For high luminosities, the trilinear
self-coupling of an intermediate mass Higgs boson can already be
measured in the first step of a linear collider, {\it i.e.}~at
c.m.~energies of 500 GeV. High $b$-tagging efficiency provided, the
experimental accuracy in $\lambda_{HHH}$ will be $\sim 20\%$.\s

The MSSM with five physical Higgs states includes six trilinear Higgs
couplings among the neutral particles and many more quadrilinear
self-couplings. The trilinear couplings are involved in a large number
of processes at $e^+e^-$ linear colliders and the LHC. In $e^+e^-$
collisions, the coupling among three light Higgs bosons $h$ can be
extracted from continuum production. Apart from the decoupling limit,
this coupling differs strongly from the corresponding SM value.  The
remaining couplings among CP-even Higgs bosons will also be
accessible, though in a smaller range of the basic input parameters
$M_A$ and $\tan\beta$. The trilinear couplings involving pseudoscalar
Higgs states are small in the MSSM and the measurement of double and
triple Higgs production processes will impose upper bounds on the size
of these couplings. At the LHC, the signal is plagued by an enormous
QCD background. Resonant decays $H\to hh$ provide a means of
extracting the $Hhh$ coupling. The measurement of the other couplings
will be more difficult since they have to be extracted from continuum
production. \s

\cleardoublepage

%% file: app.tex
\newcommand{\sla}[1]{#1\!\!\!/}
\newcommand{\ewwhh}{\eta_{\ensuremath{\mbox{\tiny WWHH}}}}
\newcommand{\ewwaa}{\eta_{\ensuremath{\mbox{\tiny WWAA}}}}
\newcommand{\ewwH}{\eta_{\ensuremath{\mbox{\tiny WWH}}}}
\newcommand{\ehhh}{\eta_{\ensuremath{\mbox{\tiny HHH}}}}
\newcommand{\ewwh}{\eta_{\ensuremath{\mbox{\tiny WWh}}}}
\newcommand{\eHHh}{\eta_{\ensuremath{\mbox{\tiny HHh}}}}
\newcommand{\ehhH}{\eta_{\ensuremath{\mbox{\tiny hhH}}}}
\newcommand{\eHaa}{\eta_{\ensuremath{\mbox{\tiny HAA}}}}
\newcommand{\ehaa}{\eta_{\ensuremath{\mbox{\tiny hAA}}}}
\newcommand{\ewgh}{\eta_{\ensuremath{\mbox{\tiny $W^+G^-H$}}}}
\newcommand{\ewghh}{\eta_{\ensuremath{\mbox{\tiny $W^+G^-h$}}}}
\newcommand{\ewhmh}{\eta_{\ensuremath{\mbox{\tiny $W^+H^-H$}}}}
\newcommand{\ewhmhh}{\eta_{\ensuremath{\mbox{\tiny $W^+H^-h$}}}}
\newcommand{\ezah}{\eta_{\ensuremath{\mbox{\tiny ZAH}}}}
\newcommand{\ezahh}{\eta_{\ensuremath{\mbox{\tiny ZAh}}}}
\newcommand{\ewhma}{\eta_{\ensuremath{\mbox{\tiny $W^+H^-A$}}}}
\newcommand{\bw}{\beta_{\ensuremath{\mbox{\tiny W}}}}
\newcommand{\bh}{\beta_{\ensuremath{\mbox{\tiny H}}}}
\newcommand{\ba}{\beta_{\ensuremath{\mbox{\tiny A}}}}
\newcommand{\bp}{\beta_+}
\newcommand{\xh}{x_{\ensuremath{\mbox{\tiny H}}}}
\newcommand{\xhpm}{x_{\ensuremath{\mbox{\tiny $H^\pm$}}}}
\newcommand{\aw}{a_{\ensuremath{\mbox{\tiny W}}}}
\newcommand{\aH}{a_{\ensuremath{\mbox{\tiny H}}}}
\newcommand{\ah}{a_h}
\newcommand{\az}{a_{\ensuremath{\mbox{\tiny Z}}}}
\newcommand{\aA}{a_{\ensuremath{\mbox{\tiny A}}}}
\newcommand{\ahpm}{a_{\ensuremath{\mbox{\tiny $H^\pm$}}}}
\newcommand{\xn}{x_0}
\newcommand{\yn}{y_0}
\newcommand{\zn}{z_0}
\newcommand{\bb}{\beta}
\newcommand{\lHh}{\lambda_{\ensuremath{\mbox{\tiny Hh}}}}
\newcommand{\mz}{M_{\ensuremath{\mbox{\tiny Z}}}}
\newcommand{\mw}{M_{\ensuremath{\mbox{\tiny W}}}}
\newcommand{\mH}{M_{\ensuremath{\mbox{\tiny H}}}}
\newcommand{\mh}{M_{\ensuremath{\mbox{\tiny h}}}}
\newcommand{\lij}{\lambda_{ij}}
\newcommand{\lia}{\lambda_{\ensuremath{\mbox{\tiny iA}}}}
\newcommand{\lhAA}{\lambda_{\ensuremath{\mbox{\tiny hAA}}}}
\newcommand{\lHAA}{\lambda_{\ensuremath{\mbox{\tiny HAA}}}}
\newcommand{\lhhphm}{\lambda_{\ensuremath{\mbox{\tiny $hH^+H^-$}}}}
\newcommand{\lHhphm}{\lambda_{\ensuremath{\mbox{\tiny $HH^+H^-$}}}}
\newcommand{\xp}{x_+}
\newcommand{\xw}{x_{\ensuremath{\mbox{\tiny W}}}}
\newcommand{\xa}{x_{\ensuremath{\mbox{\tiny A}}}}
\newcommand{\Xh}{x_{\ensuremath{\mbox{\tiny h}}}}
\newcommand{\XH}{x_{\ensuremath{\mbox{\tiny H}}}}
\newcommand{\sh}{\hat{s}}
\newcommand{\lh}{\lambda_{\ensuremath{\mbox{\tiny $hH_i H_j$}}}}
\newcommand{\lH}{\lambda_{\ensuremath{\mbox{\tiny $HH_i H_j$}}}}
\newcommand{\ct}{\cos\theta}
\newcommand{\ctt}{\cos^2 \theta}
\newcommand{\eep}{e^+ e^-}
\newcommand{\ra}{\rightarrow}



\begin{appendix}
\addcontentsline{toc}{chapter}{Appendix}

\chapter{Scalar integrals}
The integrals that turn up in the calculation of the virtual
corrections to the interference cross section in $b\bar{b}$ production
in subsection \ref{quarkfinal} are listed in this appendix. With
$n=4-2\epsilon$ and the definitions, given in subsection
\ref{quarkfinal},
\beq
C_\epsilon = \Gamma(1+\epsilon) \left( \frac{4\pi\mu^2}{m_b^2} 
\right)^\epsilon
\eeq
and
\beq
t_1 = t -m_b^2
\eeq
the scalar 2-point integrals read
\beq
B_1(s) = \frac{1}{i\pi^2} \int 
\frac{d^n q}{(q^2-m_b^2)((q+p)^2-m_b^2)} = C_\epsilon \left[
\frac{1}{\epsilon} + 2+\beta\ln\left( -\frac{1-\beta}{1+\beta} \right)
\right]
\label{b1f}
\eeq
with $p^2=s$ and $\beta = \sqrt{1-\frac{4m_b^2}{s+i\eta}}$ and
\beq
B_2(t) = \frac{1}{i\pi^2} \int
\frac{d^n q}{(q^2-m_b^2)(q+p)^2} = C_\epsilon \left[
\frac{1}{\epsilon}+2-\frac{t_1}{t}\ln\left(-\frac{t_1}{m_b^2}\right)
\right]
\label{b2f}
\eeq
with $p^2=t$. The $\epsilon$--poles in the two-point integrals are due
to UV singularities. The final result depends on two different
combinations of the above 2-point integrals:
\beq
\tilde{B}(s) = B_1(s) - B_2(m_b^2) \;, \qquad
\tilde{B}(t) = B_2(t) - B_2(m_b^2)
\eeq
$B_2(m_b^2)$ is given by
\beq
B_2(t=m_b^2) = C_\epsilon \left(\frac{1}{\epsilon}+2 \right)
\eeq
so that with (\ref{b1f}) and (\ref{b2f})
\beq
\tilde{B}(t) = -\frac{t_1}{t}\ln\left(-\frac{t_1}{m_b^2} \right)
\eeq
and
\beq
\tilde{B}(s) = \beta \ln\left( -\frac{1-\beta}{1+\beta} \right)
\eeq
Scalar 3-point integrals:
\beq
C_1(s) &=& \frac{1}{i\pi^2} \int \frac{d^n q}{(q^2-m_b^2)((q+p_1)^2-m_b^2)(q+p_1+p_2)^2} \non\\
&=& \frac{C_\epsilon}{\beta s} 
\left\{ \frac{1}{\epsilon}
\ln\left( -\frac{1-\beta}{1+\beta} \right) \right. \non\\
& & \left. + Li_2 \left(\frac{2}{1+\beta}\right) - Li_2 \left(
\frac{2}{1-\beta}\right) - 2 
\left[ Li_2\left(\frac{1}{\beta}\right) - Li_2\left(-\frac{1}{\beta}\right)
\right] \right \}
\eeq
with
$p_1+p_2+p_3 = 0$, $p_1^2 = s$, $p_2^2 = m_b^2$ and $p_3^2=m_b^2$.
\beq
C_1(t) &=& \frac{1}{i\pi^2} \int \frac{d^n q}{(q^2-m_b^2)(q+p_1)^2((q+p_1+p_2)^2-m_b^2)} \non \\
&=& \frac{1}{t-m_b^2}\left\{ \frac{\pi^2}{6}-Li_2 \left( \frac{t}{m_b^2}
\right) \right\}
\eeq
with
$p_1+p_2+p_3 = 0$, $p_1^2 = t$, $p_2^2 = m_b^2$ and $p_3^2=0$.
\beq
C(s) &=& \frac{1}{i\pi^2} \int \frac{d^n q}{(q^2-m_b^2)((q+p_1)^2-m_b^2)((q+p_1+p_2)^2-m_b^2)} \non\\
&=& \frac{1}{2s} \ln^2 \left( -\frac{1-\beta}{1+\beta} \right)
\eeq
with $p_1+p_2+p_3 = 0$, $p_1^2 = 0$, $p_2^2 = 0$ and $p_3^2=s$.\s

\noindent Scalar 4-point integral:
\beq
&& D(s,t) = \frac{1}{i\pi^2} \non\\
&& \int \frac{d^n q}{(q^2-m_b^2)((q+p_1)^2-m_b^2)(q+p_1+p_2)^2-m_b^2)(q-p_4)^2} \non\\
&& = \frac{2C_\epsilon}{\beta s(m_b^2-t)}\left 
\{ Li_2\left(\frac{1}{\beta}\right) - 
Li_2\left(-\frac{1}{\beta}\right) \right. \non\\
&& \left. + \ln \left(-\frac{1-\beta}{1+\beta}\right)
\left[ \ln \left(1-\frac{t}{m_b^2}\right) - \frac{1}{2}
\frac{1}{\epsilon} \right] \right\}
\eeq
with $ p_1^2 = 0$, $p_2^2 = 0$, $p_3^2 = m_b^2$ and $p_4^2 = m_b^2$.
The $\epsilon$--poles in $C_1(s)$ and $D(s,t)$ are IR singularities
due to the massless gluon exchange. The integrals $C_1(u)$ and
$\tilde{B}(u)$ are given by $C_1(t)$ and $\tilde{B}(t)$ after
replacing $t$ with $u$.

\chapter{Double Higgs-strahlung processes}
In this appendix the cross sections for the pair production of the
heavy MSSM Higgs bosons in the Higgs-strahlung processes, $\eep \to Z
H_i H_j $ and $ZAA$ with $H_{i,j}=h,H$, are depicted.  The process
$\eep \ra Z H_iA $ involves only gauge couplings at tree level. The
notation is the same as in subsection \ref{smzhhprod} and the
trilinear couplings have been given in section \ref{susysec}.
Modifications of the MSSM Higgs-gauge couplings with respect to the SM
are determined by the mixing parameters:
\beq
\begin{array}{l@{:\;\;}l@{=\;}l l@{:\;\;}l@{=\;}l l@{:\;\;}l@{=\;}l}
VVh & d_1 & \sin(\beta-\alpha)\;\;\;\; & VVH & d_2 & 
\cos(\beta-\alpha) \;\;\;\;& VVA & d_3 & 0 \\
VAh & c_1 & \cos(\beta-\alpha) \;\;\;\;& VAH & c_2 & 
-\sin(\beta-\alpha) \;\;\;\;& WAH & c_3 & 1
\end{array}
\eeq
for $V=Z$ and $W$. Except for $WAH$, the Higgs bosons are neutral.

\section{$\eep \to Z H_i H_j$} 
The double differential cross section of the process $e^+ e^- \to ZH_i
H_j$ for unpolarized beams can be cast into the form 
\beq
\frac{d\sigma[e^+ e^- \to ZH_i H_j]}{dx_1 dx_2} &=&  
\frac{ \sqrt{2} \, G_F^3 \, M_Z^6}{ 384\, \pi^3 s\,} 
\frac{ v_e^2+a_e^2}{(1-\mu_Z)^2} \  {\cal Z}_{ij} 
\label{eq:sigma}
\eeq
Using the variables $y_1, y_2, y_3$ defined in subsection
\ref{smzhhprod}, and the scaled masses $\mu_i=M_{H_i}^2/s,$
$\mu_{ij}=\mu_i-\mu_j,$ etc., the coefficient ${\cal Z}_{ij}$ in the
cross sections can be expressed as
\beq
{\cal Z}_{ij} &=& {\mathfrak a}^2_{ij}\,f_0 
+ \frac{{\mathfrak a}_{ij}}{2} \, \left[
  \frac{ d_i d_j \, f_3 }{y_1+\mu_{iZ}} 
+ \frac{ c_i c_j \, f_3}{y_1+\mu_{iA}} 
   \right] +\frac{ (d_i d_j)^2}{4\mu_Z (y_1+\mu_{iZ})}
\left[\frac{f_1}{y_1+\mu_{iZ}}+ \frac{f_2}{y_2+\mu_{jZ}} \right]\non \\
&+&  \!\!\!
\frac{ (c_i c_j )^2 }{4\mu_Z (y_1+\mu_{iA})}
\left[
\frac{f_1}{y_1+\mu_{iA}} 
+ \frac{f_2}{y_2+\mu_{jA}}\right] 
+ \frac{ d_i d_j c_i c_j} {2 \mu_Z (y_1+\mu_{iA})}
\left[
\frac{f_1}{y_1+\mu_{iZ}}
+\frac{f_2}{y_2+\mu_{jZ}} \right] 
\non\\
&+& \!\!\!
\Bigg\{ (y_1,\mu_i) \leftrightarrow (y_2,\mu_j) \Bigg\} 
\eeq
with 
\beq
{\mathfrak a}_{ij} = 
 \left[ \frac{ d_1 \lambda_{hH_iH_j}}{y_3 - \mu_{1Z}} + 
\frac{ d_2 \lambda_{HH_iH_j}}{y_3 - \mu_{2Z}} \right] + 
\frac{2 d_i d_j}{y_1+\mu_{iZ}} + \frac{2 d_i d_j}{y_2+\mu_{jZ}}
+ \frac{\delta_{ij}}{\mu_Z}
\eeq
The coefficients $f_{0}$ to $f_3$ are given by
\beq
f_0 &=& \mu_Z [(y_1+y_2)^2 + 8\mu_Z]/8 \non\\
f_1 &=& (y_1-1)^2(\mu_Z-y_1)^2-4\mu_iy_1(y_1+y_1\mu_Z-4\mu_Z) \non\\
& & + \mu_Z(\mu_Z-4\mu_i)(1-4\mu_i)-\mu_Z^2 + (\mu_i-\mu_j)^2 
[y_1(y_1-2)+1-4\mu_i] 
\non\\
& & + (\mu_i-\mu_j)
[8\mu_i(-y_1-\mu_Z)+2y_1 \mu_Z(y_1-2) + 2\mu_Z + 2y_1(y_1-1)^2]
\non\\
f_2 &=& [\mu_Z(1+\mu_Z - y_1 -y_2 - 8\mu_i)-(1+\mu_Z)y_1 y_2]
(2+2\mu_Z -y_1-y_2)
\non\\
& & + y_1 y_2[y_1 y_2 + \mu_Z^2+1+4\mu_i (1+\mu_Z)]
+ 4\mu_i \mu_Z(1+\mu_Z+4\mu_i)+ 
\mu_Z^2 \non\\
& & -2(\mu_i-\mu_j)^3-(\mu_i-\mu_j)^2[y_2(y_1-1)+10\mu_Z +4 \mu_j 
+3y_1 -1]\non\\
& & +(\mu_i-\mu_j)[\mu_Z(2(-y_1 y_2-y_1-8\mu_j)+6(\mu_Z+1-y_2))\non\\
& & +y_1((y_2-1)^2 -
y_1(1+y_2))+y_2(y_2-1)-4\mu_j(y_1-y_2)] \non\\
f_3 &=& y_1(y_1-1)(\mu_Z-y_1)-y_2(y_1+1)(y_1+\mu_Z)
+2\mu_Z(\mu_Z+1-4\mu_i) \non\\
& & + 2(\mu_i-\mu_j)^2 - (\mu_i-\mu_j)[y_2+y_1^2-3y_1+y_1y_2-4\mu_Z]
\label{feqs}
\eeq
Resonance contributions are accounted for by replacing the propagator
products with \\ $\pi_1(\mu_i) \pi_2(\mu_j) \to Re~\{\pi_1(\mu_i)
\pi_2(\mu_j^*) \}$ where $\mu_i \to \mu_i - i \gamma_i$ and $\gamma_i
= M_{H_i} \Gamma_{H_i}/s$. For $i=j=1$, the expressions
eqs.~(\ref{zhh1}--\ref{zhh3}) for the process $\eep \to Zhh$ are
obtained.

\section{$\eep \to ZAA$} 
The differential cross section of the process $e^+ e^- \to ZAA$ can be
obtained from Eq.~(\ref{eq:sigma}) with ${\cal Z}_{ij} = {\cal
  Z}_{33}$:
\beq
{\cal Z}_{33} &=&  {\mathfrak a}^2_{33} \,  
f_0 + \frac{{\mathfrak a}_{33}}{2} \, \left[
\frac{c_1^2}{y_1-\mu_{1A}}+\frac{ c_2^2}{y_1-\mu_{2A}} 
\right]f_3  
\non \\
& & 
+ \frac{1}{4 \mu_Z} \left[ \frac{c_1^2} {y_1-\mu_{1A}}+ 
\frac{c_2^2}{y_1-\mu_{2A}} \right] \left[ 
\frac{c_1^2}{y_2-\mu_{1A}}+ 
\frac{c_2^2}{y_2-\mu_{2A}} \right]f_2  
\non \\
& &  
+ \frac{1}{4\mu_Z} \left[ 
\frac{c_1^2}{y_1-\mu_{1A}}+ 
\frac{c_2^2}{y_1-\mu_{2A}} 
\right]^2 f_1  
+ \Bigg\{ y_1 \leftrightarrow y_2 \Bigg\} 
\eeq
where 
\beq
{\mathfrak a}_{33} = 
\left[ \frac{ d_1 \lambda_{hAA}}{y_3 - \mu_{1Z}} + 
\frac{ d_2 \lambda_{HAA}}{y_3 - \mu_{2Z}} \right]
+  \frac{1}{\mu_Z} 
\eeq 
The coefficients $f_0$ to $f_3$ are given by (\ref{feqs}) after
replacing $\mu_1,$ $\mu_2$ with $\mu_A$.



\chapter{Triple Higgs boson production}
In this appendix the cross sections for the triple Higgs boson
production of MSSM Higgs bosons, $\eep \to A H_i H_j$ and $\eep \ra
AAA$ with $H_{i,j}=h,H$, are listed . Due to CP-invariance the process
$\eep \to H_iA A$ does not occur at tree level.

\section{$\eep \to A H_i H_j$} 
In the same notation as above, the double differential cross section
of the process $e^+ e^- \to A H_i H_j$ for unpolarized beams can be
written as:
\beq 
\frac{d\sigma}{dx_1 dx_2} = 
\frac{G_F^3 M_Z^6}{768 \sqrt{2} \pi^3 s} \, 
\frac{v_e^2+a_e^2}{(1-\mu_Z)^2} \, {\mathfrak A}_{ij} 
\eeq
with the function ${\mathfrak A}_{ij}$ 
\beq
{\mathfrak A}_{ij} &=&  \left[ 
\frac{ \lambda_{hH_iH_j} c_1} {y_3-\mu_{1A}}
+ \frac{ \lambda_{H H_iH_j} c_2} {y_3-\mu_{2A}} \right]^2 g_0
+ \frac{\lambda^2_{H_jAA} c_i^2} {(y_1+\mu_{iA})^2} g_1
+ \frac{ \lambda^2_{H_iAA} c_j^2} {(y_2+\mu_{jA})^2} g_1' \non \\
&& + \frac{c_j^2 d_i^2 } {(y_1+\mu_{iZ})^2} g_2 +
\frac{c_i^2 d_j^2 } {(y_2+\mu_{jZ})^2} g_2'+
\left[ \frac{\lambda_{hH_iH_j} c_1 } {y_3-\mu_{1A}}
+ \frac{ \lambda_{H H_iH_j} c_2 } {y_3-\mu_{2A}} \right] \non \\
&& \times 
\left[ \frac{ \lambda_{H_jAA} c_i} {y_1+\mu_{iA}} g_3
+ \frac{ \lambda_{H_iAA} c_j} {y_2+\mu_{jA}} g_3' 
+ \frac{c_j d_i } {y_1+\mu_{iZ}} g_4 +
\frac{c_i d_j } {y_2+\mu_{jZ}} g_4' \right] \non \\
&&+ 
\frac{\lambda_{H_i AA} \lambda_{H_j AA} c_ic_j}{(y_1+\mu_{iA})(y_2+\mu_{jA})}g_5
+ \frac{c_ic_j d_i d_j}{(y_1+\mu_{iZ})(y_2+\mu_{jZ})}g_8 
\non \\
&& + 
\frac{ \lambda_{H_j AA} c_ic_j d_i}{(y_1+\mu_{iA})(y_1+\mu_{iZ})}g_6
+ \frac{\lambda_{H_i AA} c_ic_j d_j}{(y_2+\mu_{jA})(y_2+\mu_{jZ})}
g_6' 
\non \\
&& 
+ \frac{\lambda_{H_j AA} c_i^2 d_j}{(y_1+\mu_{iA})(y_2+\mu_{jZ})} 
g_7 
+ \frac{\lambda_{H_i AA} c_j^2 d_i}{(y_2+\mu_{jA})(y_1+\mu_{iZ})}
g_7'
\eeq
The coefficients $g_k$ read:
\beq
g_0 &=& \mu_Z[(y_1+y_2)^2-4\mu_A] \non \\
g_1 &=& \mu_Z(y_1^2-2y_1-4\mu_i+1) \non \\
g_2 &=& \mu_Z (2y_1 + y_1^2 -4y_2 +4y_2^2 + 4y_1 y_2 +1+ 4\mu_i 
- 8\mu_j - 8\mu_A) + (\mu_j-\mu_A)^2 \non\\
&& [8+(-2y_1+y_1^2-4\mu_i+1)/\mu_Z] + 2(\mu_j-\mu_A) 
(2y_1 y_2+y_1^2+2y_2-1)\non\\
g_3 &=& 2\mu_Z(y_1^2+y_1 y_2- y_1+y_2+2\mu_j-2\mu_i-2\mu_A) \non \\
g_4 &=& 2\mu_Z(y_1-y_2+y_1^2+2y_2^2+3y_1 y_2 -2\mu_j +2 \mu_i-6 
\mu_A) \non\\ 
&&+2(\mu_j-\mu_A)(-y_1+y_2+ y_1^2 + y_1y_2+2\mu_j-2\mu_A-2\mu_i) 
    \non \\
g_5 &=& 2\mu_Z(y_1+ y_2+y_1y_2+2\mu_j+ 2\mu_i-2\mu_A-1) \non \\
g_6 &=& 2\mu_Z( y_1^2+2 y_1y_2+2y_2+4\mu_j-4\mu_A -1) \non \\
 && + 2(\mu_j-\mu_A)(y_1^2-2y_1-4\mu_i+1) \non \\
g_7 &=& 2[ \mu_Z(2y_1^2+y_1y_2+y_2-3y_1+2\mu_j-6\mu_i-2\mu_A+1) 
\non \\
 && + (\mu_i-\mu_A)(y_1+y_2+y_1y_2+2\mu_j+2\mu_i-2\mu_A-1)] \non \\
g_8 &=& 2 \left\{\mu_Z(y_1+y_2+2y_1^2+2y_2^2+5y_1 y_2 -1 + 2\mu_j 
+ 2\mu_i -10\mu_A)  \right. \non \\
&& +2(\mu_i-\mu_A)(\mu_i-3\mu_j-\mu_A-2y_2+1) 
+2(\mu_j-\mu_A)(\mu_j-3\mu_i-\mu_A-2y_1+1)  \non \\
&& + [(\mu_j-\mu_A)((1+y_1)(y_2+2y_1-1)\mu_Z
+2\mu_i^2+4\mu_A^2 +\mu_A-\mu_i)\non \\
&& + (\mu_i-\mu_A) ((1+y_2) (y_1+2y_2-1)\mu_Z 
+ 2\mu_j^2+4\mu_A^2+\mu_A-\mu_j) \non \\
&& + \left. 6\mu_A(\mu_A^2-\mu_i\mu_j) + 
(\mu_i-\mu_A)(\mu_j-\mu_A) (1+y_1)(1+y_2)]/\mu_Z 
\right\}
\eeq
and 
\beq
g_k' (y_1,y_2,\mu_i,\mu_j) = g_k(y_2,y_1,\mu_j,\mu_i) 
\eeq



\chapter{Heavy Higgs production in $W_L W_L$ fusion}
The amplitudes and cross sections for pair production of CP--even
Higgs bosons in the longitudinal $W$ approximation $W_L W_L \ra H_i
H_j$, as well as for $W_L W_L \ra AA$ are presented in this appendix.
The notation is the same as in section \ref{wwsmfusion}.

\section{$W_L W_L \to H_i H_j$ } 

The amplitudes for the process $W_L W_L \to H_i H_j$ are given by: 
\beq
{\cal M}_{LL} &=& \frac{G_F \hat{s}}{\sqrt{2}} \left\{ 
(1+\beta_W^2)\left[\delta_{ij} + 
\frac{\lambda_{hH_iH_j} d_1}{(\hat{s} -M_h^2)/M_Z^2}
+ \frac{\lambda_{H H_iH_j} d_2}{(\hat{s}-M_H^2)/M_Z^2} \right] \right. 
\non \\
&&+ \frac{d_i d_j}{\beta_W \lambda_{ij}} \left[\frac{  r_W
+ (\beta_W -\lambda_{ij} \cos \theta )^2} {\cos \theta-x_W} -
\frac{ r_W + (\beta_W +\lambda_{ij} \cos \theta )^2}{\cos \theta+x_W} 
\right] \non \\
&&+ \left. \frac{c_i c_j}{\beta_W \lambda_{ij}} \left[ \frac{ 
r_+ + (\beta_W -\lambda_{ij} \cos \theta )^2 }{\cos \theta-x_+} - 
\frac{ r_+ + (\beta_W +\lambda_{ij} \cos \theta )^2 }{\cos \theta+x_+} 
\right] \right\}
\eeq
where $\mu_{i,j}= M_{H_{i,j}}^2/\hat{s}$,
$\beta_W=(1-4M_W^2/\hat{s})^{1/2}$ and $\lambda_{ij}$ is the usual
two--body phase space function, $\lambda^2_{ij} = (1-\mu_i -
\mu_j)^2-4\mu_i\mu_j$. Furthermore,
\beq
\begin{array}{l@{\quad\quad}l}
\hspace{-0.7cm}x_W= (1-\mu_i -\mu_j)/(\beta_W \lambda_{ij}) &
\hspace{-0.4cm}r_W= 1-\beta_W^4 -\beta^2_W (\mu_i-\mu_j)^2 \\[0.1cm]
\hspace{-0.7cm}x_+= (1-\mu_i -\mu_j+2 M_{H^\pm}^2/\hat{s} 
-2M_W^2/\hat{s})/(\beta_W\lambda_{ij}) &
\hspace{-0.4cm}r_+ =  -\beta^2_W (\mu_i-\mu_j)^2 
\end{array}
\eeq
After integrating over the scattering angle, the total cross section
of the subprocess can be cast into the form
\beq
\sigma_{LL} (H_i H_j) &=& \frac{1}{1+\delta_{ij}} 
\frac{G_F^2 M_W^4}{2\pi \hat{s}}
\frac{\lambda_{ij}}{ \beta_W (1-\beta_W^2)^2} \non \\
&& \Bigg\{ (1+\beta_W^2)^2 \left[ \delta_{ij} 
+ \frac{\lambda_{hH_iH_j} d_1}{(\hat{s} -M_h^2)/M_Z^2} 
+ \frac{\lambda_{H H_iH_j} d_2}{(\hat{s} -M_H^2)/M_Z^2} \right]^2  
\non \\
&& +  \frac{2(1+\beta_W^2)} {\beta_W \lambda_{ij} }  \left[
\delta_{ij}  + \frac{\lambda_{hH_iH_j}d_1}{(\hat{s} -M_h^2)/M_Z^2}
 + \frac{\lambda_{H H_iH_j}d_2}{(\hat{s} -M_H^2)/M_Z^2} \right] \left[ 
d_i d_j a_1^W  + c_i c_j a_1^+   \right] \non \\
&&+ \left( \frac{d_i d_j}{\beta_W \lambda_{ij} } \right)^2 a_2^W  
+ \left( \frac{c_i c_j}{\beta_W \lambda_{ij} } \right)^2 a_2^+  + 4 
\left(\frac{d_i d_jc_i c_j}{\beta_W^2 \lambda^2_{ij} } \right) [a_3^W 
+ a_3^+ ] \Bigg\} 
\eeq
with 
\beq
a_1^W  &=& [ (x_W \lambda_{ij} -\beta_W)^2 + r_W ] \log \frac
{x_W-1}{x_W+1} +2 \lambda_{ij} (x_W \lambda_{ij} -2\beta_W ) \non \\
a_2^W &=& \left[ \frac{1}{x_W} \log \frac{x_W-1}{x_W+1} + \frac{2}
{x_W^2-1} \right] \bigg[ x_W^2 \lambda_{ij}^2 (3 \lambda_{ij}^2 x_W^2 
+2 r_W +14 \beta_W^2) \non \\
&&  -(\beta_W^2+ r_W)^2   -4 \lambda_{ij} \beta_W x_W 
 (3 \lambda_{ij}^2 x_W^2 +\beta_W^2+r_W ) \bigg] \non \\
&& - \frac{4}{x_W^2-1} \left[ \lambda_{ij}^2 (\lambda_{ij}^2 x_W^2 
+4 \beta^2_W - 4 \lambda_{ij} x_W \beta_W) - (\beta_W^2 +r_W)^2 
\right] \non \\
a_3^W&=& \frac{1}{x_+^2 - x_W^2} \, \log \frac{x_W-1}{x_W+1} \bigg[
2 \beta_W \lambda_{ij} x_W [(\beta_W^2+x_W^2 \lambda_{ij}^2)(x_W+x_+)
+ x_W r_W +x_+ r_+]  \non \\
&&  -x_+( r_+ + r_W + \lambda_{ij}^2 x_W^2)(\beta_W^2 + 
\lambda_{ij}^2 x_W^2)
-\beta_W^2( x_+ \beta_W^2 +4\lambda_{ij}^2 x_W^3+
x_+ x_W^2 \lambda_{ij}^2 ) \non \\
&&- x_+ r_W r_+  \bigg] + \lambda_{ij}^2 
\left[ \lambda_{ij}^2 x_+ x_W -2 
\beta_W \lambda_{ij} (x_W+x_+) + 4\beta_W^2   \right] \non 
\eeq
\beq
a_i ^+ \equiv a_i^W \ (x_W \leftrightarrow x_+ \ , 
\ r_W \leftrightarrow r_+ )
\eeq

\section{$W_L W_L \to AA $}
Since there are only a few diagrams involved in the process $W_L W_L
\to AA $ and since the masses of the final state particles are equal,
the amplitude and cross section adopt a much simpler form for
pseudoscalar Higgs bosons:
\beq
{\cal M}_{LL} &=& \frac{G_F \hat{s}}{\sqrt{2}}\left\{ 
(1+\beta_W^2) \left[1 + 
\frac{\lambda_{hAA} d_1} {(\hat{s} -M_h^2)/M_Z^2}
+ \frac{\lambda_{H AA} d_2} {(\hat{s} -M_H^2)/M_Z^2} \right] \right. 
\non \\
&&+ \left. \frac{1}{\beta_W \beta_A} \left[ 
\frac{ (\beta_W - \beta_A \cos \theta )^2}{\cos \theta-x_A} -  
\frac{ (\beta_W +\beta_A \cos\theta)^2}{\cos \theta+x_A} 
\right] \right\}
\eeq
with 
\beq
\beta_A=(1-4M_A^2/\hat{s})^{1/2} \quad \mathrm{and} \quad 
x_A=(1-2 M_A^2/\hat{s} +2 M_{H^\pm}^2/\hat{s} -2 M_W^2/\hat{s})
/(\beta_W \beta_A)
\eeq
The total cross section for the subprocess $W_L W_L \to AA$ may be
written as
\beq 
\sigma_{LL} (AA) 
\!\!&=&\!\! \frac{G_F^2 M_W^4}{4\pi\hat{s}}\frac{\beta_A}
{\beta_W (1-\beta_W^2)^2} \Bigg\{ (1+\beta_W^2)^2 \left[ 1 + 
\frac{\lambda_{hAA} d_1 } {(\hat{s} -M_h^2)/M_Z^2} + 
\frac{\lambda_{HAA} d_2 }{(\hat{s} -M_H^2)/M_Z^2} \right]^2  \non \\
&&+ 2(1+\beta_W^2) \left[ 1 + 
\frac{\lambda_{hAA} d_1 }{(\hat{s} -M_h^2)/M_Z^2}
\; + \; \frac{\lambda_{HAA} d_2 }{(\hat{s} -M_H^2)/M_Z^2} 
\; \right] \ \\
&& \times \frac{1}{\beta_W \beta_A}
\bigg[ (x_A \beta_A - \beta_W)^2 \log \frac{x_A-1}{x_A+1} 
 + 2\beta_A (x_A \beta_A -2\beta_W) \bigg] + \frac{1}{\beta_A^2 \beta_W^2} 
\times \non \\
&& \Bigg(
\log \frac{x_A-1}{x_A+1} \left[ 3 \beta_A^2 x_A (\beta_A x_A-2 \beta_W)^2
+\beta_W^2(2\beta_A^2 x_A -4 \beta_W \beta_A -\beta_W^2/x_A) \right] \non \\
&& 
+ \frac{2}{x_A^2-1} \left[ (3x_A^2 \beta_A^2-2\beta_A^2+\beta_W^2)
(\beta_Ax_A -2\beta_W)^2 + \beta_W^2 (\beta_A^2 x_A^2-3 \beta_W^2) 
\right] \Bigg) \Bigg\}  \non 
\eeq

\section{Asymptotic energies}
In the high-energy limit the leading part of the $WW$ fusion cross
sections does not depend on the trilinear couplings $H_i H_j H_k$ or
$H_i H_j A$. After convolution with the $W$ luminosities, however, the
dominant contribution to the leptonic cross sections $e^+e^-$
$\longrightarrow \bar{\nu}_e \nu_e H_i H_j$ and $AA$ stems from the
threshold regions, independent of the $e^+e^-$ energies. These
processes are therefore also in the high-energy limit in leading
order sensitive to the trilinear couplings.

\end{appendix}
\cleardoublepage

%% file: lit.tex

%% file: master.bbl
\addcontentsline{toc}{chapter}{Bibliography}
\begin{thebibliography}{99}
\bibitem{wsalam}
        J.~Goldstone, A.~Salam and S.~Weinberg, Phys.\ Rev.\  
        {\bf 127} (1962) 965;
        S.~Weinberg, Phys.\ Rev.\ Lett.\ {\bf 19} (1967) 1264; 
        S.L.~Glashow, S.~Weinberg, 
        Phys.\ Rev.\ Lett.\ {\bf 20} (1968) 224;
        A.~Salam, {\it Proceedings Of The Nobel Symposium}, Stockholm 1968, 
        ed.\ N.~Svartholm
\bibitem{higgs}
        P.W.~Higgs, Phys.\ Lett.\ {\bf 12} (1964) 132; and 
        Phys.\ Rev.\ {\bf 145} (1966) 1156; F.~Englert and R.~Brout, 
        Phys.\ Rev.\ Lett.\ {\bf 13} (1964) 321; G.S.~Guralnik, 
        C.R.~Hagen and T.W.~Kibble, Phys.\ Rev.\ Lett.\ {\bf 13} 
        (1964) 585.
\bibitem{gunion}
        For a review see: J.F.~Gunion, H.E.~Haber, G.~Kane and 
        S.~Dawson, \textit{The Higgs Hunter's Guide}, Addison-Wesley, 
        1990.
\bibitem{smpub} 
        M.~Spira and P.M.~Zerwas, {\it Int. Universit\"atswochen}, 
        Schladming 1997, hep-ph/9803257;
        C.~Quigg, Acta Phys.\ Polon.\ {\bf B30} (1999) 2145-2192.
\bibitem{susy}
        D.V.~Volkov and V.P.~Alkulov, Phys.\ Lett.\ {\bf B46} (1973) 109;
        J.~Wess and B.~Zumino, Nucl.\ Phys.\ {\bf B70} (1974) 39;
        H.P.~Nilles, Phys.\ Rep.\ {\bf 110} (1984) 1; 
        H.E.~Haber, G.L.~Kane, Phys.\ Rep.\ {\bf 117} (1985) 75;
        M.F.~Sohnius, Phys.\ Rep.\ {\bf 128} (1985) 39.
\bibitem{susyprod}
        J.F.~Gunion and H.E.~Haber, Nucl.\ Phys.\ {\bf B272} (1986) 1 
        and {\bf B278} (1986) 449.
\bibitem{gut} 
        H.~Georgi and S.L.~Glashow, Phys.\ Rev.\ Lett.\ {\bf 32} (1974) 438;
        H.~Georgi, H.R.~Quinn and S.~Weinberg, Phys.\ Rev.\ Lett.\ 
        {\bf 33} (1974) 451;
        more recent reviews are, for example, G.G.~Ross, 
        \textit{Grand Unified Theories}, 
        Benjamin, New York, 1984; R.~Mohapatra, Prog.\ Part.\ Nucl.\ Phys.\
        {\bf 26} (1991) 1.
\bibitem{sinw}
        J.~Ellis, S.~Kelley and D.V.~Nanopoulos, Phys.\ Lett.\ {\bf B260} 
        (1991) 306; U.~Amaldi, W.~de Boer and H.~F\"urstenau, Phys.\ Lett.\ 
        {\bf B260} (1991) 447; P.~Langacker and M.~Luo, Phys.\ Rev.\ {\bf D44} 
        (1991) 817; G.G.~Ross and R.G.~Roberts, Nucl.\ Phys.\ {\bf B377} 
        (1992) 571.
\bibitem{atlascms}
        ATLAS Collaboration, Technical Design Report CERN-LHCC 99-14;
        CMS Collaboration, Technical Proposal, Report CERN-LHCC 94-38.
\bibitem{accomando}
        Conceptual Design Report of a 500~GeV $e^+e^-$ Linear Collider, 
        eds.\ R.~Brinkmann et al., DESY/ECFA 1997-048/182;
        E.~Accomando et al., Phys.\ Rep.\ {\bf 299} (1998) 1;
        P.M.~Zerwas, based on lectures at the Carg\`{e}se 1999 Summer 
        Institute, the Moscow 1999 Workshop on {\it Quantum Field Theory} 
        and the Lund 1999 Workshop on 
        {\it Future Electron Positron Colliders}, hep-ph/0003221.
\bibitem{ggprop}
        I.F.~Ginzburg, G.L.~Kotkin, V.G.~Serbo and V.I.~Telnov, Pizma ZhETF 
        {\bf 34} (1981) 514; JETP Lett.\ {\bf 34} (1982) 491.
\bibitem{ggvgl}
        I.F.~Ginzburg, G.L.~Kotkin, V.G.~Serbo and V.I.~Telnov: 
        Nucl.\ Inst.\ Methods {\bf 205} (1993) 47; I.F.~Ginzburg, 
        G.L.~Kotkin, S.L.~Panfil, V.G.~Serbo and V.I.~Telnov, 
        Nucl.\ Inst.\ Methods {\bf 219} (1984) 5.
%
%
%
\bibitem{kobay}
        N.~Cabibbo, Phys.\ Rev.\ Lett.\ {\bf 10} (1963) 531;  
        M.~Kobayashi and T.Maskawa, Prog.\ Theor.\ Phys.\ {\bf 49} (1973) 
        652.
\bibitem{zumino}
        J.~Wess and B.~Zumino, Nucl.\ Phys.\ {\bf B78} (1974) 1;
        S.~Ferrara and B.~Zumino, Nucl.\ Phys.\ {\bf B79} (1974) 413.
\bibitem{fayetterm}
        P.~Fayet and J.~Iliopoulos, Phys.\ Lett.\ {\bf B51} (1974) 461.
\bibitem{smallterm}
        see for example: L.~Alvarez-Gaume, J.~Polchinski and M.B.~Wise,
        Nucl.\ Phys.\ {\bf B221} (1983) 495.
\bibitem{girardello}
        L.~Girardello and M.T.~Grisaru, Nucl.\ Phys.\ {\bf B194} (1982) 65.
\bibitem{hall}
        L.~Hall, J.~Lykken and S.~Weinberg, Phys.\ Rev.\ {\bf D27} (1973) 
        2359.
\bibitem{haber}
        H.E.~Haber, to appear in 
        {\it Perspectives on Higgs Physics II}, ed.~G.L.~Kane, 
        World Scientific, Singapore 1997, hep-ph/9707213.
\bibitem{okada}
        H.E.~Haber and R.~Hempfling, Phys.\ Rev.\ Lett.\ {\bf 66} 
        (1991) 1815; Y.~Okada, M.~Yamaguchi and T.~Yanagida, Prog.\ 
        Theor.\ Phys.\ {\bf 85} (1991) 1; J.~Ellis, G.~Ridolfi and 
        F.~Zwirner, Phys.\ Lett.\ {\bf B257} (1991) 83.
\bibitem{carena}
        J.R.~Espinosa and M.~Quiros, Phys.\ Lett.\ {\bf B266} (1991) 389; 
        R.~Hempfling and A.~Hoang, Phys.\ Lett.\ {\bf B331} (1994) 99;
        J.A.~Casas, J.~Espinosa, M.~Quiros and A.~Riotto, Nucl.\ Phys.\ 
        {\bf B436} (1995) 3; (E) {\bf B439} (1995) 466;
        M.~Carena, J.R.~Espinosa, M.~Quiros and C.E.M.~Wagner, Phys.\ 
        Lett.\ {\bf B335} (1995) 209; 
        M.~Carena, J.~Espinosa, M.~Quiros and C.E.M.~Wagner, Phys.\ 
        Lett.\ {\bf B355} (1995) 209;
        M.~Carena, M.~Quiros and 
        C.E.M.~Wagner, Nucl.\ Phys.\ {\bf B461} (1996) 407; 
        H.E.~Haber, R.~Hempfling and A.H.~Hoang, Z.\ Phys.\ {\bf C75} 
        (1997) 539; S.~Heinemeyer, W.~Hollik and G.~Weiglein, 
        Eur.\ Phys.\ J.\ {\bf C9} (1999) 343; 
        R.-J.~Zhang, Phys.\ Lett.\ {\bf B447} (1999) 89.
\bibitem{djouadi1}
        V.~Barger, M.S.~Berger, A.L.~Stange and R.J.N.~Phillips, 
        Phys.\ Rev.\ {\bf D45} (1992) 4128;
        Z.~Kunszt and F.~Zwirner, Nucl.\ Phys.\ {\bf B385} (1992) 3.
\bibitem{djouadi2}
        A.~Djouadi, H.E.~Haber and P.M.~Zerwas, Phys.\ Lett.\ 
        {\bf B375} (1996) 203 and (E) in press.
%
%
%
\bibitem{lat}
        A.~Hasenfratz, K.~Jansen, C.~Lang, T.~Neuhaus and H.~Yoneyama, 
        Phys.\ Lett.\ {\bf B199} (1987) 531; J.~Kuti, L.~Liu and Y.~Shen, 
        Phys.\ Rev.\ Lett.\ {\bf 61} (1988) 678; M.~L\"uscher and P.~Weisz, 
        Nucl.\ Phys.\ {\bf B318} (1989) 705.
\bibitem{ren1}
        M.~Chanowitz, M.~Furman and I.~Hinchliffe, 
        Phys.\ Lett.\ {\bf B78} (1978) 285;
        N.~Cabibbo, L.~Maiani, G.~Parisi and R.~Petronzio, Nucl.\ Phys.\ 
        {\bf B158} (1979) 295; R.A.~Flores and M.~Sher, 
        Phys.\ Rev.\ {\bf D27} (1983) 1679; M.~Lindner, Z.\ Phys.\ 
        {\bf C31} (1986) 295; M.~Sher, Phys.\ Rep.\ {\bf 179} (1989) 273; 
        Phys.\ Lett.\ {\bf B317} (1993) 159 and addendum {\bf B331} (1994) 
        448; G.~Altarelli and G.~Isidori, Phys.\ Lett.\ {\bf B337} (1994) 
        141; J.~Casas, J.~Espinosa and M.~Quiros, Phys.\ Lett.\ {\bf B342} 
        (1995) 171.
\bibitem{ren2}
        J.~Espinosa and M.~Quiros, Phys.\ Lett.\ {\bf B353} (1995) 257.
\bibitem{exp}
        ALEPH, DELPHI, L3 and OPAL Collaborations, 
        {\it Searches for Higgs bosons: Preliminary combined 
          results using LEP data collected at energies up to 202~GeV}, 
        ALEPH 2000-028 CONF 2000-023, DELPHI 2000-050 CONF 365, 
        L3 note 2525, OPAL Technical Note TN546, March 2000.
\bibitem{abdhabil}
        For a review see:
        A.~Djouadi, Int.\ Journal of Mod.\ Phys.\ {\bf A10} (1995) 1.
\bibitem{spirahabil}
        For a review see:
        M.~Spira, Fortschr.\ Phys.\ {\bf 46} (1998) 3.
\bibitem{kinnunen}
        R.~Kinnunen and D.~Denegri, CMS-NOTE-1997/057.
\bibitem{kintalk}
        R.~Kinnunen, talk given at the ECFA/DESY Workshop on 
        {\it Physics and Detectors for a Linear Collider}, 
        Obernai 16-19 Oct.\ 1999.
\bibitem{tevspir}
        M.~Spira, contributed to {\it Physics at Run II: Workshop on 
          Supersymmetry/Higgs: Summary Meeting}, Batavia, IL, 19-21 
        Nov 1998, hep-ph/9810289.  
\bibitem{conway}
        M.~Carena, H.~Haber et al., Proc.\ workshop 
        {\it "Physics at RunII - Supersymmetry/Higgs"}, 
        Fermilab 1998 (to appear).
%
%
\bibitem{baer}
        H.~Baer et al., Phys.\ Rev.\ {\bf D47} (1993) 1062; 
        S.G.~Frederiksen, N.P.~Johnson, G.L.~Kane and J.H.~Reid, 
        preprint SSCL-577-mc, Jul 1992;
        S.G.~Frederiksen, N.P.~Johnson, G.L.~Kane and J.H.~Reid, 
        Phys.\ Rev.\ {\bf D50} (1994) 4244. 
\bibitem{mrenna}
        M.~Carena, S.~Mrenna, C.E.M.~Wagner, Phys.\ Rev.\ {\bf D60} 
        (1999) 075010.
\bibitem{hrad1}
        J.~Ellis, M.K.~Gaillard and D.V.~Nanopoulos, Nucl.\ Phys.\ 
        {\bf B106} (1976) 292; B.W.~Lee, C.~Quigg and H.B.~Thacker, 
        Phys.\ Rev.\ {\bf D16} (1977) 1519; B.L.~Ioffe and V.A.~Khoze, 
        Sov.\ J.\ Part.\  Nucl.\ {\bf 9} (1978) 50.
\bibitem{hrad2}
        V.~Barger, K.~Cheung, A.~Djouadi, B.A.~Kniehl and P.M.~Zerwas, Phys.\ 
        Rev.\ {\bf D49} (1994) 79.
\bibitem{wfus1}
        D.R.T.~Jones and S.T.~Petvoc, Phys.\ Lett.\ {\bf B84} (1979) 440.
\bibitem{wfus2}
        R.N.~Cahn and S.~Dawson, Phys.\ Lett.\ {\bf B136} (1984) 196; 
        (E) ibid.\ {\bf B138} 1984,464.
\bibitem{wfus2a} 
        G.L.~Kane, W.W.~Repko and W.B.~Rolnik, Phys.\ Lett.\ {\bf B148} 
        (1984) 367.
\bibitem{wfus3}
        G.~Altarelli, B.~Mele and F.~Pitolli, Nucl.\ Phys.\ {\bf B287} (1987) 
        205; W.~Kilian, M.~Kr\"amer and P.M.~Zerwas, Phys.\ Lett.\ {\bf B373} 
        (1996) 135.
\bibitem{schreiber}
        E.~Boos, M.~Sachwitz, H.J.~Schreiber and S.~Shichanin, Z.\ Phys.\ 
        {\bf C61} (1994) 675.
\bibitem{lohmann}
        P.~Gracia-Abia and W.~Lohmann, talk given at the 
        ECFA/DESY Workshop on 
        {\it Physics and Detectors for a Linear Collider}, Sitges 
        28 Apr-5 May 1999.  
\bibitem{djkalzer} 
        A.~Djouadi, J.~Kalinowski and P.M.~Zerwas, Z.\ Phys.\ {\bf C57} 
        (1993) 569.
\bibitem{ohmann}
        A.~Djouadi, J.~Kalinowski, P.~Ohmann and 
        P.M.~Zerwas, Z.\ Phys.\ {\bf C74} (1997) 93.
\bibitem{janot}
        P.~Janot, in Proc.\ 
        {\it Physics and Experiments with $e^+e^-$ 
          Linear Colliders}, Waikoloa/Hawaii 1993, eds.\ F.~Harris, S.~Olsen, 
        S.~Pakvasa and X.~Tata, World Scientific 1993.
\bibitem{supp0}
        J.F.~Gunion and H.E.~Haber, Phys.\ Rev.\ {\bf D48} (1993) 5109.
\bibitem{supp1}
        D.L.~Borden, V.A.~Khoze, J.~Ohnemus and W.J.~Stirling, 
        Phys.\ Rev.\ {\bf D50} (1994) 4499;
        G.~Jikia and A.~Tkabladze, Nucl.\ Inst.\ Meth.\ {\bf A355} (1995) 81.
\bibitem{supp2}
        G.~Jikia and A.~Tkabladze, Phys.\ Rev.\ {\bf D54} (1996) 2030.
\bibitem{rem}
        B.~Kamal, Z.~Merebashvili and A.P.~Contogouris, Phys.\ Rev.\ 
        {\bf D51} (1995) 4808 and (E) ibid.\ {\bf D55} (1997) 3229;
        G.~Jikia and A.~Tkabladze, hep-ph/0004068.
\bibitem{largel1}
        V.S.~Fadin, V.A.~Khoze and A.D.~Martin, Phys.\ Rev.\ {\bf D56} 
        (1997) 484;
        M.~Melles and W.J.~Stirling, Phys.\ Rev.\ {\bf D59} (1999) 094009;
        Eur.\ Phys.\ J.\ {\bf C9} (1999) 101.
\bibitem{largel2}
        M.~Melles and W.J.~Stirling, Nucl.\ Phys.\ {\bf B564} (2000) 325.
\bibitem{sudak}
        V.V.~Sudakov, Sov.\ Phys.\ JETP {\bf 3} (1956) 65.
\bibitem{graudenz}
         H.~Zheng and D.~Wu, Phys.\ Rev.\ {\bf D42} (1990) 3760;
         S.~Dawson and R.P.~Kauffman, Phys.\ Rev.\ {\bf D47} (1993) 1264;
\bibitem{nvoll}
         A.~Djouadi, M.~Spira, J.~van der Bij and P.M.Zerwas, Phys.\ Lett.\ 
         {\bf B257} (1991) 187.
\bibitem{vollqcd}
         A.~Djouadi, M.~Spira and P.M.~Zerwas, Phys.\ Lett.\ {\bf B311} 
         (1993) 255;
         K.~Melnikov and O.~Yakovlev, Phys.\ Lett.\ {\bf B312} (1993) 179;
         M.~Inoue, R.~Najima, T.~Oka and J.~Saito, Mod.\ Phys.\ Lett.\ 
         {\bf A9} (1994) 1189; 
         M.~Spira, A.~Djouadi, D.~Graudenz and P.M.~Zerwas, Nucl.
         \ Phys.\ {\bf B453} (1995) 17.
\bibitem{squark}
         A.~Djouadi, V~Driesen, W.~Hollik and J.I.~Illana, Eur.\ Phys.\ J.\ 
         {\bf C1} (1998) 149.
\bibitem{qcdcorr}
        E.~Braaten and J.P.~Leveille, Phys.\ Rev.\ {\bf D22} (1980) 715;
        N.~Sakai, Phys.\ Rev.\ {\bf D22} (1980) 2220;
        T.~Inami and T.~Kubota, Nucl.\ Phys.\ {\bf B179} (1981) 171;
        S.G.~Gorishny, A.L.~Kataev and S.A.~Larin, Sov.\ J.\ Nucl.\ Phys.\ 
        {\bf 40} (1984) 329;
        M.~Drees and K.~Hikasa, Phys.\ Rev.\ {\bf D41} (1990) 1547; 
        Phys.\ Lett.\ {\bf B240} (1990) 455 and (E) {\bf B262} (1991) 497;
        S.G.~Gorishny, A.L.~Kataev, S.A.~Larin and L.R.~Surguladze, Mod.\ 
        Phys.\ Lett.\ {\bf A5} (1990) 2703; Phys.\ Rev.\ {\bf D43} (1991) 1633;
        A.L.~Kataev and V.T.~Kim, Mod.\ Phys.\ Lett.\ {\bf A9} (1994) 1309;
        L.R.~Surguladze, Phys.\ Lett.\ {\bf 341} (1994) 61;
        K.G.~Chetyrkin, J.H.~K\"uhn and A.~Kwiatkowski, Proc.\ of the 
        Workshop {\it QCD at LEP} Aachen, 1994;
        K.G.~Chetyrkin, Phys.\ Lett.\ {\bf B390} (1997) 309;
        K.G.~Chetyrkin and A.~Kwiatkowski, Nucl.\ Phys.\ {\bf B461} (1996) 3;
        S.A.~Larin, T.~van Ritbergen and J.A.M.~Vermaseren, Phys.\ Lett.\ 
        {\bf B362} (1995) 134.
\bibitem{spirdjou}
        A.~Djouadi, M.~Spira and P.M.~Zerwas, Z.\ Phys.\ {\bf C70} 
        (1996) 427. 
\bibitem{effcoup}
        B.A.~Kniehl and M.~Spira, Z.\ Phys.\ {\bf C69} (1995) 77.
\bibitem{dawson}
        S.~Dawson, S.~Dittmaier and M.~Spira, Phys.\ Rev.\ {\bf D58} 
        (1998) 115012.
\bibitem{susycorrec}
        A.~Dabelstein, Nucl.\ Phys.\ {\bf B456} (1995) 25;
        R.A.~Jim\'{e}nez and J.~Sol\`{a}, Phys.\ Lett.\ {\bf B389} (1996) 53;
        J.A.~Coarasa, R.A.~Jim\'{e}nez and J.~Sol\`{a}, Phys.\ Lett.\ 
        {\bf B389} (1996) 312;
        H.~Eberl, K.~Hidaka, S.~Kraml, W.~Majerotto and Y.~Yamada, 
        hep-ph/9912463;
        M.~Carena, D.~Garcia, U.~Nierste and C.E.M.~Wagner, hep-ph/9912516;
        S.~Heinemeyer, W.~Hollik and G.~Weiglein, hep-ph/0003022.
\bibitem{hove}
        G.~'t Hooft and M.~Veltman, Nucl.\ Phys.\ {\bf B44} (1972) 189.
\bibitem{passa}
        G.~Passarino and M.~Veltman, Nucl.\ Phys.\ {\bf B160} (1979) 151.
\bibitem{largel3} 
        M~Melles, W.J.~Stirling and V.A.~Khoze, Phys.\ Rev.\ {\bf D61} 
        (2000) 054015. 
\bibitem{kuehn}
        J.H.~K\"uhn, E.~Mirkes and J.~Steegborn, Z.\ Phys.\ {\bf C57}
        (1993) 615.
%
%
%
\bibitem{hdecay}
        A.~Djouadi, J.~Kalinowski and M.~Spira, Comput.\ Phys.\ Comm.\ 
        {\bf 108} (1998) 56.
\bibitem{ewcorr}
        J.~Fleischer and F.~Jegerlehner, Phys.\ Rev.\ {\bf D23} (1981) 2001; 
        D.Yu.~Bardin, B.M.~Vilenski\u{i} and P.Kh.~Khristova, Sov.\ J.\ 
        Nucl.\ Phys.\ {\bf 53} (1991) 152; 
        A.~Dabelstein and W.~Hollik, Z.\ Phys.\ {\bf C53} (1992) 507; 
        B.A.~Kniehl, Nucl.\ Phys.\ {\bf B376} (1992) 3;
        A.~Djouadi, D.~Haidt, B.A.~Kniehl, B.~Mele and P.M.~Zerwas, 
        Proc.\ Workshop on {\it $e^+e^-$ Collisions at 500~GeV: The 
        Physics Potential}, ed.~P.M.~Zerwas, Report DESY 92-123A;
        B.A.~Kniehl, Phys.\ Rep.\ {\bf 240} (1994) 211.
\bibitem{zerwas}
        T.G.~Rizzo, Phys.\ Rev.\ {\bf D22} (1980) 389;
        W.-Y.~Keung and W.J.~Marciano, Phys.\ Rev.\ {\bf D30} (1984) 248;
        R.N.~Cahn, Rep.\ Prog.\ Phys.\ {\bf 52} (1989) 389.
\bibitem{zerwas1}
        A.~Djouadi, J.~Kalinowski and P.M.~Zerwas, Z.\ Phys.\ 
        {\bf C70} (1996) 435.
\bibitem{morettistir}
        S.~Moretti and W.J.~Stirling, Phys.\ Lett.\ {\bf B347} (1995) 291 
        and (E) {\bf B366} (1996) 451.
\bibitem{squarkl} 
        B.~Kileng, Z.\ Phys.\ {\bf C63} (1994) 87;
        S.~Dawson, A.~Djouadi and M.~Spira, Phys.\ Rev.\ Lett.\ {\bf 77} 
        (1996) 16. 
\bibitem{lpaper}
         W.~Kilian, M.~M\"uhlleitner, M.~Spira and P.M.~Zerwas, DESY 99-171,
         in preparation.
\bibitem{pwacc}
         M.~Melles, W.J.~Stirling and V.A.~Khoze, Phys.\ Rev.\ 
         {\bf D61} (2000) 054015.
\bibitem{brient}
         J.-C.~Brient, talk given at the 
         ECFA/DESY Workshop on 
         {\it Physics and Detectors for a Linear Collider}, Oxford  
         20-23 March 1999;
         D.~Reid, ibid.
\bibitem{bor}
         G.~Borisov and F.~Richard, talk given at the 
         ECFA/DESY Workshop on 
         {\it Physics and Detectors for a Linear Collider}, Orsay April 1998;
         M.~Battaglia, G.~Borisov and F.~Richard, talk given at the 
         ECFA/DESY Workshop on 
         {\it Physics and Detectors for a Linear Collider}, Sitges 
         28 Apr-5 May 1999;
         G.~Borisov and F.~Richard, hep-ph/9905413; M.~Battaglia, 
         hep-ph/9910271.
\bibitem{schrei}
         H.J.~Schreiber, talk given at the 
         ECFA/DESY Workshop on 
         {\it Physics and Detectors for a Linear Collider}, Sitges 
         28 Apr-5 May 1999.
\bibitem{desch}
         K.~Desch and N.~Meyer, talk given at the 
         ECFA/DESY Workshop on 
         {\it Physics and Detectors for a Linear Collider}, Obernai 
         16-19 October 1999.
\bibitem{bpriv}
         M.~Battaglia, private communication.
\bibitem{dkz}
         A,~Djouadi, J.~Kalinowski and P.M.~Zerwas, Mod.\ Phys.\ Lett.\ 
         {\bf A7} (1992) 1765 and Z.\ Phys.\ {\bf C54} (1992) 255.
\bibitem{reina}
         S.~Dawson and L.~Reina, Phys.\ Rev.\ {\bf D60} (1999) 015003.
\bibitem{liao}
         S.~Dittmaier, M.~Kr\"amer, Y.~Liao, M.~Spira, P.M.~Zerwas,
         Phys.\ Lett.\ {\bf B478} (2000) 247.
\bibitem{ours}
        W.~Kilian and P.M.~Zerwas, Proceedings, {\it XXIX 
        Int. Conference on High Energy Physics}, Vancouver 1998, 
        hep-ph/9809486;
        A.~Djouadi, W.~Kilian, M.~M\"uhlleitner and P.M.~Zerwas, Eur.\ Phys.\ 
        J.\ {\bf C10} (1999) 27; 
        A.~Djouadi, W.~Kilian, M.~M\"uhlleitner and P.M.~Zerwas, Eur.\ Phys.\ 
        J.\ {\bf C10} (1999) 45;
        A.~Djouadi, W.~Kilian, M.~M\"uhlleitner and P.M.~Zerwas, DESY 99-171, 
        PM/99-55, TTP99-48, hep-ph/0001169.
\bibitem{dubinin}
        M.N.~Dubinin and A.V.~Semenov, SNUTP-98-140, hep-ph/9812246.
\bibitem{gounaris}
        G.~Gounaris, D.~Schildknecht and F.~Renard, Phys.\ Lett.\ 
        {\bf B83} (1979) 191 and (E) {\bf B89} (1980) 437.
\bibitem{bargerhan} 
        V.~Barger, T.~Han and R.J.N.~Phillips, Phys.\ Rev.\ {\bf D38} 
        (1988) 2766.
\bibitem{ilyin}
        V.A.~Ilyin, A.E.~Pukhov, Y.~Kurihara, Y.~Shimizu and 
        T.~Kaneko, Phys.\ Rev.\ {\bf D54} (1996) 6717.
\bibitem{boudjema}
        F.~Boudjema and E.~Chopin, Z.\ Phys.\ {\bf C73} (1996) 85.
\bibitem{bargerhan2}
        V.~Barger and T.~Han, Mod.\ Phys.\ Lett.\ {\bf A5} (1990) 667.
\bibitem{dicus}
        D.A.~Dicus, K.J.~Kallianpur and S.S.D.~Willenbrock, Phys.\ 
        Lett.\ {\bf B200} (1988) 187; A.~Abbasabadi, W.W.~Repko, 
        D.A.~Dicus and R.~Vega, Phys.\ Rev.\ {\bf D38} (1988) 2770.
\bibitem{abba}
        A.~Abbasabadi, W.W.~Repko, D.A.~Dicus and R.~Vega, 
        Phys.\ Lett. {\bf B213} (1988) 386.
\bibitem{kalli}
        K.J.~Kallianpur, Phys.\ Lett.\ {\bf B215} (1988) 392.
\bibitem{Novi}
        A.~Dobrovolskaya and V.~Novikov, Z.\ Phys.\ {\bf C52} (1991) 
        427.
\bibitem{eboli}
        O.J.P.~Eboli, G.C.~Marques, S.F.~Novaes and 
        A.A.~Natale, {Phys.~Lett.} {\bf B197} (1997) 269.
\bibitem{glover}
        E.W.N.~Glover and J.J.~van der Bij, Nucl.\ Phys.\ {\bf B309} 
        (1988) 282.
\bibitem{plehn}
        T.~Plehn, M.~Spira and P.M.~Zerwas, Nucl. Phys. {\bf B479} 
        (1996) 46; (E) Nucl.\ Phys.\ {\bf B531} (1998) 655.
\bibitem{dreeseal}
        A.~Belyaev, M.~Drees, O.J.P~Eboli, J.K.~Mizukoshi and
        S.F.~Novaes, {Phys.~Rev.}  {\bf D60} (1999) 075008; A.~Belyaev,
        M.~Drees and J.K.~Mizukoshi, preprint SLAC-PUB-8249, September 1999,
        hep-ph/9909386.
\bibitem{jikia}
        G.~Jikia, Nucl. Phys. {\bf B412} (1994) 57.
\bibitem{moretti}
        D.J.~Miller and S.~Moretti, 
        Eur.\ Phys.\ J.\ {\bf C13} (2000) 459 and hep-ph/0001194.
\bibitem{houches}
        R.~Lafaye, D.J.~Miller, M.~M\"uhlleitner and S.~Moretti, 
        hep-ph/0002238. 
\bibitem{prochouches}
        A.~Djouadi et al., Workshop on 
        {\it Physics at TeV Colliders}, Les Houches, France, 7-18 Jun 1999, 
        hep-ph/0002258.
\bibitem{gay}
        P.~Lutz, talk given at the ECFA/DESY Workshop on 
        {\it Physics and Detectors for a Linear Collider}, Oxford 
        20-23 March 1999;
        P.~Gay, talk given at the ECFA/DESY Workshop on 
        {\it Physics and Detectors for a Linear Collider}, Obernai 
        16-19 October 1999. 
\bibitem{newhdecay}
        See \cite{hdecay} and M.~Spira, private communication.
\bibitem{osland}
         P.~Osland and P.N.~Pandita, Phys.\ Rev.\ {\bf D59} (1999) 
        055013.    
\bibitem{repko}
        S.~Dawson, Nucl.\ Phys.\ {\bf B249} (1985) 42.
\bibitem{boos}
        E.E.~Boos, M.N.~Dubinin, V.A.~Ilyin, A.E.~Pukhov and 
        V.I.~Savrin, Report SNUTP-94-116, hep-ph/9503280;
        P.A.~Baikov et al., \textit{Proceedings of the Workshop 
          QFTHEP-96}, eds. B.~Levtchenko and V.~Savrin, Moscow 1996, 
        hep-ph/9701412.
\bibitem{lutzcom}
        P.~Lutz, private communication.
\bibitem{martin}
        A.~Martin, R.~Roberts and W.~Stirling, Phys.\ Lett.\ {\bf B354} 
        (1995) 155. 
\bibitem{singleh}
        A.~Djouadi, M.~Spira and P.M.~Zerwas, Phys.\ Lett.\ {\bf B264} 
        (1991) 440; S.~Dawson, Nucl.\ Phys.\ {\bf B359} (1991) 283;
        D.~Graudenz, M.~Spira and P.M.~Zerwas, 
        Phys.\ Rev.\ Lett.\ {\bf 70} (1993) 1372;
        M.~Spira, A.~Djouadi, D.~Graudenz and P.M.~Zerwas, 
        Nucl.\ Phys.\ {\bf B453} (1995) 17.
\bibitem{froide}
        E.~Richter-Was and D.~Froidevaux, Z.\ Phys.\ {\bf C76} (1997) 665;
        E.~Richter-Was et al., Int.\ J.\ Mod.\ Phys.\ {\bf A13} (1998) 1371.
\bibitem{daigu}
        J.~Dai, J.F.~Gunion and R.~Vega, Phys.\ Lett.\ {\bf B371} (1996) 71 
        and {\it ibid.}~{\bf 378} (1996) 801.
\end{thebibliography}
